\documentstyle[12pt,graphicx,axodraw,epsfig,epsf,feynmp,rotating]{article}

\catcode`@=11
\def\citer{\@ifnextchar
[{\@tempswatrue\@citexr}{\@tempswafalse\@citexr[]}}

\def\@citexr[#1]#2{\if@filesw\immediate\write\@auxout{\string\citation{#2}}\fi
  \def\@citea{}\@cite{\@for\@citeb:=#2\do
    {\@citea\def\@citea{--\penalty\@m}\@ifundefined
       {b@\@citeb}{{\bf ?}\@warning
       {Citation `\@citeb' on page \thepage \space undefined}}%
\hbox{\csname b@\@citeb\endcsname}}}{#1}}
\catcode`@=12

\newcommand{\nn}{\noindent}
\newcommand{\non}{\nonumber}
\newcommand{\ra}{\rightarrow}
\newcommand{\s}{\\ \vspace*{-3mm}}
\newcommand{\lsim}{\raisebox{-0.13cm}{~\shortstack{$<$ \\[-0.07cm] $\sim$}}~}
\newcommand{\gsim}{\raisebox{-0.13cm}{~\shortstack{$>$ \\[-0.07cm] $\sim$}}~}
\newcommand{\gae}{\stackrel{\scriptscriptstyle>}{\scriptscriptstyle\sim}}
\newcommand{\beq}{\begin{eqnarray}}
\newcommand{\eeq}{\end{eqnarray}}
\newcommand{\bq}{\begin{equation}}
\newcommand{\eq}{\end{equation}}
\newcommand{\be}{\begin{equation}}
\newcommand{\ee}{\end{equation}}
\newcommand{\sla}[1]{/\!\!\!#1}
\begin{document}


\begin{flushright}
PM/00--03\\
February 2000 
\end{flushright}

\begin{center}
\vspace*{.5cm}

{\Large\sc \bf THE HIGGS WORKING GROUP: }

\vspace*{0.4cm}

{\Large\sc \bf Summary Report} 

\vspace*{1.cm} 

Conveners: \\[0.2cm] 
{\sc A. Djouadi$^1$, R. Kinnunen$^2$, E. Richter--W\c{a}s$^{3,4}$ and H.U. 
Martyn$^5$}

\vspace*{0.6cm}

 Working Group: \\[0.2cm]
{\sc 

K.A.~Assamagan$^6$, 
C.~Bal\'azs$^7$, 
G.~B\'elanger$^8$, 
E.~Boos$^9$, 
F.~Boudjema$^8$,
M.~Drees$^{10}$,
N.~Ghodbane$^{11}$, 
M.~Guchait$^5$, 
S.~Heinemeyer$^5$, 
V.~Ilyin$^{9}$,
J.~Kalinowski$^{12}$, 
J.L.~Kneur$^1$, 
R.~Lafaye$^8$, 
D.J.~Miller$^5$, 
S.~Moretti$^{13}$,
M.~M\"uhlleitner$^5$,  
A.~Nikitenko$^{2,3}$, 
K.~Odagiri$^{13}$, 
D.P.~Roy$^{14}$,
M.~Spira$^{15}$,
K.~Sridhar$^{14}$ and
D.~Zeppenfeld$^{16,17}$.
}

\vspace*{1.cm}

{\small

$^1$ LPMT, Universit\'e Montpellier II, F--34095 Montpellier Cedex 5, France.\\
$^2$ Helsinki Institute of Physics, Helsinki, Finland. \\
$^3$ CERN, IT Division, 1211 Geneva 23, Switzerland. \\
$^4$ Institute of Computer Science, Jagellonian University,  
and Institute of Nuclear Physics, \\ 30--059 Krakow, ul. Nawojki 26a, Poland. \\
$^5$ DESY, Notkestrasse 85, D--22603 Hamburg, Germany. \\
$^6$ Hampton University, Hampton, VA 23668, USA. \\
$^7$ Department of Physics and Astronomy, University of Hawaii, Honolulu, 
HI 96822. \\
$^8$ LAPP, BP 110, F--74941 Annecy le Vieux Cedex, France. \\
$^{9}$ Institute of Nuclear Physics, MSU, 11 9899 Moscow, Russia. \\ 
$^{10}$ Physik Department, TU M\"unchen, James Franck Str., D--85748 Garching, 
Germany. \\
$^{11}$ IPNL, Univ. Claude Bernard, F--69622 Villeurbanne Cedex, France.\\
$^{12}$ Institute of Theoretical Physics, Warsaw University, PL--00681 Warsaw, 
Poland.  \\
$^{13}$ Rutherford Appleton Laboratory, Chilton, Didcot, Oxon OX11 OQX, U.K. \\
$^{14}$ Theoretical Physics Department, TIFR, Homi Bhabha Road, Bombay 400 
005, India. \\
$^{15}$ II. Inst. Theor. Physik, Universit\"at Hamburg, 
D--22761 Hamburg, Germany. \\
$^{16}$ CERN, Theory Division, CH--1211, Geneva, Switzerland. \\
$^{17}$ Department of Physics, University of Wisconsin, Madison, WI 53706, USA.
}

\vspace*{1cm}

{\it Report of the HIGGS working group for the Workshop \\[0.1cm]
``Physics at TeV Colliders", Les Houches, France 8--18 June 1999.}

\end{center} 

\newpage


\vspace*{1cm} 
\begin{center}
{\bf \large CONTENTS} 
\end{center} 

\vspace*{0.8cm} 

\nn {\bf \ \ \ \ SYNOPSIS} \hfill 3 \\

\nn {\bf 1. Measuring Higgs boson couplings at the LHC} 
\hfill 4 \\[0.2cm] \hspace*{0.5cm}
D.~Zeppenfeld, R.~Kinnunen, A.~Nikitenko and E.~Richter--W\c{a}s. \\

\nn {\bf 2. Higgs boson production at hadron colliders at NLO} 
\hfill 20 \\[0.2cm] \hspace*{0.5cm}
C. Bal\'azs, A. Djouadi, V. Ilyin and M. Spira. \\

\nn {\bf 3. Signatures of Heavy Charged Higgs Bosons at the LHC}
\hfill 36 \\[0.2cm]  \hspace*{0.5cm}
K.A. Assamagan, A. Djouadi, M. Drees, M. Guchait, R. Kinnunen, 
J.L. Kneur, \\ \hspace*{0.5cm} 
D.J. Miller, S. Moretti, K. Odagiri and D.P. Roy. \\

\nn {\bf 4. Light stop effects and Higgs boson searches at the LHC.}
\hfill 54 \\[0.2cm]  \hspace*{0.5cm}
G.~B\'elanger, F.~Boudjema, A.~Djouadi, V.~Ilyin, 
J.L.~Kneur, S.~Moretti, \\ \hspace*{0.5cm} E. Richter--W\c{a}s and  
K.~Sridhar. \\

\nn {\bf 5. Double Higgs production at TeV Colliders in the MSSM}
\hfill 67 \\[0.2cm] \hspace*{0.5cm}
R. Lafaye, D.J. Miller, M. M\"uhlleitner and  S.~Moretti. \\

\nn {\bf 6. Programs and Tools for Higgs Bosons} 
\hfill 88 \\[0.2cm] \hspace*{0.5cm}
E. Boos, A. Djouadi, N. Ghodbane, S. Heinemeyer,
V. Ilyin, J. Kalinowski, \\ \hspace*{0.5cm} J.L. Kneur and M. Spira. 

\newpage


\begin{center}
{\large\sc {\bf SYNOPSIS}} 
\end{center}

\vspace*{0.1cm}

During this Workshop, the Higgs working group has addressed the prospects 
for searches for Higgs particles at future TeV colliders [the Tevatron RunII, 
the LHC and  a future high--energy $e^+e^-$ linear collider] in the context 
of the Standard Model (SM) and its supersymmetric extensions such as the 
Minimal Supersymmetric Standard Model (MSSM).  

In the past two decades, the main focus in Higgs physics at these colliders
was on the assessment of the discovery of Higgs particles in the simplest 
experimental detection channels. A formidable effort has been devoted to 
address this key issue, and there is now little doubt that a Higgs particle 
in both the SM and the MSSM cannot escape detection at the LHC or at the
planed TeV linear $e^+e^-$ colliders.  

Once Higgs particles will be found, the next important step and challenge would
be to make a detailed investigation of their fundamental properties and to
establish in all its facets the electroweak symmetry breaking mechanism.  To
undertake this task, more sophisticated analyses are needed since for instance,
one has to include the higher--order corrections [which are known to be rather
large at hadron colliders in particular] to the main detection channels to
perform precision measurements and to consider more complex Higgs production
and decay mechanisms [for instance the production of Higgs bosons with other
particles, leading to multi--body final states] to pin down some of the Higgs
properties such as the self--coupling or the coupling to heavy states.  

We have addressed these issues at the Les Houches Workshop and initiated a few 
theoretical/experimental analyses dealing with the measurement of Higgs boson
properties and higher order corrections and processes. This report summarizes 
our work.  

The first part of this report deals with the measurements at the LHC of the SM
Higgs boson couplings to the gauge bosons and heavy quarks. In part 2, the
production of the SM and MSSM neutral Higgs bosons at hadron colliders,
including the next--to--leading order QCD radiative corrections, is discussed.
In part 3, the signatures of heavy charged Higgs particles in the MSSM are
analyzed at the LHC. In part 4, the effects of light top squarks with large
mixing on the search of the lightest MSSM Higgs boson is analyzed at the LHC. 
In part 5, the double Higgs production is studied at hadron and $e^+e^-$
colliders in order to measure the trilinear Higgs couplings and to reconstruct 
the scalar potential of the MSSM. Finally, part 6 summarizes the work performed
on the programs and tools which allow the determination of the Higgs boson decay
modes and production cross sections at various colliders.  \bigskip


\noindent {\bf Acknowledgements}: \smallskip

\noindent We thank the organizers of this Workshop, and in particular ``le
Grand Ordonateur" Patrick Aurenche, for the warm, friendly and very stimulating
atmosphere of the meeting.  We thank also our colleagues of the QCD and SUSY
working groups for the nice and stimulating, strong and super, interactions
that we had.  Thanks also go to the ``personnel" of the Les Houches school for
allowing us to do physics late at night and for providing us with a hospitable
environment for many hot or relaxed discussions.  

\newpage


\begin{center}
{\large\sc {\bf Measuring Higgs boson couplings at the LHC}} 

\vspace{0.5cm}

{\sc
D.~Zeppenfeld, R.~Kinnunen, A.~Nikitenko and E.~Richter--W\c{a}s}
\end{center}

\begin{abstract}
For an intermediate mass Higgs boson with SM-like couplings the LHC
allows observation of a variety of decay channels in production by gluon 
fusion and weak boson fusion. Cross section ratios provide measurements 
of various ratios of Higgs couplings, with accuracies of order 15\%
for 100~fb$^{-1}$ of data in each of the two LHC experiments.
For Higgs masses above 120~GeV, minimal assumptions on the Higgs sector 
allow for an indirect measurement of the total Higgs boson width with an 
accuracy of 10 to 20\%, and of the $H\to WW$ partial width with an accuracy
of about 10\%.
\end{abstract} 

\section{Introduction}

Investigation of the symmetry breaking mechanism of the electroweak 
$SU(2)\times U(1)$ gauge symmetry will be one of the prime tasks of the
LHC. Correspondingly, major  efforts have been concentrated on devising 
methods for Higgs boson discovery, for the entire mass range 
allowed within the Standard Model (SM) (100~GeV$\lsim m_H \lsim 1$~TeV,
after LEP2), and for Higgs boson search in extensions of the SM,
like its minimal supersymmetric extension the MSSM~\cite{ATLAS,CMS}. 
While observation of one or more Higgs 
scalar(s) at the LHC appears assured, discovery will be followed by
a more demanding task: the systematic investigation of Higgs boson 
properties. Beyond observation of the various CP even and CP odd scalars 
which nature may have in store for us, this means the determination of 
the couplings of the Higgs boson to the known fermions and gauge bosons,
i.e. the measurement of $Htt$, $Hbb$, $H\tau\tau$ and $HWW$, $HZZ$, 
$H\gamma\gamma$ couplings, to the extent possible. 

Clearly this task very much depends on the expected Higgs boson mass. For
$m_H>200$~GeV and within the SM, only the $H\to ZZ$ and $H\to WW$ channels 
are expected to be observable, and the two gauge boson modes are related 
by SU(2). Above $m_H\approx 250$~GeV, where detector effects will no longer
dominate the mass resolution of the $H\to ZZ\to 4\ell$ resonance,
additional information is expected from a direct measurement of the total 
Higgs boson width, $\Gamma_H$. A much richer spectrum of decay modes
is predicted for the intermediate mass range, i.e. if a SM-like
Higgs boson has a mass between the reach of LEP2 ($\lsim 110$~GeV) and 
the $Z$-pair threshold. The main reasons for focusing on this range are 
present indications from electroweak precision
data, which favor $m_H<250$~GeV~\cite{EWfits}, as well as expectations
within the MSSM, which predicts the lightest Higgs boson to have a mass
$m_h\lsim 130$~GeV~\cite{mhMSSM}.

Until recently, the prospects of detailed and model independent 
coupling measurements at the LHC were considered somewhat 
remote~\cite{snowmass_H}, because few promising search channels were
known to be accessible, for any given Higgs boson mass. 
Taking ATLAS search scenarios as an example, these were~\cite{ATLAS}
\beq 
\label{eq:ggHAA}
gg\to H\to \gamma\gamma\;, \qquad 
{\rm for}\quad m_H \lsim 150~{\rm GeV}\;, \\
\label{eq:ggHZZ}
gg\to H\to ZZ^*\to 4\ell\;,\qquad 
{\rm for}\quad m_H \gsim 130~{\rm GeV}\;, 
\eeq
and
\bq
\label{eq:ggHWW}
gg\to H\to WW^*\to \ell\bar\nu\bar\ell\nu\;, \qquad 
{\rm for}\quad m_H \gsim 150~{\rm GeV}\;, 
\eq
with the possibility of obtaining some additional information from processes 
like $WH$ and/or $t\bar tH$ associated production with subsequent 
$H\to\bar bb$ and $H\to\gamma\gamma$ decay for Higgs boson masses near 
100~GeV. Throughout this contribution, ``$gg\to H$'' stands for inclusive 
Higgs production, which is
dominated by the gluon fusion process for a SM-like Higgs boson.

This relatively pessimistic outlook is changing considerably now, due to the
demonstration that weak boson fusion is a promising Higgs production channel
also in the intermediate mass range. Previously, this channel 
had only been explored for Higgs masses above 300~GeV. 
Specifically, it was recently shown in parton 
level analyses that the weak boson fusion channels, with subsequent Higgs
decay into photon pairs~\cite{RZ_gamgam,R_thesis},
\bq
\label{eq:wbfHAA}
qq\to qqH,\;H\to \gamma\gamma\;, \qquad 
{\rm for}\quad m_H \lsim 150~{\rm GeV}\;,
\eq
into $\tau^+\tau^-$ pairs~\cite{R_thesis,RZ_tautau_lh,RZ_tautau_ll},
\bq
\label{eq:wbfHtautau}
qq\to qqH,\;H\to \tau\tau\;, \qquad 
{\rm for}\quad m_H \lsim 140~{\rm GeV}\;,
\eq
or into $W$ pairs~\cite{R_thesis,RZ_WW}
\bq
\label{eq:wbfHWW}
qq\to qqH,\;H\to WW^{(*)}\to e^\pm \mu^\mp /\!\!\!{p}_T\;, 
\qquad {\rm for}\quad m_H \gsim 120~{\rm GeV}\;,
\eq
can be isolated at the LHC. Preliminary analyses, which try to extend
these parton level results to full detector simulations, look 
promising~\cite{sasha}. The weak boson fusion channels utilize the 
significant background reductions which are expected from 
double forward jet tagging~\cite{Cahn,BCHP,DGOV} and central jet vetoing 
techniques~\cite{bjgap,bpz_mj}, and promise low
background environments in which Higgs decays can be studied in detail.  
The parton level results predict highly significant signals with 
(substantially) less than 100~fb$^{-1}$.

The prospect of observing several Higgs production and decay 
channels, over the entire intermediate mass range, suggests a reanalysis of 
coupling determinations at the LHC~\cite{snowmass_H}. 
This contribution attempts a first such 
analysis, for the case where the branching fractions of an intermediate
mass Higgs resonance are fairly similar to the SM case, i.e. we analyze
a SM-like Higgs boson only. We make use of the previously published analyses
for the inclusive Higgs production channels~\cite{ATLAS,CMS} and of the 
weak boson fusion 
channels~\cite{RZ_gamgam,R_thesis,RZ_tautau_lh,RZ_tautau_ll,RZ_WW}.
The former were obtained by the experimental collaborations and include 
detailed detector simulations. The latter are based on parton level results,
which employ full QCD tree level matrix elements for all signal and 
background processes. We will not discuss here differences in the 
performance expected for the ATLAS and CMS detectors nor details in the 
theoretical assumptions which lead to different estimates for expected 
signal and background rates. The reader is referred to the original 
publications from which numbers are extracted.
In Section~\ref{sec2} we summarize expectations for the various 
channels, including expected accuracies for cross section measurement of
the various signals for an integrated luminosity of 100~fb$^{-1}$. 
Implications for the determination of coupling ratios and the measurement 
of Higgs boson (partial) decay widths are then obtained in
Section~\ref{sec3}. A final summary is given in Section~\ref{sec4}.

\section{Survey of intermediate mass Higgs channels}
\label{sec2}

The various Higgs channels listed in Eqs.~(\ref{eq:ggHAA}--\ref{eq:wbfHWW})
and their observability at the LHC have all been discussed in the 
literature. Where available, we give values as presently quoted by the 
experimental collaborations. In order to compare
the accuracy with which the cross sections of different Higgs production
and decay channels can be measured, we need to unify these results. For
example, $K$-factors of unity are assumed throughout. Our goal 
in this section is to obtain reasonable estimates for the relative errors,
$\Delta\sigma_H/\sigma_H$, which are expected after collecting 
100~${\rm fb}^{-1}$ in each the ATLAS and the CMS detector, i.e. we estimate 
results after a total of 200~${\rm fb}^{-1}$ of data have been collected 
at the LHC. Presumably these data will be taken with a mix of both low 
and high luminosity running.

\begin{table*}[thb]
\caption{Number of expected events for the inclusive SM $H\to \gamma\gamma$ 
signal and expected backgrounds, assuming an integrated luminosity of 
100~${\rm fb}^{-1}$ and high luminosity performance. Numbers correspond 
to optimal $\gamma\gamma$ invariant mass windows for CMS and ATLAS.
The expected relative statistical errors on the signal cross section 
are given for the individual experiments and are combined in the last line.
}
\begin{center}
\vspace{0.15in}
\label{table:gg.AA}
\begin{tabular}{c|c|cccccc}
    & $m_H$ & 100  & 110  & 120  & 130  & 140  & 150  \\
\hline
CMS~\protect\cite{CMS_ecal_tdr,katri_lassila} 
    & $N_S$ &   865 &  1038 &  1046 &   986 &  816 &  557 \\
    & $N_B$ & 29120 & 22260 & 16690 & 12410 & 9430 & 7790 \\
    & $\Delta\sigma_H/\sigma_H$ 
            & 20.0\% & 14.7\% & 12.7\% & 11.7\% & 12.4\% & 16.4\% \\
\hline
ATLAS~\protect\cite{ATLAS} 
      & $N_S$ &  1045 &  1207 &  1283 &  1186 &   973 &   652 \\
      & $N_B$ & 56450 & 47300 & 39400 & 33700 & 28250 & 23350 \\
    & $\Delta\sigma_H/\sigma_H$ 
              & 22.9\% & 18.2\% &15.7\% & 15.7\% & 17.6\% & 23.8\% \\
\hline 
Combined & $\Delta\sigma_H/\sigma_H$ 
              & 15.1\% & 11.4\% & 9.9\% & 9.4\% & 10.1\% & 13.5\% \\
\end{tabular}
\end{center}
\end{table*}

We find that the measurements are largely dominated by statistical errors.
For all channels, event rates with 200~${\rm fb}^{-1}$ of data will be large 
enough to use the Gaussian approximation for statistical errors. The 
experiments measure the signal cross section by separately determining the 
combined signal + background rate, $N_{S+B}$, and the expected number of 
background events, $\langle N_B\rangle$. The signal cross section is then 
given by
\bq
\sigma_H = {N_{S+B}-\langle N_B\rangle \over \epsilon\int{\cal L}dt }=
{N_S\over \epsilon\int{\cal L}dt }\;,
\eq
where $\epsilon$ denotes efficiency factors.
Thus the statistical error is given by
\bq
{\Delta\sigma_H\over\sigma_H} = {\sqrt{N_{S+B}}\over N_S} = 
{\sqrt{N_S+N_B}\over N_S}\;,
\eq
where in the last step we have dropped the distinction between the 
expected and the actual number of background events.
Systematic errors on the background rate are added in quadrature to the
background statistical error, $\sqrt{N_B}$, where appropriate.

Well below the $H\to WW$ threshold, the search for $H\to\gamma\gamma$ events
is arguably the cleanest channel for Higgs discovery. LHC detectors have 
been
designed for excellent two-photon invariant mass resolution, with this 
Higgs signal in mind. We directly take the expected signal and background
rates for the inclusive  $H\to\gamma\gamma$ search from the detailed studies
of the CMS and ATLAS collaborations~\cite{CMS_ecal_tdr,katri_lassila,ATLAS}, 
which were performed 
for an integrated luminosity of 100~${\rm fb}^{-1}$ in each detector. 
Expectations are summarized in Table~\ref{table:gg.AA}. Rates correspond to 
not including a $K$-factor for the expected signal and background cross 
sections in CMS and ATLAS. 
Cross sections have been determined with the set MRS (R1) of parton
distribution functions (pdf's) for CMS, while ATLAS numbers are based on 
the set CTEQ2L of pdf's. 

The inclusive $H\to\gamma\gamma$ signal will be observed as a narrow
$\gamma\gamma$ invariant mass peak on top of a smooth background 
distribution.
This means that the background can be directly measured from the very high
statistics background distribution in the sidebands. We expect any 
systematic errors on the extraction of the signal event rate to be 
negligible compared to the statistical 
errors which are given in the last row of Table~\ref{table:gg.AA}.
With  100~${\rm fb}^{-1}$ of data per experiment 
$\sigma(gg\to H)\cdot B(H\to\gamma\gamma)$ can be determined with a relative
error of 10 to 15\% for Higgs masses between 100 and 150~GeV. Here we 
do not include additional systematic errors, e.g. from the luminosity
uncertainty or from higher order QCD corrections, because we will mainly 
consider cross section ratios in the final analysis in the next Section. 
These systematic errors largely cancel in the cross section ratios. 
Systematic errors common to several channels will be considered later, where
appropriate.

A Higgs search channel with a much better signal to background ratio, 
at the price of lower statistics, however, is available 
via the inclusive search for $H\to ZZ^*\to 4\ell$ events. Expected event
numbers for 100~${\rm fb}^{-1}$ in both ATLAS~\cite{ATLAS} and 
CMS~\cite{CMS_HZZ} are listed in 
Table~\ref{table:gg.ZZ}. These numbers were derived using CTEQ2L pdf's 
and are corrected to contain no QCD K-factor. 
For those Higgs masses where no ATLAS or CMS prediction is available, we
interpolate/extrapolate the results for the nearest Higgs mass, taking 
the expected $H\to ZZ^*$ branching ratios into account for the signal.
Similar to the case of
$H\to \gamma\gamma$ events, the signal is seen as a narrow peak in the 
four-lepton invariant mass distribution, i.e. the background can be 
extracted directly from the signal sidebands.
The combined relative error on the measurement
of $\sigma(gg\to H)\cdot B(H\to ZZ^*)$ is listed in the last line of 
Table~\ref{table:gg.ZZ}. For Higgs masses in the 130--150~GeV range, and
above $Z$-pair threshold, a 10\% statistical error on the 
cross section measurement is possible.
In the intermediate range, where $H\to WW$ dominates, and for lower Higgs 
masses, where the Higgs is expected to dominantly decay into $\bar bb$,
the error increases substantially.

\begin{table*}[t]
\caption{Number of expected events for the inclusive SM 
$H\to ZZ^*\to \ell^+\ell^-\ell^+\ell^-$ 
signal and expected backgrounds, assuming an integrated luminosity of 
100~${\rm fb}^{-1}$ and high luminosity performance. Numbers correspond to 
optimal four-lepton invariant mass windows for CMS and ATLAS and to the 
combined total. Rates in parentheses correspond to numbers interpolated,
according to $H\to ZZ^*$ branching ratios for the signal. 
The expected relative statistical errors on the signal cross section are 
given for each experiment and are combined in the last line.}
\vspace{0.15in}
\label{table:gg.ZZ}
\begin{center}
\begin{tabular}{cc|ccccccc}
    & $m_H$   &   120  &    130  & 140 &  150  & 160 & 170   & 180  \\
\hline
CMS~\protect\cite{CMS_HZZ}
      & $N_S$ &  19.2  &   55.3  & (99)& 131.4 & (48)& 29.4  & (76.5)\\
      & $N_B$ &  12.9  &   17.1  & (20)&  22.5 & (26)& 27.5  & (27)  \\
    & $\Delta\sigma_H/\sigma_H$ 
            & 29.5\% & 15.4\% &  11.0\% & 9.4\% & 17.9\% & 25.7\% & 13.3\% \\
\hline
ATLAS~\protect\cite{ATLAS}
      & $N_S$ &  10.3  &   28.7  & (51)& 67.6  & (31)& 19.1  &  49.7 \\
      & $N_B$ &  4.44  &   7.76  & (8) & 8.92  & (8) & 8.87  &  8.81 \\
    & $\Delta\sigma_H/\sigma_H$ 
     & 37.3\% & 21.0\% &  15.1\%& 12.9\% & 20.1\%& 27.7\% & 15.4\% \\
\hline 
Combined 
&$\Delta\sigma_H/\sigma_H$  
            & 23.1\% & 12.4\%  & 8.9\%&  7.6\% & 13.4\% & 18.8\% & 10.1\% \\
\end{tabular}
\end{center}
\end{table*}

\begin{table}[thb]
\vspace{-0.15in}
\caption{Number of expected events for the inclusive SM 
$H\to WW^*\to \ell^+\nu\ell^-\bar\nu$ 
signal and expected backgrounds, assuming an integrated luminosity of 
30~${\rm fb}^{-1}$. Numbers correspond to optimized cuts, varying with the 
mass of the Higgs boson being searched for. The expected relative errors 
on the signal cross section are given for each experiment, 
separating the statistical error, the effect of a systematic 5\% error of the 
background level, and the two added in quadrature. The combined error for the
two experiments assumes 100\% correlation of the systematic errors on the 
background determination.
}
\vspace{0.15in}
\label{table:gg.WW}
\begin{small}
\begin{tabular}{cc|cccccccc}
      & $m_H$ & 120  &  130   &   140  &   150  
                                       &   160  &   170  &   180  &   190  
\\
\hline
CMS%
    & $N_S$ &   44 &  106   &   279  &   330  & 468 & 371 & 545 & \\
\protect\cite{DittDrein}
    & $N_B$ &  272 &  440   &   825  &   732  & 360 & 360 & 1653 & \\
    & $\Delta\sigma_H/\sigma_H$(stat.)     
 & 40.4\% & 22.0\% & 11.9\% & 9.9\% & 6.1\%& 7.3\%&8.6\%&  \\
    & $\Delta\sigma_H/\sigma_H$(syst.)     
 & 30.9\% & 20.8\% & 14.8\% & 11.1\% & 3.8\%& 4.9\%&15.2\%&  \\
    & $\Delta\sigma_H/\sigma_H$(comb.)     
 & 50.9\% & 30.3\% & 19.0\% & 14.9\% & 7.3\%& 8.8\%&17.4\%& 20.6\% \\
\hline
ATLAS%
    & $N_S$ & & &  &  240   &   400  &   337  &   276  &   124  \\
\protect\cite{ATLAS}
    & $N_B$ & & &  &  844   &   656  &   484  &   529  &   301  \\
    & $\Delta\sigma_H/\sigma_H$ (stat.)     
     &  & &  & 13.7\% & 8.1\% & 8.5\% & 10.3\% & 16.6\% \\
    & $\Delta\sigma_H/\sigma_H$ (syst.)     
     &  & &  & 17.6\% & 8.2\% & 7.2\% & 9.6\% & 12.1\% \\
    & $\Delta\sigma_H/\sigma_H$ (comb.)     
 & 50.9\% & 30.3\% & 19.0\% & 22.3\% & 11.5\% & 11.1\% & 14.1\% & 20.6\% \\
\hline 
Com
    &$\Delta\sigma_H/\sigma_H$  (comb.) & 
 42.1\% & 26.0\% & 17.0\% & 14.8\% & 7.0\% & 8.0\% & 13.6\% & 16.9\% \\
\end{tabular}
\end{small}
\end{table}

Above $m_H\approx 135$~GeV, $H\to WW^{(*)}$ becomes the dominant SM Higgs
decay channel. The resulting inclusive $WW\to\ell^+\nu\ell^-\bar\nu$ signal
is visible above backgrounds, after exploiting the characteristic
lepton angular correlations for spin zero decay into $W$ pairs near 
threshold~\cite{DittDrein}. The inclusive channel, which is dominated by 
$gg\to H\to WW$, has been analyzed by ATLAS for $m_H\geq 150$~GeV and for
integrated luminosities of 30 and 100~${\rm fb}^{-1}$~\cite{ATLAS}
and by CMS for $m_H\geq 120$~GeV and 
30~${\rm fb}^{-1}$~\cite{DittDrein}. The expected event numbers for 
30~${\rm fb}^{-1}$ are listed in Table~\ref{table:gg.WW}.
The numbers are derived without QCD K-factors and use CTEQ2L for ATLAS and 
MRS(A) pdf's for CMS results. 

Unlike the two previous modes, the two missing neutrinos in the $H\to WW$ 
events do not allow for a reconstruction of the narrow Higgs mass peak. 
Since the Higgs signal is only seen as a broad enhancement of the expected 
background rate in lepton-neutrino transverse mass distributions, with 
similar shapes of signal and background after application of all cuts, a 
precise determination
of the background rate from the data is not possible. Rather one has to rely
on background measurements in phase space regions where the signal is weak,
and extrapolation to the search region using NLO QCD predictions. The 
precise
error on this extrapolation is unknown at present, the assumption of a 5\%
systematic background uncertainty appears optimistic but attainable. 
It turns out that with 30~${\rm fb}^{-1}$ already, the systematic error
starts to dominate, because the background exceeds the signal rate by
factors of up to 5, depending on the Higgs mass. Running at high 
luminosity makes matters worse, because the less efficient reduction of 
$\bar tt$ backgrounds, due to less stringent $b$-jet veto criteria, 
increases the 
background rate further. Because of this problem we only present results
for 30~${\rm fb}^{-1}$ of low luminosity running in Table~\ref{table:gg.WW}.
Since neither of the LHC collaborations has presented predictions for the
entire Higgs mass range, we take CMS simulations below 150~GeV and ATLAS
results at 190~GeV, but divide the resultant statistical errors by a 
factor $\sqrt{2}$, to take account of the presence of two experiments.
Between 150 and 180~GeV we combine both experiments, assuming 100\%
correlation in the systematic 5\% normalization error of the background.

\begin{table*}[htb]
\caption{Number of expected $\gamma\gamma jj$ events from the
$qq\to qqH,\; H\to \gamma\gamma$ weak boson fusion
signal and expected backgrounds, assuming an integrated luminosity of 
100~${\rm fb}^{-1}$. Numbers correspond to optimal $\gamma\gamma$ invariant 
mass windows for CMS and ATLAS and to the combined total, as projected
from the parton level analysis of Refs.~\protect\cite{RZ_gamgam,R_thesis}. 
The expected relative statistical errors on the signal cross section are 
given for each experiment and are combined in the last line.}
\vspace{0.15in}
\label{table:WBF.AA}
\begin{center}
\begin{tabular}{c|c|cccccc}
    & $m_H$ &       100  & 110  & 120  & 130  & 140  & 150  \\
\hline
projected CMS & $N_S$ &       37   &  48  &  56  &  56  &  48  &  33 \\
performance   & $N_B$ &       33   &  32  &  31  &  30  &  28  &  25 \\
    & $\Delta\sigma_H/\sigma_H$     
            & 22.6\% & 18.6\% & 16.7\% & 16.6\% & 18.2\%& 23.1\% \\
\hline
projected ATLAS & $N_S$ &     42   &  54  &  63  &  63  &  54  &  37 \\
performance     & $N_B$ &     61   &  60  &  56  &  54  &  51  &  46 \\
    & $\Delta\sigma_H/\sigma_H$     
            & 24.2\% & 19.8\% & 17.3\% & 17.2\% & 19.0\% & 24.6\% \\
\hline 
combined 
& $\Delta\sigma_H/\sigma_H$  
            & 16.5\% & 13.6\% & 12.0\% & 11.9\% & 13.1\% & 16.8\% \\
\end{tabular}
\end{center}
\vspace*{-3mm}
\end{table*}

The previous analyses are geared towards measurement of the inclusive 
Higgs production cross section, which is is dominated by the gluon fusion
process. 15 to 20\% of the signal sample, however, is expected to arise
from weak boson fusion, $qq\to qqH$ or corresponding antiquark initiated 
processes. The weak boson fusion component can be isolated 
by making use of the two forward tagging jets which are present in these
events and by vetoing additional central jets, which are unlikely to arise
in the color singlet signal process~\cite{bjgap}. A more detailed discussion
of these processes can be found in Ref.~\cite{R_thesis} from which most of 
the following numbers are taken.

\begin{table*}[thb]
\caption{Number of expected signal and background events for the
$qq\to qqH\to\tau\tau jj$ channel, for 100~${\rm fb}^{-1}$ and two
detectors. Cross sections are added for
$\tau\tau\to \ell^\pm h^\mp\sla p_T$ and $\tau\tau \to e^\pm \mu^\mp\sla 
p_T$
events as given in Refs.~\protect\cite{R_thesis,RZ_tautau_ll}. 
The last line gives the expected statistical relative error on the 
$qq\to qqH,\; H\to \tau\tau$ cross section. }
\label{table:WBF.tautau}
\vspace{0.15in}
\begin{center}
\begin{tabular}{c|cccccc}
    $m_H$ &       100  & 110  & 120  & 130  & 140  & 150  \\
\hline
         $N_S$ &  211  & 197  & 169  & 128  &  79  & 38   \\
         $N_B$ &  305  & 127  & 51   &  32  &  27  & 24   \\
$\Delta\sigma_H/\sigma_H$ &
                  10.8\% & 9.1\% & 8.8\% & 9.9\% & 13.0\% & 20.7\% \\
\end{tabular}
\end{center}
\vspace*{-3mm}
\end{table*}

The $qq\to qqH,\;H\to\gamma\gamma$ process was first analyzed in 
Ref.~\cite{RZ_gamgam}, where cross sections for signal and background
were obtained with full QCD tree level matrix elements. The parton level
Monte Carlo determines all geometrical acceptance corrections. Additional
detector effects were included by smearing parton and photon 4-momenta with 
expected detector resolutions and by assuming trigger, identification and 
reconstruction efficiencies of 0.86 for each of the two tagging jets and 
0.8 for each photon. Resulting cross sections were presented 
in Ref.~\cite{R_thesis} for a fixed $\gamma\gamma$ invariant mass window of 
total width $\Delta m_{\gamma\gamma}=2$~GeV. We correct these numbers for 
$m_H$ dependent mass resolutions in the experiments. We take $1.4\sigma$ mass
windows, as given in Ref.~\cite{ATLAS} for high luminosity running,
which are expected to contain 79\% of the signal events for ATLAS.
The 2~GeV window for $m_H=100$~GeV at CMS~\cite{CMS_ecal_tdr,katri_lassila} is 
assumed 
to scale up like the ATLAS resolution and assumed to contain 70\% of the 
Higgs signal. The expected total signal and background rates for
100~${\rm fb}^{-1}$ and resulting relative errors for the extraction of the
signal cross section are given in Table~\ref{table:WBF.AA}. Statistical 
errors only are considered for the background subtraction, since the 
background level can be measured 
independently by considering the sidebands to the Higgs boson peak.

The next weak boson fusion channel to be considered is $qq\to qqH,\;
H\to \tau\tau$. Again, this channel has been analyzed at the parton level,
including some estimates of detector effects, as discussed for the 
$H\to \gamma\gamma$ case. Here, a lepton identification efficiency of 0.95
is assumed for each lepton $\ell=e,\;\mu$. 
Two $\tau$-decay modes have been considered so 
far: $H\to\tau\tau \to \ell^\pm h^\mp\sla p_T$~\cite{RZ_tautau_lh}
and $H\to\tau\tau \to e^\pm \mu^\mp\sla p_T$~\cite{RZ_tautau_ll}. These 
analyses were performed for low luminosity running. Some deterioration at 
high luminosity is expected, as in the analogous $H/A\to\tau\tau$ channel
in the MSSM search~\cite{ATLAS}. At high luminosity, pile-up effects
degrade the $\sla p_T$ resolution significantly, which results in a 
worse $\tau\tau$ invariant mass resolution. At a less significant level, 
a higher $p_T$ threshold for the minijet veto technique will 
increase the QCD and $t\bar t$ backgrounds. The $\tau$-identification 
efficiency is similar at high and low luminosity. 
We expect that the reduced performance at high 
luminosity can be compensated for by considering the additional channels 
$H\to\tau\tau \to e^+e^-\sla p_T,\;\mu^+\mu^-\sla p_T$. $Z+$jets and
$ZZ+$jets backgrounds (with $ZZ\to \ell^+\ell^-\nu\bar\nu$) are strongly 
suppressed by rejecting same flavor lepton pairs which are compatible 
with $Z$ decays ($m_{\ell\ell}= m_Z\pm 6$~GeV). Drell-Yan plus jets 
backgrounds are further reduced by requiring significant $\sla p_T$.
Since these analyses have not yet been performed, we 
use the predicted cross sections for only those two channels which have 
already been discussed in the literature 
and scale event rates to a combined 200~${\rm fb}^{-1}$ of data. 
Results are given in Table~\ref{table:WBF.tautau}.

\begin{table*}[t]
\vspace{-0.2in}
\caption{Number of events expected for 
$qq\to qqH,\;H\to WW^{(*)}\to\mu^\pm e^\mp\sla p_T$ 
in 200~fb$^{-1}$ of data, and corresponding 
backgrounds~\protect\cite{RZ_WW}.
The expected relative statistical error on the signal cross section 
is given in the last line.}
\vspace{0.15in}
\label{table:WBF.WW}
\begin{center}
\begin{tabular}{c|cccccccc}
$m_H$ &  120  &  130  &  140  &  150  &  160  &  170  &  180  &  190 \\
\hline
$N_S$ &  136  &  332  &  592  &  908  & 1460  & 1436  & 1172  &  832 \\
$N_B$ &  136  &  160  &  188  &  216  &  240  &  288  &  300  &  324 \\
$\Delta\sigma_H/\sigma_H$ &
       12.1\% & 6.7\% & 4.7\% & 3.7\% & 2.8\% & 2.9\% & 3.3\% & 4.1\%  \\
\end{tabular}
\end{center}
\vspace*{-6mm}
\end{table*}

The previous two weak boson channels allow reconstruction of the Higgs
resonance as an invariant mass peak. This is not the case for 
$H\to WW\to \ell^+\nu\ell^-\bar\nu$ as discussed previously for the 
inclusive search. The weak boson fusion 
channel can be isolated separately by employing forward jet tagging and
color singlet exchange isolation techniques in addition to tools like
charged lepton angular correlations which are used for the inclusive
channel. The corresponding parton level analysis for $qq\to qqH$, 
$H\to WW^{(*)}\to\mu^\pm e^\mp\sla p_T$ has been performed in
Ref.~\cite{RZ_WW} and we here scale the results to a total integrated 
luminosity of 200~${\rm fb}^{-1}$, which takes into account the availability 
of two detectors. As for the tau case, the analysis was done for low 
luminosity running conditions and somewhat higher backgrounds are expected
at high luminosity. On the other hand the $WW^{(*)}\to\mu^+\mu^-\sla p_T$ 
and $WW^{(*)}\to e^+e^-\sla p_T$ modes should roughly double the 
available statistics since very few signal events have lepton pair 
invariant masses compatible with $Z\to\ell\ell$ decays. Therefore our 
estimates are actually conservative. Note that the expected background 
for this weak boson fusion process is much smaller than for the 
corresponding
inclusive measurement. As a result modest systematic uncertainties will
not degrade the accuracy with which $\sigma(qq\to qqH)\cdot B(H\to 
WW^{(*)})$
can be measured. A 10\% systematic error on the background, double the error
assumed in the inclusive case, would degrade the statistical accuracy by, 
typically, a factor 1.2 or less. As a result, we expect that a very precise
measurement of  $\sigma(qq\to qqH)\cdot B(H\to WW^{(*)})$ can be performed 
at the LHC, with a statistical accuracy of order 5\% or even better in the 
mass range $m_H\geq 140$~GeV. Even for $m_H$ as low as 120~GeV a 12\% 
measurement is expected.

\section{Measurement of Higgs properties}
\label{sec3}

One would like to translate the cross section measurements of the various
Higgs production and decay channels into measurements of Higgs boson 
properties, in particular into measurements of the various Higgs boson
couplings to gauge fields and fermions. This translation requires knowledge
of NLO QCD corrections to production cross sections, information on the
total Higgs decay width and a combination of the measurements discussed
previously. The task here is to find a strategy for combining the 
anticipated
LHC data without undue loss of precision due to theoretical uncertainties
and systematic errors.

For our further discussion it is convenient to rewrite all Higgs boson 
couplings in terms of partial widths of various Higgs boson decay channels. 
The Higgs-fermion couplings $g_{Hff}$, for example, which in the SM are 
given
by the fermion masses, $g_{Hff} = m_f(m_H)/v$, can be traded for the
$H\to \bar ff$ partial widths,
\bq
\Gamma_f = \Gamma(H\to \bar ff) = c_f {g_{Hff}^2\over 8\pi}
\biggl( 1-{4m_f^2\over m_H^2} \biggr)^{3\over 2}\;m_H\;.
\eq
Here $c_f$ is the color factor (1 for leptons, 3 for quarks). Similarly the 
square of the $HWW$ coupling ($g_{HWW}=gm_W$ in the SM) or the $HZZ$ 
coupling is proportional to the partial widths $\Gamma_W=\Gamma(H\to WW^*)$ 
or $\Gamma_Z=\Gamma(H\to ZZ^*)$~\cite{Keung:1984hn}. Analogously we trade 
the squares of the effective $H\gamma\gamma$ and $Hgg$ couplings for 
$\Gamma_\gamma=\Gamma(H\to\gamma\gamma)$ and $\Gamma_g=\Gamma(H\to gg)$.
Note that the $Hgg$ coupling is essentially proportional to $g_{Htt}$, the
Higgs boson coupling to the top quark.
 
The Higgs production cross sections are governed by the same squares of
couplings. This allows to write e.g. the $gg\to H$ production cross section
as~\cite{Barger:1987nn}
\bq
\sigma(gg\to H) = \Gamma(H\to gg){\pi^2\over 8m_H^3}\tau
\int_{\tau}^1 {dx\over x}g(x,m_H^2)g({\tau\over x},m_H^2)\;,
\eq
where $\tau = m_H^2/s$. Similarly the $qq\to qqH$ cross sections 
via $WW$ and $ZZ$ fusion are proportional to $\Gamma(H\to WW^*)$ and 
$\Gamma(H\to ZZ^*)$, respectively. In the narrow width approximation,
which is appropriate for the intermediate Higgs mass range considered here,
these production cross sections need to be multiplied by the branching
fractions for final state $j$, $B(H\to j)=\Gamma_j/\Gamma$, where $\Gamma$
denotes the total Higgs width. This means that the various cross section 
measurements discussed in the previous Section provide measurements of
various combinations $\Gamma_i\Gamma_j/\Gamma$. 

The production cross sections are subject to QCD corrections, which 
introduces theoretical uncertainties. While the $K$-factor for the gluon 
fusion process is large~\cite{HggNLO}, which suggests a sizable theoretical 
uncertainty on the production cross section, the NLO corrections to the weak 
boson fusion cross section are essentially identical to the ones encountered 
in deep inelastic scattering and are quite small~\cite{wbfNLO}. Thus we can 
assign a small theoretical uncertainty to the latter, of order 5\%, while 
we shall use a larger theoretical error for the gluon fusion process, of 
order 20\%~\cite{HggNLO}. 
The problem for weak boson fusion is that it consists of a mixture of
$ZZ\to H$ and $WW\to H$ events, and we cannot distinguish between the two 
experimentally. In a large class of models the ratio of $HWW$ and $HZZ$ 
couplings is identical to the one in the SM, however, and this includes the
MSSM. We therefore make the following $W,Z$-universality assumption:
\begin{itemize}
\item{}
The $H\to ZZ^*$ and $H\to WW^*$ partial widths are related by SU(2) as
in the SM, i.e. their ratio, $z$, is given by the SM value,
\bq
\Gamma_Z = z\; \Gamma_W = z_{SM}\;\Gamma_W \;.
\eq
\end{itemize}
Note that this assumption can be tested, at the 15-20\% level for 
$m_H>130$~GeV, by forming the ratio 
$B\sigma(gg\to H\to ZZ^*)/B\sigma(gg\to H\to WW^*)$, in which QCD 
uncertainties cancel (see Table~\ref{table:XYratio}).

With  $W,Z$-universality, the three weak boson fusion cross sections give us
direct measurements of three combinations of (partial) widths,
\beq
X_\gamma = {\Gamma_W\Gamma_\gamma \over  \Gamma}\qquad &{\rm from}&\;\;
qq\to qqH,\; H\to\gamma\gamma \;, \\
X_\tau = {\Gamma_W\Gamma_\tau \over \Gamma}\qquad &{\rm from}&\;\;
qq\to qqH,\; H\to\tau\tau \;, \\
X_W = {\Gamma_W^2\over \Gamma}\quad\qquad &{\rm from}&\;\;
qq\to qqH,\; H\to WW^{(*)} \;,
\eeq
with common theoretical systematic errors of 5\%. In addition the three gluon 
fusion channels provide measurements of 
\beq
Y_\gamma = {\Gamma_g\Gamma_\gamma\over \Gamma}\qquad &{\rm from}&\;\;
gg\to H\to\gamma\gamma \;, \\
Y_Z = {\Gamma_g\Gamma_Z\over \Gamma}\qquad &{\rm from}&\;\;
gg\to H\to ZZ^{(*)} \;, \\
Y_W = {\Gamma_g\Gamma_W\over \Gamma}\qquad &{\rm from}&\;\;
gg\to H\to WW^{(*)} \;,
\eeq
with common theoretical systematic errors of 20\%. 

The first precision test of the Higgs sector is provided by taking ratios 
of the $X_i$'s and ratios of the $Y_i$'s. In these ratios the QCD 
uncertainties, and all other uncertainties related to the initial state, 
like luminosity and pdf errors, cancel.  Beyond testing $W,Z$-universality,
these ratios provide useful 
information for Higgs masses between 100 and 150~GeV and 120 to 150~GeV, 
respectively, where more than one channel can be observed in the weak 
boson fusion and gluon fusion groups. Typical errors on these cross 
section ratios are expected to be in the 15 to 20\% range (see 
Table~\ref{table:XYratio}).
Accepting an additional systematic error of about 20\%, a measurement 
of the ratio $\Gamma_g/\Gamma_W$, which determines the $Htt$ to $HWW$ 
coupling ratio, can be performed, by measuring the cross section 
ratios $B\sigma(gg\to H\to\gamma\gamma)/\sigma(qq\to 
qqH)B(H\to\gamma\gamma)$
and  $B\sigma(gg\to H\to WW^*)/\sigma(qq\to qqH)B(H\to WW^*)$. Expected
accuracies are listed in Table~\ref{table:XYratio}. In these estimates 
the systematics coming from understanding detector acceptance is not 
included. 

\begin{table}[th]
\caption{Summary of the accuracy with which various ratios of 
partial widths can be determined with 200 fb$^{-1}$ of data. The first two 
columns give the ratio considered and indicate the method by which it is 
measured. $Y_Z/Y_W$, for example, indicates a measurement of 
$\sigma B(H\to ZZ^*)/\sigma B(H\to WW^*)$ in gluon fusion, while $X_i$ 
ratios correspond to weak boson fusion (see text for details). The 
statistical combination of several channels for a given width ratio is 
indicated by $\oplus$. 5\% and 20\% theoretical uncertainties for weak 
boson and gluon fusion cross sections affect the mixed gluon/weak boson 
fusion ratios only, which are needed for a measurement of 
$\Gamma_g/\Gamma_W$. The effect of this systematic error is indicated 
in the last line.
}
\vspace{0.15in}
\label{table:XYratio}
\begin{tabular}{c|l|ccccccccc}
$m_H$ & & 100  &  110  &  120 &  130 &  140 &  150 &  160 &  170 &  180  \\
\hline
$z=\Gamma_Z/\Gamma_W$ & $Y_Z/ Y_W$ &
               &       & 48\% & 29\% & 19\% & 17\% & 15\% & 20\% & 17\%   \\
 & ${Y_Z\over Y_\gamma}{X_\gamma\over X_W}$ &
               &       & 30\% & 21\% & 19\% & 23\% &      &      &       \\
 & ${Y_Z\over Y_W}\oplus {Y_Z\over Y_\gamma}{X_\gamma\over X_W}$ &
               &       & 29\% & 19\% & 15\% & 14\% & 15\% & 20\% & 17\%   \\
\hline
$\Gamma_\gamma/\Gamma_W$ & ${Y_\gamma\over Y_W}\oplus{X_\gamma\over X_W}$ &
               &       & 16\% & 12\% & 11\% & 13\% &      &      &       \\
$\Gamma_\tau/\Gamma_W$ & ${X_\tau\over X_W}$ &
               &       & 15\% & 12\% & 14\% & 21\% &      &      &       \\
$\Gamma_\tau/\Gamma_\gamma$ & ${X_\tau\over X_\gamma}$ &
          20\% &  16\% & 15\% & 16\% & 18\% & 27\% &      &      &       \\
\hline
$\Gamma_g/\Gamma_W$ & ${Y_\gamma\over X_\gamma}\oplus{Y_W\over X_W}$ &
          22\% &  18\% & 15\% & 13\% & 12\% & 13\% &  8\% &   9\% & 14\%  \\
       & ${Y_\gamma\over X_\gamma}\oplus{Y_W\over X_W}\oplus 21\%$ &
          30\% &  27\% & 25\% & 24\% & 24\% & 24\% & 22\% & 22\% & 25\%  \\
\end{tabular}
\end{table}

Beyond the measurement of coupling ratios, minimal additional assumptions 
allow an indirect measurement of the total Higgs width. First of all, the 
$\tau$ partial width, properly normalized, is measurable with an accuracy of 
order 10\%. The $\tau$ is a third generation fermion with isospin 
$-{1\over 2}$, just like the $b$-quark. In all extensions of the SM with 
a common source of lepton and quark masses, even if generational symmetry 
is broken, the ratio of $b$ to $\tau$ Yukawa couplings is given by 
the fermion mass ratio. We thus assume, in addition to $W,Z$-universality,
that
\begin{itemize}
\item{}
The ratio of $b$ to $\tau$ couplings of the Higgs is given by their mass 
ratio, i.e.
\bq
y = {\Gamma_b\over \Gamma_\tau} = 3c_{QCD}{g_{Hbb}^2\over g_{H\tau\tau}^2}
= 3c_{QCD}{m_b^2(m_H)\over m_\tau^2}\;,
\eq
where $c_{QCD}$ is the known QCD and phase space correction factor.

\item{}
The total Higgs width is dominated by decays to $\bar bb$, $\tau\tau$, 
$WW$, $ZZ$, $gg$ and $\gamma\gamma$, i.e. the branching ratio for 
unexpected channels is small:
\beq
\epsilon = 1-\biggl(&& B(H\to b\bar b)+B(H\to \tau\tau)+B(H\to WW^{(*)})+
\nonumber \\
&& B(H\to ZZ^{(*)})+B(H\to gg)+B(H\to \gamma\gamma) \ \biggr) \qquad \ll 1\;.
\eeq
\end{itemize}
Note that, in the Higgs mass range of interest, these two assumptions are 
satisfied for both CP even Higgs bosons in most of the MSSM parameter space.
The first assumption holds in the MSSM at tree level, but
can be violated by large squark loop contributions, in particular for small
$m_A$ and large $\tan\beta$~\cite{carena,MSSMcalc}. 
The second assumption might be 
violated, for example, if the $H\to\bar cc$ partial width is exceptionally 
large. However, a large up-type Yukawa coupling would be noticeable in 
the $\Gamma_g/\Gamma_W$ coupling ratio, which measures the $Htt$ coupling.

\begin{figure}[thb]
\vspace*{-.7cm}
\begin{center}
\includegraphics[width=9.0cm,angle=0]{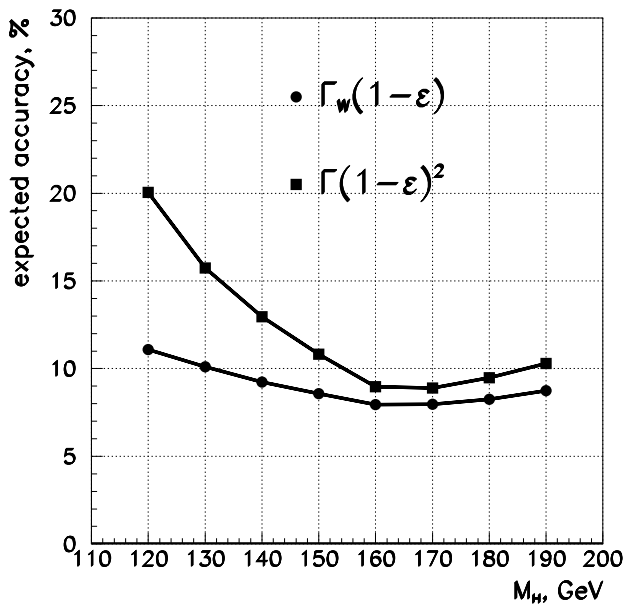}
\end{center}
\vspace*{-0.3cm}
\caption{Expected accuracy with which the Higgs boson width can be measured 
at the LHC, with 100~fb$^{-1}$ of data in each experiment. Results are shown
for the extraction of the the $H\to WW$ partial width, $\Gamma_W$, and
and the total Higgs boson width, $\Gamma$. $\epsilon$ is the sum of the
residual (small) branching ratios of unobserved channels, mainly 
$H\to c\bar c$ (see text).}
\label{fig:Widthaccurcy}
\end{figure}

With these assumptions consider the observable
\begin{eqnarray}
\tilde\Gamma_W &=& X_\tau(1+y) + X_W(1+z) + X_\gamma + \tilde X_g 
\nonumber \\
&=& \biggl(\Gamma_\tau+\Gamma_b +\Gamma_W +\Gamma_Z+
\Gamma_\gamma+\Gamma_g\biggr){\Gamma_W\over\Gamma}
=(1-\epsilon)\Gamma_W  \;,
\end{eqnarray}
where $\tilde X_g = \Gamma_g\Gamma_W/\Gamma$ is determined by combining 
$Y_W$
and the product $Y_\gamma X_W/X_\gamma$. $\tilde\Gamma_W$ provides a lower 
bound on $\Gamma(H\to  WW^{(*)})=\Gamma_W$. Provided $\epsilon$ is small 
($\epsilon < 0.1$ suffices for practical purposes), the determination 
of $\tilde\Gamma_W$ provides a direct measurement of the $H\to WW^{(*)}$ 
partial width. Once $\Gamma_W$ has been determined, the total width of the 
Higgs boson is given by
\bq
\label{eq:Gamma_tot}
\Gamma = {\Gamma_W^2\over X_W}={1\over X_W}
\biggl(X_\tau(1+y) + X_W(1+z) + X_\gamma + \tilde X_g \biggr)^2
{1\over (1-\epsilon)^2}\; .
\eq
For a SM-like Higgs boson the Higgs width is dominated by the $H\to\bar bb$
and $H\to WW^{(*)}$ channels. Thus, the error on $\tilde\Gamma_W$ is
dominated by the uncertainties of the $X_W$ and $X_\tau$ measurements 
and by the theoretical uncertainty on the $b$-quark mass, which enters 
the determination of $y$ quadratically. According to the Particle Data Group, 
the present 
uncertainty  on the $b$ quark mass is about $\pm 3.5\%$~\cite{pdg98}. 
Assuming a luminosity error of $\pm 5\%$ in addition to the theoretical 
uncertainty of the weak boson fusion cross section of $\pm 5\%$, the 
statistical
errors of the $qq\to qqH,\;H\to \tau\tau$ and $qq\to qqH, H\to WW$
cross sections of Tables~\ref{table:WBF.tautau} and \ref{table:WBF.WW} 
lead to an expected accuracy of the $\tilde\Gamma_W$ determination 
of order 10\%. 
More precise estimates, as a function of the Higgs boson mass, are shown
in Fig.~\ref{fig:Widthaccurcy}.

The extraction of the total Higgs width, via Eq.~(\ref{eq:Gamma_tot}),
requires a measurement of the $qq\to qqH, H\to WW^{(*)}$ cross section,
which is expected to be available for $m_H\gsim 115$~GeV~\cite{RZ_WW}.
Consequently, errors are large for Higgs masses close to this lower limit
(we expect a relative error of $\approx 20\%$ for $m_H=120$~GeV and 
$\epsilon<0.05$). But for Higgs boson masses around the $WW$ threshold, 
$\Gamma(1-\epsilon)^2$ can be determined with an error of about 10\%. 
Results are shown in Fig.~\ref{fig:Widthaccurcy} and look highly promising.

\section{Summary}
\label{sec4}

In the last section we have found that various ratios of Higgs partial 
widths
can be measured with accuracies of order 10 to 20\%, with an integrated 
luminosity of 100 fb$^{-1}$ per experiment. This translates into 5 to 10\%
measurements of various ratios of coupling constants. The 
ratio $\Gamma_\tau/\Gamma_W$ measures the coupling of down-type fermions
relative to the Higgs couplings to gauge bosons. To the extent that the
$H\gamma\gamma$ triangle diagrams are dominated by the $W$ loop,
the width ratio $\Gamma_\tau/\Gamma_\gamma$ measures the same relationship.
The fermion triangles leading to an effective $Hgg$ coupling are expected 
to be dominated by the top-quark, thus, $\Gamma_g/\Gamma_W$ probes the 
coupling of up-type fermions relative to the $HWW$ coupling.
Finally, for Higgs boson masses above $\approx 120$~GeV,
the absolute normalization of the $HWW$ coupling is accessible
via the extraction of the $H\to WW^{(*)}$ partial width
in weak boson fusion.

Note that these measurements test the crucial aspects of the Higgs sector.
The $HWW$ coupling, being linear in the Higgs field, 
identifies the observed Higgs boson as the scalar 
responsible for the spontaneous breaking of $SU(2)\times U(1)$: a 
scalar without a vacuum expectation value couples to gauge bosons only 
via $HHWW$ or $HHW$ vertices at tree level, i.e. the interaction is 
quadratic in scalar fields.
The absolute value of the $HWW$ coupling, as compared to the SM expectation,
reveals whether $H$ may be the only mediator of spontaneous symmetry breaking
or whether additional Higgs bosons await discovery. Within the framework of 
the MSSM this is a measurement of $|\sin (\beta-\alpha)|$, at the 
$\pm 0.05$ level. 
The measurement of the ratios of $g_{Htt}/g_{HWW}$ and 
$g_{H\tau\tau}/g_{HWW}$ then probes the mass generation of both up and down
type fermions. 

The results presented here constitute a first look only at the issue of
coupling extractions for the Higgs. This is the case for the weak boson fusion
processes in particular, which prove to be extremely valuable if not 
essential. Our analysis
is mostly an estimate of statistical errors, with some rough estimates
of the systematic errors which are to be expected for the various 
measurements of (partial) widths and their ratios. A number of issues need 
to be addressed in further studies, in particular with regard to the weak
boson fusion channels.

\begin{itemize}
\item[(a)]
The weak boson fusion channels and their backgrounds have only been studied 
at the parton level, to date. Full detector level simulations, and 
optimization of strategies with more complete detector information is 
crucial for further progress. 

\item[(b)] 
A central jet veto has been suggested as a powerful tool to suppress 
QCD backgrounds to the color singlet exchange processes which we call 
weak boson fusion. The feasibility of this tool and its reach need to be 
investigated in full detector studies, at both low and high luminosity.

\item[(c)]
In the weak boson fusion studies of $H\to WW$ and $H\to\tau\tau$ decays,
double leptonic $e^+e^-\sla p_T$ and $\mu^+\mu^-\sla p_T$ signatures have 
not yet been considered. Their inclusion promises to almost double the 
statistics available for the Higgs coupling measurements, at the price
of additional $ZZ+$jets and Drell-Yan plus jets backgrounds which are
expected to be manageable.

\item[(d)]
Other channels, like $WH$ or $t\bar tH$ associated production with 
subsequent
decay $H\to \bar bb$ or $H\to\gamma\gamma$, provide additional information
on Higgs coupling ratios, which complement our analysis at small Higgs mass
values, $m_H\lsim 120$~GeV~\cite{CMS,snowmass_H}. These channels need to be 
included in the analysis.

\item[(e)]
Much additional work is needed on more reliable background determinations. 
For the $H\to WW^{(*)}\to \ell^+\ell^{'-}\sla p_T$ channel in particular, 
where no narrow Higgs resonance peak can be reconstructed, a precise
background estimate is crucial for the measurement of Higgs couplings. 
Needed
improvements include NLO QCD corrections, single top quark production 
backgrounds, the combination of shower Monte Carlo programs with higher
order QCD matrix element calculations and more.

\item[(f)]
Both in the inclusive and WBF analyses any given channel contains a 
mixture of events from $gg\to H$ and $qq\to qqH$ production processes.
The determination of this mixture adds another source of systematic 
uncertainty, which was not included in the present study.
In ratios of $X$ observables (or of different $Y_i$) these uncertainties
largely cancel, except for the effects of acceptance variations due to
different signal selections. Since an admixture from the wrong production 
channel is expected at the 10 to 20\% level only, these systematic errors 
are not expected to be serious. 

\item[(g)]
We have only analyzed the case of a single neutral, CP even Higgs 
resonance with couplings which are close to the ones predicted in the SM.
While this case has many applications, e.g. for the large $m_A$ region of 
the MSSM, more general analyses, in particular of the MSSM case, are 
warranted and highly promising. 

\end{itemize}

While much additional work is needed, our study clearly shows that the 
LHC has excellent potential to provide detailed and accurate information 
on Higgs boson interactions. The observability of the Higgs boson at the LHC
has been clearly established, within the SM and extensions like the MSSM.
The task now is to sharpen the tools for accurate measurements of 
Higgs boson properties at the LHC.

\subsubsection*{Acknowledgements}

We would like to thank the organizers of the 
Les Houches Workshop for getting us together in an inspiring atmosphere. 
Useful discussions with M.~Carena, A.~Djouadi, K.~Jakobs and G.~Weiglein
are gratefully acknowledged.
We thank CERN for the hospitality extended to all of us during various 
periods of this work. 
The research of E.~R.-W. was partially supported by the Polish Government 
grant KBN 2P03B14715, and by the Polish-American Maria Sk\c lodowska-Curie 
Joint Fund II in cooperation with PAA and DOE under project PAA/DOE-97-316. 
The work of D.~Z. was supported in part by the University of Wisconsin Research
Committee with funds granted by the Wisconsin Alumni Research Foundation and
in part by the U.~S.~Department of Energy under Contract
No.~DE-FG02-95ER40896.


\bibliographystyle{plain}

\setcounter{figure}{0}
\setcounter{table}{0}
\setcounter{section}{0}
\setcounter{equation}{0}
\newpage

\begin{center}

{\large\sc {\bf Higgs boson production at hadron colliders at NLO}}

\vspace{0.5cm}

{\sc C. Bal\'azs, A. Djouadi, V. Ilyin and M. Spira} 
\end{center}

\begin{abstract}
We discuss the production of neutral Higgs bosons at the hadron colliders 
Tevatron and LHC, in the context of the Standard Model and its minimal
supersymmetric extension. The main focus will be on the next--to--leading 
order QCD radiative corrections to the main Higgs production mechanisms and 
on Higgs production in processes of higher order in the strong coupling 
constant. 
\end{abstract}

\section{Introduction}

One of the most important missions of future high--energy colliders will be the
search for scalar Higgs particles and the exploration of the electroweak
symmetry breaking mechanism. In the Standard Model (SM), one doublet of
complex scalar fields is needed to spontaneously break the symmetry, leading
to a single neutral Higgs particle $H^0$ \cite{mssm}. In the SM, the Higgs 
boson mass is a free parameter and can have a value anywhere between 100 GeV 
and 1 TeV. 
In contrast, a firm prediction of supersymmetric extensions of the SM 
is the existence of a light scalar Higgs boson \cite{mssm}. In the Minimal
Supersymmetric Standard Model (MSSM) the Higgs sector contains a quintet of
scalar particles [two CP-even $h$ and $H$, a pseudoscalar $A$ and two charged
$H^\pm$ particles] \cite{mssm}, the Higgs boson $h$ of which should be light,
with a mass $M_{h} \lsim 135$ GeV. If this particle is not found at
LEP2, it will be produced at the upgraded Tevatron (where a large luminosity, 
$\int {\cal L} \sim 20$ fb$^{-1}$, is expected) \cite{tev,tev-michael} or at 
the LHC \cite{ATLAS2,CMS2,habil}, if the MSSM is indeed realized in Nature.

Since Higgs boson production at hadron colliders involves strongly interacting
particles in the initial state, the lowest order cross sections are in general
affected by large uncertainties arising from higher order corrections. If the
next-to-leading QCD corrections to these processes are included, the total
cross sections can be defined properly and in a reliable way in most of the
cases.  In this contribution, we will discuss the next--to--leading order 
(NLO) QCD radiative corrections to the main neutral Higgs production mechanisms
as well as neutral Higgs boson production in processes of higher order in the 
strong coupling constant. 

The contribution is organized as follows. In the next section \cite{subsec1},
we summarize the main processes for the production of the neutral Higgs bosons
of the MSSM at hadron colliders and discuss the effects of their
next--to--leading order QCD corrections; we will then discuss the recently
evaluated SUSY--QCD corrections to some of these processes. In section 3
\cite{subsec2}, we will concentrate on Higgs boson production in association
with heavy quarks which in the MSSM might have the largest cross sections due a
possible strong  enhancement of the Yukawa couplings of third generation
quarks; we will discuss in particular the next--to--leading order QCD
corrections to Higgs production in heavy quark fusion. In section 4
\cite{subsec3}, we will analyze the detection of the SM and lightest MSSM [in
the decoupling regime] Higgs boson in the channel $\gamma \gamma$+jet at the
LHC [where the Higgs boson is produced in the gluon--gluon fusion mechanism and
decays into two photons].  

\section{MSSM neutral Higgs production at hadron colliders: \\ 
Next--to--Leading--Order QCD corrections}

\subsection{Summary of standard NLO QCD corrections}

At hadron colliders, the production of the neutral Higgs bosons in the MSSM 
is provided by the following processes: \s

(a) The gluon--gluon fusion, mediated by heavy quark loops, is the
dominant production mechanism for neutral Higgs particles, $gg \ra \Phi$ with
$\Phi=h,H$ or $A$ \cite{ggh}. Since the Higgs particles in the mass range of
interest, $M_\Phi \lsim 135$ GeV, dominantly decay into bottom quark pairs,
this process is rather difficult to exploit at the Tevatron because of the huge
QCD background \cite{tev}.  In contrast, at the LHC rare decays of the 
$h$ boson to two photons or decays of the $H,A$ bosons to $\tau$ and $\mu$
lepton pairs make this process very useful \cite{ATLAS2,CMS2}.  \s

(b) Higgs--strahlung off $W$ or $Z$ bosons for the CP-even Higgs particles
[due to CP--invariance, the pseudoscalar $A$ particle does not couple to the
massive gauge bosons at tree level]: $q\bar{q} \ra V^* \ra \Phi V$ with
$\Phi=h,H$ and $V=W,Z$ \cite{vh}. At the Tevatron, the process $q\bar{q}' \ra
hW$ [with the $h$ boson decaying into $b\bar{b}$ pairs] develops a cross
section of the order of a fraction of a picobarn for a SM--like $h$ boson with a
mass below $\sim 135$ GeV, making it the most relevant mechanism to study
\cite{tev}. At the LHC, both the $b\bar{b}$ and $\gamma \gamma$ decay modes of
the $h$ boson may be exploited \cite{ATLAS2}. \s

(c) If the heavier $H,A, H^\pm$ bosons are not too massive, the pair 
production of two Higgs particles in the Drell--Yan type process, $q\bar{q}
\ra \Phi_1 \Phi_2$ \citer{pair,krause}, might lead to a variety of final 
states [$hA, HA, H^\pm h, H^\pm H, H^\pm A, H^+ H^-$] with reasonable cross 
sections [in particular for $M_A \sim M_H \sim M_{H^\pm} \lsim 250$ GeV and 
small values of tan$\beta$, the ratio of the vacuum expectation values of 
the two Higgs doublets] especially at the LHC. Moreover, neutral and
charged Higgs boson pairs will be produced in gluon fusion $gg\to \Phi_1\Phi_2$
\citer{9a,9b}. \s

(d) The production of CP--even Higgs bosons via vector boson fusion,
$q q \ra qqV^*V^* \ra qq\Phi$ \cite{vvh}. In the case of a SM-like $h$ boson,
this process has a sizeable cross section at the LHC.
While decays of the Higgs boson into heavy quark pairs are problematic to be
detected in the jetty environment of the LHC, decays into $\tau$ lepton
pairs make this process useful at the LHC as discussed recently \cite{zepp}. \s

(e) The production of neutral Higgs bosons via radiation off 
heavy bottom and top quarks [$q\bar{q},gg\ra b\bar{b}\Phi,
t\bar{t}\Phi$] might play an important role in SUSY theories \cite{tth}.
In particular, because the couplings of the Higgs boson to $b$ quarks can be
strongly enhanced for large values of tan$\beta$, Higgs production in 
association with $b\bar{b}$ pairs can give rise to large production
rates. \s

It is well known that for processes involving strongly interacting particles,
as is the case for the ones discussed above, the lowest order cross 
sections are affected by large uncertainties arising from higher
order corrections. If the next-to-leading QCD corrections to these processes 
are included, the total cross sections can be defined properly and in a 
reliable way in most of the cases. 

For the standard QCD corrections, the next-to-leading corrections are available
for most of the Higgs boson production processes\footnote{The small NLO QCD 
corrections to the important Higgs decays into photons are also 
available \cite{photons}.}. They are parameterized by 
the K-factors [defined as the ratios of the next-to-leading order cross 
sections to the lowest order ones]: \s

-- For Higgs boson production via the gluon fusion processes, the 
K--factors have been calculated a few years ago in the SM \cite{ggSM} and in 
the MSSM \cite{ggSUSY}; the [two-loop] QCD corrections to the heavy top and to
the bottom quark loops [which gives the dominant contributions to the 
cross section for large $\tan \beta$ values] have been found to be significant 
since they increase the cross sections by up to a factor of two. \s

-- The K--factors for Higgs production in association with a gauge boson $(b)$
and for Drell--Yan--like Higgs pair production $(c)$, can be inferred from 
the one the Drell--Yan production of weak vector bosons and increase the cross 
section by approximately 30\% \cite{vhqcd}. \s

-- The QCD corrections to pair production $gg\to \Phi_1\Phi_2$ are only known
in the limit of light Higgs bosons compared with the loop--quark mass. This
is a good approximation in the case of the lightest $h$ boson which, due to 
phase space, has the largest cross section in which the top quark loop is 
dominant for small values of $\tan \beta$ or in the decoupling limit. The
corrections enhance the cross sections by up to a factor of two \cite{9b}. \s

-- For Higgs boson production in the weak boson fusion process $(d)$, the QCD
corrections can be derived in the structure function approach from
deep-inelastic scattering; they turn out to be rather small, enhancing the
cross section by about 10\% \cite{vvhqcd}. \s

-- Finally, the full QCD corrections to the associated Higgs production with
heavy quarks $(e)$ are not yet available; they are only known in the limit of
light Higgs particles compared with the heavy quark mass \cite{tthqcd} which is
only applicable to $t\bar{t}h$ production; in this limit the QCD corrections
increase the cross section by about 20--60\%.  

\subsection{SUSY QCD corrections}

Besides these standard QCD corrections, additional SUSY-QCD corrections must be
taken into account in SUSY theories; the SUSY partners of quarks and gluons,
the squarks and gluinos, can be exchanged in the loops and contribute to the
next-to-leading order total cross sections. In the case of the gluon fusion
process, the QCD corrections to the squark loop contributions have been
calculated in the limit of light Higgs bosons and heavy gluinos; the K--factors
were found to be of about the same size as the ones for the quark loops
\cite{gghqcdsusy}. 

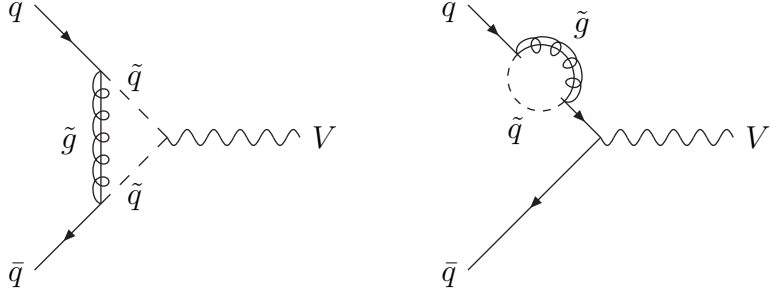
\begin{figure}[hbt]
\begin{center}
\begin{picture}(180,100)(-50,0)

\ArrowLine(0,100)(25,75)
\ArrowLine(25,25)(0,0)
\Line(25,75)(25,25)
\Gluon(25,75)(25,25){-3}{5}
\DashLine(25,75)(50,50){5}
\DashLine(50,50)(25,25){5}
\Photon(50,50)(100,50){-3}{5}
\put(105,46){$V$}
\put(-10,96){$q$}
\put(-10,-4){$\bar q$}
\put(35,70){$\tilde{q}$}
\put(35,25){$\tilde{q}$}
\put(10,46){$\tilde{g}$}

\end{picture}
\begin{picture}(180,100)(-30,0)

\ArrowLine(0,100)(20,80)
\ArrowLine(35,65)(50,50)
\ArrowLine(50,50)(0,0)
\Photon(50,50)(100,50){-3}{5}
\GlueArc(27.5,72.5)(12.5,-45,135){3}{4}
\CArc(27.5,72.5)(12.5,-45,135)
\DashCArc(27.5,72.5)(12.5,135,315){3}
\put(105,46){$V$}
\put(-10,96){$q$}
\put(-10,-4){$\bar q$}
\put(15,50){$\tilde{q}$}
\put(40,90){$\tilde{g}$}
\end{picture}
\vspace*{5mm}
\caption[ ]{\label{fg:dia}  Generic diagrams contributing to the SUSY-QCD 
corrections to the $q\bar q V$ vertex [$V=\gamma,Z,W$] at next--to--leading 
order.}
\end{center}
\vspace*{-3mm}
\end{figure}

During this workshop, we studied the SUSY--QCD corrections to the Higgs
production cross sections for Higgs--strahlung, Drell--Yan like Higgs pair
production and weak boson fusion processes \cite{paper}.  This analysis
completes the theoretical calculation of the NLO production cross sections of
these processes in the framework of supersymmetric extensions of the Standard
Model.  These corrections originate from $q\bar{q}V$ one--loop vertex
corrections, where squarks of the first two generations and gluinos are
exchanged, and the corresponding quark self-energy counterterms,
Fig.~\ref{fg:dia}.  \s

Including these SUSY--particle loop corrections, the lowest order partonic 
cross section for the Drell--Yan type processes will be shifted by
\beq 
\hat{\sigma}_{\rm LO} \ra \hat{\sigma}_{\rm LO} \left[ 1+ \frac{2}{3} 
\frac{\alpha_s (\mu)}{\pi} \Re e C (\hat{s}, m_{\tilde{q}}, m_{\tilde{g}})
\right] 
\eeq
For degenerate unmixed squarks [as is approximately the case for the first two 
generation squarks], the expression of the factor $C$ is simply given by
\beq
C (s, m_{\tilde{q}}, m_{\tilde{g}})
= 2 \int_0^1 x{\rm d}x \int_0^1 {\rm d} y {\rm log} \frac{
m_{\tilde{g}}^2 +( m_{\tilde{q}}^2 -m_{\tilde{g}}^2)x}
{-s x^2 y(1-y) +( m_{\tilde{q}}^2 -m_{\tilde{g}}^2)x
+m_{\tilde{g}}^2 - i\epsilon}
\eeq

\smallskip

For the fusion processes, the standard QCD corrections have been calculated 
within the structure function approach \cite{vvhqcd}. Since at lowest order, the
proton remnants are color singlets, at NLO no color will be exchanged between
the first and the second incoming (outgoing) quark line and hence the QCD
corrections only consist of the well-known corrections to the structure
functions $F_i(x,M^2)~(i=1,2,3)$. The final result for the QCD-corrected
cross section can be obtained from the replacements
\begin{eqnarray}
F_i(x,M^2)  \to  F_i(x,M^2) + \Delta F_i(x,M^2,Q^2) \hspace*{1cm} (i=1,2,3)
\end{eqnarray}
with $\Delta F_i(x,M^2,Q^2)$ the standard QCD corrections \cite{vvhqcd}. 
The typical renormalization and factorization scales
are fixed by the corresponding vector-boson momentum transfer
$\mu^2=M^2=-q_i^2$ for $x=x_i$ ($i=1,2$). \s

Including the SUSY--QCD correction at both $q_j q_j V$ vertices, the LO order 
structure functions $F_i(x_j, M^2)$ ($i=1,\ldots,3$ and $j=1,2$) have to be
shifted to: 
\beq
F_i(x_j, M^2) \ra F_i(x_j, M^2) \bigg[ 1+ \frac{2}{3} \frac{\alpha_s (\mu)}
{\pi} \Re e C (q_j^2, m_{\tilde{q}}, m_{\tilde{g}}) \bigg]
\eeq

To illustrate the size of these corrections, we perform a numerical analysis 
for the light scalar Higgs boson $h$ in the decoupling limit of large 
pseudoscalar masses, $M_A \sim 1$ TeV. In this case the light $h$ boson 
couplings to standard particles approach the SM values. The only relevant 
processes are then the Higgs--strahlung process $q\bar q \to hV$, the vector 
boson fusion mechanism $qq\to qqV^*V^*\to qqh$ and the gluon--gluon
fusion mechanism $gg \to h$. 

\begin{figure}[hbtp]
\vspace*{0.4cm}
\hspace*{-1.5cm}
\begin{turn}{-90}%
\epsfxsize=7cm \epsfbox{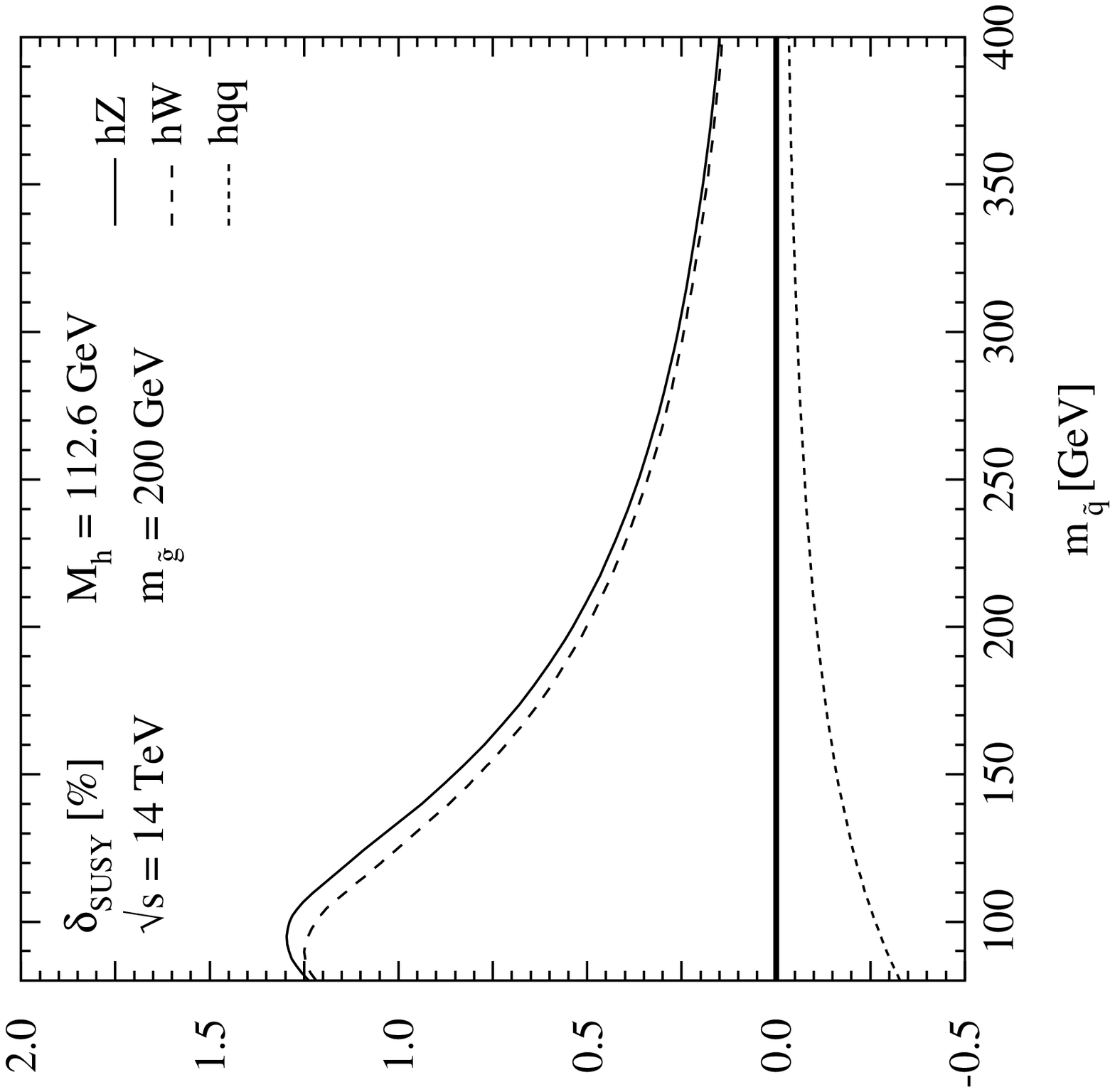}
\end{turn}
\hspace*{-1.8cm}
\begin{turn}{-90}%
\epsfxsize=7cm \epsfbox{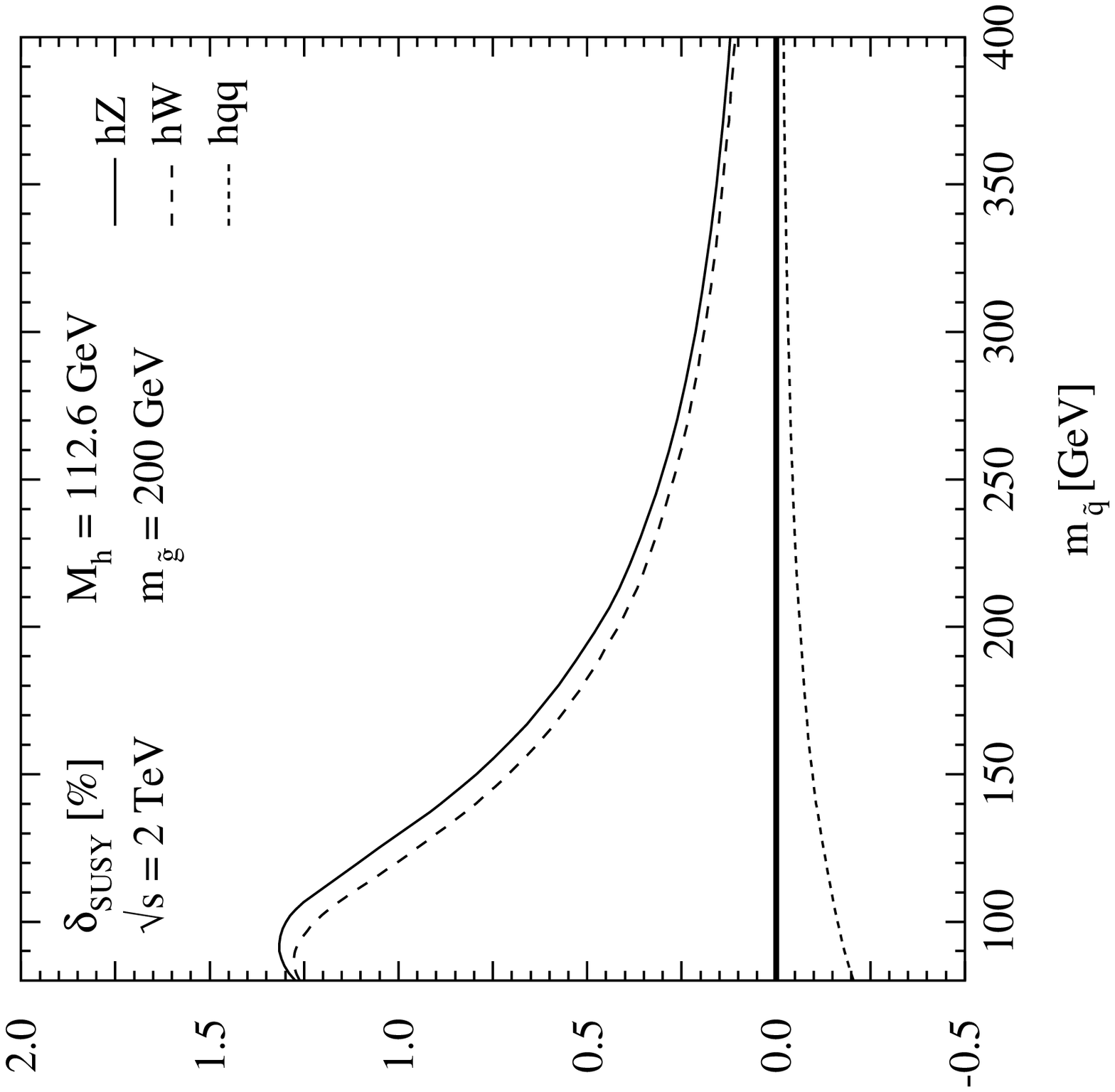}
\end{turn}
\vspace*{-0.1cm}
\caption[ ]{ Relative corrections due to  virtual squark and
gluino exchange diagrams to Higgs boson production via Higgs-strahlung
$q\bar q\to h+W/Z$ and vector boson fusion $qq\to qqV^*V^*\to qqh$
[$V=W,Z$] at the LHC (left) and the Tevatron (right). }
\label{fg:susy}
\end{figure}

We evaluated the Higgs mass for $\tan \beta=30$, $M_A=1$ TeV and vanishing
mixing in the stop sector; this yields a value $M_h=112.6$ GeV for the light
scalar Higgs mass. For the sake of simplicity we decompose the $K$ factors
$K=\sigma_{NLO}/\sigma_{LO}$ into the usual QCD part $K_{QCD}$ and the
additional SUSY correction $\delta_{SUSY}$: $K = K_{QCD} +\delta_{SUSY}$.  The
NLO (LO) cross sections are convoluted with CTEQ4M (CTEQ4L) parton densities
\cite{cteq4} and NLO (LO) strong couplings $\alpha_s$.  The additional SUSY-QCD
corrections $\delta_{SUSY}$ are presented in Fig.~\ref{fg:susy} as a function
of a common squark mass for a fixed gluino mass $m_{\tilde g}=200$ GeV [for the
sake of simplicity we kept the stop mass fixed for the determination of the
Higgs mass $M_h$ and varied the loop-squark mass independently].  

The SUSY-QCD corrections increase the
Higgs-strahlung cross sections by less than 1.5\%, while they decrease the
vector boson fusion cross section by less than 0.5\%. The maximal shifts
are obtained for small values of the squark masses of about 100 GeV, which are 
already ruled out by present Tevatron analyses \cite{PDG}; for more reasonable 
values of these masses, the corrections are even smaller. Thus, the additional
SUSY-QCD corrections, which are of similar size at the LHC and the Tevatron,
turn out to be small. For large squark/gluino masses they become even 
smaller due to the decoupling of these particles, as can be inferred from the 
upper squark mass range in Fig.~\ref{fg:susy}. \s

In summary, the SUSY--QCD corrections to Higgs boson production in
these channels are very small and can be safely neglected.

\section{Associated Higgs production with $b\bar{b}$ pairs} 

\subsection{Constraints on the MSSM parameter space}

In the MSSM, the Yukawa couplings between the Higgs bosons and the down--type
fermions, in particular the relatively heavy bottom quarks, are enhanced for
large $\tan\beta$ values. This enhancement can be so significant that it
renders the cross section of the associated production channel ($p{\bar p},pp
\to \phi^0 b \bar{b}$, with $\phi^0 =h^0,A^0,H^0$) the highest at the Tevatron
and the LHC, along with the cross section  of the gluon fusion mechanism $pp
\to gg ({\rm via~heavy~(s)fermion~loop}) \to \phi^0 X$ \cite{habil}.  The Higgs
bosons in this regime decay mainly into $b\bar{b}$ pairs, leading to 4
$b$--jets which can be tagged experimentally \cite{cdfwh}. Due to the lack of
phase space and the reduced couplings, the associated production with top
quarks is not feasible at the Tevatron, and is difficult at the LHC. This makes
it possible for the Tevatron RunII and LHC to discover Higgs bosons in the
$\phi^0 b \bar{b}$ process and to impose stringent constraints on the
SUSY--Higgs sector in a relatively model independent way.  [At the LHC, the
associated $H/A+ b\bar{b}$ production with the $\tau^+ \tau^-$ and $\mu^+
\mu^-$ Higgs decay channels is very important \cite{ATLAS2,CMS2} and allows to
cover most of the parameter space for large $\tan \beta$.]

In Ref.~\cite{Balazs-Cruz-He-Tait-Yuan}, an effective search strategy was
presented for the extraction of the signal from the backgrounds [which have
been calculated]. Using HDECAY \cite{hdecay2} to calculate the Higgs [and SUSY]
spectrum and branching fractions, and combining signals from the search of more
than one scalar boson [provided their masses differ by less than a resolution
$\Delta m_{exp}$ which can be chosen as the total Higgs decay width], contours
in the $\tan \beta$-$m_A$ plane of the MSSM, for which the Tevatron and LHC are
sensitive, can be derived. When scanning over the parameter space, the set of
soft breaking input parameters should be compatible with the current data from
LEP~II and the Tevatron while, preferably, not exceeding 1 TeV. The most
important parameters here are the masses and mixing of top squarks, and the
value and sign of the Higgsino mass parameter $\mu$.

\begin{figure}[t] 
\vspace*{-1.cm}
\begin{center}
\hskip  -1.0cm \psfig{figure=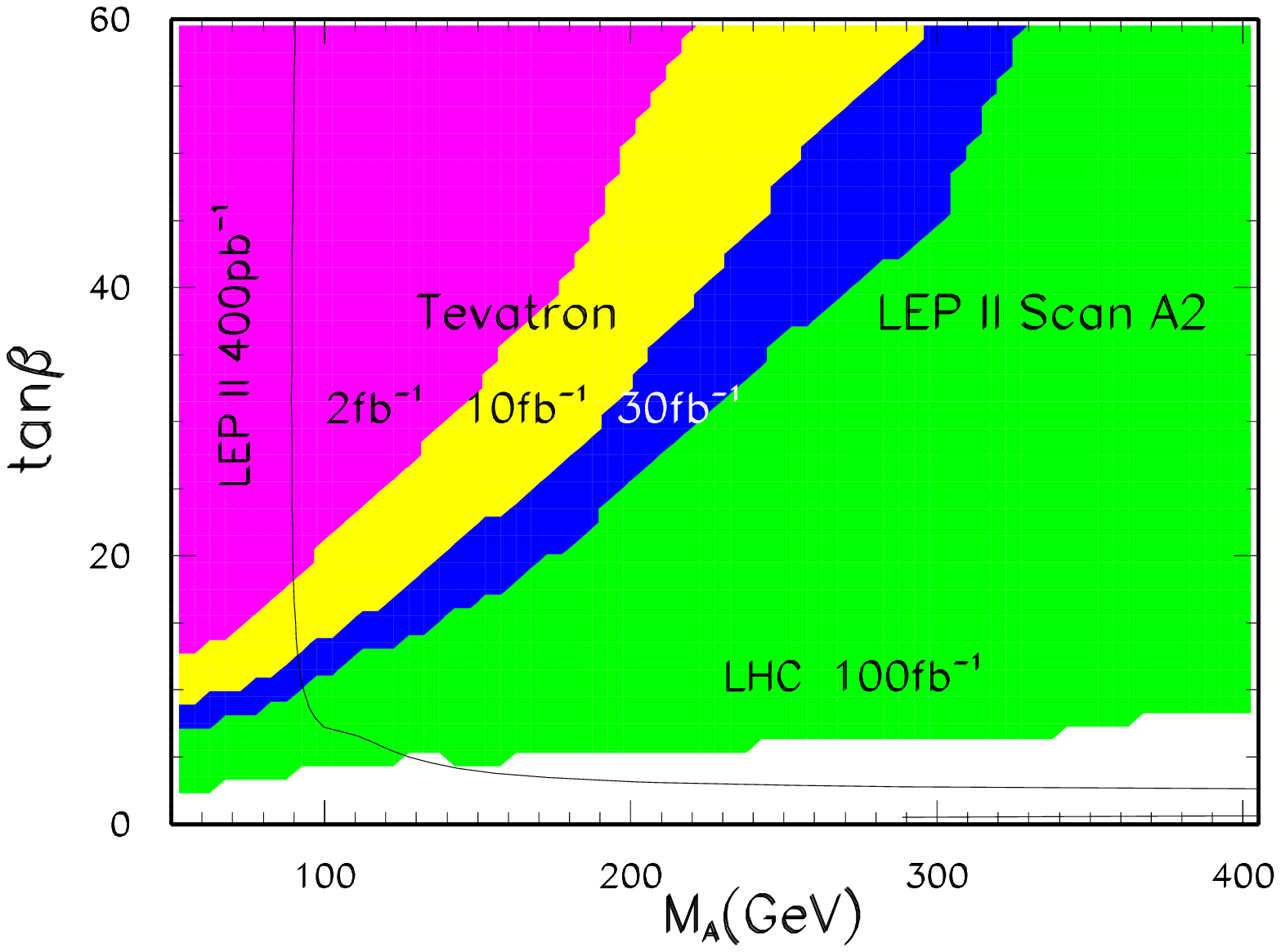,height=3.0in} 
\hskip -16.7cm \psfig{figure=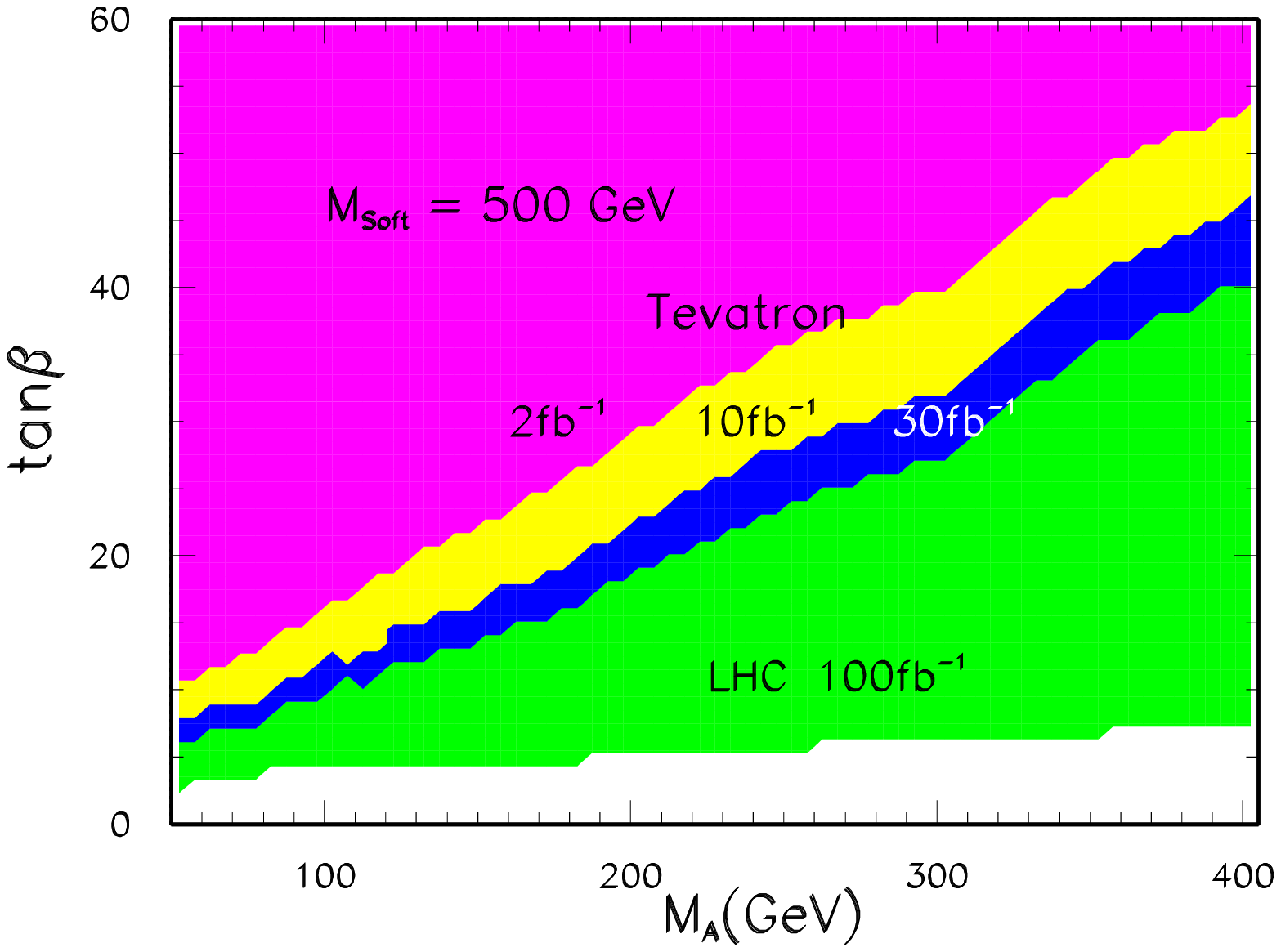,height=3.0in} 
\end{center}
\vspace*{-1.2cm} 
\caption{
$95\%$~C.L. theoretical estimates of sensitivity contours in the 
$\tan\beta$-$m_A$ planes of the MSSM. The areas above the four boundaries 
can be excluded by the Tevatron RunII and the LHC;
$M_{\rm soft}=500$ GeV (a) and the ``LEP~II Scan~A2'' (b) are shown.
From Ref.~\cite{Balazs-Cruz-He-Tait-Yuan}.}
\label{fig:Exclusion}
\vspace*{-.3cm}
\end{figure}

For soft breaking parameters $M_{\rm soft}= \mu = 500$ GeV,
Fig.~\ref{fig:Exclusion}a shows the $95\%$~C.L. exclusion contours in the
$\tan\beta$-$m_A$ plane, derived from the measurement of $\sigma( p\bar{p},pp
\to \phi^0 b \bar{b} \to b \bar{b} b \bar{b})$. The areas above the four
boundaries are accessible at the Tevatron RunII with the indicated luminosities
and for the LHC with 100 fb$^{-1}$. The potential of hadron colliders with
these processes is compared in Fig.~\ref{fig:Exclusion}b with that of LEPII
[where Higgs bosons are searched for in the $Zh$ and $hA$ production channels]
for the ``benchmark'' parameter scan ``LEPII Scan~A2'' discussed in
\cite{Abbiendi:1999sy} for $\sqrt{s}=200$ GeV and a luminosity of
$100$~pb$^{-1}$ per experiment. As can be seen, the Tevatron can already cover
a substantial region with only a 2 fb$^{-1}$ luminosity.  Furthermore, for $m_A
\gae 100$ GeV, Tevatron and LEPII are complementary.  The LHC can further probe
the MSSM down to values $\tan \beta \sim$ 7~(15) for $m_A < 400~(1000)$ GeV.

In conclusion, detecting the $\phi^0 b \bar{b}$ signal at hadron colliders
could effectively probe the MSSM Higgs sector, especially for large $\tan
\beta$ values\footnote{Note that so far, existing experimental studies are not
confirming the potential of this channel at the LHC \cite{ATLAS2}, while the
results seem to be more promising at the Tevatron Run II \cite{tev}.}. Similar
conclusions are reached in Ref.~\cite{dai} for the LHC and in
Ref.~\cite{tev,Carena:1998gk}. The results given here show a substantial
improvement compared to Ref.~\cite{dressetal}, where only the $p{\bar p}\to
\phi^0 b \bar{b} \to \tau^+ \tau^- b \bar b$ process is discussed at the
Tevatron RunI. Detailed interpretation of the above results in the MSSM and
other models [such as composite Higgs models with strong dynamics associated
with heavy quarks] can be found in Ref.~\cite{Balazs-Cruz-He-Tait-Yuan}. The
analyses can be improved in many ways, for instance with a better $b$--trigger,
which bears central significance for the detection of the $b$--jets.


\subsection{QCD corrections to Higgs production in heavy quark fusion}

Recently it was proposed that, due to the top-mass enhanced flavor mixing
Yukawa coupling of the charm and bottom to charged scalar or pseudoscalar
bosons ($\phi^\pm$), the $s$-channel partonic process $c\bar{b}, \bar{c}
b\to\phi^\pm$ can be an important mechanism for the production of $\phi^\pm$
\cite{HY}. This mechanism is also important for $s$-channel neutral scalar
production via $b {\bar b}$ fusion\footnote{Note that the subprocess $b\bar{b}
\to \phi^0$ alone overestimates the complete cross section via bottom fusion;
one has to add consistently the cross sections for $bg \to b \phi^0$ and $gg
\to b\bar{b} \phi^0$ to have a reliable value.}.  In this section, we describe
the complete NLO QCD corrections to these processes. The results were
originally calculated in Ref.~\cite{Balazs-He-Yuan}, to which we refer for
details. The QCD corrections for the SM Higgs production $b\bar{b} \to H$ has
been also discussed in Ref.~\cite{note-added}. The overlapping parts of the two
calculations are in  agreement.

\begin{figure}[ht!]
\vspace{-.5cm}
\begin{center}
\psfig{figure=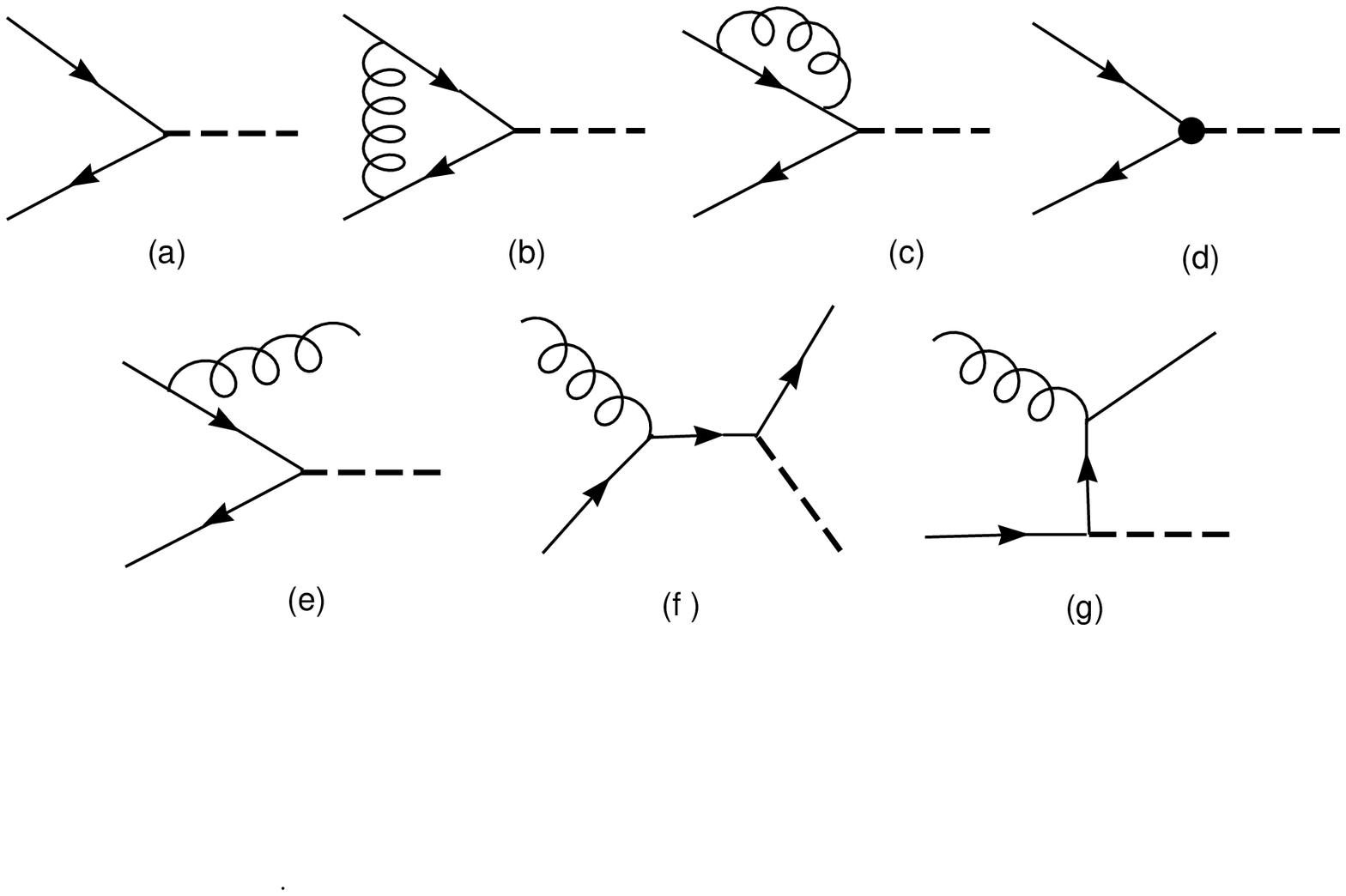,width=11.6cm}
\end{center}
\vspace{-4.2cm}
\caption{ 
Representative diagrams for charged or neutral (pseudo-)scalar
(dashed line) production from quark-anti quark and quark-gluon collisions at 
$O(\alpha_s^0)$ and $O(\alpha_s^1)$:
(a)~leading order contribution; 
(b-d)~self-energy and vertex corrections (with counter term);
(e)~real gluon radiation in $q\bar{q'}$-fusion;
(f-g)~$s$- and $t$-channel gluon-quark fusion.
}
\label{fig:qQPhi}
\vspace*{-3mm}
\end{figure}

The NLO contributions to the process $b\bar{b} \to \phi^0$ contain three 
parts: (i)~the one-loop Yukawa vertex and quark self-energy corrections 
(Fig.~\ref{fig:qQPhi}b-d); (ii) real gluon emission in  
$q\bar{q}^{\prime}$ annihilation (Fig.~\ref{fig:qQPhi}e); (iii)~$s$- 
and $t$-channel gluon-quark fusion (Fig.~\ref{fig:qQPhi}f-g). In addition, 
the renormalization of the fermion--higgs--fermion Yukawa coupling has to be 
performed. Since the factorization scale $\mu_F = m_\phi$ is much larger 
than the mass of the bottom quark, when computing the Wilson coefficient 
functions the $b$-quarks were treated as massless partons in the proton or 
anti-proton, similarly to Ref.~\cite{Haber}. The only effect of the heavy 
quark mass is to determine at which scale $\mu_F$ this heavy parton 
becomes active. (This is the Collins-Wilczek-Zee (CWZ) \cite{CWZ} 
scheme). The CTEQ4 PDFs \cite{cteq4} are used to calculate the rates, 
because they are consistent with the scheme used in the current study 
\cite{ACOT}.

\smallskip

The $\alpha_s$ corrections involve the contributions from the emission of 
real gluons, and as a result the scalar particle will acquire a non-vanishing
transverse momentum $Q_T$. When the emitted gluons are soft, they
generate large logarithmic contributions of the (lowest order) form 
$\alpha _s$$\ln^m\left(Q^2/Q_T^2\right)$$/Q_T^2$, where $Q$ is
the invariant mass of the scalar and $m=0,1$. These large
logarithms spoil the convergence of the perturbative series, and falsify
the $O(\alpha _s) $ prediction of the transverse momentum when $Q_T\ll
Q$. To predict the $Q_T$ distribution one can use the Collins--Soper--Sterman 
(CSS) formalism \cite{CSS}, resumming the logarithms of the
type $\alpha _s^n$ $\ln ^m\left( Q^2/Q_T^2\right)/Q_T^2$, to all orders
$n$ in $\alpha _s$ ($m=0,...,2n-1$). The resummation calculation is performed 
along the same lines as for vector boson production 
(cf. \cite{Balazs-YuanWZ}). To recover the ${\cal O}(\alpha_s)$ cross 
section, the Wilson coefficients $C_{i\alpha }^{(1)}$ are included in the 
resummed calculation in \cite{Balazs-He-Yuan}. The non-perturbative sector of 
the CSS resummation is assumed to be the same as for vector boson production
in Ref.~\cite{Balazs-YuanWZ}.

The resummed total rate is the same as the $O(\alpha _s)$ rate, when we include
$C^{(1)}_{i\alpha}$ and the usual fixed order NLO corrections at high $Q_T$,
and switch from the resummed distribution to the fixed order one at $Q_T=Q$. 
When calculating the total rate, we have applied this matching prescription. In
the case of the scalar production, the matching takes place at high $Q_T \sim
Q$ values, and the above matching prescription is numerically irrelevant when
calculating the total rate, since the cross sections around $Q_T \sim Q$ are
negligible.  Thus, as expected, the resummed total rate differs from the
$O(\alpha _s)$ rate only by a few percent.  Since the difference of the
resummed and fixed order rates and the $K$--factors (c.f. Fig.~\ref{fig:K-A0}) 
is small, we can conclude that for inclusive scalar production once the
resummation is performed, the $O(\alpha _s^2)$ corrections are likely to be
smaller than the uncertainty  from the PDF's.  

\smallskip

\begin{figure}[ht!]
\vspace*{-1.3cm}
\begin{center}
\psfig{figure=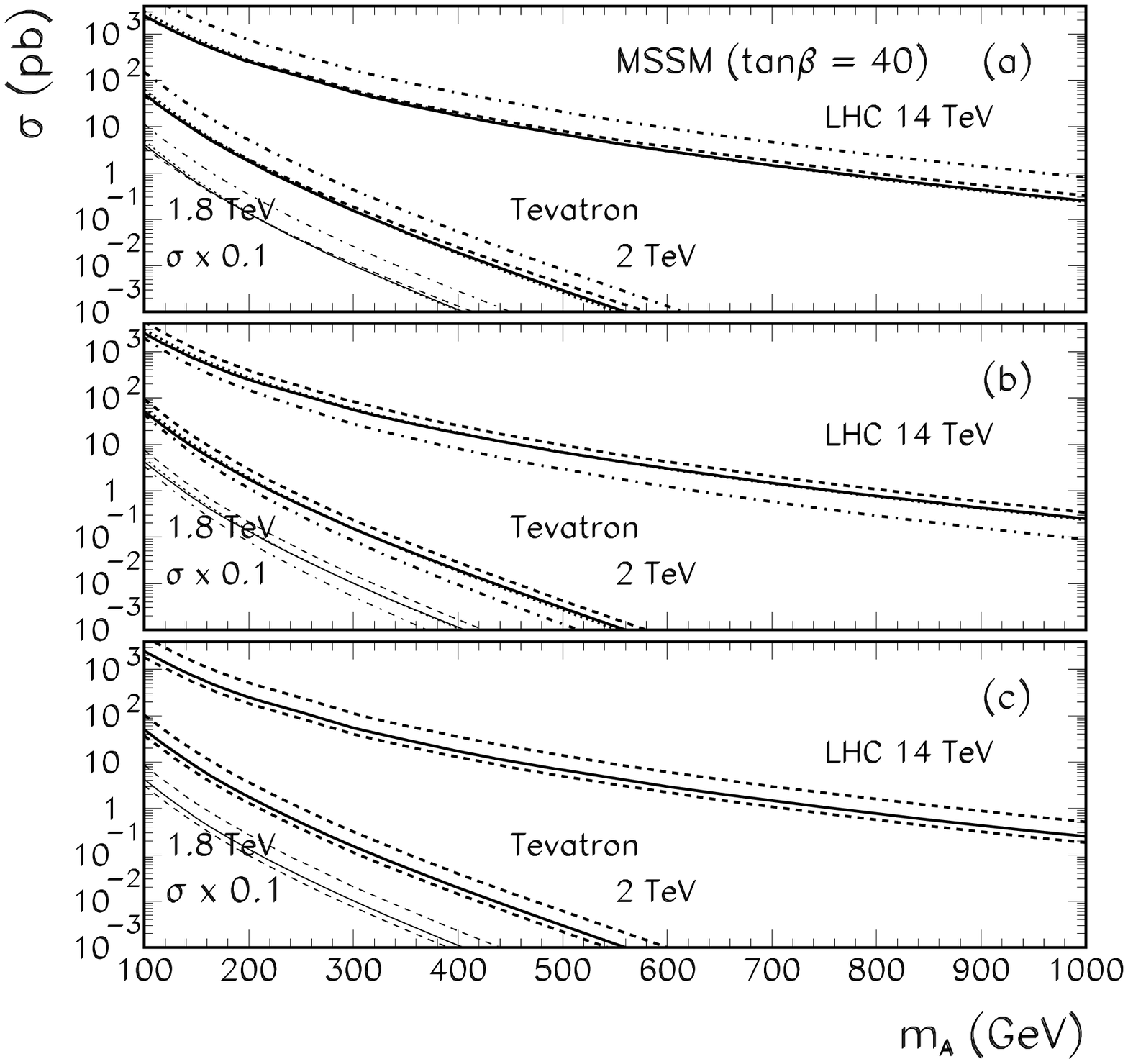,width=11cm,height=10.6cm}
\end{center}
\vspace*{-6mm}
\caption{ 
Cross sections for $A^0$ production in the MSSM with $\tan \beta =40$ at the 
Tevatron and the LHC. (a) The NLO cross sections with the resummed running 
(solid) and one-loop Yukawa coupling (dashed), as well as the LO cross 
sections with resummed running (dotted) and tree-level 
Yukawa coupling (dash-dotted) are shown. The cross sections 
at $\sqrt{s} = 1.8$~TeV (thin set of lowest curves) are multiplied by 0.1 
not to overlap with the $\sqrt{s} = 2$ TeV curves. (b) The NLO
(solid), the $b\bar b$ (dashed) and $bg$ (dash-dotted)
sub-contributions, and the LO (dotted) contributions. 
The (negative) $bg$ cross sections are multiplied by $-1$.
(c) The NLO cross sections with QCD running Yukawa
coupling (solid curves) and those with additional SUSY corrections 
(top/bottom dashed lines for $\mu =+/-500$\,GeV). }
\label{fig:Sigma_MSSMbb}
\vspace*{-.4cm}
\end{figure}

Since the QCD corrections are universal, the application to the production 
of neutral scalar or pseudo-scalar $\phi^0$ via the $b\bar{b}$ fusion is 
straightforward. In the following, we will consider only the production of 
the pseudo-scalar $A^0$ within the context of the MSSM. The total LO and 
NLO cross sections for the inclusive processes $p\bar{p},pp\to A^0X$ at 
the Tevatron and the LHC are shown in Fig.~\ref{fig:Sigma_MSSMbb} for 
$\tan \beta =40$. For other values the cross sections can be obtained by 
scaling with the factor $(\tan \beta /40)^2$.

Fig.~\ref{fig:Sigma_MSSMbb}a shows a significant improvement from the pure 
LO results (dash-dotted curves) due to the resummation of the large 
logarithms of $m_\phi^2/m_b^2$ into the running coupling. The good 
agreement between the LO results with running coupling and the NLO 
results is due to a non-trivial, and process-dependent, cancellation 
between the individual $O(\alpha_s)$ contributions of the $b\bar{b}$ and 
$bg$ sub-processes (which are connected via mass factorization). 

\begin{figure}[ht!]
\begin{center}
\vspace*{-1.cm}
\psfig{figure=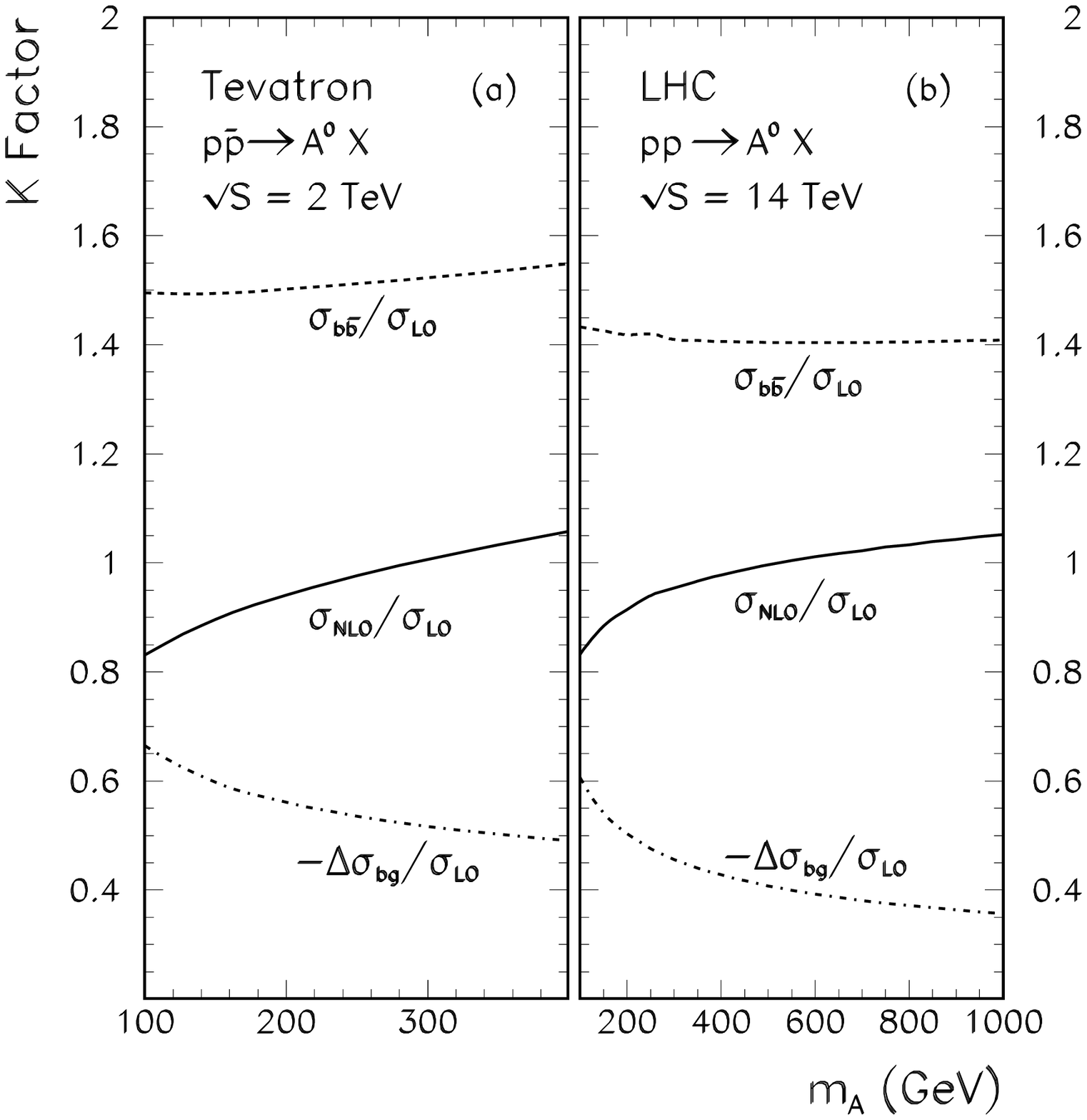,width=11cm,height=7.6cm}
\vspace*{-6mm}
\end{center}
\caption{ 
The $K$-factors for $A^0$ production in the MSSM with 
$\tan \beta =40$ for
the NLO ($K=\sigma_{\rm NLO}/\sigma_{\rm LO}$, solid lines),
$b\bar{b}$ ($K=\sigma_{b \bar{b}}/\sigma_{\rm LO}=
(\sigma_{\rm LO}+\Delta\sigma_{b \bar{b}})/\sigma_{\rm LO}$, 
dashed lines), and $bg$ ($K=-\Delta\sigma_{bg}/\sigma_{\rm LO}$, 
dash-dotted lines) contributions, 
at the Tevatron (a) and LHC (b).
}
\label{fig:K-A0}
\vspace*{-3mm}
\end{figure}

For large $\tan \beta$, the SUSY correction to the running 
$\phi^0$-$b$-$\bar{b}$ 
Yukawa coupling can be significant \cite{large-tanb}, and can be included in a 
similar way as it is done for the $\phi^0b\bar{b}$ associate 
production~\cite{cbhbb}. To illustrate the effects of these corrections, all 
MSSM soft-breaking parameters and $\mu$ were set to $500$~GeV. Depending 
on the sign of $\mu$, the correction to the coupling can take either the 
same or opposite sign as the full NLO QCD correction \cite{cbhbb}. In 
Fig.~\ref{fig:Sigma_MSSMbb}c, the solid curves represent the NLO cross 
sections with QCD correction alone, while the results including the SUSY 
corrections to the running bottom Yukawa coupling are shown for $\mu 
=+500$\,GeV (top dashed curves) and $\mu =-500$\,GeV (bottom dashed 
curves). These partial SUSY corrections can change the cross sections by 
about a factor of 2. 

\begin{figure}[t]
\begin{center}
\vspace*{-1.cm}
\psfig{figure=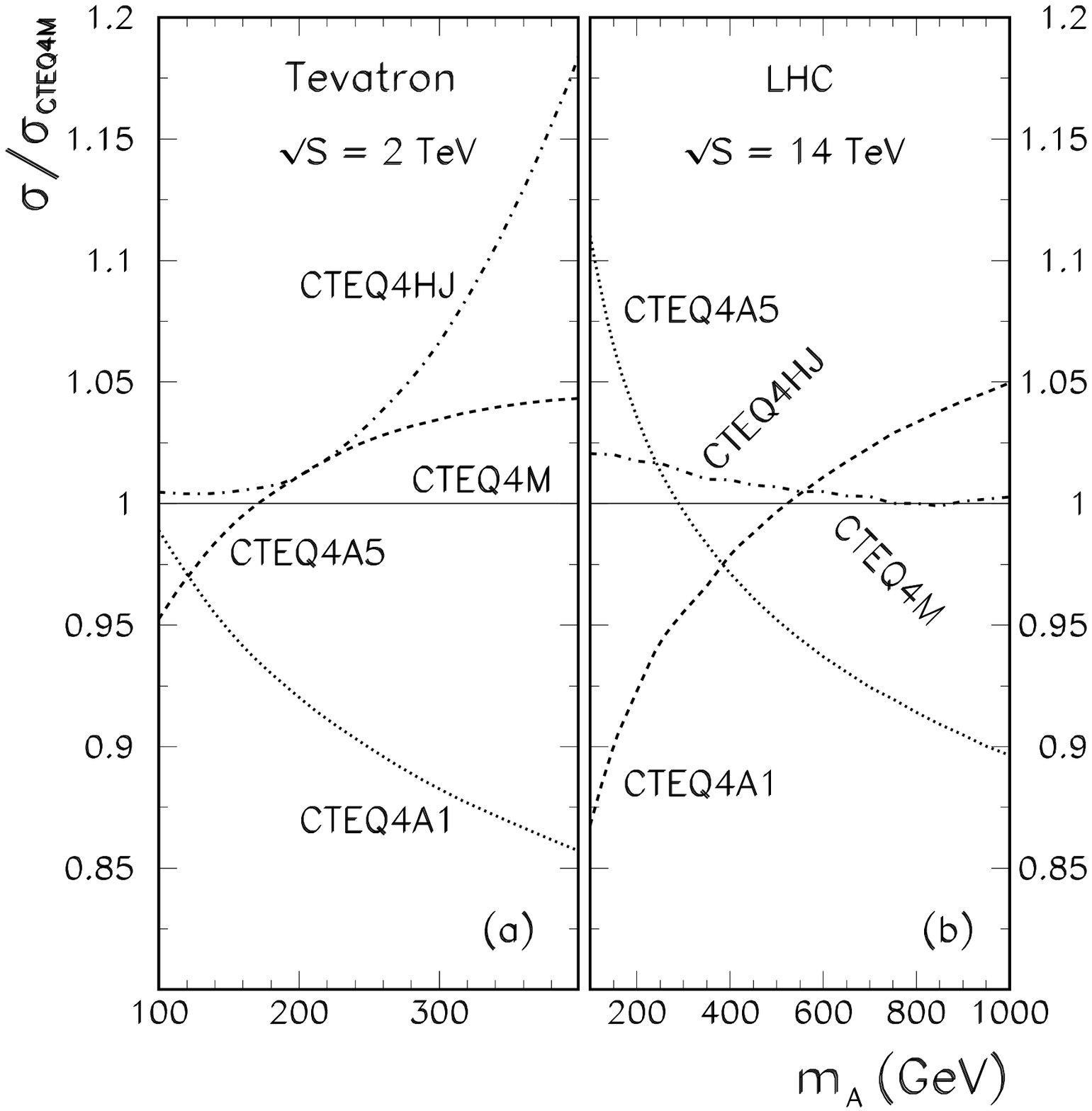,width=11cm,height=7.4cm}
\vspace*{-6mm}
\end{center}
\caption{
The ratios of NLO cross sections computed by four different sets of
CTEQ4 PDFs to the cross section computed by CTEQ4M for neutral
pseudo-scalar ($A^0$) production in the MSSM with $\tan \beta =40$, at the
upgraded Tevatron (a) and the LHC (b).}
\label{fig:PDF-bb}
\vspace*{-.2cm}
\end{figure}
\begin{figure}[ht!]
\begin{center}
\vspace*{-.7cm}
\psfig{figure=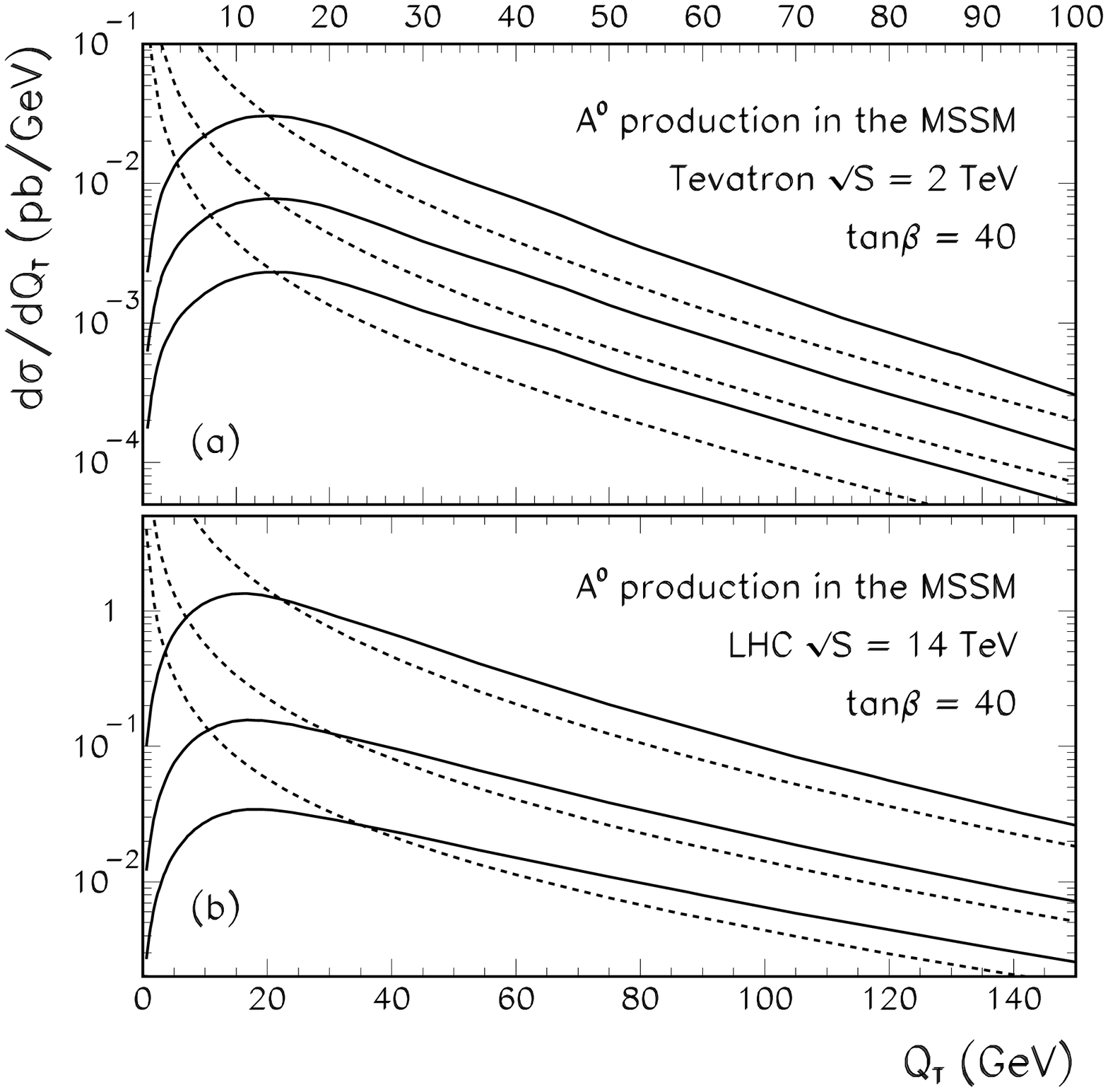,width=12cm,height=9.4cm}
\vspace*{-8mm}
\end{center}
\caption{
Transverse momentum distributions of pseudo-scalar $A^0$ produced
via hadron collisions, calculated in the MSSM with $\tan \beta =40$. 
The resummed (solid) and $O(\alpha _s)$ (dashed) curves 
are shown for $m_A =$ 200, 250, and
300\,GeV at the upgraded Tevatron (a), and 
for $m_A =$ 250, 400, and 550\,GeV at the LHC (b).
}
\label{fig:QT-MSSM}
\vspace*{-.2cm}
\end{figure}

The $K$-factors, the ratios of the NLO versus LO cross sections as defined
in Ref.~\cite{Balazs-He-Yuan}, for the 
$p\bar{p},pp\to A^0X$ processes are presented in Fig.~\ref{fig:K-A0} for 
the MSSM with $\tan \beta =40$. Depending on the $A^0$ mass, they range from 
about $-(16$$\sim$17)\% to +5\% at the Tevatron and the LHC. The 
uncertainties of the CTEQ4 PDFs for $A^0$-production at the Tevatron and 
the LHC are summarized in Fig.~\ref{fig:PDF-bb}. 

The transverse momentum distributions of $A^0$, produced at the upgraded 
Tevatron and at the LHC, are shown in Fig.~\ref{fig:QT-MSSM} for various 
$A^0$ masses with $\tan \beta = 40$. The solid curves are the result of 
the multiple soft-gluon resummation, and the dashed ones are from the 
$O(\alpha_s)$ calculation. The fixed order distributions are singular as 
$Q_T \to 0$, while the resummed ones have a maximum at some finite $Q_T$, 
and vanish at $Q_T = 0$. When $Q_T$ becomes large, of the order of $m_A$, 
the resummed curves merge into the fixed order ones. The average resummed 
$Q_T$ varies between 25 and 30 (40 and 60) GeV in the 200 to 300 (250 to 
550) GeV mass range of $m_A$ at the Tevatron (LHC).

In summary, the overall NLO corrections to the $p\bar{p},pp\to A^0X$ 
processes are found to vary between $-(16$$\sim$17)\% and +5\% at the 
Tevatron and the LHC in the relevant range of the $A^0$ mass. The 
uncertainties of the NLO rates due to the different PDFs also have been 
systematically examined, and found to be around $20\%$. The QCD 
resummation, including the effects of multiple soft-gluon radiation, was 
also performed to provide a better prediction of the transverse momentum 
distribution of the scalar $\phi^0$. This latter is important when 
extracting the experimental signals. Similar results can be easily 
obtained for the other neutral higgs bosons ($h^0$ and $H^0$) by properly 
rescaling the coupling. These QCD corrections can also be applied to the 
generic two higgs-doublet model (called type-III 2HDM\cite{2HDM3}), in 
which the two higgs doublets $\Phi_1$ and $\Phi_2$ couple to both up- and 
down-type quarks.

\newpage

\section{Higgs search in the $\gamma\gamma$+jet channel at LHC}

The observation of a Higgs boson with a mass $M_H<140$ GeV at the LHC in the
inclusive channel $pp\to\gamma\gamma+X$ is not easy \cite{atlasTDR,cmsTDR} as 
it is necessary to separate a rather elusive Higgs boson signal from the
continuum background. In Ref.~\cite{Hjet} the reaction $pp\to H(\to\gamma
\gamma)$+jet, when the Higgs boson is produced with large transverse momentum 
recoiling against a hard jet, was analyzed as a discovery channel. The signal 
rate is much smaller, but there remains enough events to discover the 
Higgs boson at a low luminosity LHC. It is important to note that the situation
with the background is undoubtedly much better in the case of Higgs production 
at high $p_T$. Thus, one has  $S/B\sim 1/2-1/3$ for CMS and ATLAS 
correspondingly, providing a discovery significance of 5 already with an 
integrated luminosity of 30 fb$^{-1}$. Furthermore, recent achievements in 
calculations of QCD next--to--leading corrections have shown an enhancement of 
the signal against the background. This circumstance together with the 
possibility to exploit the event kinematics in a more efficient way allow the
hope that this reaction will be the most reliable discovery channel for Higgs
bosons with masses $M_H=110-135$ GeV.

Typical acceptances of the LHC detectors ATLAS and CMS were taken into account
in the analysis: two photons are required with $p_t^\gamma>40$ GeV  for each
photon (harder than for the inclusive channel), and  $|\eta_\gamma|<2.5$, while
a jet was required with $E_t^{jet}>30$ GeV and $|\eta_{jet}|<4.5$, thus
involving the forward parts of the hadronic calorimeter. The isolation cut
$\Delta R>0.3$ was applied for each $\gamma \gamma$ and  $\gamma g(q)$ pair.

There are three QCD subprocesses giving a signal from the Higgs boson in the
channel under discussion in QCD leading order: $gg\to H+g$, $gq\to H+q$ and
$q\bar q\to H+g$. It was found that the $gg\to H+g\to\gamma\gamma+g$ subprocess
gives the main contribution to the signal rate. In total, the QCD signal
subprocesses give 5.5, 10.6 and 9.8 fb for $M_H=100$, 120 and 140 GeV,
correspondingly within the kinematical cuts described above.

Another group of signal subprocesses includes the electroweak reactions of
Higgs production through $WW$ or $ZZ$ fusion and in association with $W$ or $Z$
boson, where one should {\it veto} the second quark jet.  The EW signal rate is
at the level of 10\% of the QCD signal.

Both the reducible and irreducible backgrounds,  $pp \to\gamma\gamma$+jet 
have been discussed in the QCD section of these Proceedings. It was found that 
in total it is about  19, 31 and 32 fb in the 1 GeV bin for $M_H=100$, 120 
and 140 GeV, correspondingly.

\begin{figure}[hb]
\vspace*{-.5cm}
\begin{center}
\unitlength=1cm
\begin{picture}(9,9)
\put(-3,8.5){\mbox{fb/deg}}
\put(-5,0){\epsfxsize=9cm \epsfysize=9cm \leavevmode \epsfbox{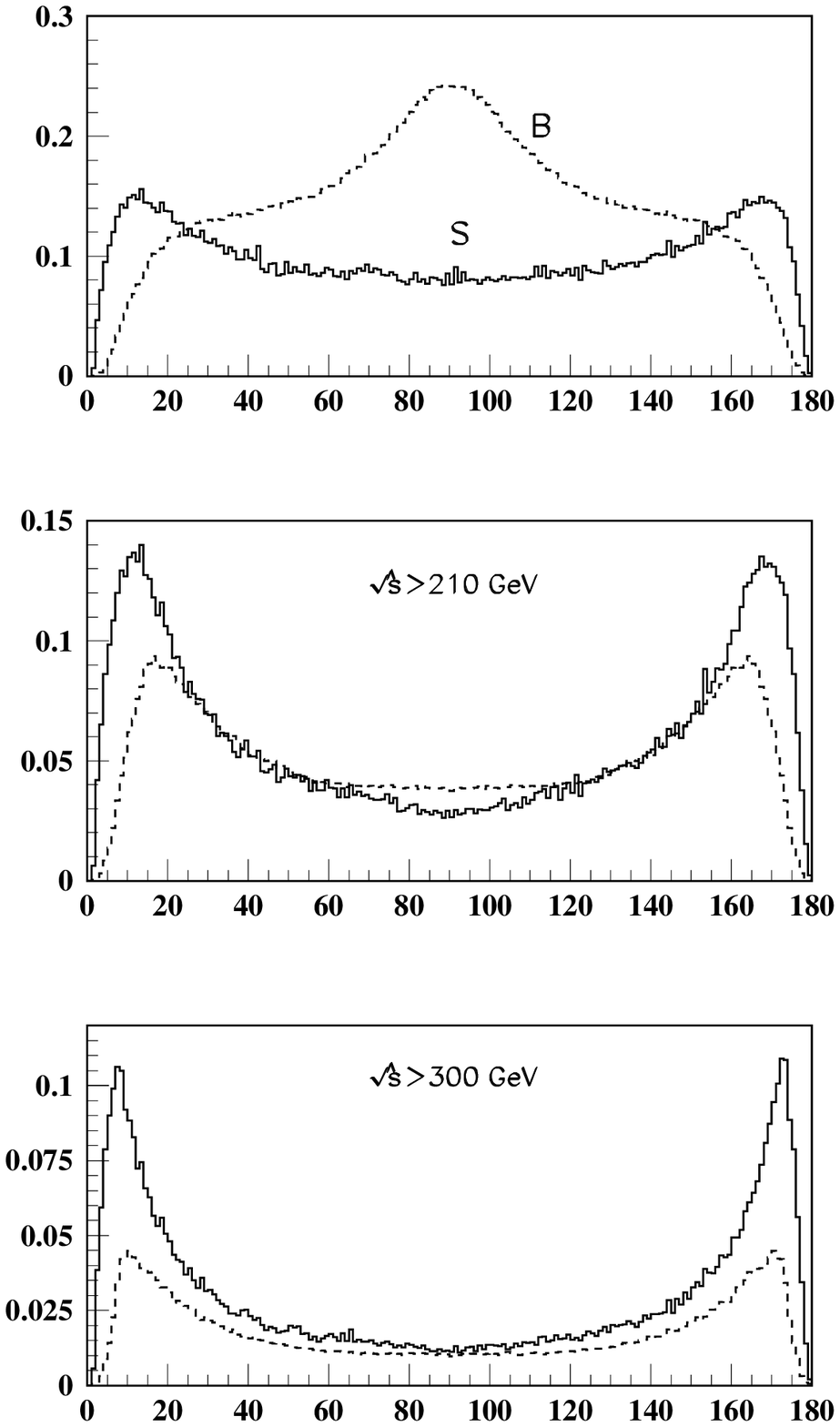}}
\put(2.5,-0.2){\mbox{ $\protect \vartheta^*_{jet}$}}
\put(3,0){\epsfxsize=9cm \epsfysize=9cm \leavevmode \epsfbox{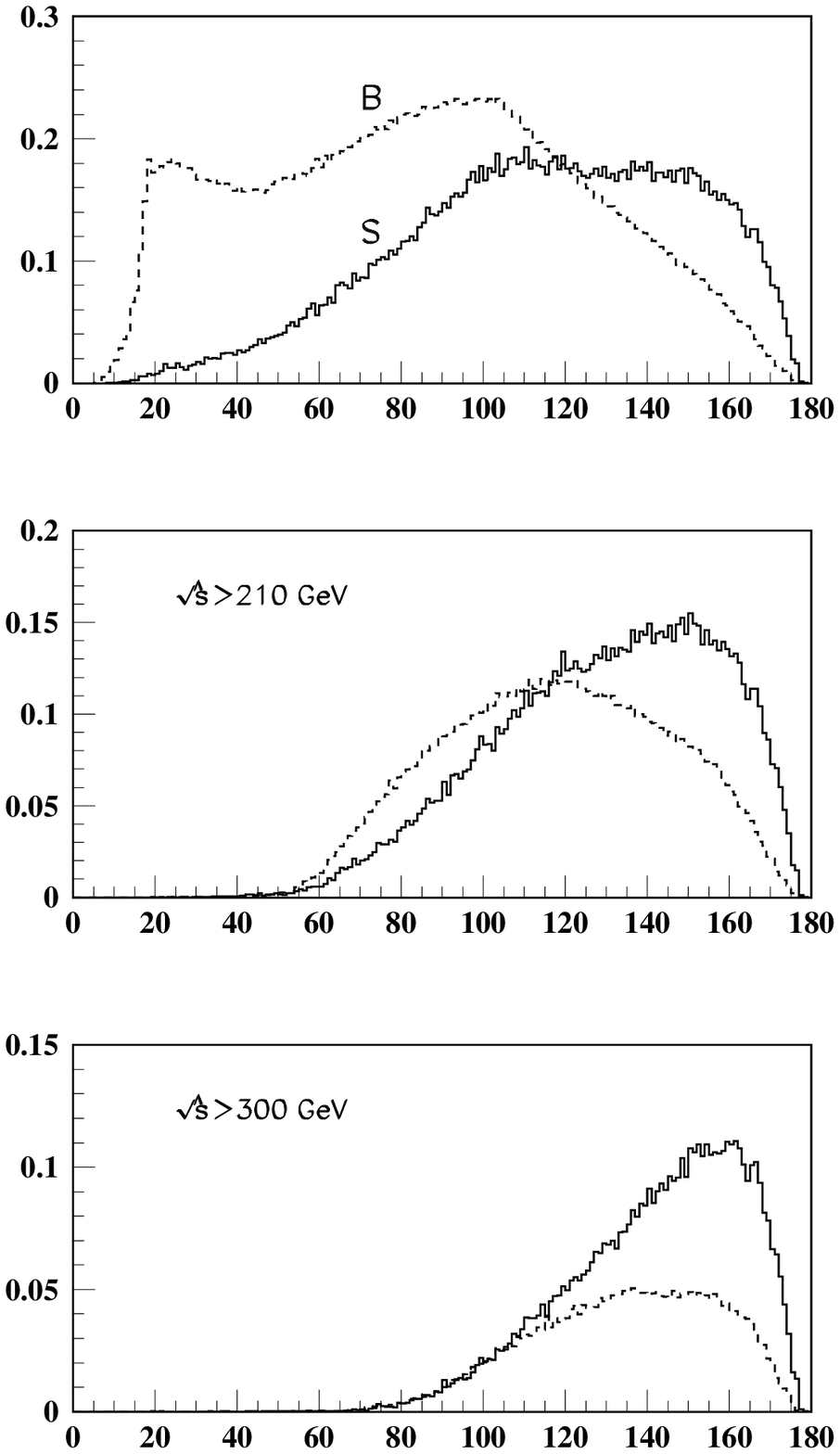}}
\put(10.5,-0.2){\mbox{ $\protect \vartheta^*_{j\gamma}$}} 
\end{picture}
\vspace*{-.2cm}
\end{center}
\caption{Distributions in the jet production angle
$\protect\vartheta^*_{jet}$
and the angle $\protect\vartheta^*_{j\gamma}$ 
between jet and the photon with smaller $p_T$
in partonic c.m.s. for the QCD signal (S)
and background (B). The Higgs mass is taken to be $M_H=120$ GeV. 
Upper plots -- no $\protect \sqrt{\hat s}$ cut, in others
$\protect \sqrt{\hat s}>210$ and 300 GeV correspondingly.
The $M_{\gamma\gamma}$ bin is equal to 1 GeV.
\label{fig:jstar}}
\vspace*{-.5cm}
\end{figure}

Further improvement of the $S/B$ ratio can be obtained by studying the 
kinematical distributions of the 3--body final states in the subprocesses 
under discussion. The background processes contribute at a smaller 
$\sqrt{\hat s}$ in comparison to the QCD signal processes. So, the 
corresponding cut improves the {\it S/B} ratio: e.g., the cut $\sqrt{\hat s}
>300$ GeV suppresses the background by a factor of 8.7 while the QCD signal 
is suppressed only by a factor of 2.6. This effect is connected with the 
different shapes, Fig.~\ref{fig:jstar}, of the jet angular distributions in 
the partonic c.m.s. for the signal and background. Indeed, for the
dominant signal subprocess $gg\to H+g$, a set of possible {\it in} spin states
does not include spin 1, while the spin of the {\it out} state is determined by
the gluon. It means, in particular, that the S--wave does not contribute here.
At the same time, in the dominant background subprocesses $gq\to
\gamma\gamma+q$ and $q\bar q\to \gamma\gamma+g$, the same spin configurations
are possible for both {\it in} and {\it out} states. It was found that the cut
on the partonic collision energy $\sqrt{\hat s}$ matches this {\it spin-states}
effect, and the best $S/B$ ratio is obtained at $\sqrt{\hat s}>300$ GeV. One
can try to exploit this effect to enhance the signal significance with the same
level of the $S/B$ ratio. Indeed, if one applies the cut on the angle between
the jet and the photon in partonic c.m.s. $\cos{\vartheta_{j-\gamma}^*}<-0.87$
for $\sqrt{\hat s}<300$ GeV and add such events to the events respecting the
only cut $\sqrt{\hat s}>300$ GeV, then the $S/B$ change is rather small, while
the significance is improved by a factor of about 1.3. The same effect can be
observed with the cut on the jet production angle in the partonic c.m.s.
$\vartheta_{jet}^*$, but one should note that  the two variables,
$\theta_{j-\gamma}^*$ and $\vartheta_{jet}^*$, are correlated. It is
desirable to perform a multivariable optimization of the event selection. 

Note that this is a result of a LO analysis, the task for the next step is to
understand how this effect will work in presence of NLO corrections to both the
signal and background.

In the analysis performed in Ref.~\cite{Hjet} the factor $K^{NLO}=1.6$ was 
used to take into account the QCD next--to--leading corrections for both 
the signal and background subprocesses. In Ref.~\cite{Schmidt,brems,FGK}, 
this assumption was confirmed by an accurate evaluation of NLO corrections 
to the signal subprocesses (where for the evaluation of the two--loop 
diagrams, the effective point--like vertices were used in the limit $M_H\ll 
m_{t}$ \cite{ggSM}). For the background, the corresponding
analysis \cite{KunFlor} has shown that the NLO corrections are not larger than
50\%. Thus, an attractive feature of the $pp\to H(\to\gamma\gamma)$+jet channel
is that theoretical uncertainties related to higher order QCD corrections can 
be under control.

\setcounter{figure}{0}
\setcounter{table}{0}
\setcounter{section}{0}
\setcounter{equation}{0}
\newpage


\begin{center}
{\large\sc {\bf
Signatures of Heavy Charged Higgs Bosons at the LHC}} 

\vspace*{.5cm}

{\sc K.A. Assamagan, A. Djouadi, M. Drees, M. Guchait, R. Kinnunen} 

\vspace*{.2cm}

{\sc J.L. Kneur, D.J. Miller, S. Moretti, K. Odagiri and D.P. Roy}

\end{center}

\begin{abstract}
We analyze the signatures of the charged Higgs particles of the Minimal
Supersymmetric extension of the Standard Model at the LHC. We will mainly focus
on the large $M_{H^\pm}$ range where the charged Higgs boson is produced
through the gluon--bottom or gluon--gluon mechanisms. The resulting $H^\pm$
signal is analyzed in its dominant $H^+ \to tb$ as well as subdominant decay
channels. Simulations for the detection of the charged Higgs boson signals in
the decay channels $H^\pm \to \tau^\pm \nu$ and $H^\pm \to cs, W^\pm h$ or
$tb$ are performed in the framework of the CMS  and ATLAS detectors, 
respectively.  
\end{abstract}

\section{Introduction} 

The minimal supersymmetric Standard Model (MSSM) contains two complex Higgs
doublets, $\phi_1$ and $\phi_2$, corresponding to eight scalar states.  Three
of these are absorbed as Goldstone bosons leaving five physical states -- the
two neutral scalars $(h^0,H^0)$, a pseudo-scalar $(A^0)$ and a pair of charged
Higgs bosons $(H^\pm)$.  All the tree-level masses and couplings of these
particles are given in terms of two parameters, $m_{H^\pm}$ and $\tan\beta$,
the latter representing the ratio of the two vacuum expectation values
\cite{HHGs3}.  While any one of the above neutral Higgs bosons may be hard to
distinguish from that of the Standard Model, the $H^\pm$ carries a distinctive
hall-mark of the SUSY Higgs sector.  Moreover the couplings of the $H^\pm$ are
uniquely related to $\tan\beta$, since the physical charged Higgs boson
corresponds to the combination 
\be
H^\pm = -\phi^\pm_1 \sin\beta + \phi^\pm_2 \cos\beta.
\label{one}
\ee
Therefore the detection of $H^\pm$ and measurement of its mass and couplings 
are expected to play a very important role in probing the SUSY Higgs sector. 

The search for charged Higgs bosons is one of the major tasks of present and
future high--energy colliders. In a model independent way, LEP2 has set a lower
limit on the $H^\pm$ mass, $m_{H^\pm} \gsim 74$ GeV, for any value of $\tan
\beta$ \cite{LEP2}. At the Tevatron, the CDF and D{\O} collaborations searched
for $H^\pm$ bosons in top decays through the process $p\bar{p}\rightarrow t
\bar{t}$, with at least one of the top quarks decaying via $t\rightarrow H^\pm
b$, leading to a surplus of $\tau$'s  due to the $H^\pm \to \tau \nu$ decay;
they excluded the low and high $\tan\beta$ regions [where the branching ratios
for this decay is large] almost up to the $M_{H^\pm} \sim m_t$ limit
\cite{Tevatron}. Detailed analyses at the LHC  have shown that the entire range
of $\tan \beta$ values should be covered for $M_{H^\pm} \lsim m_t$ \cite{LHC}
using this process.  

At this workshop, we focused on the large mass region, $M_{H^\pm} > m_t$, where
the previous production process is not at work and for which only a few
preliminary studies have been performed. We summarize our work in this
contribution. After a brief summary of the $H^\pm$ decay modes [both in the
MSSM and in some of its extensions], we will discuss in section 3, the various
signals for a heavy charged Higgs boson at the LHC. We will then present, in
sections 4 and 5, two simulations for the detection of the $H^\pm$ signals in
the decay channels $H^\pm \to \tau^\pm \nu$ in the CMS and $H^\pm \to cs, 
W^\pm h, tb$ in the ATLAS detectors. 

\section{Production and decay modes of the $H^\pm$ bosons}

The decays of the charged Higgs bosons are in general controlled by their
Yukawa couplings to up-- and down--type fermions $u,d$ given by \cite{HHGs3}:
\be
{g V_{ij} \over \sqrt{2} M_W} H^+ \left[\cot \beta \; m_{u} \; \bar u_i d_{jL}
+ \tan \beta \; m_d \; \bar u_i d_{jR} \right],
\label{coupling}
\ee
For values $\tan \beta >1$, as is the case in the MSSM, the couplings to
down--type fermions are enhanced. The coupling $H^-tb$, which is of utmost
importance in the production and the decays\footnote{It should be mentioned
that most analyses of the $H^\pm$ boson decay modes and detection signals at
colliders are based on the lowest order vertex, represented by the Yukawa
coupling of eq.~(\ref{coupling}), but improved by standard QCD corrections
\cite{QCDs3} by using the running quark masses. One loop electroweak corrections
to this vertex can give a large variation in the signal cross--section
at high or low $\tan\beta$, as recently shown in \cite{EWloop}. The
corresponding correction from SUSY--QCD loops is possibly large \cite{QCDloop}
depending on the SUSY parameters [but for the production, they are not yet
completely available].  The inclusion of these corrections is evidently
important for a quantitative evaluation of the $H^\pm$ signal.} of the $H^\pm$
bosons, is large for $\tan\beta \sim 1$ and $\sim m_t/m_b$. Interestingly these
two regions of $\tan\beta$ are favored by $b$--$\tau$ unification for a related
reason: i.e.  one needs a large $tbH^\pm$ Yukawa coupling contribution to the
RGE to control the rise of $m_b$ as one goes down from the GUT to the low
energy scale \cite{GUT}.  

\begin{figure}[htb]
\vspace*{-4.5cm}
\centerline{\mbox{\psfig{figure=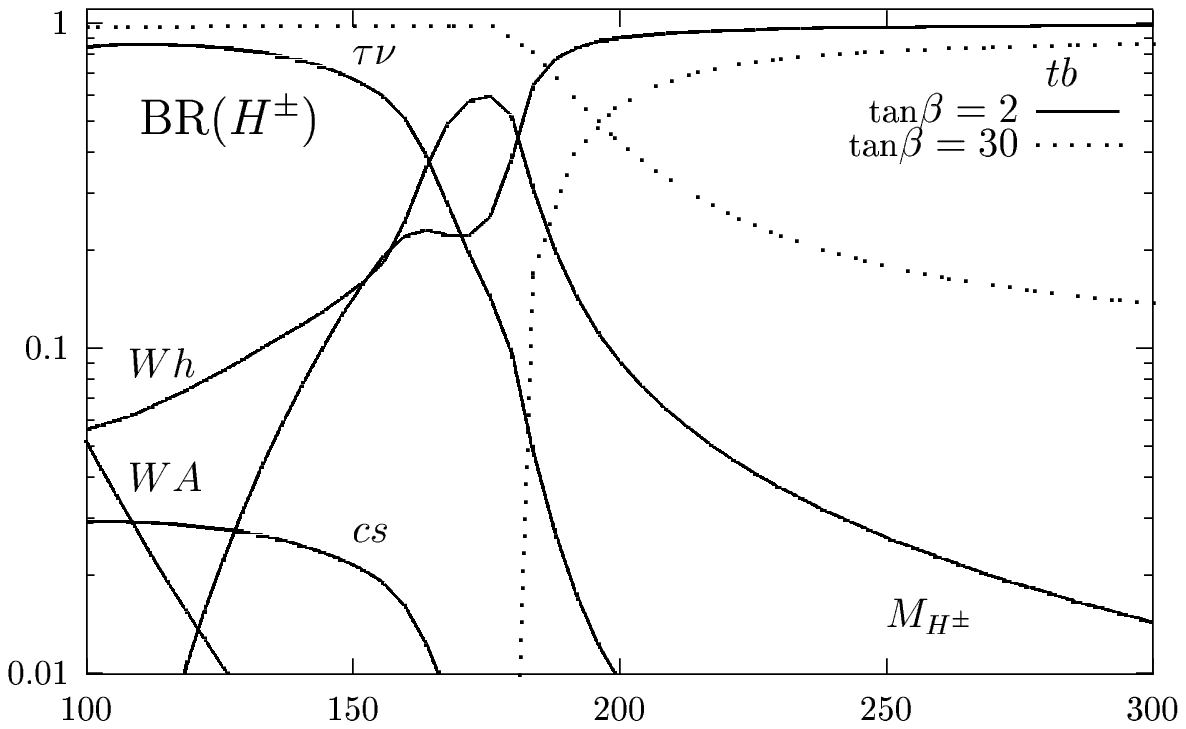,width=20cm}}}
\vspace*{-16.5cm}
\caption{Branching ratios of the charged Higgs boson decays for $\tan \beta=2$ 
and 30. They are obtained using the program HDECAY \cite{HDECAY3}. }
\end{figure}

The value of $\tan \beta$ determines to a large extent the decay pattern
\cite{decays} of the charged Higgs bosons. For large $\tan \beta$ values the
pattern is simple, a result of the strong enhancement of the couplings  to
down--type fermions: below the top--bottom threshold, $H^\pm$ bosons will decay
into $\tau \nu_\tau$ pairs while above this threshold, they will decay into $tb$
pairs with BR $\sim 85\%$ and $\tau \nu_\tau$ pairs with BR $\sim 15\%$ for
large enough $M_{H^\pm}$ values. For small $\tan \beta$ values, $\tan \beta
\lsim 5$, the pattern is more complicated, in particular around and below the
$tb$ threshold.  Decays into $Wh$ final states play an important role since
they reach the level of several ten percent leading to a significant reduction
of the dominant branching ratio into $\tau\nu$ states. Note that the off--shell
three body decays \cite{decays} $H^\pm \to bt^* \to b\bar{b} W^\pm$ and $H^\pm 
\to hW^*, AW^* \to h f\bar{f},A f\bar{f}$ [the latter being kinematically 
forbidden at the two--body level] can be rather important. The $H^\pm$ branching
ratios are summarized in Fig.~1 for the values $\tan \beta=2$ and 30.  \bigskip

In the MSSM, the charged and pseudoscalar Higgs boson masses are related 
\cite{HHGs3}, 
\beq
M_{H^\pm}^2 =M_A^2+M_W^2
\eeq
and the LEP limit on the lightest scalar and pseudoscalar Higgs masses,  
$m_{h_0} (m_{A_0}) \gsim 90 - 100 ~{\rm GeV}$ implies first, that $M_{H^\pm} 
\gsim 120$ GeV $[M_{H^\pm} \gsim 200$ GeV for $\tan \beta=2$] and second, that 
the $H^\pm \rightarrow Wh^0 (WA^0)$ decay 
channel has as high a threshold as the $t\bar b$ channel, while the latter 
has a more favorable coupling. Consequently the $H^\pm \rightarrow Wh^0 
(WA^0)$ decay $BR$ is restricted to be $\lsim 5\%$ over the LEP allowed 
region [Fig.~1].  However the constraints discussed above do not hold in 
singlet extensions of the MSSM like the NMSSM \cite{NMSSM}.  Consequently  
$H^\pm \rightarrow Wh^0 (WA^0)$ can be the dominant decay mode for $M_{
H^\pm} \sim 160~{\rm GeV}$ in the low $\tan\beta$ region and lead to a 
spectacular signal at the LHC, as illustrated in Table 1. This decay channel 
will be analyzed in detail in the next sections. \bigskip

\noindent {\small Table 1. Maximal branching fractions for $H^\pm \rightarrow
W(h^0_1,A^0_1)$ decay in the NMSSM for fixed input values of $\tan\beta$
and output $H^\pm$ mass of $\sim 160$ GeV.  The values of the
$h^0_1,A^0_1$ masses and branching fractions are shown along with the
corresponding model parameters.   Also shown are the $t 
\rightarrow b H^\pm$ branching fraction and the size of the resulting $H^\pm
\rightarrow W(h^0_1,A^0_1)$ decay signal at LHC.} \smallskip

\[
\begin{tabular}{|c|c|c|c|c|c|c|c|c|}
\hline
$\tan\beta$ & $M_{H^\pm}$ & $B_{H^\pm}$ & $\langle N\rangle$ &
$\lambda, k$ & $A_\lambda,A_k$ & $m_{h1},m_{A1}$ & $B_{h1},B_{A1}$ &
$\sigma_{H^\pm}$ \\
& (GeV) & (\%) & (GeV) & & (GeV) & (GeV) & (\%) & (fb) \\
\hline
&&&&&&&& \\
 & 164 & 0.4 & 147 & .39,--.25 & --158,--59 & 56,36 & 51,43 & 2 \\
2&     &     &     &          &          &       &       & \\
 & 160 & 0.8 & 273 & .40,--.73 & 12, 8    & 115,15& 0,97  & -- \\
&&&&&&&& \\
\hline
&&&&&&&& \\
 &     &     & 231 & .21,--.41 & --101,111 & 51,137& 86,0  & 2.2 \\
2.5 & 160 & 0.5&   &          &          &       &       &     \\
 &     &     & 278 & .33,--.72 & 16,8     & 113,15& 0,95  & --  \\    
&&&&&&&& \\
\hline
&&&&&&&& \\
 &    &    &   196 & .14,--.33 & --184,--8 & 54,27 & 69,16 & 1.6 \\
3 & 160 & 0.4 & & & & & & \\
 &      &     & 341 & .22,--.62 & 23, 6 & 110,19 & 0,90 & -- \\
&&&&&&&& \\
\hline
\end{tabular}
\]
\bigskip

An important point which should be mentioned, is that in most of the analyses of
the $H^\pm$ signals, it is always assumed that it decays only into 
standard particles and that the SUSY decay modes are shut. But for large 
values of $M_{H^\pm}$, at least the decays into the lightest neutralinos and 
charginos [and possibly into light sleptons and $\tilde{t}, \tilde{b}$
squarks] can be kinematically allowed. These modes could have large decays 
widths, and could thus suppress the $H^\pm \ra tb$ branching ratio in a 
drastic way \cite{SUSYdecays}. 

In Fig.~2, the branching fraction  BR$(H^\pm \to \chi^0_i \chi_j^\pm$) [with
$i=$1--4 and $j$=1--2] are shown as function of $M_{H^\pm}$ for the four values
$\tan \beta=2, 5,10$ and 30. The choice of the gaugino and higgsino mass
parameters $M_2=\mu=200$ GeV has been made leading to the lightest chargino and
neutralino masses $m_{\chi_1^0} \sim 80$--90 GeV and $m_{\chi_1^+} \sim$
125--150 GeV depending on the value of $\tan \beta$ [small masses are obtained
with small $\tan \beta$ input]. The values of the scalar masses are such that 
sleptons and squarks are too heavy to appear in the decay products of the 
$H^\pm$ boson. As can be seen, for small and large values of $\tan
\beta$, the $H^\pm tb$ couplings are enhanced and the chargino/neutralino
decays are important only for large $H^\pm$ masses where many $\chi_i^0
\chi_j^0$ channels are open.  For intermediate $\tan \beta$ values,  the $H^\pm
tb$ Yukawa couplings are suppressed, and the chargino/neutralino decays are
dominant for charged Higgs boson masses of a few hundred GeV.  

In scenarii where sleptons and squarks [in particular stop and sbottom squarks]
are also light, $H^\pm$ bosons decays into these states might be kinematically
possible as well and would be dominant. This will again suppress in a
dramatic way the branching ratio for the $H^\pm \to tb$ signature
\cite{SUSYdecays}. These SUSY decays, although discussed in the literature,
have not been analyzed experimentally up to now. They should, however, not be
overlooked for heavy charged Higgs bosons, as they might jeopardize the
detection of these particles at the LHC.  

\begin{figure}[htb]
\vspace*{-4.6cm}
\centerline{\mbox{\psfig{figure=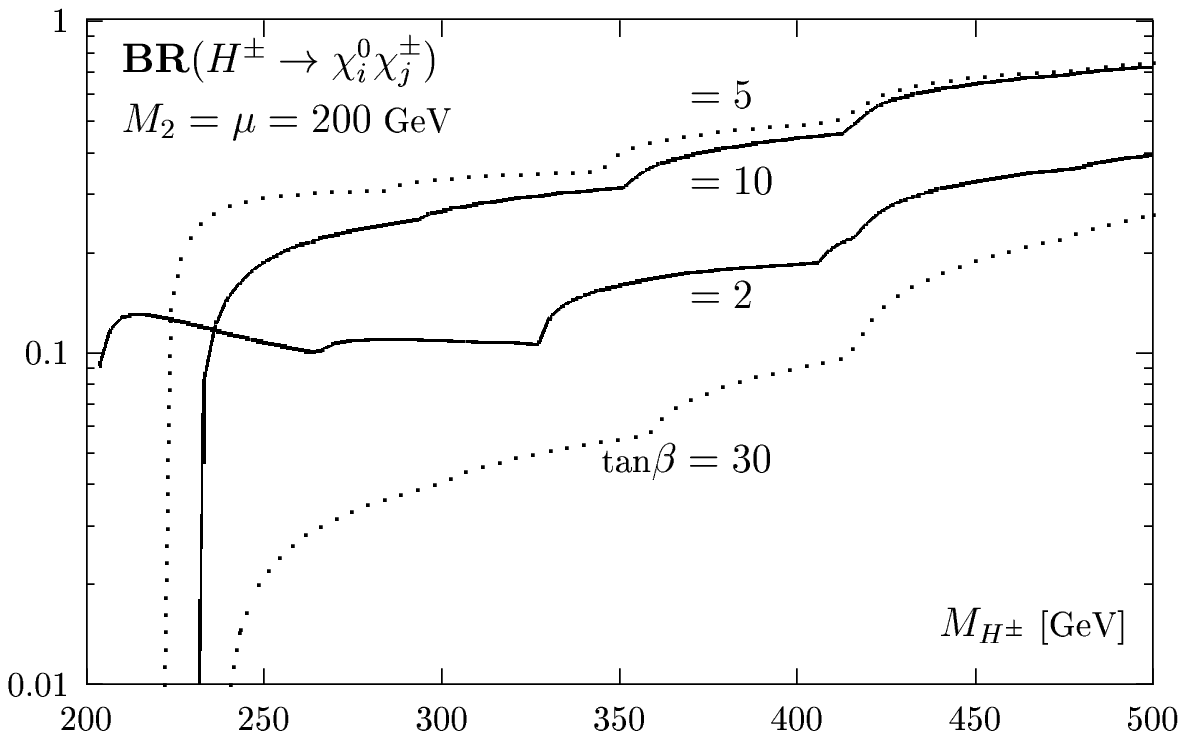,width=20cm}}}
\vspace*{-16.5cm}
\caption{Branching ratios of the charged Higgs boson decays into charginos
and neutralinos as a function of $M_{H^\pm}$ for a set of $\tan \beta$ values;
$M_2$ and $\mu$ are fixed to 200 GeV.}
\end{figure}

Finally, we briefly discuss the production modes of a heavy charged
Higgs boson, with $m_{H^\pm} >m_t$, at the LHC. The two mechanisms which
have sizeable cross sections are:
\beq
pp \to & gb (g\bar{b}) \to tH^- \; (\bar{t}H^+) & \cite{gb,[2]} 
\nonumber \\
pp \to & gg/qq' \to tH^- \bar{b} + \bar{t}H^+ b & \cite{[3],[8],[5]} \ \ 
\eeq
The signal cross-section from the $2 \to 2$ mechanism $gb \to tH^-$ [where the 
$b$ quark is obtained from the proton] is 2--3 times larger than the $2 \to 3$ 
process $gg/qq \to t\bar{b} H^-$ [where the $H^-$ boson is radiated from a 
heavy quark line]. This is shown in Fig.~3a at LHC energies for the values 
$\tan \beta=2$ and 40. When the decays $H^- \to t\bar{b}$ and $t \to Wb$ take 
place, the first process gives rise to 3 $b$--quarks in the final state while 
the second one gives 4 $b$--quarks. Both processes contribute to the inclusive 
production where at most 3 final $b$--quarks are tagged. However, the two 
processes have to be properly combined \cite{dicus} to avoid double counting 
of the contribution where a gluon gives rise to a $b\bar{b}$ pair that is 
collinear to the initial proton. The cross section of the inclusive process in 
this case is shown in Fig.~3b, and is mid--way between the two cross sections 
eqs.~(4) \cite{[8]}. 

\begin{figure}[htb]
\vspace*{-4.3cm}
\centerline{\mbox{
\psfig{figure=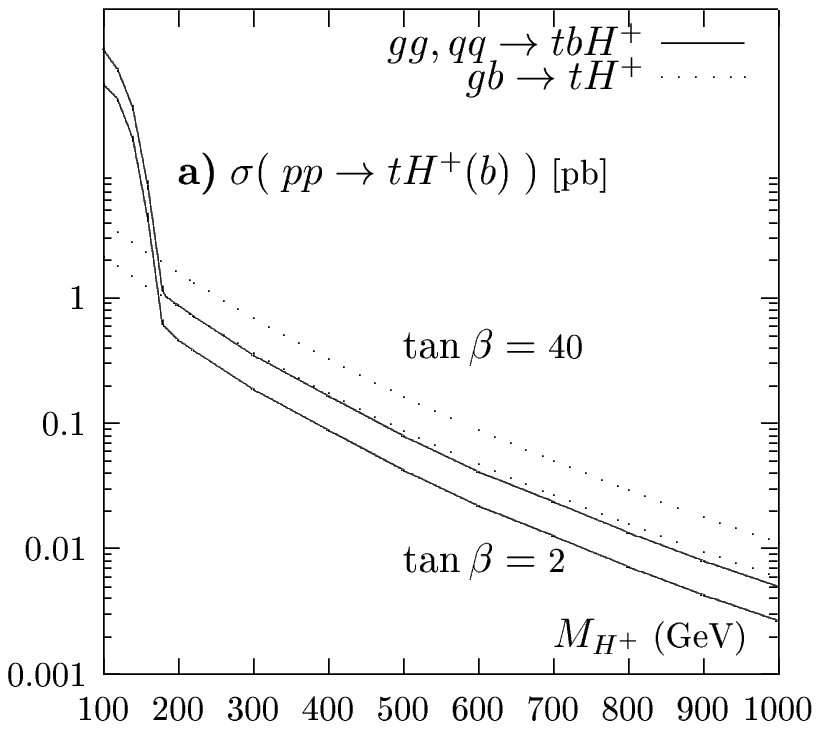,width=20cm} -\hspace*{-12cm}
\psfig{figure=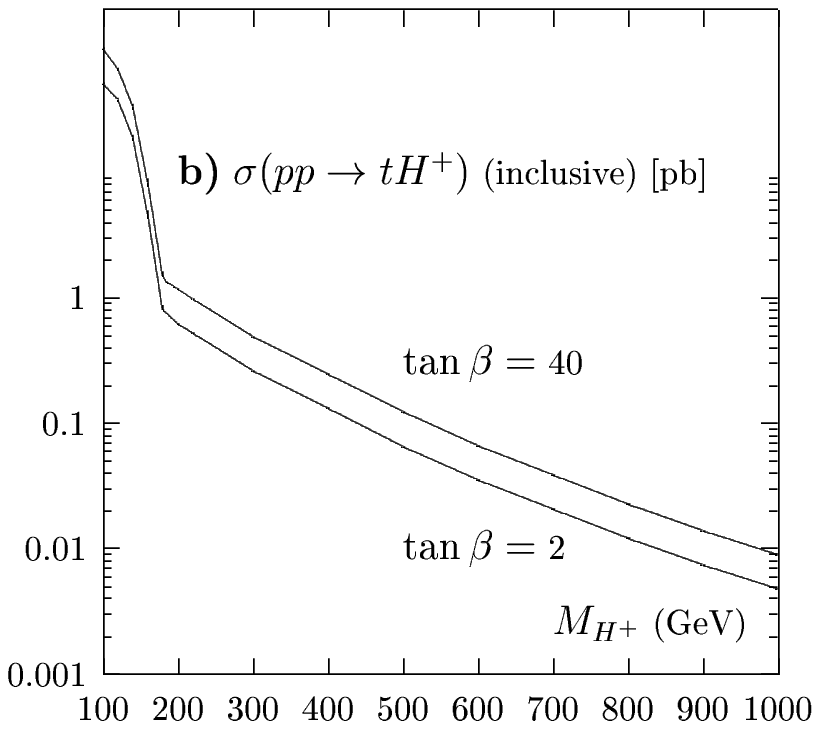,width=20cm} -\hspace*{-3cm}}}
\vspace*{-16.3cm}
\caption{Production cross sections for charged Higgs bosons at the LHC for
$\tan \beta=2$ and 40. (a) Individual cross sections from the $gg/qq$ and
$gb$ processes and (b) combination of the two processes with the subtraction
of the common piece.}
\end{figure}

Other mechanisms for $H^\pm$ production at hadron colliders are the Drell--Yan 
type process for pair production, $qq \to H^+H^-$, the associated production 
process with $W$ bosons, $qq \to H^\pm W^\mp $ \cite{[4]} and the gluon--gluon 
fusion process for pair production, $gg \to H^+H^-$ \cite{ggs3}. However, the 
rates are rather small at the LHC, in the first case because of the weak 
couplings and the low quark luminosities at high energies and in the
second case because the process is induced by loops of heavy quarks and is
thus suppressed by additional electroweak coupling factors. We will thus focus
in this study on the two processes eq.~(4).

\newpage
\section{Signatures of $H^\pm$ bosons at the LHC} 

The $t \rightarrow bH^+$ decay is known to provide a promising signature for
charged Higgs boson search at the LHC for $M_H < m_t$.  But it is hard to extend
the $H^\pm$ search beyond $m_t$, because in this case the combination of
dominant production and decay channel, $tH^- \rightarrow t\bar tb$, suffers
from a large QCD background \cite{[2],[3]}.  Moreover the subdominant production
channels of $H^\pm W^\mp$ and $H^\pm H^\mp$ have been found to give no viable
signature at LHC \cite{[4]}.  In view of this we have undertaken a systematic study of
a heavy $H^\pm$ signature at the LHC from its dominant production channel $ gb
(g\bar{b}) \to t H^- (\bar{t} H^+)$, followed by the decays $H^- \rightarrow
\bar tb,\tau\bar\nu$ and $W^- h^0$.  While the first decay represents the
dominant channel of charged Higgs bosons, the $\tau\nu$ and $Wh^0$ are the
largest subdominant channels in the high and low $\tan\beta$ regions
respectively, with [see also Fig.~1]
\be
B_{\tau\nu} (\tan\beta \gsim 10) \sim 15\% ~~{\rm and}~~ B_{Wh^0}
(\tan\beta = 1 - 5) \lsim 5\%.
\label{two}
\ee
The signature for the dominant decay channel of $H^- \rightarrow \bar
tb$ has been analyzed separately assuming three and four $b$--jet tags.
The analyses presented in this section are based on parton level Monte 
Carlo programs with a Gaussian smearing of lepton and jet momenta for 
simulating the detector resolution.

\bigskip

\noindent {\bf (i) $H^\pm \rightarrow tb$ Signature with Four $b$-tags
\cite{[5]}}\footnote{While this work was initiated earlier, some of the
issues analyzed during the workshop led to the final version presented
here.}: 
\medskip

The dominant signal and background processes are
\be
gg \rightarrow tH^- \bar b + {\rm h.c.} \rightarrow t\bar t b\bar b,
\label{three}
\ee
\be
gg \rightarrow t\bar t b\bar b,
\label{four}
\ee
followed by the leptonic decay of one top and hadronic decay of the
other, i.e.
\be
t\bar t b\bar b \rightarrow b\bar b b\bar b W^+ W^- \rightarrow b\bar
b b\bar b \ell\nu q\bar q.
\label{five}
\ee

A basic set of kinematic and isolation cuts,
\be
p_T > 20 ~{\rm GeV}, ~|\eta| < 2.5, ~\Delta R = \left[(\Delta \phi)^2
+ (\Delta \eta)^2\right]^{1/2} > 0.4
\label{six}
\ee
is imposed on all the jet and lepton momenta.  The $p_T$ cut is also
imposed on the missing-$p_T$, obtained by vector addition of the
$p_T$'s after resolution smearing.  This is followed by the mass
reconstruction of the $W$ and the top quark pair, so that one can
identify the pair of $b$-jets accompanying the latter.  While the
harder of these two $b$-jets $(b_1)$ comes from $H^\pm$ decay in the
signal, both of them come mainly from gluon splitting in the
background.  Consequently the $S/B$ ratio is improved by imposing the
following cuts on this $b$-jet pair:
\be
M_{bb} > 120 ~{\rm GeV}, ~E_{b_1} > 120 ~{\rm GeV~~and}~~
\cos\theta_{bb} < 0.75.
\label{seven}
\ee
Then each of this $b$-jet pair is combined with each of the
reconstructed pair of top to give 4 entries for the invariant mass
$M_{tb}$ per event.  One of these 4 entries corresponds to the $H^\pm$
mass for the signal event, while the others constitute a combinatorial
background.  Fig.~4 shows this $tb$ invariant mass distribution for
the signal (\ref{three}) and background (\ref{four}).  The right hand
scale corresponds to the cross-section for $\epsilon^4_b = 0.1$ --
i.e. an optimistic $b$-tagging efficiency of $\epsilon_b = 0.56$.
Reducing it to a more conservative value of $\epsilon_b = 0.4$ would
reduce both the signal and background by a factor of 4 each. 

\hrule width 0pt
\vspace {3.5in}
\includegraphics{sigfig1.eps} 
\label{fig:sigfig1}
\begin{enumerate}
\item[{Fig.~4:}] The reconstructed $tb$ invariant mass distribution of
the $H^\pm$ signal (\ref{three}) and the QCD background (\ref{four})
in the isolated lepton plus multi-jet channel with 4 $b$-tags.  The
scale on the right corresponds to a $b$-tagging efficiency factor
$\epsilon^4_b = 0.1$.
\end{enumerate}

\noindent {\small Table 2. Number of signal and background events in the 4 
$b$-tagged channel per $100 ~{\rm fb}^{-1}$ luminosity in a mass window of
$M_{H^\pm} \pm 40~{\rm GeV}$ at $\tan\beta = 40 ~(\epsilon_b =0.4)$.} 
\[
\begin{tabular}{|c|c|c|c|}
\hline
&&& \\
$M_{H^\pm}({\rm GeV})$ & $S$ & $B$ & $S/\sqrt{B}$ \\
&&& \\
\hline 
310 & 32.7 & 26.9 & 6.3 \\
407 & 22.7 & 17.3 & 5.5 \\
506 & 13.2 & ~9.9 & 4.2 \\
605 & ~7.5 & ~5.5 & 3.2 \\
\hline
\end{tabular}
\]
\vspace*{5mm} 

Table 2 lists the number of signal and background events for a typical
annual luminosity of $100~{\rm fb}^{-1}$, expected from the high
luminosity LHC run, assuming $\epsilon_b = 0.4$.  While the $S/B$
ratio is $> 1$, the viability of the signal is limited by the signal
size\footnote{Increasing the $p_T$ cut of $b$-jets from 20 to 30 GeV
would reduce the signal (background) size by a factor of about 3(4), 
hence reducing the viability of this signal.}.  
One expects a $> 3\sigma$ signal up to $M_{H^\pm} = 600~{\rm
GeV}$ at $\tan\beta = 40$. The signal size is very similar at $\tan\beta = 
1.5$, but smaller in between [the signal process (\ref{three}) is 
controlled by the $tbH^\pm$ Yukawa coupling, eq.~(\ref{coupling}), which 
is large for $\tan\beta \sim 1$ and $\sim m_t/m_b$, as discussed previously].
\bigskip

\noindent {\bf (ii) $H^\pm \rightarrow tb$ Signature with Three
$b$-tags \cite{[7]}:}
\medskip

The contributions to this signal come from (\ref{three}) as well as
\be
gb \rightarrow tH^- + {\rm h.c.} \rightarrow t\bar tb + {\rm h.c.},
\label{nine}
\ee
followed by the leptonic decay of one top and hadronic decay of the
other.  The signal cross-section from (\ref{nine}) is 2--3 times larger
than from (\ref{three}) [Fig.~3], while their kinematic distributions are very 
similar. Combining the two cross sections and subtracting
the overlapping piece to avoid double counting results in a signal 
cross-section, which is mid--way between the two; see Fig.~3. 

The background comes from (\ref{four}) as well as 
\be
gb \rightarrow t\bar tb + {\rm h.c.} ~~{\rm and}~~ gg \rightarrow
t\bar tg,
\label{ten}
\ee
where the gluon jet in the last case is mis-tagged as a $b$-jet.
Assuming the standard mis-tagging factor of 1\% this contribution turns
out in fact to be the largest source of the background, as we see
below. 

The basic kinematic cuts are as in (\ref{six}) except for a harder
$p_T$-cut, 
\be
p_T > 30 ~{\rm GeV},
\label{eleven}
\ee
since the 3 $b$-jets coming from $H^\pm$ and $t\bar t$ decays are all
reasonably hard.  This is followed by the mass reconstruction of the
top quark pair as before, so that one can identify the accompanying
(3rd) $b$-jet.  We impose a
\be
p_T > 80 ~{\rm GeV}
\label{twelve}
\ee
cut on this $b$-jet to improve the $S/B$ ratio.  Finally this $b$-jet
is combined with each of the reconstructed top pair to give two
entries of $M_{tb}$ per event.  One of them corresponds to the $H^\pm$
mass for the signal while the other constitutes the combinatorial
background.  Fig.~5 shows this $tb$ invariant mass distribution of the
signal along with the above mentioned backgrounds, including a
$b$-tagging efficiency factor of
\be
\epsilon_b = 0.4.
\label{thirteen}
\ee
While the $S/B$ ratio is $< 1$ the signal cross-section is much larger
than the previous case.  Table 3 lists the number of signal and
background events for a luminosity of $100~{\rm fb}^{-1}$ at
$\tan\beta = 40$.  The results are very similar at $\tan\beta = 1.5$.
Comparing this with Table 2 we see that the $S/\sqrt{B}$ ratio is very
similar in the two channels.  One should bear in mind however the
larger $p_T$ cut (\ref{eleven}) assumed for the 3 $b$-tagged channel.
The cross-sections in both the cases were calculated with the MRS-LO
structure functions \cite{[9]}. 

\newpage 

\hrule width 0pt
\vspace {3.2in} 
\includegraphics{sigfig2.eps} 
\label{fig:sigfig2}
\vspace*{5mm}
\begin{enumerate}
\item[{Fig.~5:}] The reconstructed $tb$ invariant mass distribution of
the $H^\pm$ signal and different QCD backgrounds in the isolated
lepton plus multi-jet channel with 3 $b$-tags.
\end{enumerate}

\noindent {\small Table 3. Number of signal and background events in the 3 
$b$-tagged channel per $100~{\rm fb}^{-1}$ luminosity in a mass window of
$M_{H^\pm} \pm 40~{\rm GeV}$ at $\tan\beta = 40 ~(\epsilon_b =0.4)$.} 
\[
\begin{tabular}{|c|c|c|c|}
\hline
&&& \\
$M_{H^\pm} ~({\rm GeV})$ & $S$ & $B$ & $S\sqrt{B}$ \\
&&& \\
\hline
310 & 133 & 443 & 6.2 \\
407 & 111 & 403 & 5.6 \\
506 & ~73 & 266 & 4.5 \\
605 & ~43 & 156 & 3.4 \\
\hline
\end{tabular}
\]
\bigskip

\noindent {\bf (iii) $H^\pm \rightarrow \tau\nu$ Signature \cite{[10]}:}
\medskip

Following the analysis of Ref.~\cite{[10]} a more exact simulation of a heavy
$H^\pm$ signature in the $\tau\nu$ decay channel was done for the CMS
detector using PYTHIA \cite{pythias3}.  The results will be presented in 
the next section.  By exploiting the distinctive $\tau$ polarization one 
can get at least as good a $H^\pm$ signature here as in the $t\bar b$ 
channel for the large $\tan\beta$ region.
\bigskip

\noindent {\bf (iv) $H^\pm \rightarrow W^\pm h^0$ Signature \cite{[12]}:} 
\medskip

For simplicity we have estimated the signal cross-section from 
\be
gb \rightarrow tH^- + {\rm h.c.} \rightarrow bW^+ W^- h^0 + {\rm
h.c.},
\label{fourteen}
\ee
followed by $h^0 \rightarrow b\bar b$, $W^\pm \rightarrow \ell\nu$ and
$W^\mp \rightarrow q\bar q$.  Thus the final state consists of the
same particles as the dominant decay mode of eq. (\ref{nine}).  Thus
we have to consider the background from the $H^- \rightarrow t\bar b$
decay (\ref{nine}) along with those from the QCD processes of
eq. (\ref{ten}). 

We require 3 $b$-tags along with the same basic cuts as in section
(ii).  This is followed by the mass reconstruction of $W^\pm$ and the
top, which helps to identify the accompanying $b$-pair and the $W$.
The resulting $bb$ and $Wb$ invariant masses are then subjected to the
constraints,
\be
M_{bb} = m_{h^0} \pm 10~{\rm GeV} ~{\rm and}~ m_{Wb} \neq m_t \pm 20
~{\rm GeV}.
\label{fifteen}
\ee
The $h^0$ mass constraint and the veto on the second top helps to
separate the $H^\pm \rightarrow W^\pm h^0$ signal from the
backgrounds.  However the former is severely constrained by the signal
size as well as the $S/B$ ratio.  Consequently one expects at best a
marginal signal in this channel and only in a narrow strip of
the $M_{H^\pm}$--$\tan\beta$ parameter space, at the boundary of the
LEP exclusion region.  Fig.~6 shows the signal (\ref{fourteen}) along
with the backgrounds from (\ref{nine}) and (\ref{ten}) against the
reconstructed $H^\pm$ mass at one such point -- $M_{H^\pm} = 220~{\rm
GeV}$ and $\tan\beta = 2$. Note that, as discussed in section 2, in 
extensions of the MSSM, the $H^\pm \rightarrow Wh^0 (WA^0)$ can
be the dominant decay mode for $M_{H^\pm} \sim 160~{\rm GeV}$ in the
low $\tan\beta$ region and lead to a spectacular signal at the LHC;
see Table 1. \bigskip

\hrule width 0pt
\vspace {3.0in}
\includegraphics{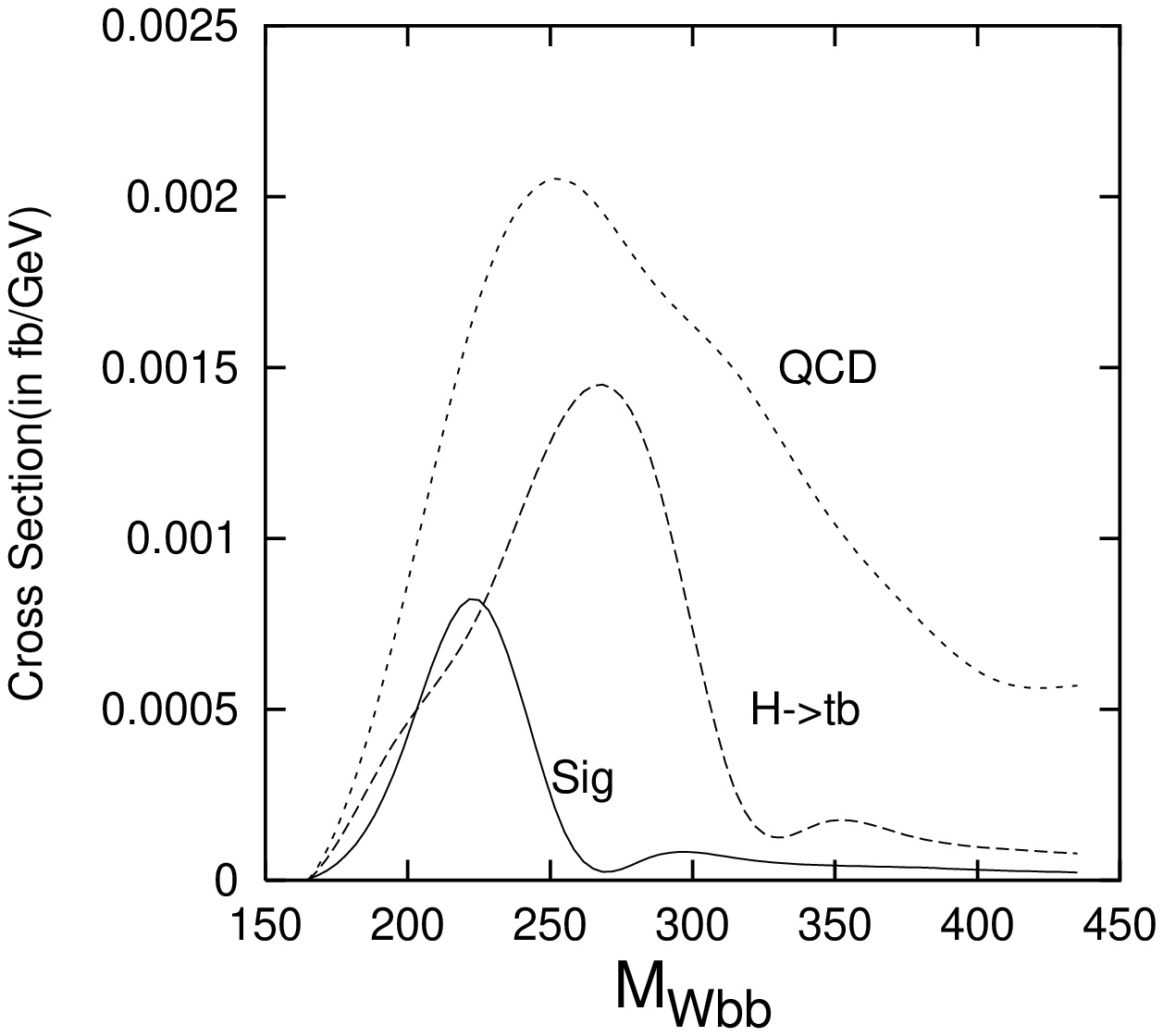} 
\label{fig:sigfig3}
\vspace*{-2mm}
\noindent Fig.~6 The $H^\pm \rightarrow Wh^0$ signal cross-section is
shown against the reconstructed $H^\pm$ mass for $M_{H^\pm} = 220~{\rm
GeV}$ and $\tan\beta = 2$ along with the $H^\pm \rightarrow tb$ and
the QCD backgrounds.
\vspace{.5cm}

It should be mentioned here that these parton level Monte Carlo
analyses of the $H^\pm$ signature in $tb$ and $Wh^0$ decay channels
need to be followed up by detailed simulation with PYTHIA, including
detector acceptance, as in the case of the $\tau\nu$ channel discussed
in the next section. Some work has started here along this line for the 
ATLAS detector; this is summarized in section 5. One should also bear 
in mind  the possibility of large radiative corrections to the Yukawa
coupling eq.~(\ref{coupling}); it is evidently important to include these
corrections for a quantitative evaluation of this signal.  

\setcounter{figure}{6}
\section{The $H^+ \rightarrow \tau \nu$ mode in CMS} 

\subsection{Introduction} 

As mentioned in the previous section, the hadronic $\tau$ signature of a heavy
charged Higgs boson from $pp \ra tH^{\pm}$ at the LHC is useful. In this 
contribution, we study the search of heavy $H^\pm$ bosons in the CMS detector 
with a
realistic simulation using the procedure of Ref.~\cite{[10]} to select the
events and to exploit the $\tau$ polarization effects. The main backgrounds are
due to $t\overline{t}$ and $W$+jet events. The $W$+jet background can be
effectively reduced with $W$ and top mass reconstruction and $b$--tagging.
Although for $t\overline{t}$ and $W$+jet events the transverse mass
reconstructed from the $\tau$--jet and the missing transverse energy is bounded
from above by the $W$ mass, some leaking of the backgrounds into the signal
region can be expected due to the experimental resolution of the $E_t^{miss}$
measurement.

\subsection{Event selection and expectations for CMS} 

Events are generated with PYTHIA \cite{pythias3} using the process $bg\ra
tH^{\pm}$. The results from Ref.~\cite{[7]} with a subtraction of double
counting between the $g\overline{b}\ra \overline{t}H^{\pm}$ and $gg \ra
\overline{t}bH^{\pm}$ processes are used to normalize the PYTHIA cross
sections. The $H^{\pm}\ra \tau\nu$ branching ratio is calculated with the
HDECAY program \cite{HDECAY3} and used in the simulation. A heavy SUSY particle
spectrum (1 TeV) is assumed with no stop mixing. The decay matrix elements with
polarization effects \cite{[10],monojet} are added in PYTHIA. For
$m_{H^{\pm}}$ = 400 GeV and $\tan\beta$ = 40 about 1700 signal events, including
only one-prong hadronic $\tau$ decays, are expected for an integrated
luminosity of 30 fb$^{-1}$. The jets and the missing transverse energy are
reconstructed with a fast simulation package CMSJET \cite{cmsjet}. For
b-tagging, results obtained from a full simulation and reconstruction of the CMS
tracker are used \cite{tracker}.  
 
The real $\tau$ jet is chosen as the $\tau$ jet candidate requiring $E_t>$ 100
GeV and $|\eta|<$2.5. The events can be triggered with a multi-jet trigger and
a higher level $\tau$ trigger even in the high luminosity running conditions. 
The $\tau$ selection is performed here using only the tracker information. The
algorithm of ref. \cite{[10]} to remove the transverse components of the
$\tau$ polarization is used requiring r = $p^{\pi}$ / $E^{\tau jet} >$ 0.8,
where $p^{\pi}$ is the momentum of a hard pion from $\tau$ decay in a cone of
$\Delta R <$ 0.1 around the calorimeter jet axis and $E^{\tau jet}$ is the
hadronic energy of the $\tau$ jet ($E_t>$ 100 GeV) reconstructed in the
calorimeters (electromagnetic and hadronic) in a cone of $\Delta R <$ 0.4. The
efficiency of this $\tau$ selection for the signal events is 20\% while for the
$t\overline{t}$ events the efficiency is only 0.4\% (including the $E_t$
threshold for jet). A reconstruction efficiency of 95\% is assumed for the hard
isolated track from $\tau$.  

A large missing transverse energy is expected in the signal events due to the
neutrino from $H^{\pm}$ decay. The $E_t^{miss}$ is reconstructed with the
CMSJET package, where the calorimeter response is parametrized including the
effects of the detector cracks and the volumes of degraded response. Efficiency
of the cut $E_t^{miss} >$ 100 GeV is about 75\% for the signal events and about
39\% for the $t\overline{t}$ background.  

A visible signal for the Higgs can be obtained in the transverse mass
reconstructed from the $\tau$-jet and the missing transverse energy if the
hadronic decay of the associated top quark is selected. For the reconstruction
of the $W$ and top masses the 
events with at least three jets with $E_t >$ 20 GeV, in
addition to the $\tau$ jet, are selected. The $W$ and top masses are
reconstructed minimizing the variable $\chi^2$ = $(m_{jj}-m_{W})^2
+(m_{jjj}-m_{t})^2$, where $m_{W}$ and $m_{t}$ are the nominal $W$ and top
masses.  A Gaussian resolution of 13.6 GeV is found for the reconstructed top
mass. The fraction of events where the three jets are found and the
reconstructed $W$ mass is within $m_{W} \pm$ 15 GeV and the reconstructed top
mass within $m_{t} \pm$ 20 GeV is 54\% for the signal, 59\% for the
$t\overline{t}$ background and 8\% for the $W$+jet events.
 
After the $W$ and top mass reconstruction and the mass window cuts b-tagging is
applied on the jet not assigned to the $W$. This jet is required to be harder
with $E_t^{jet}>$ 30 GeV. The tagging efficiencies based on the impact
parameter method obtained from a full simulation and track reconstruction in
the CMS tracker are used \cite{tracker}. At least two tracks with $p_t >$ 1 GeV
and impact parameter significance $\sigma^{ip}>$ 2 are required inside the jet
reconstruction cone of 0.4. For b-jets with $E_t$ = 50 GeV the efficiency is
found to be $\sim$ 50\% averaged over the full $\eta$ range ($|\eta|<2.5$). The
mis-tagging rate for the corresponding light quark and gluon jets is 1.3\%.

\begin{figure}[h]
\vfill \begin{minipage}{.495\linewidth}
\begin{center}
\mbox{\epsfig{file=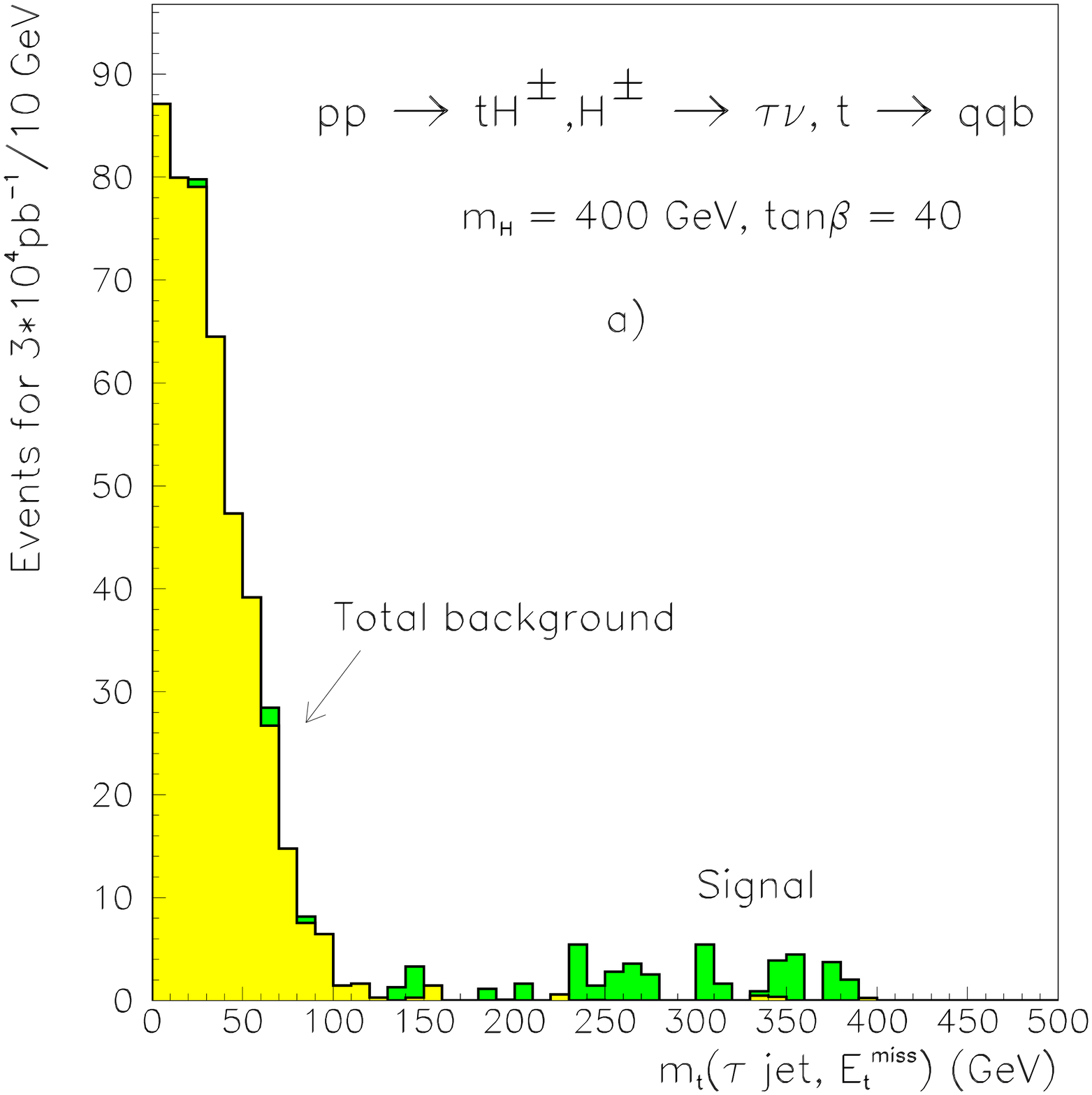,width=.9\linewidth}}
\end{center}
\end{minipage}\hfill
\begin{minipage}{.495\linewidth}
\begin{center}
\mbox{\epsfig{file=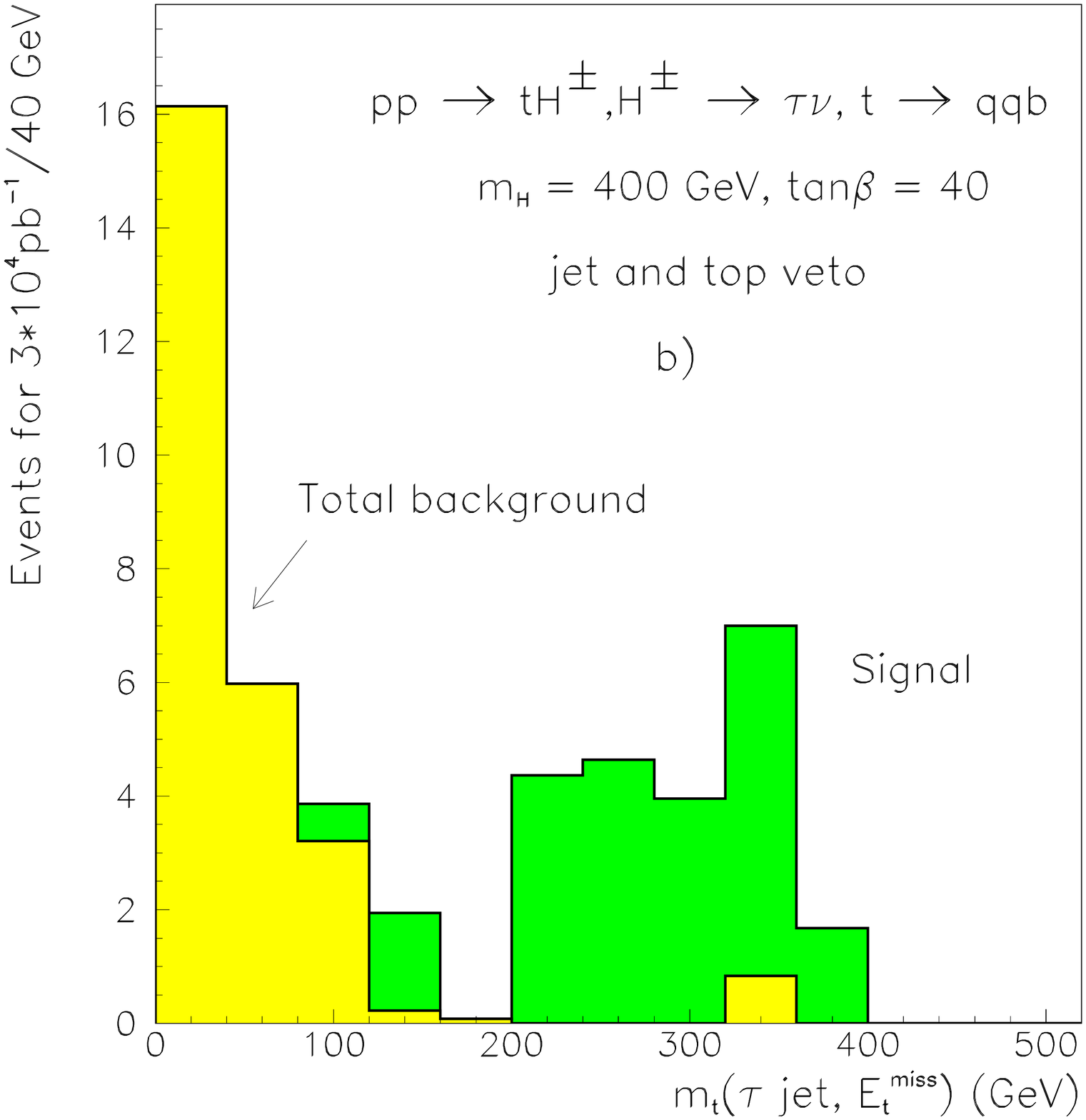,width=.9\linewidth}}
\end{center}
\end{minipage}
\vspace{ 3mm}
\caption{a) Transverse mass reconstructed from $\tau$ jet and $E_t^{miss}$ for 
$H^{\pm}\ra \tau\nu$ from $pp\ra tH^{\pm}$ with $m_{H^{\pm}}$ = 400 GeV and 
$\tan\beta$ = 40 over the total background from $t\overline{t}$ and $W$+jet 
events. b) the same as in a) but with a veto on a central jet and a second 
top.}
\label{mt1}
\end{figure}

The reconstructed transverse mass $m_T^{\tau\nu}$ over the total background is
shown in Fig.~7a for $m_{H^{\pm}}$ = 400 GeV and $\tan\beta$ = 40 for 30
fb$^{-1}$. For $m_T^{\tau\nu}>$100 GeV about 44 signal events are expected for
$m_{H^{\pm}}$ = 400 GeV and $\tan\beta$ = 40 and about 25 events for $m_{H^{
\pm}}$ = 200 GeV and $\tan\beta$ = 30, for an integrated luminosity of 
30 fb$^{-1}$. About 5 background events from $t\overline{t}$ and $W+jet$ are 
expected for $m_T^{\tau\nu}>$100 GeV. 
Further reduction of the $t\overline{t}$ background is still
possible using a jet veto cut and a veto on a second top in the event. Since a
soft and a relatively forward spectator b-jet from the production process is
expected in the signal events, a central and hard jet veto with $|\eta^{jet}|<$
2 and $E_t^{jet} >$ 50 GeV is used. For the reconstruction of the second top
from the $\tau$ jet, missing energy and one of the remaining jets, the
longitudinal component of the missing energy is first resolved from the $W$
mass constraint selecting the smaller of the two solutions. The reconstructed
top mass is required to fall outside the window of $m_{t} \pm$ 60 GeV. The
central jet veto and the second top veto, being closely correlated cuts, reduce
$t\overline{t}$ background by a factor of $\sim$7. The efficiency for the
signal is 54\%. The transverse mass $m_T^{\tau\nu}$ distribution over the total
background including the jet and second top veto is shown in Fig.~7b for
$m_{H^{\pm}}$ = 400 GeV and $\tan\beta$ = 40 and in Fig.~8a for $m_{H^{\pm}}$ =
200 GeV and $\tan\beta$ = 30. 

\begin{figure}[h]
\vfill \begin{minipage}{.495\linewidth}
\begin{center}
\mbox{\epsfig{file=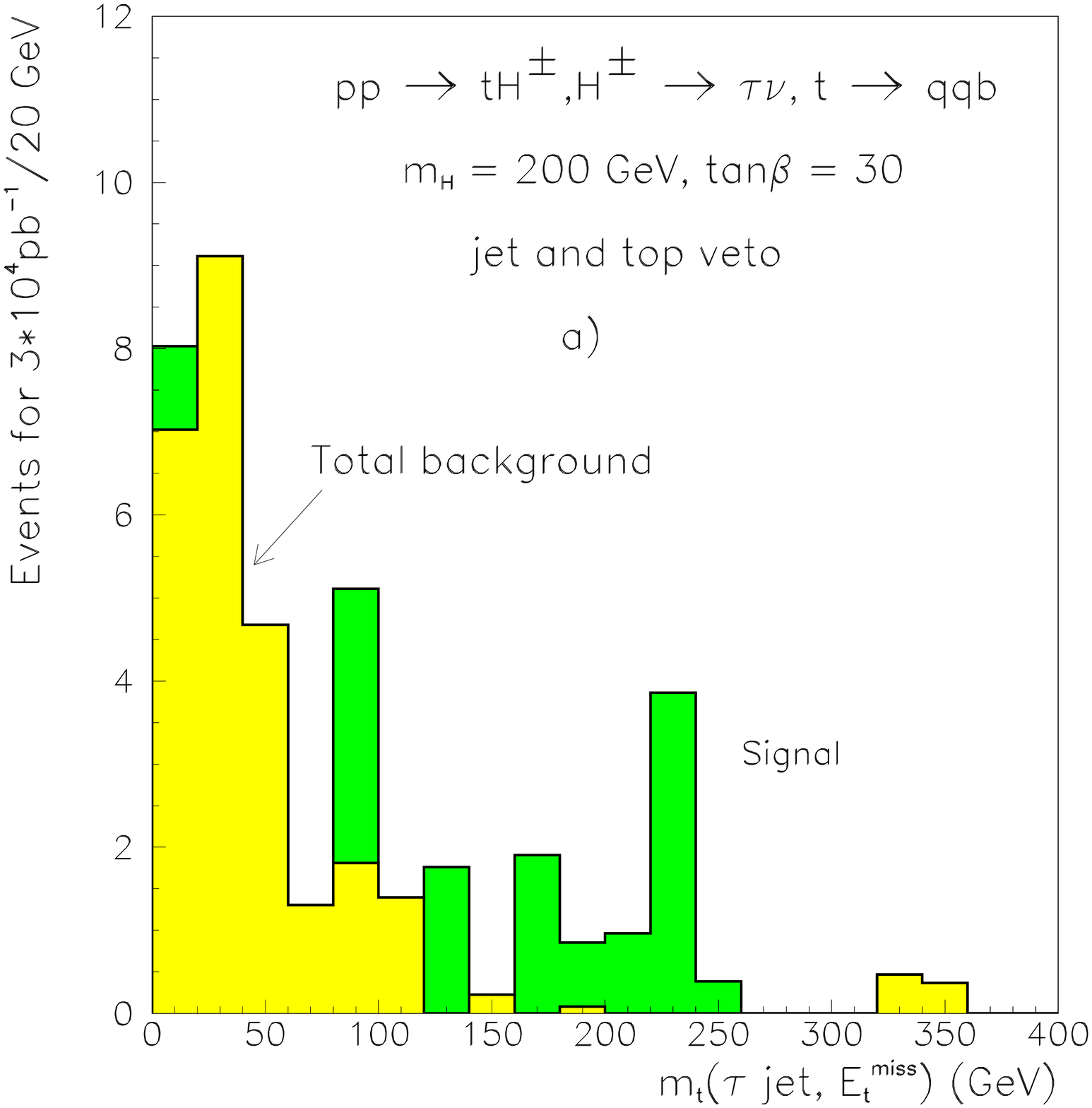,width=.9\linewidth}}
\end{center}
\end{minipage}\hfill
\begin{minipage}{.495\linewidth}
\begin{center}
\mbox{\epsfig{file=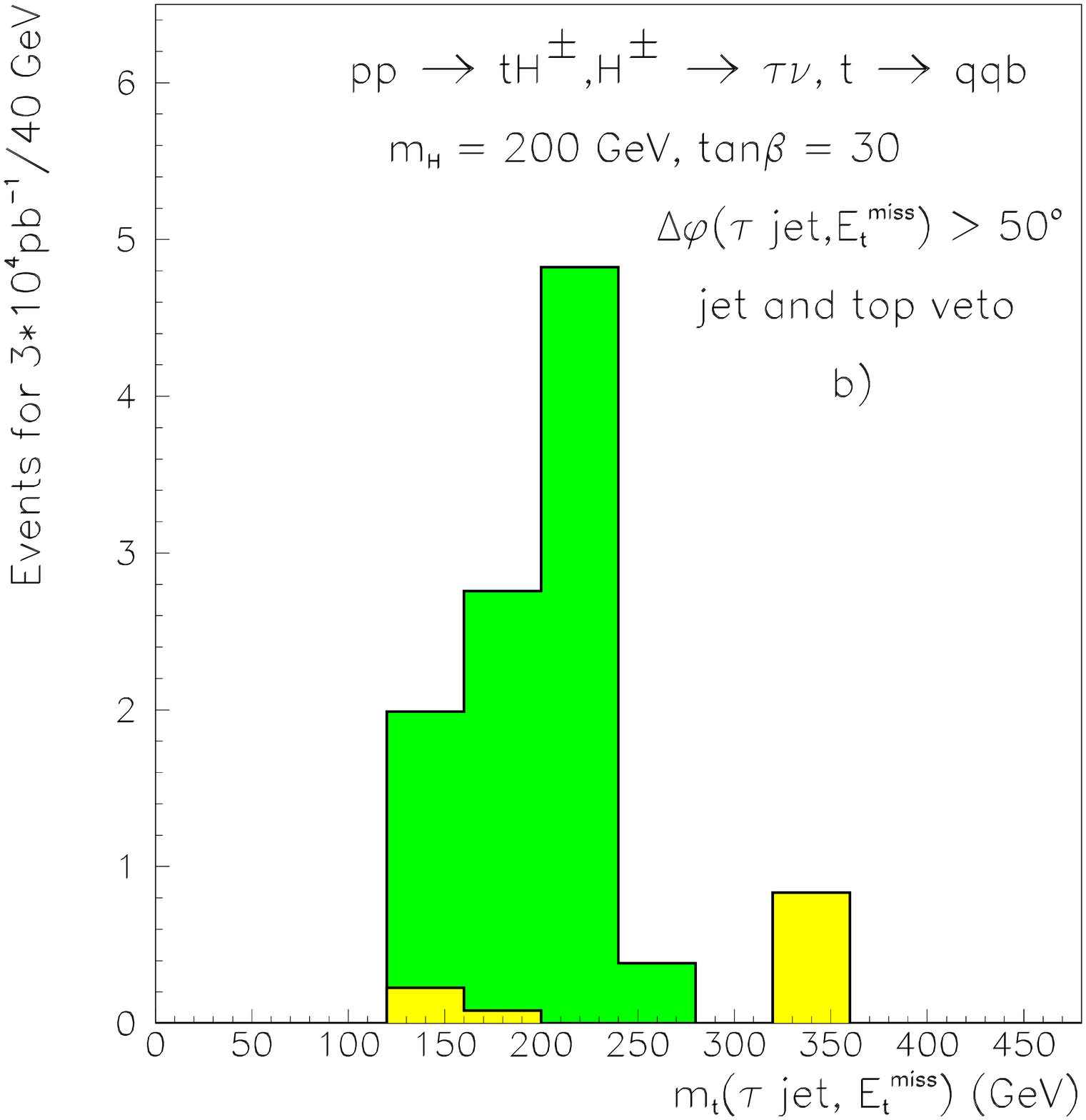,width=.9\linewidth}}
\end{center}
\end{minipage}
\vspace{ 3mm}
\caption{a) Transverse mass reconstructed from $\tau$ jet and $E_t^{miss}$ for 
$H^{\pm}\ra \tau\nu$ from $pp\ra tH^{\pm}$ with $m_{H^{\pm}}$ = 200 GeV and 
$\tan\beta$ = 30 over the total background from $t\overline{t}$ and $W$+jet 
events with central jet and second top veto. b) the same as in a) but for 
$\Delta\phi > 50^o$ where $\Delta\phi$ is the angle between the $\tau$ jet and 
the $E_t^{miss}$ vector in the transverse plane.}
\label{mt2}
\end{figure}

The visibility of the signal can be significantly improved, especially at
$m_{H^{\pm}}$ = 200 GeV, with a cut on the $\Delta\phi$ angle between the
$\tau$ jet and the $E_t^{miss}$. Although $\Delta\phi$ is directly proportional
to $m_T^{\tau\nu}$, a cut in $\Delta\phi$ suppresses the background efficiently
at the lower end of the expected signal region as can be seen from Fig.~8b
showing the signal over the total background with $\Delta\phi > 50^o$ for
$m_{H^{\pm}}$= 200 GeV and $\tan\beta$ = 30.

\subsection{Conclusion} 

Our preliminary study leads to the conclusion that $H^{\pm}\ra \tau\nu$ from
$pp\ra tH^{\pm}$ is a promising discovery channel for charged Higgs bosons at
the LHC. For the evaluation of the final discovery reach in the $m_A$,
$\tan\beta$ parameter space a detailed simulation of the $E_t^{miss}$
measurement for the background events is needed.  The study can be extended to
high luminosity but some additional loss of efficiency should be expected due
to the harder $E_t^{jet}$ cuts due to trigger requirements.  

\section{The $H^+ \rightarrow cs, Wh, tb$ modes in ATLAS}  

\subsection{Introduction} 

In this section we describe the charged Higgs boson discovery potential of the
ATLAS detector in the ($m_{H^\pm}$, $\tan\beta$) parameter space which has been
investigated using the ATLFAST~\cite{K6} and PYTHIA 5.7~\cite{pythias3} simulation
packages. This is a particle--level simulation performed at $\sqrt{s}=14$~TeV,
but with the detector resolutions and efficiencies parametrized from full
detector simulations. It is assumed that the mass scale of supersymmetric
partners of ordinary matter is above the charged Higgs bosons so that $H^\pm$
decays into supersymmetric partners are forbidden~\cite{K8}. A central value
175~GeV is used for the top-quark mass. 

The decays $H^\pm\rightarrow tb$ and $H^\pm\rightarrow\tau\nu$ are the dominant
channels in most of the parameter space~\cite{HDECAY3}. The decay channel
$H^\pm\rightarrow\tau\nu$ has been studied extensively for ATLAS  for
$m_{H^\pm} < m_t$, and the signal appears as an excess of $\tau$
leptons~\cite{K10}. The channel $H^\pm\rightarrow Wh^0$ is only relevant in a
tiny range of MSSM parameter space but it constitutes a unique test for MSSM
and may be sensitive to the singlet extension to MSSM, i.e., NMSSM. The
$H^\pm\rightarrow c\bar{s}$ channel is studied as a complement to the
$\tau$-lepton channel: if the charged Higgs is detected by observing the excess
of $\tau$-leptons over the SM prediction, then the $c\bar{s}$
channel could be used to measure $m_{H^\pm}$. Discovery is possible through
the $H^+ \to t\bar{b}$ channel for low $(\lsim 3)$ and large $(\gsim 25)$ 
$\tan \beta$ values up to masses $m_{H^\pm} \sim 400$ GeV. In the following, 
a brief description of the analysis is presented; details can be found
elsewhere~\cite{K11,K12}.

\begin{figure}[!htbp]
\vspace*{-.2cm}
\begin{center}
\mbox{
     \epsfxsize=7.5cm
     \epsffile{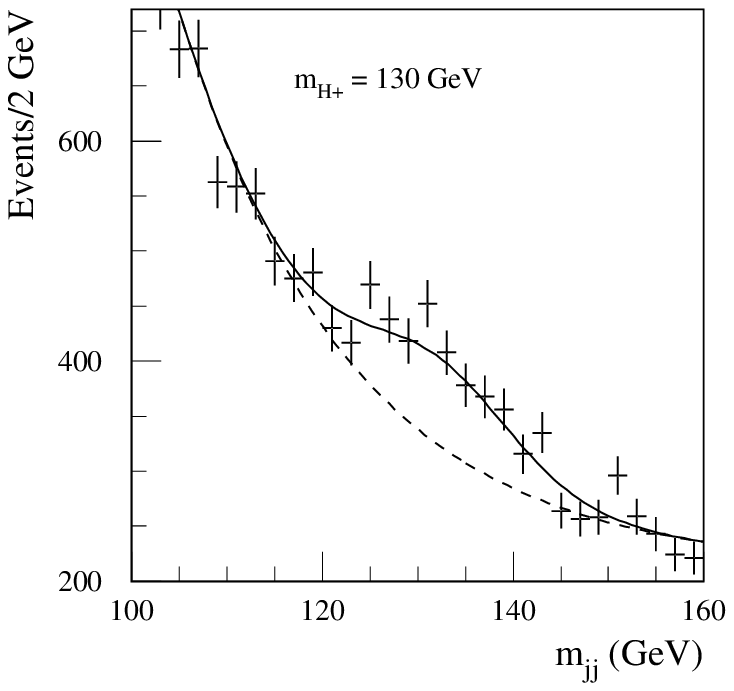}}
\caption{\small For the $H^\pm\rightarrow cs$ channel, the expected $m_{jj}$
distribution from signal and background events (solid) and from the background
(dashed) for $m_{H^\pm}=130$~GeV and $\tan\beta=1.5$ and for an integrated
luminosity of 30~fb$^{-1}$. Errors are statistical only.}
\label{fig:fig1}
\end{center}
\vspace*{-8mm}
\end{figure}

\subsection{$H^\pm$ Discovery Potential}

\ \ \ \textbf{(i)} $\mathbf{t\rightarrow bH^\pm\rightarrow bc\bar{s}}$,
$\mathbf{m_{H^\pm} < m_t}$: $t\bar{t}$ events are generated through
\mbox{$gg,\, q\bar{q}\rightarrow t\bar{t}$} with one top-quark decaying into
the charged Higgs, and the other into $W$, \mbox{$\bar{t}\rightarrow
Wb\rightarrow l\nu b$}.  The major background is $t\bar{t}$ production itself
with both top-quark decaying into $W$'s; one of the $W$'s goes to jets and the
other to leptons. This process is studied for $\tan\beta=1.5$ and $m_{H^\pm}=
110$ and 130~GeV. The events with a final state consisting of two b-tagged jets
($|\eta|<2.5$, and $p_T>15$~GeV), and a single isolated lepton ($|\eta| < 2.5$,
$p_T^e> 20$ and $p_T^\mu>6$~GeV) are selected and the charged mass peak is
searched for the di-jet mass distribution $m_{jj}$. The combinatorial
background is reduced by applying a b-jet veto and a jet-veto on extra jets.
Fig.~\ref{fig:fig1} shows the di-jet mass distribution for both the signal and
the background. This channel complements the $H^\pm \rightarrow \tau \nu$
channel in that if the $H^\pm$ is detected by observing the excess of
$\tau$-leptons, the $H^\pm\rightarrow c\bar{s}$ channel can be used to
determine $m_{H^\pm}$. \bigskip

\begin{figure}[!htbp]
\vspace*{-.7cm}
\begin{center}
\mbox{
     \epsfxsize=7.5cm
     \epsffile{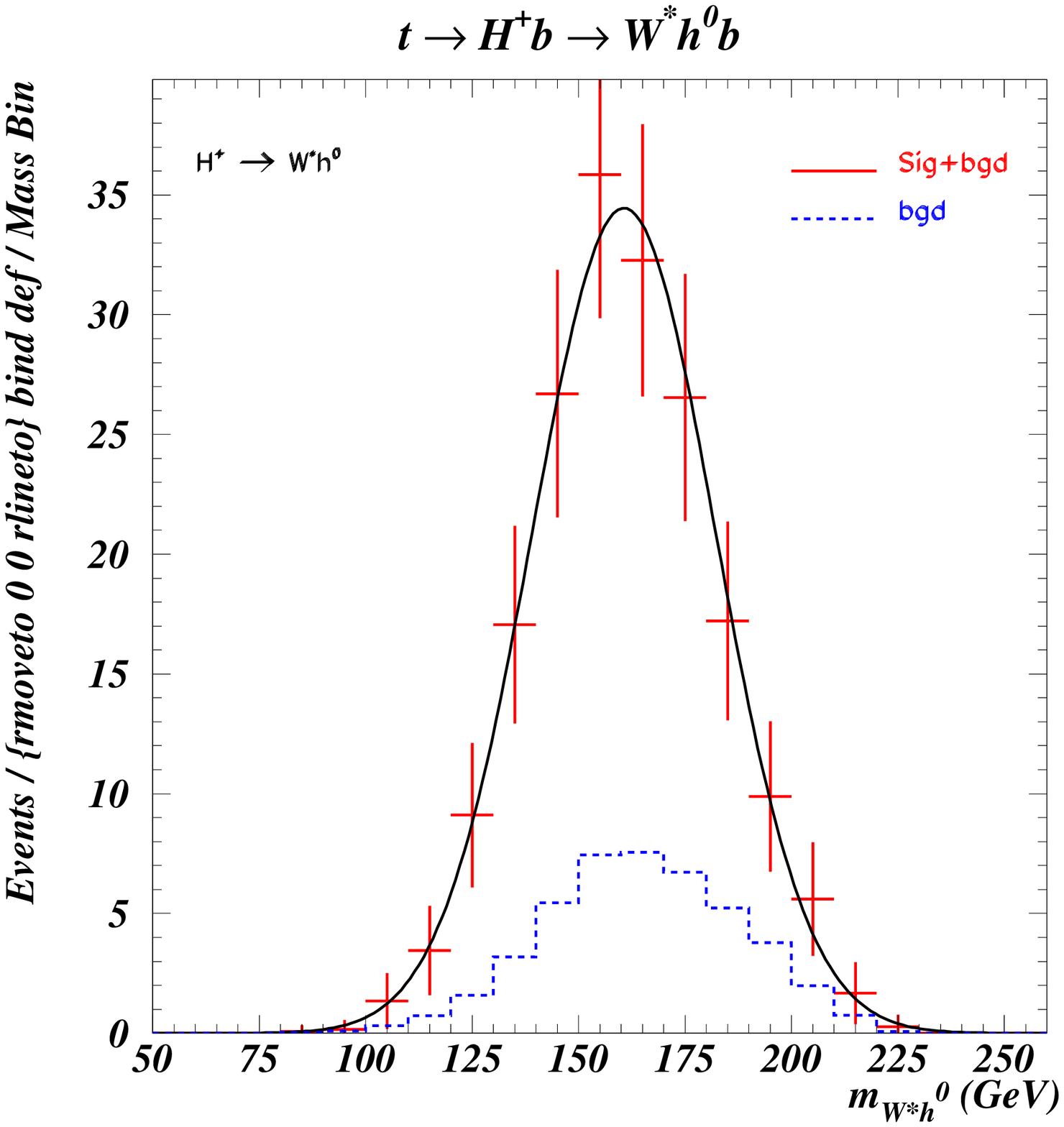}}
\caption{\small For the $H^\pm\rightarrow W^*h^0$ channel, the reconstructed 
mass distribution from signal$+$background events (solid) and from background 
events (dashed) for $m_{H^\pm}=152$~GeV, $\tan\beta=2$, and for and integrated 
luminosity of 300~fb$^{-1}$. Errors are statistical.}
\label{fig:fig2}
\end{center}
\vspace*{-3mm}
\end{figure}

\textbf{(ii)} $\mathbf{t\rightarrow bH^\pm}$, $\mathbf{H^\pm\rightarrow
W^*h^0}$, $\mathbf{m_{H^\pm} < m_t}$: The production mechanism is the same as
in the previous case, but here,  $H^\pm\rightarrow W^*h^0$, with
$h^0\rightarrow b\bar{b}$. The final state contains two $W$'s, one of which is
off-shell and one of which decays to leptons and the other to jets. The major
backgrounds are $t\bar{t}b\bar{b}$ and $t\bar{t}q\bar{q}$ followed by the
decays of the top-quarks as described above. The present channel is studied for
$m_{H^\pm}=152$~GeV and for $\tan\beta=2$ and 3 corresponding to $m_{h^0}=83.5$
and 93.1~GeV respectively. We search for an isolated lepton, four b-tagged jets
($p_T^b > 30$~GeV) and at least two non b-jets with $p_T^j > 30$~GeV. The
details of this analysis can be found in~\cite{K12}. It suffices to say that
although the backgrounds are over two orders of magnitude higher that the
signal at the start, we propose a reconstruction method which permits  the
extraction of the signal with a significance exceeding $5\sigma$ in the low
$\tan\beta\;(1.5-2.5)$ region. At high $\tan\beta$, though the reconstruction
remains comparable the signal rate decreases so significantly that discovery
potential vanishes in this region.  Fig.~\ref{fig:fig2} shows the charged Higgs
mass reconstruction for $\tan\beta=2$. \bigskip

\begin{figure}[!t]
\vspace*{-.5cm}
\begin{center}
\mbox{
     \epsfxsize=7.5cm
     \epsffile{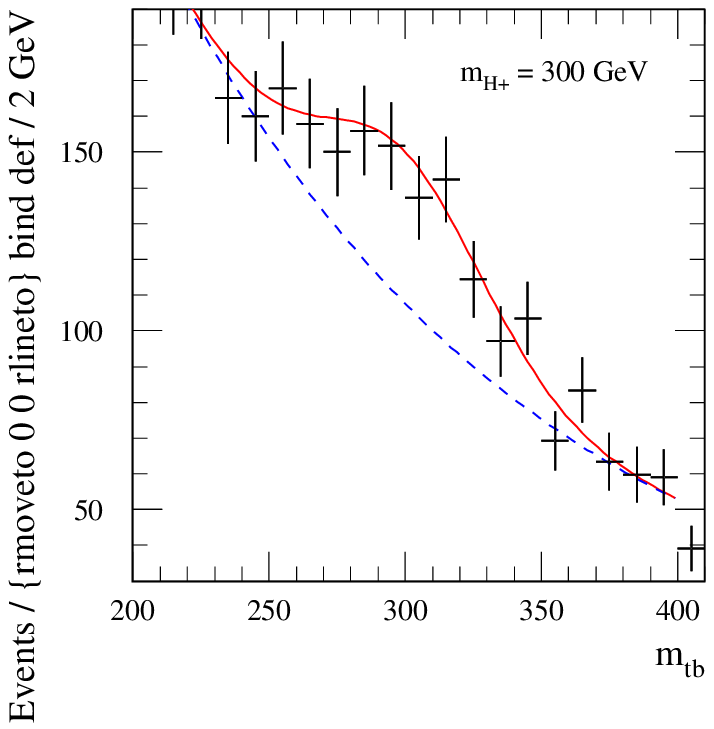}}
\caption{\small Signal and background distributions for the reconstructed 
invariant mass $m_{tb}$ for $m_{H^\pm}=300$ GeV, $\tan\beta=1.5$ 
and an integrated luminosity of 30~fb$^{-1}$. Errors are statistical only.}
\label{fig:fig3}
\end{center}
\vspace*{-7mm}
\end{figure}

\textbf{(iii)} $\mathbf{m_{H^\pm} > m_t}$: Above the top-quark mass, we
consider the production of $H^\pm$ through the $2\rightarrow 2$ process
$gb\rightarrow tH^\pm$. Two decay channels of $H^\pm$ are examined in details,
$H^\pm\rightarrow tb$ and $H^\pm\rightarrow Wh^0\rightarrow Wb\bar{b}$. In both
cases the major background comes from $t\bar{t}b$ and $t\bar{t}q$ events. In
either case, we search for an isolated lepton, three b-tagged jets and at least
two non b-jets. The details of these analyses can be found elsewhere
\cite{K11,K12}. Discovery is possible through the $H^\pm\rightarrow tb$ channel
for low ($<3$) and for high ($>25$) $\tan\beta$ up to $m_{H^\pm}\sim
400$~GeV~\cite{K11}. Fig.~\ref{fig:fig3} shows the charged Higgs mass
reconstruction for $\tan\beta=1.5$ and $m_{H^\pm}=300$~GeV.  On the other hand,
the $H^\pm\rightarrow Wh^0$ channel presents no discovery potential for the
charged Higgs in the MSSM.  Initially, the total background is at least three 
orders of magnitude higher than the signal in the most favorable case studied 
($\tan\beta=3$).
We propose a reconstruction technique which improves the signal-to-background
ratios by two orders of magnitude. However, this improvement is still not
enough to observe a clear signal; for example, at $\tan\beta=3$, a significance
of only 3.3 can be expected after three years of high luminosity
operation~\cite{K12}. 

\subsection{Conclusions}

The possibility of detecting the charged Higgs through the decay channels
$H^\pm\rightarrow c\bar{s}$, $H^\pm\rightarrow Wh^0$, and
$H^\pm\rightarrow tb$ with the ATLAS detector has been studied as a
function of $\tan\beta$, below and above the top-quark mass. Below the
top-quark mass and at low $\tan\beta$, both channels \mbox{$H^\pm\rightarrow
c\bar{s}$} and \mbox{$H^\pm\rightarrow Wh^0$} present significant discovery
potential. These two channels would complement the \mbox{$H^\pm\rightarrow
\tau\nu$} searches in that if the latter is observed through the excess of
$\tau$-leptons, the former channels can be used to measure the mass of the
charged Higgs. Above the top-quark mass, the process \mbox{$H^\pm\rightarrow
tb$} presents a significant discovery potential in the low and the high
$\tan\beta$ regions up to 400~GeV.

\subsubsection*{Acknowledgements}

R.K. would like to thank Daniel Denegri for helpful discussions.
K.A.A. expresses gratitude to E.~Richter-W\c{a}s for fruitful 
discussions and constructive criticisms; his work is supported by a grant 
from the USA National Science Foundation (grant number 9722827).


\setcounter{figure}{0}
\setcounter{table}{0}
\setcounter{section}{0}
\setcounter{equation}{0}
\newpage

\newcommand{\beqn}{\begin{eqnarray}}
\newcommand{\eeqn}{\end{eqnarray}}
\def\sinb{\sin\beta}
\def\cosb{\cos\beta}
\def\sinbb{\sin (2\beta)}
\def\cosbb{\cos (2 \beta)}
\def\tgb{\tan \beta}
\def\tgbt{$\tan \beta\;\;$}
\def\tgbsq{\tan^2 \beta}
\def\sinal{\sin\alpha}
\def\cosal{\cos\alpha}
\def\stop{\tilde{t}}
\def\sto{\tilde{t}_1}
\def\stt{\tilde{t}_2}
\def\stl{\tilde{t}_L}
\def\str{\tilde{t}_R}
\def\msto{m_{\sto}}
\def\mstosq{m_{\sto}^2}
\def\mstt{m_{\stt}}
\def\msttsq{m_{\stt}^2}
\def\mt{m_t}
\def\mtsq{m_t^2}
\def\sint{\sin\theta_{\stop}}
\def\sintt{\sin 2\theta_{\stop}}
\def\cost{\cos\theta_{\stop}}
\def\sintsq{\sin^2\theta_{\stop}}
\def\costsq{\cos^2\theta_{\stop}}
\def\mqtt{\M_{\tilde{Q}_3}^2}
\def\mutt{\M_{\tilde{U}_{3R}}^2}
\def\msbo{m_{\tilde{b}_1}}
\def\neut{\chi_0}
\def\mh{M_h}
\def\mchi{m_\chi^+}
\newcommand{\mw}{M_{W}}
\newcommand{\mz}{M_{Z}}
\newcommand{\mzz}{M_{Z}^{2}}
\newcommand{\cw}{\cos\theta_W}
\newcommand{\sw}{\sin\theta_W}
\newcommand{\tw}{\tan\theta_W}
\def\cww{\cos^2\theta_W}
\def\sww{\sin^2\theta_W}
\def\tww{\tan^2\theta_W}

\begin{center}
{\large\sc {\bf Light stop effects and Higgs boson searches at the LHC.}}

\vspace*{.5cm}

{\sc G.~B\'elanger, F.~Boudjema, A.~Djouadi, V.~Ilyin,}

\vspace*{2mm}

{\sc J.L.~Kneur, S.~Moretti, E. Richter--W\c{a}s 
and  K.~Sridhar}
\end{center} 

\begin{abstract}
We analyze the effects of light top squarks with large mixing on the search 
of the lightest Higgs boson of the Minimal Supersymmetric extension of the 
Standard Model at the LHC. We discuss both the stop loop effects in the main 
production and decay processes, and the associated production of top squarks 
with the lightest Higgs boson. 
\end{abstract} 


\section{Introduction}

The third generation fermions, and especially the top quark because of its
large Yukawa coupling, play an important r\^ole in the mechanism of electroweak
symmetry breaking and the properties of the Higgs bosons \cite{HHG}. Recall
that if the top quark were rather light, the Minimal Supersymmetric extension
of the Standard Model (MSSM) would have been already discarded since the
lightest Higgs boson $h$ that it predicts would have been lighter than the $Z$
boson, $M_h \leq M_Z$ \cite{HHG}, and would have not escaped detection at LEP2.
The contribution of the top quark and its SUSY partners to the radiative
corrections to $M_h$ can push the mass value up to $M_h \sim 135$ GeV
\cite{rcmass}, beyond the reach of LEP2. The mixing in the stop sector is also
important since large values of the mixing parameter $\tilde{A}_t = A_t + \mu /
\tgb$ [where $A_t$ is the trilinear coupling, $\mu$ the higgsino mass parameter
and $\tgb$ the ratio of the vev's of the two Higgs doublets which break the
electroweak symmetry; see Ref.~\cite{HK} for the SUSY parameters] can increase
the $h$ boson mass for a given value of $\tgb$ \cite{rcmass}.  

On the other hand, while the sfermions of the two first generations can be very
heavy, naturalness  arguments suggest that the SUSY particles that couple
substantially to the Higgs bosons [stops, sbottoms for large $\tgb$, and the
electroweak gauginos and higgsinos] could be relatively light. In this respect,
the case of the stop sector is special: because of the large $m_t$ value, the
mixing in this sector can be very strong, leading to a mass eigenstate
$\tilde{t}_1$ lighter that all other squarks, and possibly lighter than the top
quark itself. At the same time, again because of the large mixing, this
particle can couple very strongly to the MSSM Higgs bosons and in particular to
the lightest CP--even particle $h$.  
 
At the LHC, a light stop with large couplings to Higgs bosons can contribute to
both the $h$ production in the main channel, the gluon--gluon fusion mechanism
$gg \ra h$, and to the main detection channel, the two--photon decay $h \ra
\gamma \gamma$. The effects can be extremely large, making this discovery
channel possibly useless at the LHC [4--6]. On the other hand, because of the
enhanced couplings and phase--space, associated production of stops and the $h$
boson at the LHC, $pp \to q\bar{q}/gg \to \tilde{t}_1 \tilde{t}_1h$, might have
sizeable cross sections [7--10].  
 
It is thus crucial to investigate how and when this scenario occurs and what
other consequences then follow at the LHC. The purpose of our working group
contribution is to update and complement the various analyses [5--10] which
have been made on this subject.  
 
\section{Stop parameters and phenomenological constraints} 

We start our discussion by recalling the parameters that define the stop
masses, mixing angle and the $\sto \sto h$ coupling. The stop mass eigenstates
are defined through the mixing angle $\theta_{\tilde{t}}$, with  the lightest
stop $\sto$ being $\sto=\cost \; \stl - \sint \; \str$. With the effective
trilinear mixing parameter, $\tilde{A}_t= A_t + \mu / \tgb$, one has for the
masses and the mixing angle\footnote{The sign conventions for $A_t$ here is 
opposite to the one adopted in Refs.~\cite{HK} and \cite{stophiggs_LHC}. 
Accordingly, the sign convention for the mixing angle is opposite to the one 
of Ref.~\cite{stophiggs_LHC} where $\sto=\cost \; \stl + \sint \; \str$.}
\beqn
\label{s2t}
\tan (2 \theta_{\stop} )= \frac{ -2 m_t \tilde{A}_t }
{ \tilde{m}_{\tilde{Q}_3}^2 - \tilde{m}_{\tilde{U}_{3R}}^2+ 
\frac{1}{2} M_Z^2 \cos 2 \beta (1 - \frac{8}{3}
s_W^2) } \;\;\;\;\; {\rm or} \;\;\;\; \sin (2 \theta_{\stop})=
\frac{-2 m_t \tilde{A}_t}{\mstosq-\msttsq}
\eeqn
\beq
m_{\tilde{t}_{1,2}}^2 = m_t^2 + \frac{1}{2} \left[ m_{\tilde Q_3}^2 + 
m_{\tilde U_{3R}}^2 +\cdots \mp \sqrt{ (m_{\tilde Q_3}^2 - m_{\tilde U_{3R}}^2 
+\cdots)^2 + 4m_t^2 \tilde{A}_t^2 } \right]\;, 
\eeq
where $m_{\tilde Q_3}$, $m_{\tilde U_{3R}}$ are the soft-SUSY breaking scalar 
masses and the dots stand for the $D$--terms $\propto M^2_Z \cos 2\beta$. 
Note that in order to enhance the mixing,  $\sin (2 \theta_{\stop})\sim 1$, 
one needs to make $\tilde{A}_t$ large and/or  have the soft-SUSY masses almost 
equal: $\tilde{m}_{\tilde{Q}_3} \simeq \tilde{m}_{\tilde{U}_{3R}}$. The $\sto 
\sto h$ vertex writes
\beqn
\label{stopstophcoupling} 
V_{\sto \sto h}&= & -g \frac{\mt}{\mw} \frac{\cos \alpha}{\sin \beta} 
\bigg[ (A_t-\mu \tan \alpha) \sint \; \cost \;-\; \mt \nonumber \\ 
&+& \frac{\mzz}{\mt} \frac{\sin \beta}{\cos \alpha} \sin(\alpha+\beta)
\biggr[ (\frac{1}{2}-\frac{2}{3} \sww) \costsq +\frac{2}{3} \sww
\sintsq \biggr)  \biggr] \nonumber \\ 
&\simeq & \frac{g}{\mw} \biggr[ \frac{1}{4} \sin^2(2 \theta_{\stop})
 (\mstosq-\msttsq) \;+\; \mt^2\biggr] 
\eeqn
where in the last line we neglected the $D$--term contributions and assumed 
the limit of large $M_A$ to be in the decoupling regime. As can be seen, in 
the presence of large mixing with large splitting between the two stop 
eigenstates, the $\tilde{t}_1 \tilde{t}_1 h$ coupling can be particularly 
large. In the case of no mixing, only the top contribution survives and the 
coupling $\sto \sto h$ is of the order of $t\bar t h$ coupling. Taking this 
limit as a reference point, the strength of the $\sto \sto h$ vertex can be 
normalized through $R_{\sto}=\left[ \mw V_{\sto\sto h}/(g \mt^2)\right]^2$. 

\smallskip 

We now summarize the constraints which can be imposed on the stop 
parameters:
\begin{itemize} 
\item The model independent mass limit on the lightest stop is obtained from
direct searches at LEP, $m_{\tilde{t}_1} \ge 90$ GeV \cite{mh_limit-nov99}. 
However, if the $\tilde{t}_1$ and the $\chi_0$ LSP are not too close in mass, a
stronger limit, $m_{\tilde{t}_1} \ge 120$ GeV \cite{stoplimit99}, is available
from Tevatron analyses. For bottom squarks, a limit $m_{\tilde{b}} \ge 250$ GeV
is available from Tevatron data in the case of no--mixing \cite{stoplimit99}.  
\item If stops are too light, the radiative corrections to the 
$h$ boson mass are not large enough and the limit $M_h \ge 90$ GeV 
\cite{mh_limit-nov99} from LEP searches plays an important role. 
\item As in the case of top/bottom splitting in the Standard Model, the 
stop/sbottom doublet can contribute significantly to electroweak precision 
observables through the $\rho$ parameter. In particular, if stops strongly 
mix and have large couplings, the contributions to $\Delta \rho$ can exceed
the value $\Delta\rho\le 0.0013$ imposed by data \cite{Deltarho}. 
\item Some values of the stop parameters might induce color and charge breaking 
minima (CCB). Since the naive constraints based on the global minima may be 
too restrictive, we will take into account the tunneling rate [for wide range 
of parameters, the global CCB minimum becomes irrelevant on the ground that 
the time required to reach the lowest energy state exceeds the present age of 
the universe], which leads to a milder constraint which may be approximated by
\cite{CCB}: $A_t^2 +3\mu^2 < 7.5 (M_{\tilde{Q}_3}^2 +M_{\tilde{U}_{3R}
}^2)$. 
\end{itemize}
\begin{figure*}[htbp]
\vspace*{-0.7cm}
\begin{center}
\mbox{
\includegraphics[width=8cm,height=8cm]{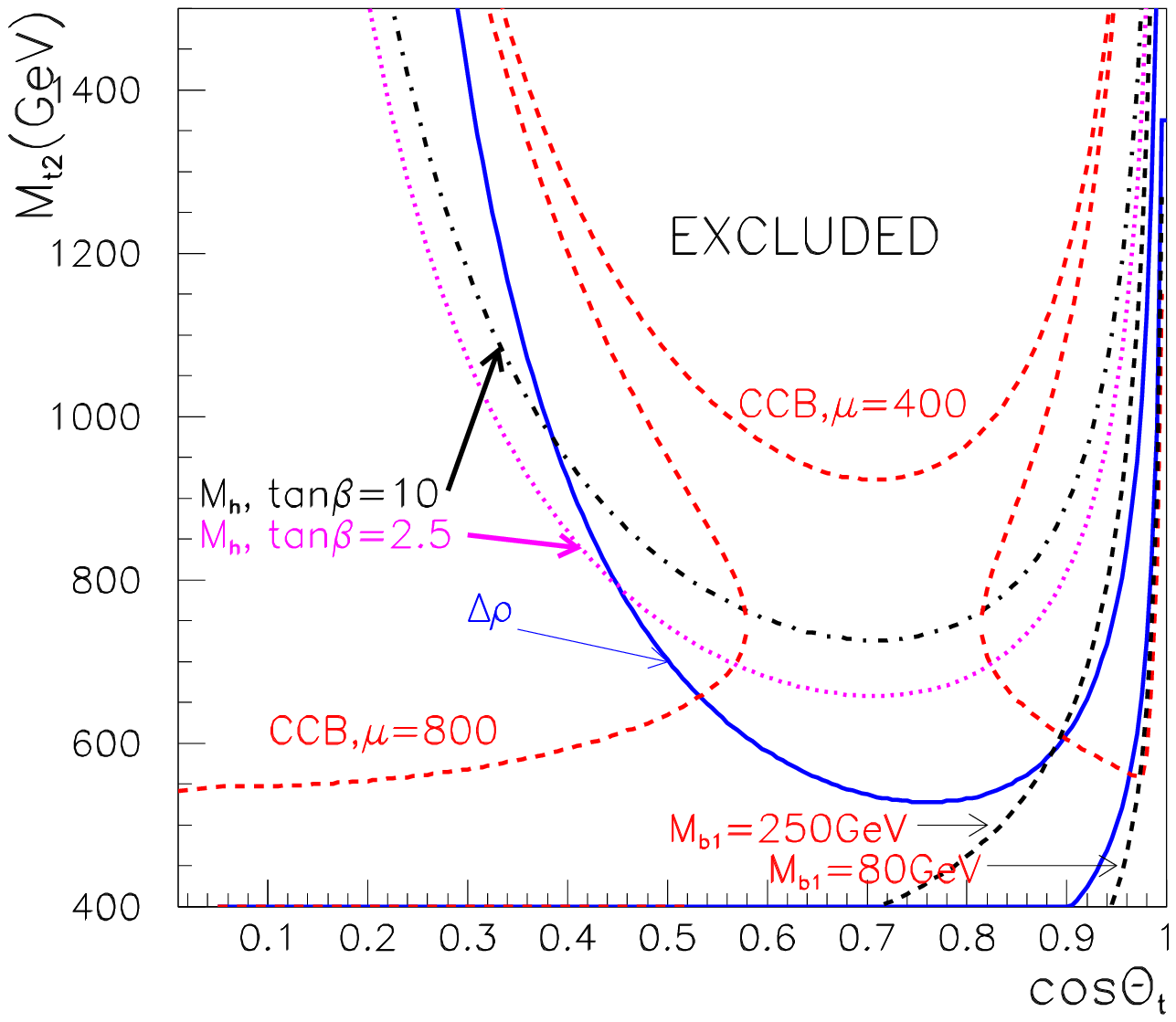}
\includegraphics[width=8cm,height=8cm]{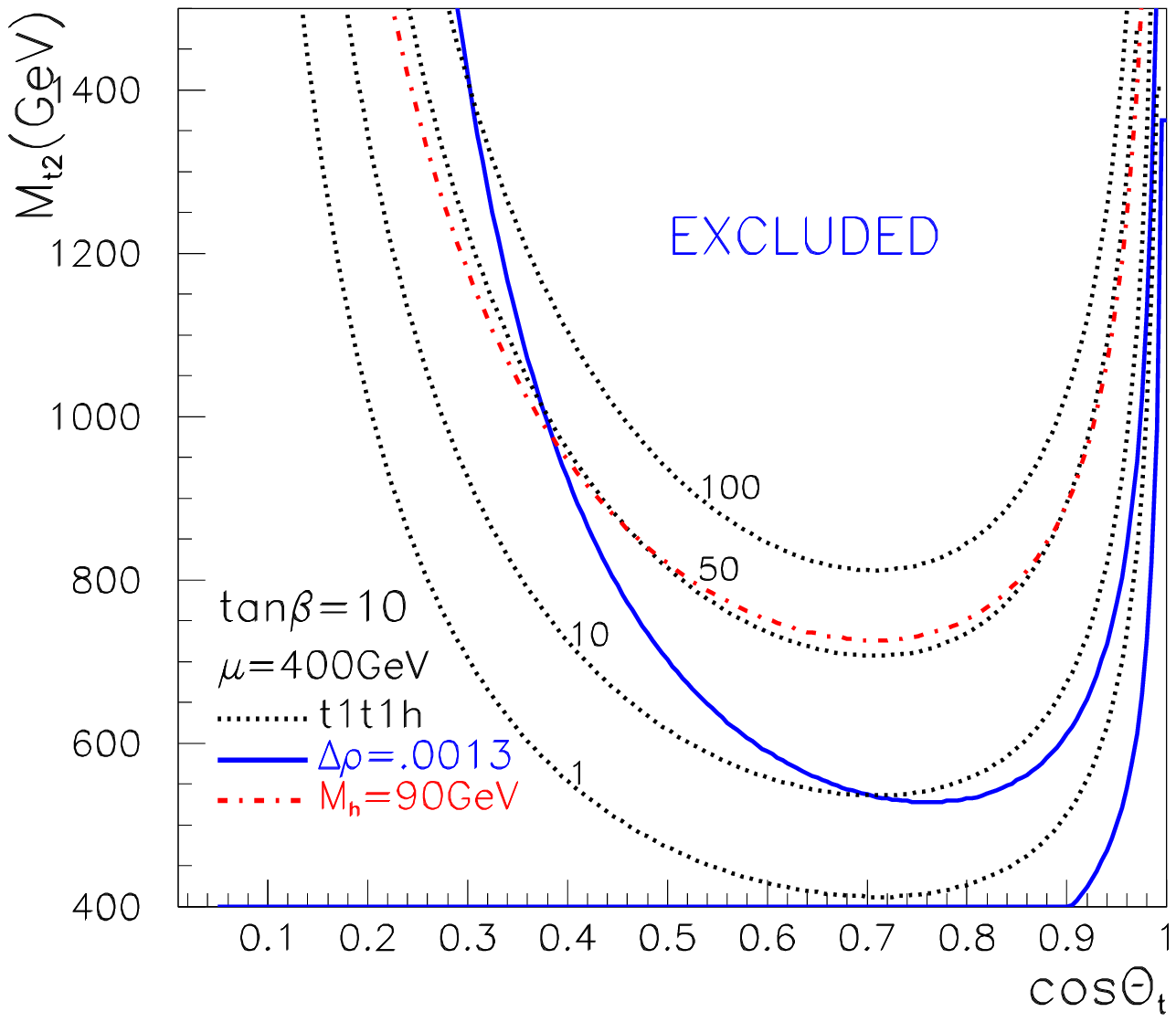}}
\caption{\label{constraints}{\small ({\bf a}) Constraint from $\Delta\rho\le
0.0013$ (full line), $\mh\ge 90$~GeV (dash-dot), CCB (dash) and $\msbo$ (dash)
for $\tgb=10$, $\mu=400$~GeV, $m_{\sto}=120$~GeV and $M_A=1$~TeV; the $\mh$
constraint for $\tgb=2.5$ is also shown (dot). ({\bf b}) Equipotential lines
(dotted) for the normalized coupling $R_{\sto}=1,10,50,100$  with $\tgb=10$ and
$\mu=400 $~GeV. The exclusion regions corresponding to $\Delta\rho\le.0013$ and
$\mh\le 90$~GeV are also reproduced.}}
\end{center}
\vspace*{-0.3cm}
\end{figure*}

Fig.~1 shows how the parameter space is restricted by the previous constraints
and which values of the ratio $R_{\sto}$ are allowed. In Fig.~1a, the excluded
region in the plane $(\cos \theta_{\tilde t}, m_{{\tilde t}_2}$) is within the
respective boundaries indicated. Note that for $\cost \approx 1$, the
$\Delta\rho$ constraint also excludes the region to the right of the second
branch of the $\Delta\rho$ curve where the present limit on the mass of the
sbottom is contained. Requiring $\msbo\ge 250$~GeV excludes the region to the
right of the curve. The CCB constraint for $\mu=800 $~GeV is also displayed,
the excluded region lies between the two ``CCB, $\mu=800$" curves.  In Fig.~1b,
we show the equipotential lines for the normalized coupling $R_{\sto}$. The
exclusion regions corresponding to $\Delta\rho\le.0013$ and $\mh\le 90$~GeV are
also reproduced.  In all cases $M_2=-\mu$ and a common gaugino mass at the GUT
scale are assumed. Note that one has to make sure that the lightest stop is not
the LSP, as has been always verified in our analysis.  Considering that the CCB
constraint is rather uncertain, it is also worth pointing out that the one used
in our analysis hardly precludes points which are not already excluded  by the
$\Delta \rho$ and $m_h$ constraints.  

\section{Higgs boson signals at the LHC}

In this section, we will discuss what might happen to the search for the
lightest MSSM Higgs boson $h$ at the LHC, if one allows all sparticles but the
stops (and to a lesser extent the charginos and neutralinos) to be rather
heavy. We will first discuss the effects of stop loops in the gluon--gluon
fusion mechanism, $gg \ra h$, and in the main Higgs detection channel, the
two--photon decay $h \ra \gamma \gamma$, and then discuss the associated
production of stops with the light Higgs boson $h$ and possibly $A$.  

\subsection{Stop loop effects} 

Since the $h \tilde{t}_1 \tilde{t}_1$ vertex eq.~(2) does not have a definite
sign [for no mixing the positive $m_t^2$ component dominates while for maximal
mixing the negative component $\frac{1}{4} \sin^2(2 \theta_{\stop}) (\mstosq-
\msttsq)$ is the leading one], the stop loop contributions can interfere either
destructively or constructively with the top loop contributions in the $gg \ra
h$ and $h \ra \gamma \gamma$ processes. Noting that while for $gg \ra h$ only
top/stop loops are present, for the decay $h \ra \gamma \gamma$, the additional
contributions from $W$ loops are dominant and have a destructive interference
with the top contributions. This means that if the rate for $h \ra gg$ is
suppressed, there will be a slight increase in $h\ra \gamma \gamma$ decay width
and vice versa. Therefore either the rate for the inclusive channel $gg \ra h
\ra \gamma \gamma$ is enhanced {\em or} the rate for the associated Higgs
production $pp \to Wh,Zh,t\bar t h$ \cite{xsections} with $h \ra \gamma \gamma$ is enhanced. It
is important to stress that, in any case, the rate for the associated $t \bar t
h$ production with the subsequent decay $h \ra b \bar b$ is hardly affected by
stop loops and will always help in these scenarii, as will be discussed later.  

We begin our analysis by defining the ratio $R_{\gamma \gamma} \equiv R_{h \ra
\gamma \gamma}$ which is the branching ratio of the lightest SUSY Higgs boson
decay into two photons over that of the SM for the same Higgs mass. In the
decoupling regime, $M_A \gg M_Z$, this ratio is affected only by SUSY--particle
loops; in this case the ratio is also sensibly the same as the ratio for
associated production of the $h$ boson with $W,Z$ bosons and/or with $t\bar{t}$
pairs, with $h$ decaying into $\gamma \gamma$. We also define $R_{gg \gamma
\gamma}$ as the ratio for the signal in the direct production channel $gg \ra
h$ times the branching ratio for the $h \ra \gamma \gamma$ decay in the two
models. The $gg$ and $\gamma \gamma$ decay widths are obtained\footnote{Note
that the ratios of $gg$ decay widths and production cross sections are almost
the same: large QCD corrections cancel out in the ratios when the dominant 
contribution comes from the top loops, and the corrections to the top and
stop contributions are practically the same; see Ref.~\cite{QCD}.} with the
help of the program {\tt HDECAY} \cite{HDECAY}.  

Fig.~\ref{hsignal_equalmass} summarizes the contribution of stop loops to
these ratios, for $\tgb=2.5, \mu=-M_2=250$ GeV and $M_A=1$ TeV. To maximize 
the effect of stop mixing, $\sin (2 \theta_{\stop}) \simeq 1$, we assume that 
$\tilde{m}_{\tilde{Q}_3} \simeq \tilde{m}_{\tilde{U}_{3R}}$. From this figure, 
one can see that: 

\begin{figure*}[htbp]
\vspace*{-1.1cm} 
\begin{center}
\includegraphics[width=12cm,height=15cm]{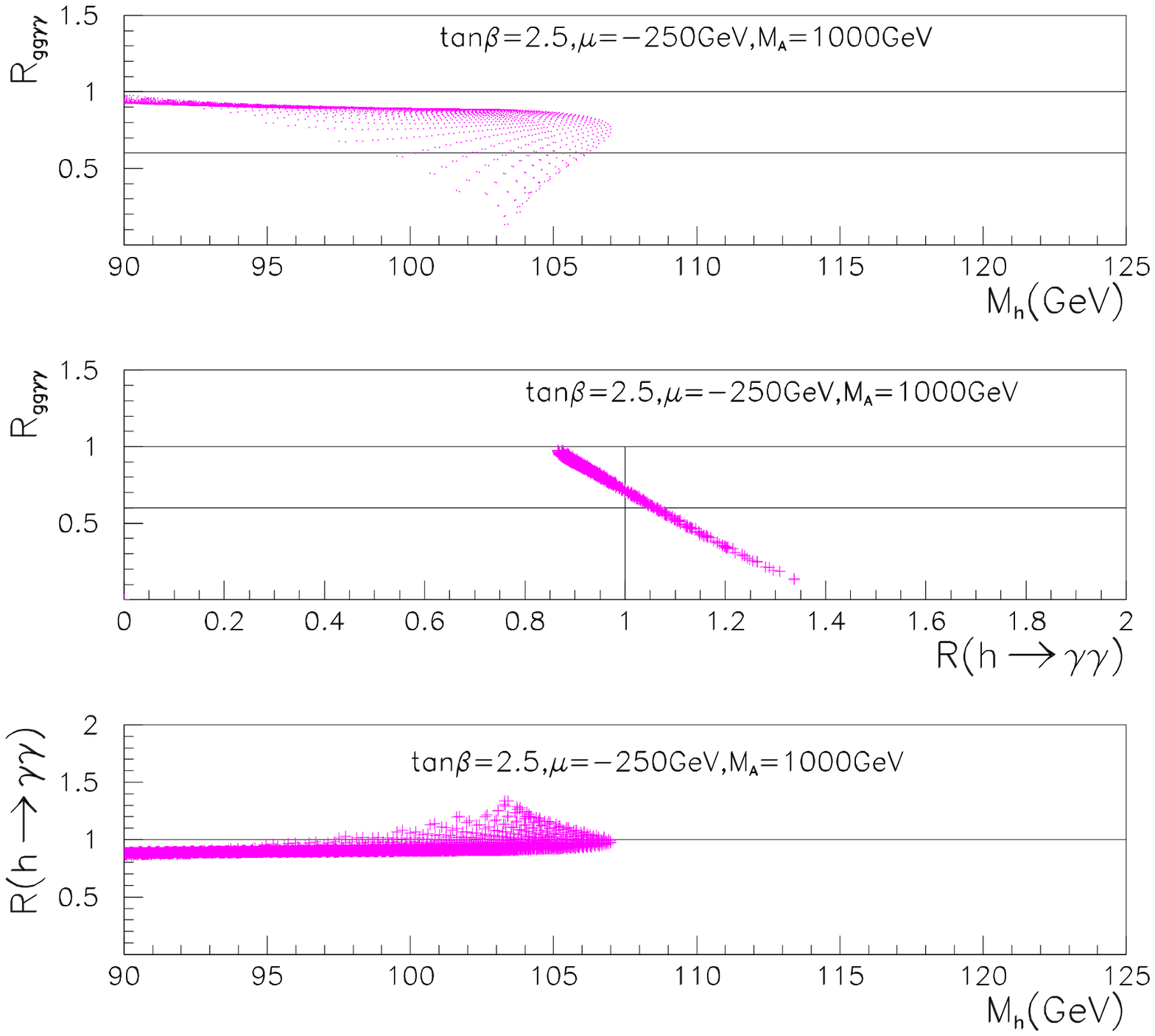}
\includegraphics[width=12cm,height=5cm]{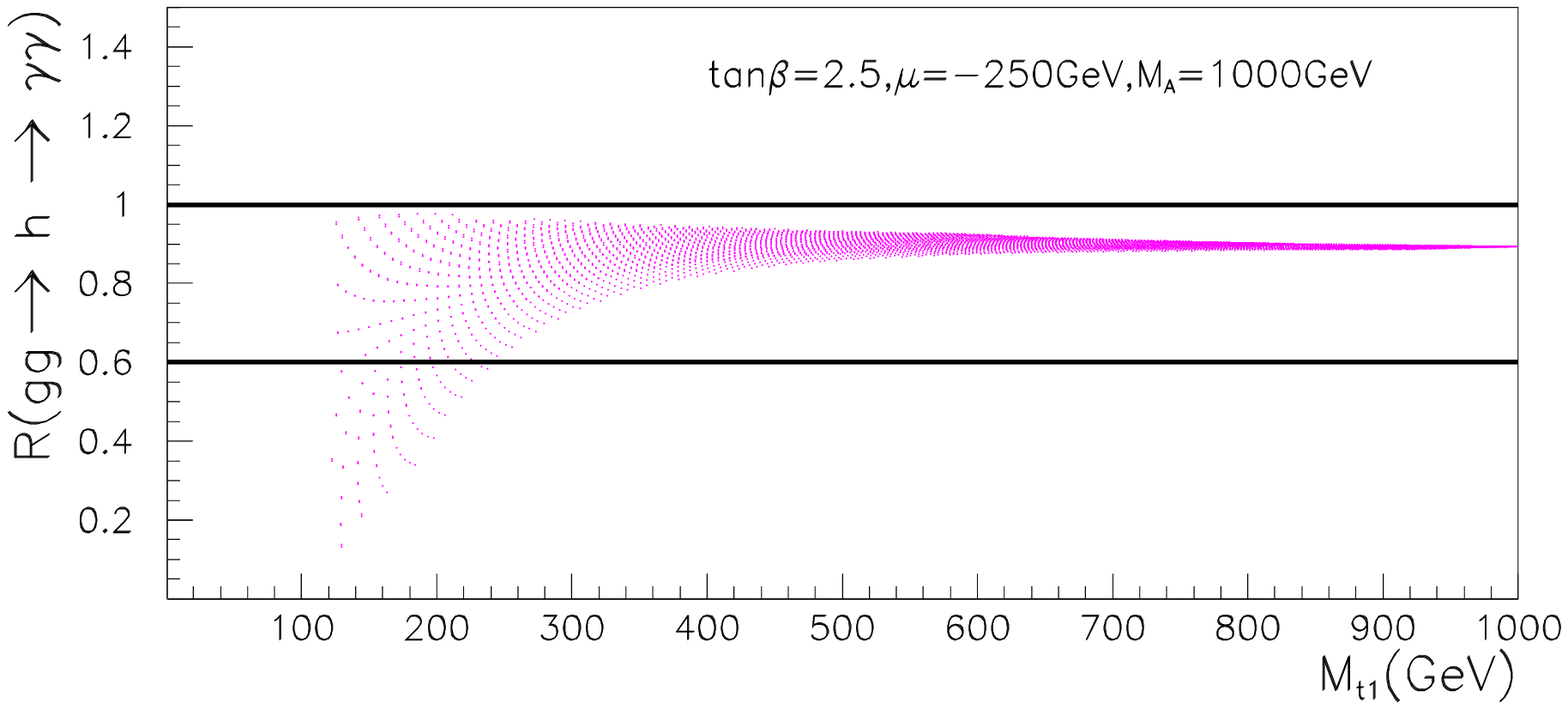}
\caption{\label{hsignal_equalmass}{\small Higgs boson $(h)$ production 
and decay ratios at the LHC for $\tilde{m}_{\tilde{Q}_3} \simeq \tilde{m}_
{\tilde{U}_{3R}}$ at $\tgb=2.5$ and large $M_A$. Figures are scanned over 
$\tilde{m}_{\tilde{Q}_3}$ and  $A_t$ within the constraints discussed above.}}
\vspace*{-0.4cm}
\end{center}
\vspace*{-0.5cm}
\end{figure*}

-- The $h \ra \gamma \gamma$ branching ratio is only mildly affected [less 
than $\sim 30\%$] by the contributions of the stop loops which can be of 
either sign. This is mainly due to the fact that the $W$ contribution to the
$h \gamma \gamma$ vertex is largely dominant in the decoupling limit. 

-- The $hgg$ coupling is always reduced compared to the SM case for large stop 
mixing and rather light stops can lead to a strong reduction in the rate of the 
inclusive production channel $gg \ra h$. The suppression factor can be as low 
as $1/10$ whereas a benchmark for discovery is about $\sim 1/2$ [although this 
benchmark depends slightly on the Higgs boson mass]. The suppression occurs 
for rather large, though not maximal, Higgs boson masses where the efficiencies
are better than for smaller Higgs masses.

-- For very heavy stops which should decouple from the $hgg$ and $h \gamma
\gamma$ vertices, the ratio $R_{gg \gamma \gamma}$ could be different from
unity since charginos could be also light and might give small contributions to
the $h \to \gamma \gamma$ decay width in the MSSM.  

\subsection{Associated Higgs production with stops} 

If the mixing in the stop sector is large, one of the top squarks can be 
rather light and at the same time, its couplings to the Higgs boson can 
be strongly enhanced. The associated production process $pp \to q\bar{q}/gg 
\to h \sto \sto$ might then be favored by phase space and the cross sections 
might be significantly large. This process is thus worth investigating at the 
LHC. 

In view of the implementation of the process $pp \to q\bar{q}/gg \to \sto \sto
h$ into an event generator, it is useful to give a ``model independent"
description of  the production cross section in the continuum, in terms of the
parameters $\msto, M_h$ [besides $\alpha_s$, $m_t$ etc...]. One can tabulate,
in a way which can be read externally, the cross section according to selected
values\footnote{Of course, in reality, the situation is slightly more
complicated since the two masses $\msto, M_h$ and the coupling $V_{\sto \sto h}$
depend on the mixing and are thus inter-related} of $M_h,\msto$ together with
the coupling $V_{\sto \sto h}$ [for simplicity and as a first step, one can
take the vertex $V_{\sto \sto h}$ such that $R_{\sto}=1$, i.e.  in the large
$M_A$ limit, no $\tilde{t}$ mixing and $D$--terms].  

The generator of partonic events for $pp \to \sto \sto h$ can be created by
using the package {\tt CompHEP} \cite{comphep} and may be down-loaded at this
http address \cite{slava-t1t1h}.  The events can be used as an external process
input in {\tt PYTHIA} \cite{pythia} or {\tt ISAJET} for further decay and
hadronization to simulate full events at the level of detectable particles. The
$\sto \sto h$ coupling is evaluated as a user's function thus allowing for an
interface with any SUSY model. The generator also includes, as an option, the
event generation of the SM process $pp \to q\bar{q}/gg \to t\bar{t}$+Higgs.

\begin{figure*}[htbp]
\vspace*{-4mm}
\begin{center}
\mbox{
\includegraphics[width=8cm,height=7.5cm]{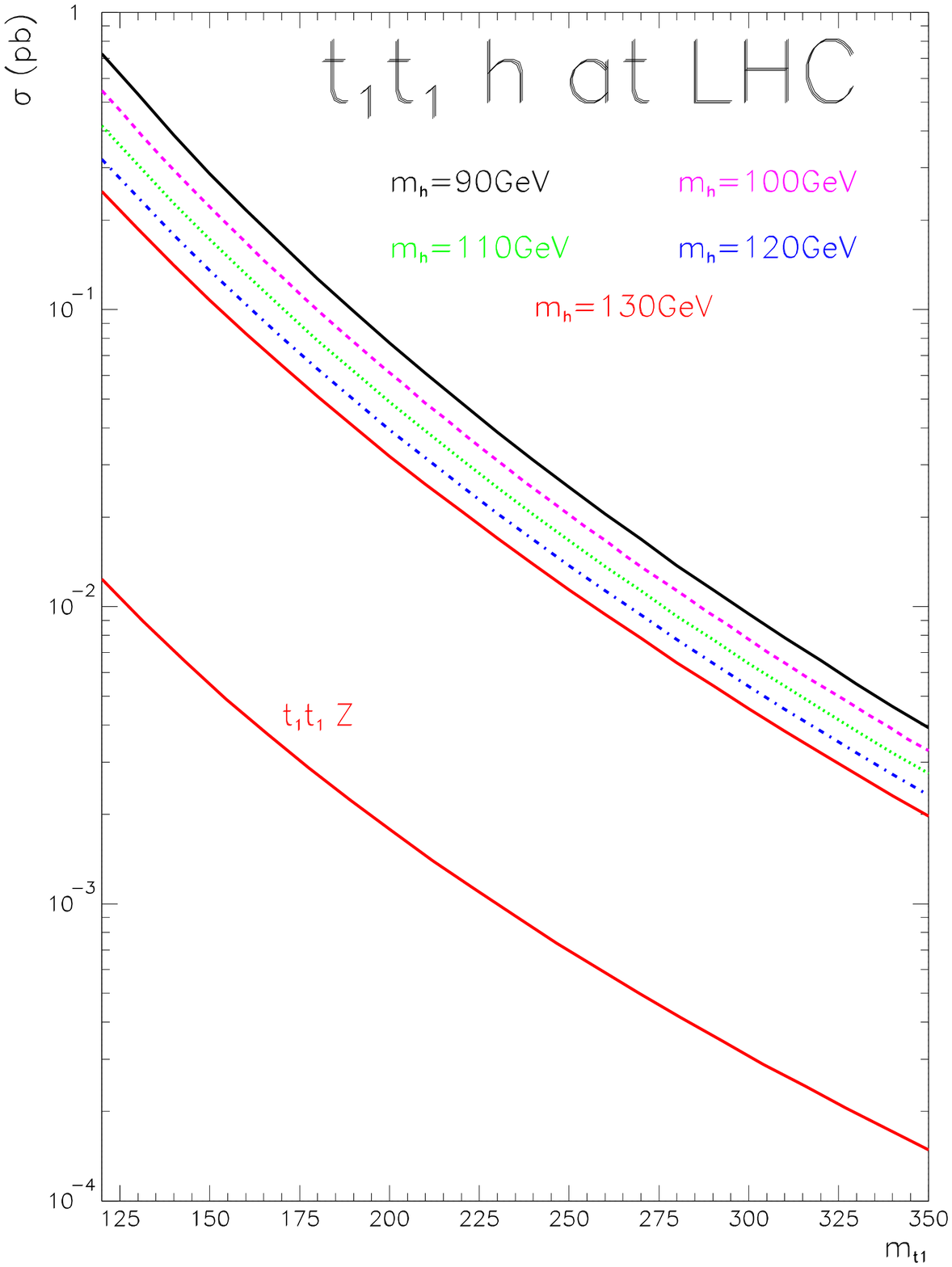}
\includegraphics[width=8cm,height=7.5cm]{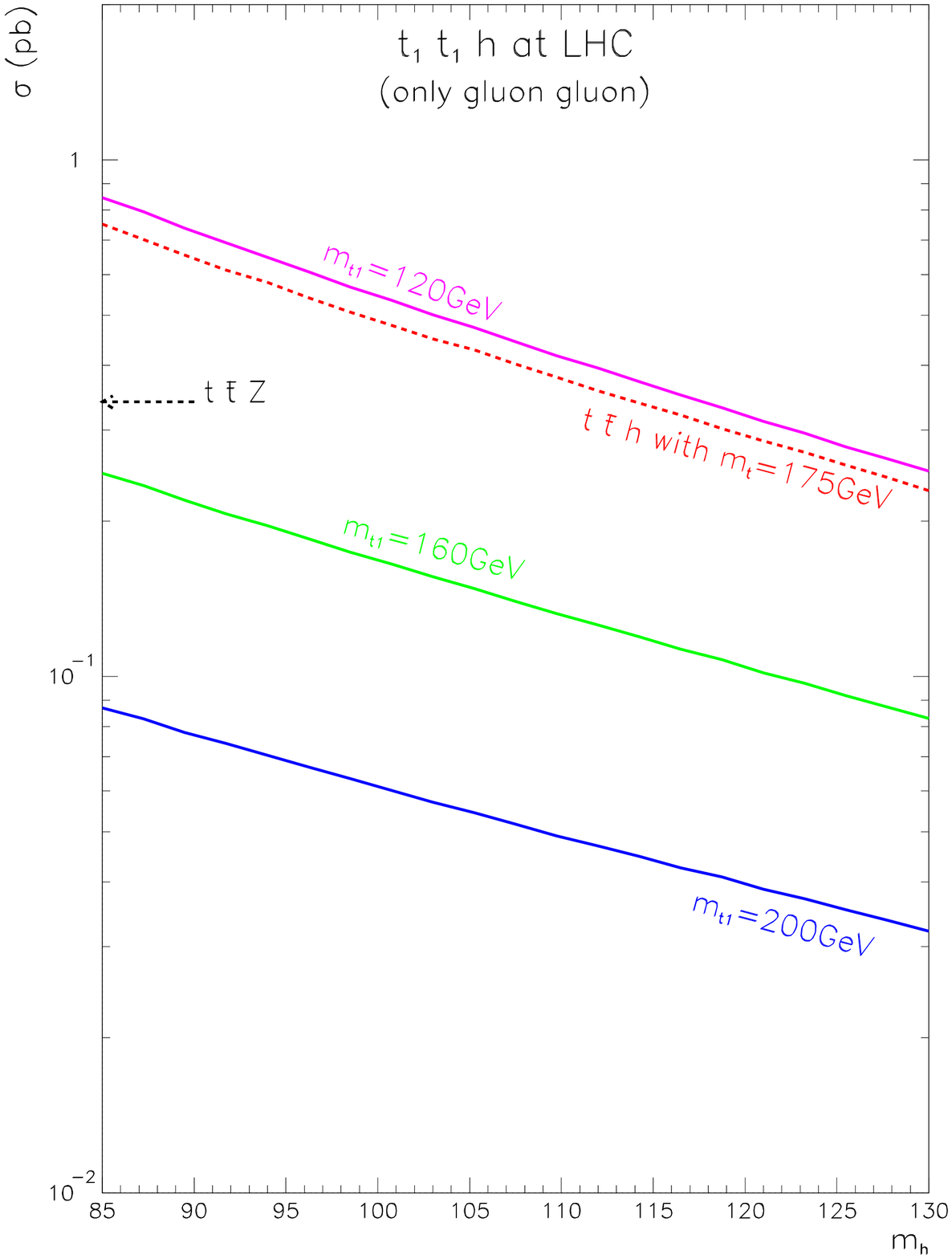}}
\caption{\label{t1t1h_mt1}{\small The cross section $pp \to \sto \sto h$ (and
similar processes) at the LHC as functions of $\msto$ (left) and $M_h$ (right) 
for a range of $M_h$ and $\msto$ values. See text for details.}} 
\end{center}
\vspace*{-6mm}
\end{figure*}

As an illustration, defined reference cross sections\footnote{Note that the
complete analytical expressions of the $pp \to gg/q\bar{q} \to \tilde{q}
\tilde{q}$+Higgs are given in Ref.~\cite{stophiggs_LHC}.} calculated with the
help of {\tt CompHEP} are displayed in Fig.~\ref{t1t1h_mt1}. The cross sections
are shown as functions of $M_h (m_{\sto})$ for given values of $m_{\sto}(M_h)$,
for a $\sto \sto h$ vertex in the limit of large $M_A$, no mixing and no
D--terms, as discussed above. Also shown are the cross sections for the
processes $pp \to t\bar{t}h , t \bar{t}Z $ [where only the dominating
contributions of the $gg$ initiated subprocesses are included] and $\sto \sto
Z$ [where the vertex has been computed with $\cos^2 \theta_{\stop}= 1/2$, i.e.
maximal mixing, and has to be rescaled by a factor $(\cos^2\theta_{\stop}/2
-2/3s_W^2)$ for other mixing values,]. We have used the CTEQ4 structure
functions \cite{CTEQ} with a scale set at the invariant mass of the subprocess.

As can be seen, the $pp \to \sto \sto h$ cross section can be large for small 
values of the stop and the Higgs masses, but drops precipitously with $\msto$ 
and to a lesser extent with $M_h$. The cross section is more than order of 
magnitude larger than $\sigma( pp \to \sto \sto Z)$ and can exceed the one for 
the SM--like process $pp \to t \bar{t} h$ for strong enough mixing 
$R_{\tilde t} \gg 1$ and  light $\tilde{t}_1$.  

\begin{figure*}[htbp]
\vspace*{-.8cm}
\begin{center}
\includegraphics[width=15cm,height=8.9cm]{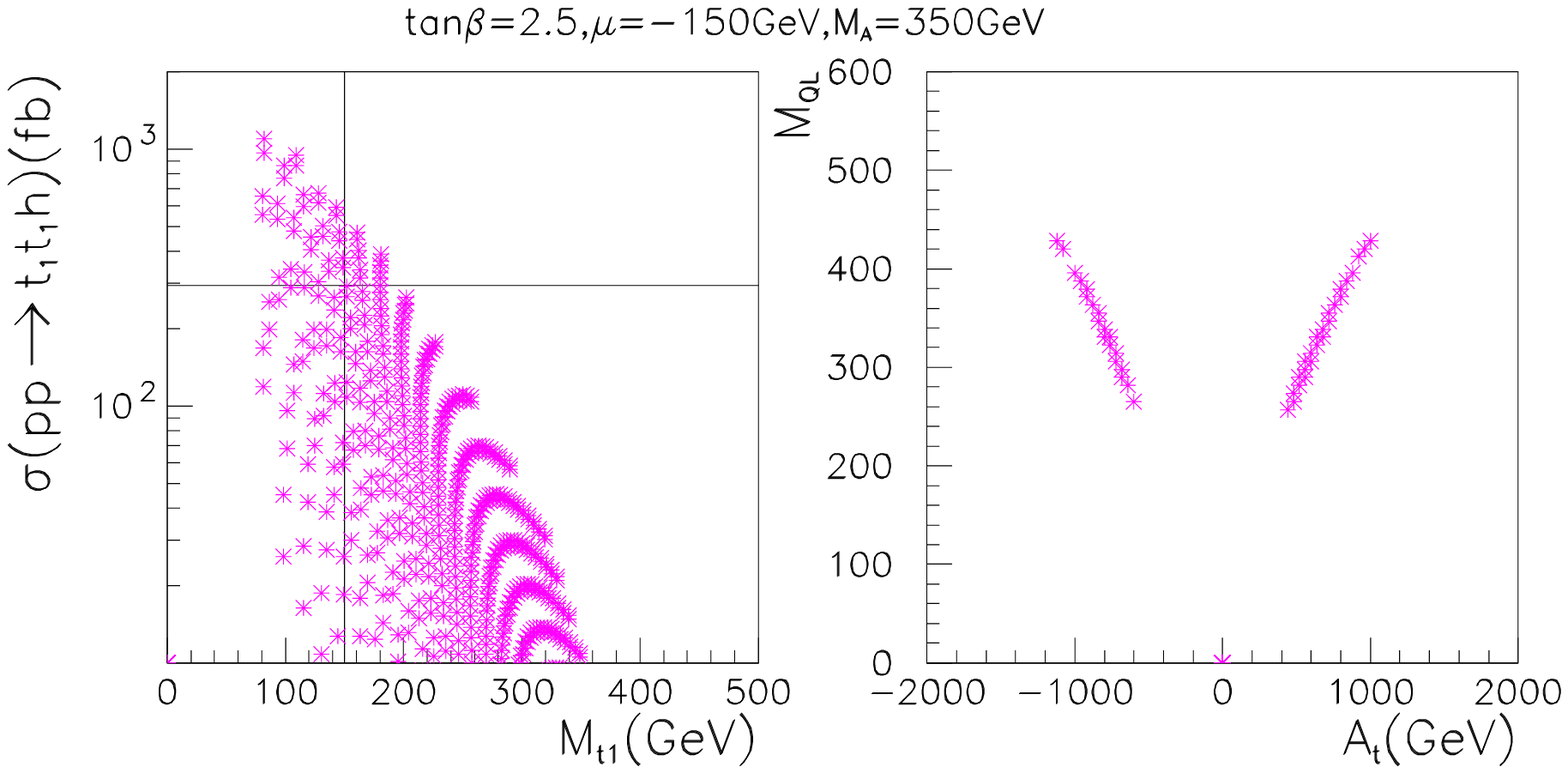}
\includegraphics[width=15cm,height=8.9cm]{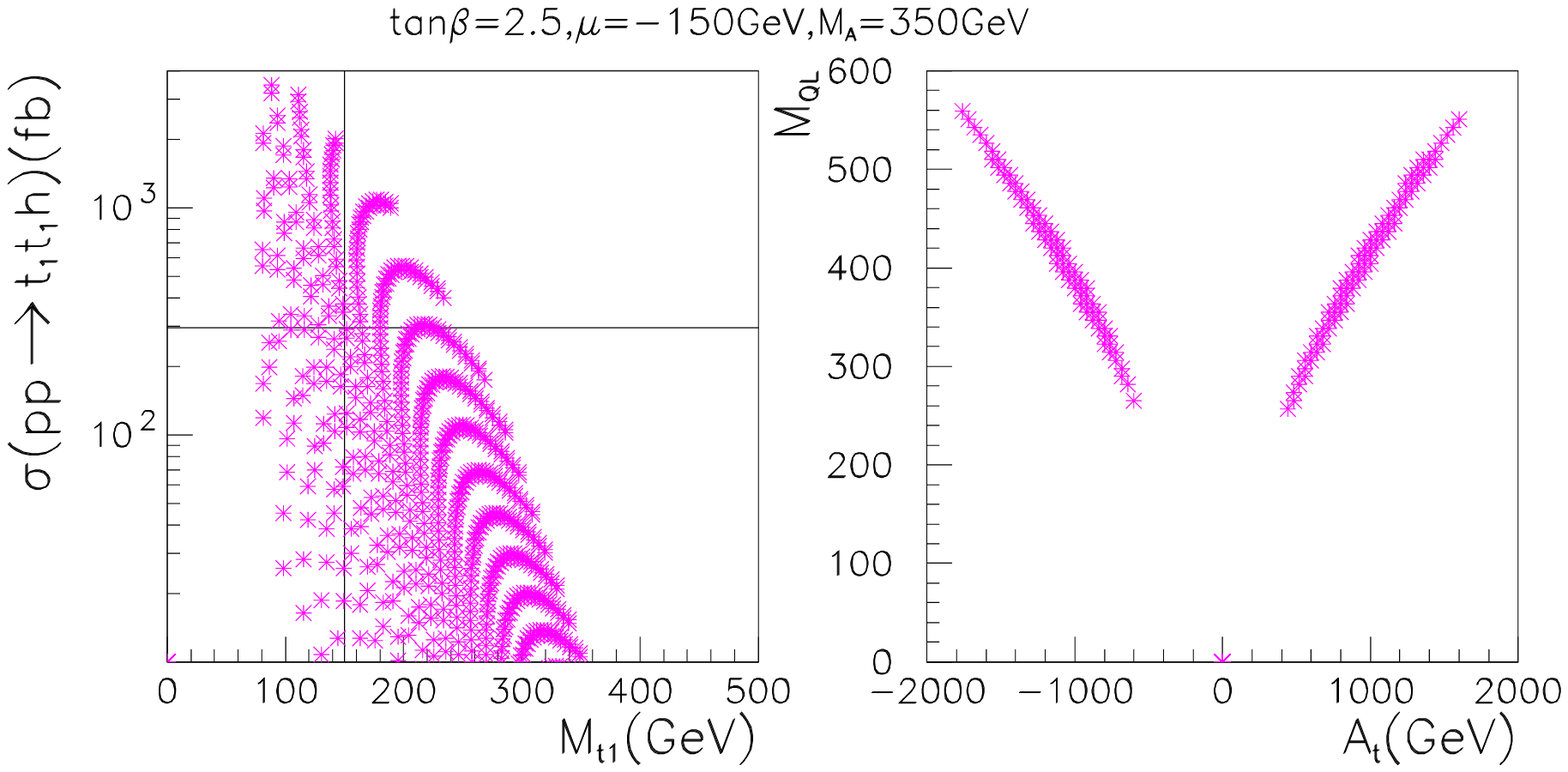}
\vspace*{-3mm}
\caption{\label{maxisigma}{\small
The $pp \to \sto \sto h$ cross sections at the LHC 
as a function of $\msto$ (left) and a scan on $m_{\tilde{Q}}$ and $A_t$ (right).
Shown also are points that pass $\sigma_{\sto \sto h}>300$ fb and $\msto \ge 
150$ GeV, imposing $\Delta\rho\le.0013$ (top) or $\Delta\rho\le.0026$ 
(bottom).}}
\end{center}
\vspace*{-8mm}
\end{figure*}

If one takes the value $\sigma(\sto \sto h) >300$ fb as a benchmark cross
section value for observing this process at the LHC, and using the
constraint on the maximum values of the $\sto \sto h$ coupling, values of
$\msto \ge 250$ GeV are hardly accessible at the LHC. This is shown in
Fig.~\ref{maxisigma} where the $pp \to \sto \sto h$ cross section is shown as
a function of $\msto$ taking all soft squark masses equal for $\tgb=2.5$ 
and imposing $\Delta\rho\le.0013$. A scan on the common soft breaking scalar 
mass and $A_t$ has also been performed; shown are points that pass the
criteria $\sigma_{\sto \sto h}>300$ fb and for which $\msto \ge 150$ GeV.
Larger values of the stop mass can be reached if ones allows a $2\sigma$ 
variation on the $\Delta \rho$ constraint, as shown in the figures
at the bottom.
 
\subsection{Comparison of inclusive and associated production processes}

\begin{figure*}[htbp]
\vspace*{-.4cm} 
\begin{center}
\includegraphics[width=12cm,height=5.5cm]{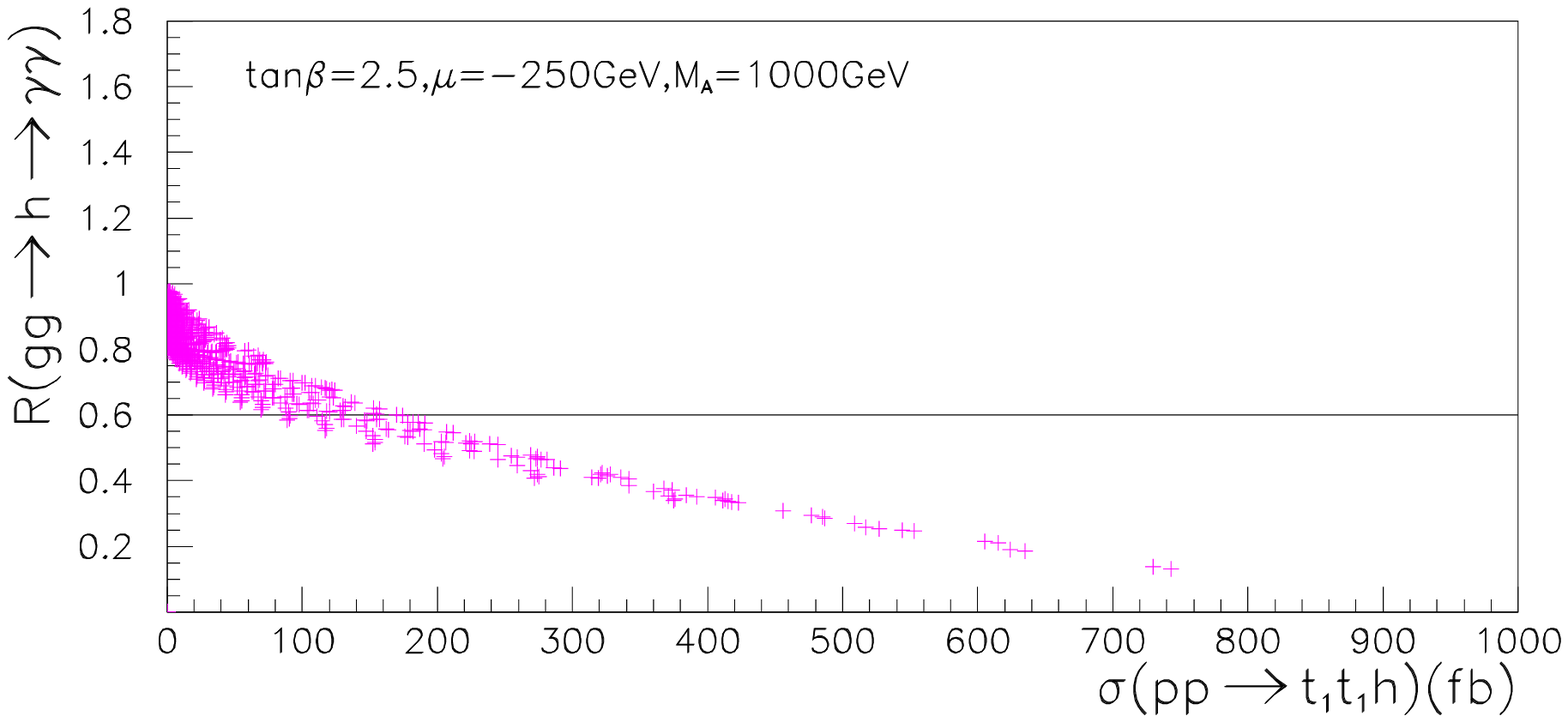}
\vspace*{-7mm}
\caption{\label{h_equal}{\small The cross sections for the inclusive 
and associated Higgs production at the LHC for $\tilde{m}_{\tilde{Q}_3} \simeq 
\tilde{m}_{\tilde{U}_{3R}}$ at $\tgb=2.5$ and large $M_A$. Figures are scanned 
over $\tilde{m}_{\tilde{Q}_3}$ and  $A_t$.}}
\end{center}
\vspace*{-0.5cm}
\end{figure*}

Let us now make a global discussion on the stop effects in both type of
processes for Higgs boson production, $gg \ra h \to \gamma \gamma$ and 
$pp \to \sto \sto h$. The two cross sections are shown in Fig.~\ref{h_equal}
in the decoupling limit for $\tgb=2.5$ and equal soft breaking scalar masses.
As can be seen, the suppression of the rate in the inclusive production channel
is compensated by a rate increase in the associated production channel.
When the suppression factor in the inclusive production is below $0.5$ 
making discovery in this channel difficult, the cross section for the process
$\sto \sto h$ is above $200$ fb. As discussed previously, a benchmark value 
for the cross section allowing the discovery of the Higgs boson in the $pp \to 
\sto \sto h$ channel has been estimated to $\sigma \sim 300$ fb. Therefore for 
some values of the parameters, neither the inclusive nor the $pp \to \sto \sto
h$ channels can be accessed. However, one also sees that for these same points
one can without difficulty use the usual $ pp \to Wh/Zh, t \bar t h$ search 
modes. 

Therefore with the remark that the process $ pp \to t \bar t h$ [with $h \ra b
\bar b$] should allow for Higgs boson discovery at the LHC within this
scenario, one should salvage the detection the $h$ boson with the bonus that
the stop should also be observed. Even though one may have to wait for the
higher luminosity stage, the scenario with light stops and large couplings
offer much better prospects than previously thought.

\begin{figure*}[htbp]
\vspace*{-1.2cm}
\begin{center}
\includegraphics[width=15cm,height=10cm]{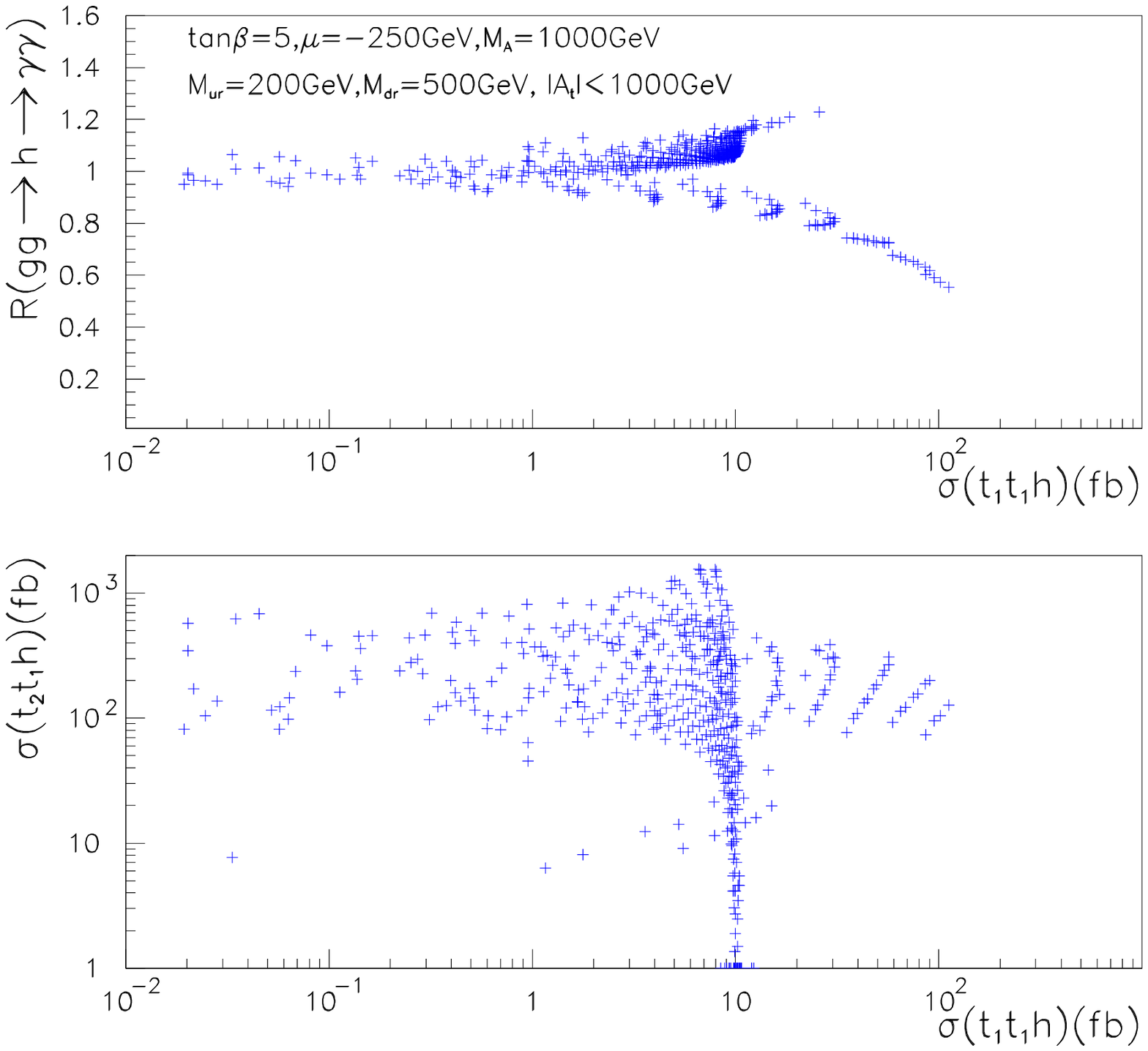}
\caption{\label{notequalmasses}{\small 
Higgs boson production rates at the LHC for unequal soft breaking scalar masses
in the decoupling limit $M_A=$1TeV and $\sigma(pp \to \stt \sto h)$ {\it vs}  
$\sigma(pp \to \sto \sto h)$.}}
\end{center}
\vspace*{-0.8cm}
\end{figure*}

The assumption of an equal value for the soft scalar masses at the weak scale
is rather unnatural [see later in mSUGRA] and could be relaxed. To illustrate
the fact that large suppression factors in the inclusive production channel,
though not as dramatic as in the previous case, still occur we show in
Fig.~\ref{notequalmasses} typical $R$ ratios for unequal values of the soft
masses. What is most interesting is that, as soon as $\sin
(2\theta_{\tilde{t}}) \neq 1$, the non--diagonal decay channel $\stt \ra \sto
h$ opens up and can have an appreciable branching ratio. This can be seen by
inspection of the $ V_{\sto \stt h}$ coupling, for which the leading component
is proportional to: $V_{\sto \stt h} \propto g/(4 \mw) \sin 4 \theta_{\stop} \,
(\mstosq-\msttsq)$.  Considering that if the $\stt$ mass is not excessively
large, $\stt$ is produced in abundance and this cascade decay can provide more
Higgs bosons than through the continuum $ p p \to \sto \sto h$ production.  

\begin{figure*}[htbp]
\vspace*{-1.3cm}
\begin{center}
\includegraphics[width=15cm,height=10cm]{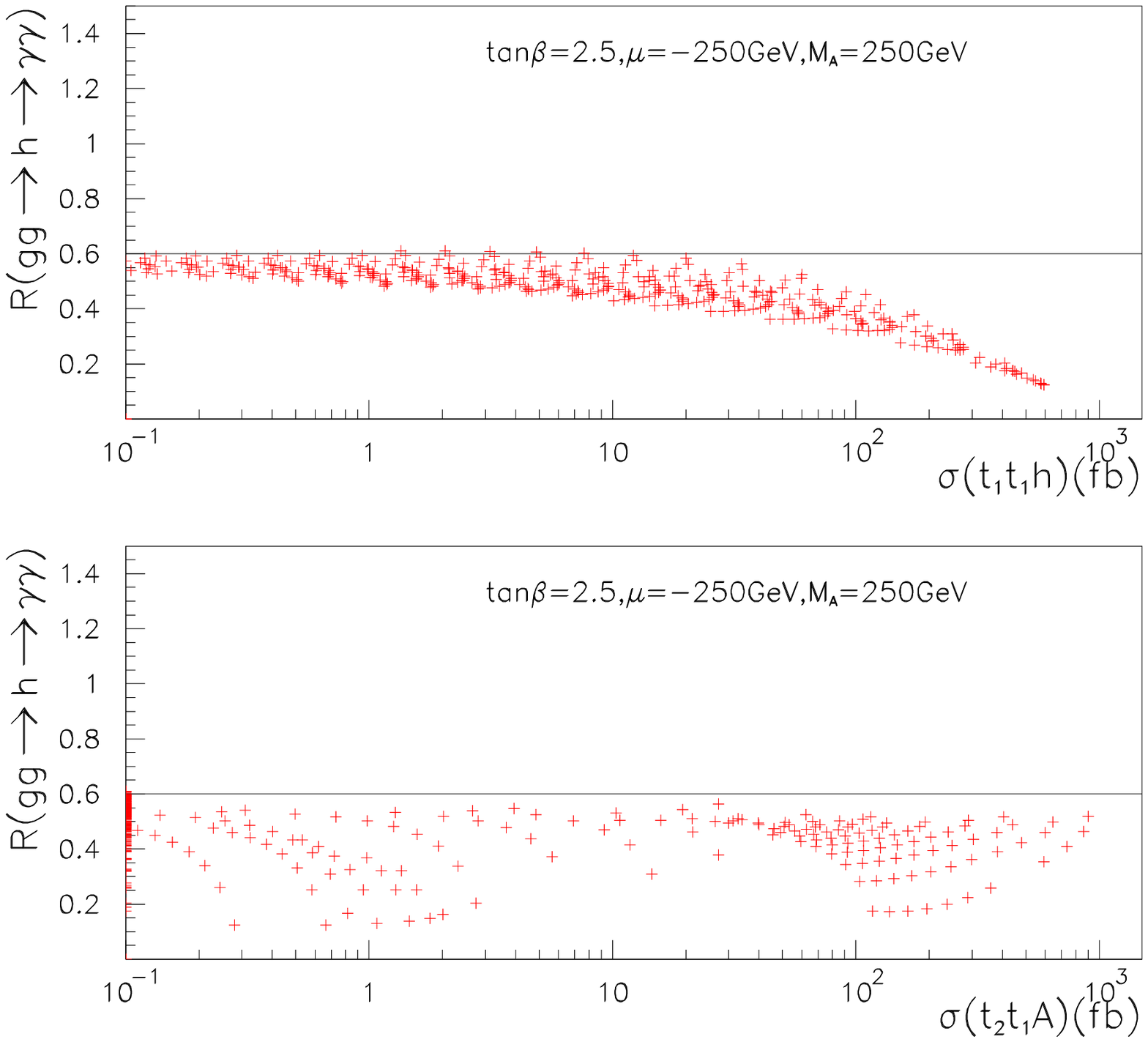}
\caption{\label{smallma}{\small 
Higgs boson production and decay rates at the LHC for low values of $M_A$.}} 
\end{center}
\vspace*{-0.8cm}
\end{figure*}

Perhaps even more interesting, is the case when $M_A$ is not too large. For
large values of $A_t$, and even when $\sin 2\theta_{\stop} \simeq 1$, one can
have a large decay rate $\stt \ra \sto A$ since the $ A \sto \stt$ coupling can
be large $V_{\sto \stt A} \propto g m_t/(2\mw) (A_t/\tgb -\mu)$, as shown in
Fig.~\ref{smallma}. This coupling is generally larger than the $\stt \sto H$
coupling and hence, within these scenarii, the decay $\stt \to \sto A$ is most
likely to occur than the decay into the heavier $H$ boson, $ \stt \to H\sto$.  

Finally, let us make a few comments on the case of the minimal Supergravity 
model
\cite{mSUGRA}, where the only input parameters are the universal scalar mass
$m_0$, the universal gaugino mass parameter $m_{1/2}$, the trilinear coupling
$A_0$, $\tgb$ and the sign of the $\mu$ parameter. The parameters $m_{0},
m_{1/2}$ and $A_0$ are chosen at the GUT scale and their evolution down to the
weak scale is given by the RGE's \cite{rge}. Proper breaking of the electroweak
symmetry is also assumed, which fixes the parameter $|\mu|$. In what follows,
the RGE's and the proper EW symmetry breaking are solved using the program {\tt
SUSPECT} \cite{SUSPECT}.  

In the mSUGRA case the cross section can be as large as in the case of the
unconstrained MSSM, but in a relatively smaller area of the SUSY parameter
space.  This is essentially due to the fact that it is generically very
difficult to have almost degenerate $\tilde{t}_L$ and $\tilde{t}_R$ in mSUGRA, 
so that the stop mixing angle which is controlled by the
ratio $\tilde A_t/(\tilde m^2_{\tilde{t}_L} -\tilde m^2_{\tilde{t}_R})$ can
become large only for very large $\tilde A_t$.  Moreover in the RG evolution
\cite{rge} $\vert A_t\vert$ tends to decrease when the energy scale is
decreasing from GUT to low-energy. This makes a large $A_t$ value at low energy
less likely, since $A_0 =A_t({\rm GUT})$ would have to be even larger, which
may conflict with e.g. the CCB constraints.  The only way to have an increasing
$\vert A_t\vert $ when running down to low energy is if $A_0 <0 $  with $A_0$ 
small enough.  This requires a large $m_{1/2}$ value, which implies not too
small $m_{\tilde t_1}$. 

Thus the mixing in the stop sector is, in general, not as large as in the 
unconstrained MSSM 
and the $\sto \sto h$ coupling for instance is, in general, smaller than in the
previous case. This implies that the rate for the inclusive production and 
detection channel $gg \ra h \ra \gamma \gamma$ [in the decoupling limit] is not
as dramatically different from the rate in the SM, as it can be in the 
unconstrained MSSM. Furthermore, the milder mixing results in a smaller cross 
sections for the process $pp \to \sto \sto h$ as is shown in Fig.~8 for
LHC energies. However, for large $\tgb$ values, the pseudoscalar $A$ boson 
tends to be rather light in mSUGRA, opening the possibility for the 
decay $\tilde{t}_2 \to \tilde{t}_1 A$ to occur with an appreciable rate as 
shown in Fig.~8b. 

Note that one should also easily observe the pseudoscalar $A$ boson in the 
loop mediated process $gg \to A$ since the rate is strongly enhanced for large 
$\tgb$ and, because of CP--invariance, light stop [or sbottom] loops cannot 
contribute to the  process.
 
\begin{figure*}[htbp]
\vspace*{-8mm}
\begin{center}
\mbox{
\includegraphics[width=8cm,height=8cm]{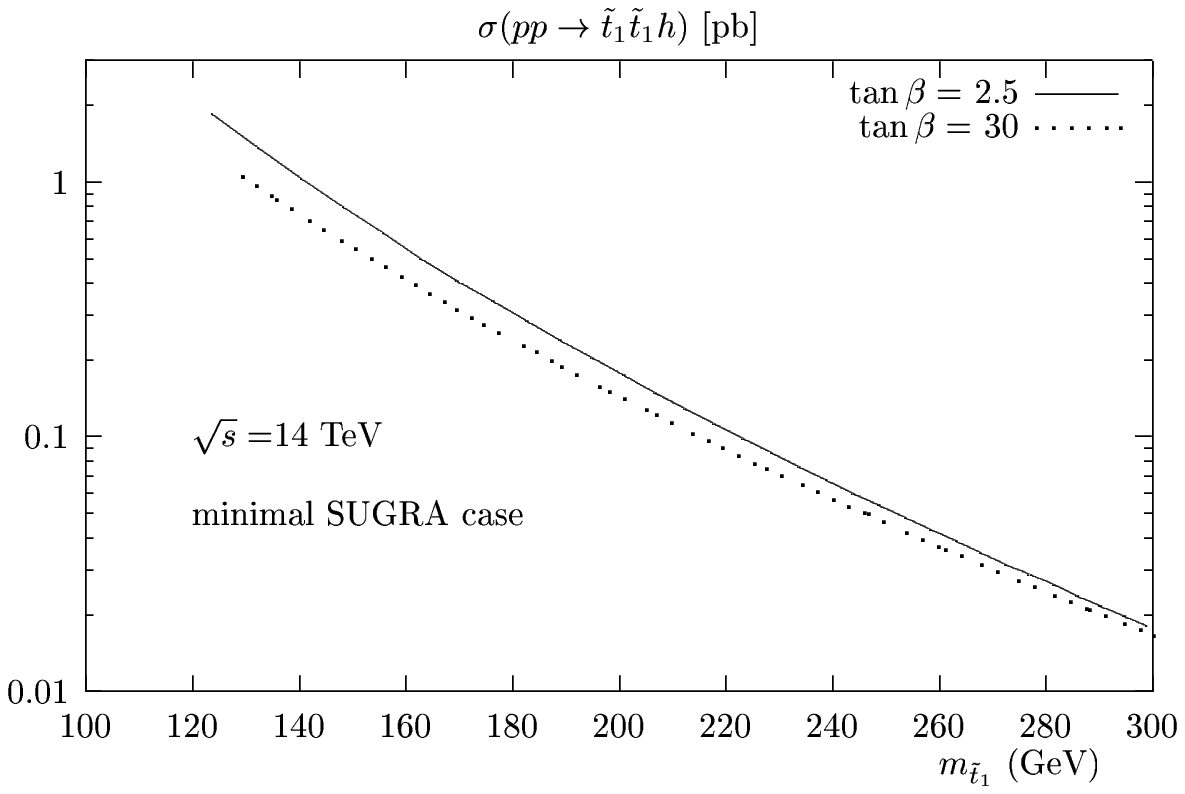}
\includegraphics[width=7.2cm,height=8cm,angle=90]{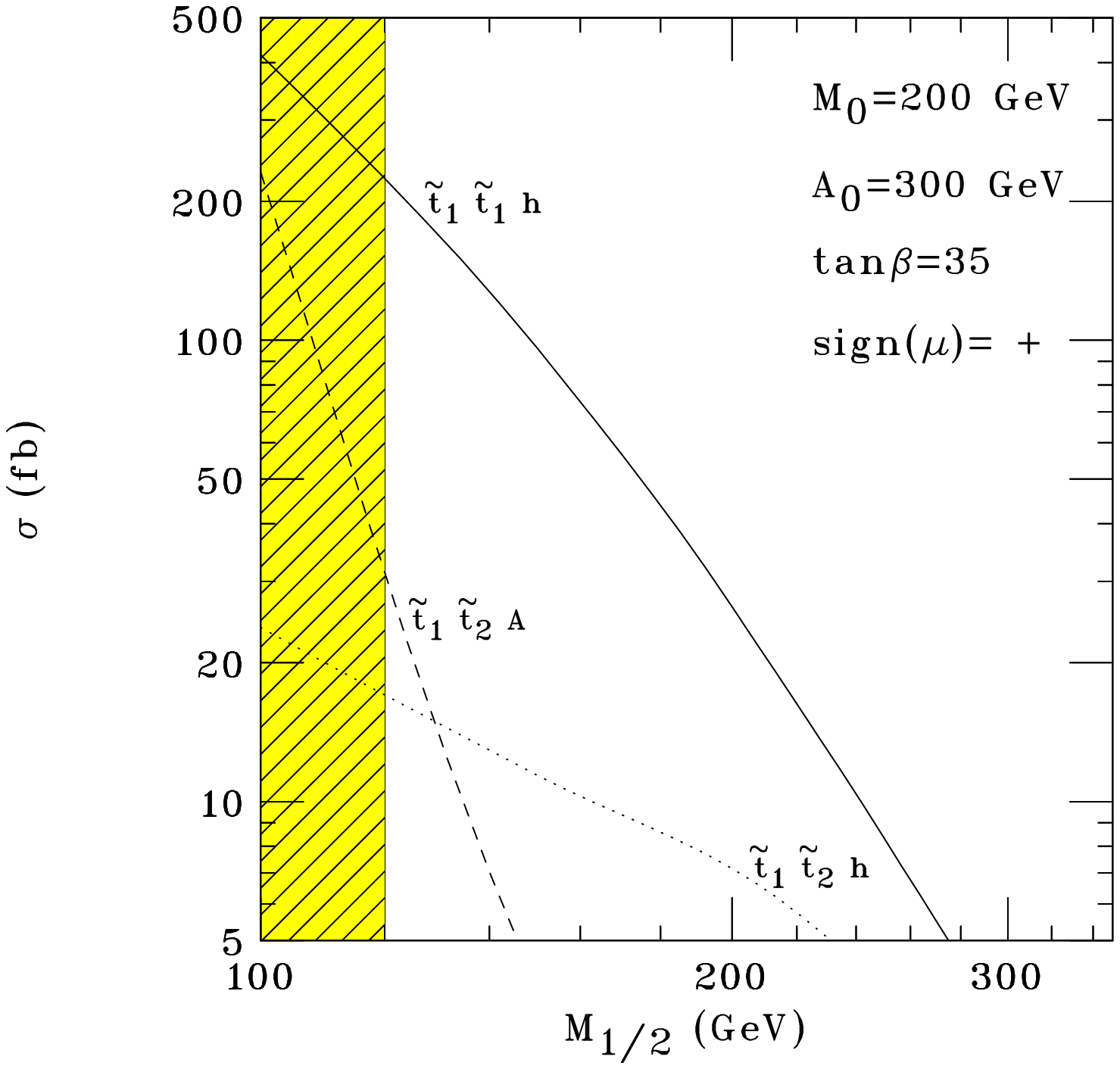}}
\caption{\label{stophiggs-msugra}{\small Cross sections in mSUGRA at LHC:
{\bf a)} $\sigma(pp \to \sto \sto h)$ for $m_{1/2}=0.3$ TeV, $A_0=2$ TeV, 
$\tgb=2.5, 30$. {\bf b)} $\sigma(pp \to \sto \sto h, \sto \stt h, \sto \stt 
A)$ for $m_{0}=0.2$ TeV, $A_0=0.3$ TeV and $\tgb=35$.
For the spectrum, SUSPECT is used in {\bf a)} and ISAJET in {\bf b)}.}}
\end{center}
\vspace*{-1.3cm}
\end{figure*}

\setcounter{figure}{0}
\setcounter{table}{0}
\setcounter{section}{0}
\setcounter{equation}{0}
\newpage

\newcommand{\br}{\begin{eqnarray}}
\newcommand{\er}{\end{eqnarray}}
\newcommand{\ba}{\begin{array}}
\newcommand{\ea}{\end{array}}
\newcommand{\bi}{\begin{itemize}}
\newcommand{\ei}{\end{itemize}}
\newcommand{\bn}{\begin{enumerate}}
\newcommand{\en}{\end{enumerate}}
\newcommand{\bc}{\begin{center}}
\newcommand{\ec}{\end{center}}
\newcommand{\ul}{\underline}
\newcommand{\ol}{\overline}
\def\epem{\ifmmode{e^+ e^-} \else{$e^+ e^-$} \fi}
\newcommand{\eeww}{$e^+e^-\rightarrow W^+ W^-$}
\newcommand{\qqQQ}{$q_1\bar q_2 Q_3\bar Q_4$}
\newcommand{\eeqqQQ}{$e^+e^-\rightarrow q_1\bar q_2 Q_3\bar Q_4$}
\newcommand{\eewwqqqq}{$e^+e^-\rightarrow W^+ W^-\ar q\bar q Q\bar Q$}
\newcommand{\eeqqgg}{$e^+e^-\rightarrow q\bar q gg$}
\newcommand{\eeqloop}{$e^+e^-\rightarrow q\bar q gg$ via quark loops}
\newcommand{\eeqqqq}{$e^+e^-\rightarrow q\bar q Q\bar Q$}
\newcommand{\eewwjjjj}{$e^+e^-\rightarrow W^+ W^-\rightarrow 4~{\rm{jet}}$}
\newcommand{\eeqqggjjjj}{$e^+e^-\rightarrow q\bar 
q gg\rightarrow 4~{\rm{jet}}$}
\newcommand{\eeqloopjjjj}{$e^+e^-\rightarrow q\bar 
q gg\rightarrow 4~{\rm{jet}}$ via quark loops}
\newcommand{\eeqqqqjjjj}{$e^+e^-\rightarrow q\bar q Q\bar Q\rightarrow
4~{\rm{jet}}$}
\newcommand{\eejjjj}{$e^+e^-\rightarrow 4~{\rm{jet}}$}
\newcommand{\jjjj}{$4~{\rm{jet}}$}
\newcommand{\qqbar}{$q\bar q$}
\newcommand{\ww}{$W^+W^-$}
\newcommand{\ar}{\rightarrow}
\newcommand{\sm}{${\cal {SM}}$}
\newcommand{\Dir}{\kern -6.4pt\Big{/}}
\newcommand{\Dirin}{\kern -10.4pt\Big{/}\kern 4.4pt}
\newcommand{\DDir}{\kern -7.6pt\Big{/}}
\newcommand{\DGir}{\kern -6.0pt\Big{/}}
\newcommand{\wwqqqq}{$W^+ W^-\ar q\bar q Q\bar Q$}
\newcommand{\qqgg}{$q\bar q gg$}
\newcommand{\qloop}{$q\bar q gg$ via quark loops}
\newcommand{\qqqq}{$q\bar q Q\bar Q$}
\newcommand{\ord}{{\cal O}}
\newcommand{\Ecm}{E_{\mathrm{cm}}}

\def\l{\left\langle}
\def\r{\right\rangle}
\def\aem{\alpha_{\rm em}}
\def\as{\alpha_{\rm s}}
\def\MW{m_{W^\pm}}
\def\MZ{m_{Z}}
\def\ycut{y_{\rm cut}}
\def\Ord{\lower .7ex\hbox{$\;\stackrel{\textstyle <}{\sim}\;$}}
\def\OOrd{\lower .7ex\hbox{$\;\stackrel{\textstyle >}{\sim}\;$}}

\begin{center}
{\large\sc {\bf
Double Higgs production at TeV Colliders \\[0.35cm]
in the Minimal Supersymmetric Standard Model}}

\vspace{0.5cm}

{\sc
R. Lafaye, D.J. Miller, M. M\"uhlleitner and  S.~Moretti}
\end{center} 

\begin{abstract}
  The reconstruction of the Higgs potential in the Minimal
  Supersymmetric Standard Model (MSSM) requires the measurement of the
  trilinear Higgs self-couplings.  The `double Higgs production'
  subgroup has been investigating the possibility of detecting
  signatures of processes carrying a dependence on these vertices at
  the Large Hadron Collider (LHC) and future Linear Colliders (LCs).
  As reference reactions, we have chosen $gg\to hh$ and $e^+e^-\to h h
  Z$, respectively, where $h$ is the lightest of the MSSM Higgs
  bosons.  In both cases, the $Hhh$ interaction is involved.  For
  $m_H\OOrd2m_h$, the two reactions are resonant in the $H\to hh$
  mode, providing cross sections which are detectable at both
  accelerators and strongly sensitive to the strength of the trilinear
  coupling involved. We explore this mass regime of the MSSM in the
  $h\to b\bar b$ decay channel, also accounting for irreducible
  background effects.
\end{abstract}

\section{Introduction}
\label{sec_intro}

Considerable attention has been devoted to double Higgs boson
production at future $e^+e^-$ and hadron colliders, both in the
Standard Model (SM) and the MSSM \cite{review,pp,ee}.  For the SM,
detailed signal-to-background studies already exist for a LC
environment \cite{ee}, for both `reducible' and `irreducible'
backgrounds \cite{Lutz,noi}, which have assessed the feasibility of
experimental analyses.  At the LHC, since here the typical SM signal
cross sections are of the order of 10 fb \cite{pp}, high integrated
luminosities would be needed to generate a statistically large enough
sample of double Higgs events. These would be further obscured by an
overwhelming background, making their selection and analysis in a
hadronic environment extremely difficult.  Thus, in this contribution we will
concentrate only on the case of the MSSM.

In the Supersymmetric (SUSY) scenario, the phenomenological potential
of these reactions is two-fold. Firstly, in some specific cases, they
can furnish new discovery channels for Higgs bosons.  Secondly, they
are all dependent upon several triple Higgs self-couplings of the theory,
which can then be tested by comparing theoretical predictions with
experimental measurements. This is the first step in the
reconstruction of the Higgs potential itself\footnote{The
  determination of the quartic self-interactions is also required, but
  appears out of reach for some time: see Refs.~\cite{pp,ee} for some
  cross sections of triple Higgs production.}.

The Higgs Working Group (WG) has focused much of its attention in assessing
the viability of these reactions at future TeV colliders. However, the
number of such processes is very large both at the LHC and a LC
\cite{pp,ee}, so only a few `reference' reactions could be studied in
the context of this Workshop.  Work is in progress for the
longer term, which aims to cover most of the double Higgs production
and decay phenomenology at both accelerators~\cite{more}.

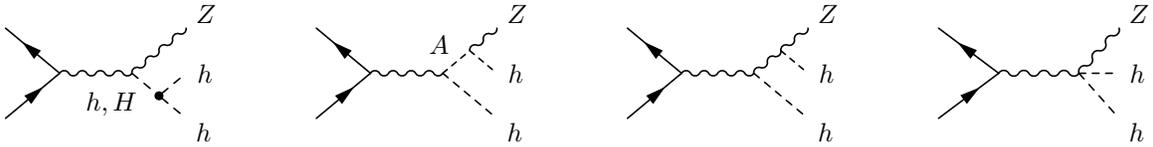
\begin{figure}[!t]
\begin{flushleft}
\underline{$gg$ to double Higgs fusion at the LHC: $gg\to hh$}
\\[1.5\baselineskip]
{\footnotesize
\unitlength1mm
\hspace{5mm}
\begin{fmfshrink}{0.7}
\begin{fmfgraph*}(30,12)
  \fmfstraight
  \fmfleftn{i}{2} \fmfrightn{o}{2}
  \fmflabel{$g$}{i1}  \fmflabel{$g$}{i2}
  \fmf{gluon,tens=2/3}{i1,v1} \fmf{phantom}{v1,v2,v3,o1}
  \fmf{gluon,tens=2/3}{w1,i2} \fmf{phantom}{w1,w2,w3,o2}
  \fmffreeze
  \fmf{fermion}{w1,x2,v1}
  \fmf{dashes, lab=$h,,H$}{x2,x3}
  \fmf{dashes}{o1,x3,o2}
  \fmffreeze
  \fmf{fermion,label=$t,,b$,label.s=left}{v1,w1}
  \fmflabel{$h$}{o1}  \fmflabel{$h$}{o2}
  \fmfdot{x3}
\end{fmfgraph*}
\hspace{15mm}
\begin{fmfgraph*}(30,12)
  \fmfstraight
  \fmfleftn{i}{2} \fmfrightn{o}{2}
  \fmf{gluon}{i1,v1} \fmf{phantom}{v1,v3} \fmf{dashes}{v3,o1}
  \fmf{gluon}{w1,i2} \fmf{phantom}{w1,w3} \fmf{dashes}{w3,o2}
  \fmffreeze
  \fmf{fermion}{v1,w1,w3,v3,v1}
  \fmflabel{$h$}{o1}  \fmflabel{$h$}{o2}
\end{fmfgraph*}
\end{fmfshrink}
}
\\[2\baselineskip]
\underline{double Higgs-strahlung at a LC: $e^+e^-\to hhZ$}
\\[1.5\baselineskip]
{\footnotesize
\unitlength1mm
\hspace{5mm}
\begin{fmfshrink}{0.7}
\begin{fmfgraph*}(24,12)
  \fmfstraight
  \fmfleftn{i}{3} \fmfrightn{o}{3}
  \fmf{fermion}{i1,v1,i3}
  \fmf{boson,tens=3/2}{v1,v2}
  \fmf{boson}{v2,o3} \fmflabel{$Z$}{o3}
  \fmf{phantom}{v2,o1}
  \fmffreeze
  \fmf{dashes,lab=$h,,H$,lab.s=right}{v2,v3} \fmf{dashes}{v3,o1}
  \fmffreeze
  \fmf{dashes}{v3,o2} 
  \fmflabel{$h$}{o2} \fmflabel{$h$}{o1}
  \fmfdot{v3}
\end{fmfgraph*}
\hspace{15mm}
\begin{fmfgraph*}(24,12)
  \fmfstraight
  \fmfleftn{i}{3} \fmfrightn{o}{3}
  \fmf{fermion}{i1,v1,i3}
  \fmf{boson,tens=3/2}{v1,v2}
  \fmf{dashes}{v2,o1} \fmflabel{$H$}{o1}
  \fmf{phantom}{v2,o3}
  \fmffreeze
  \fmf{dashes,lab=$A$,lab.s=left}{v2,v3} 
  \fmf{boson}{v3,o3} \fmflabel{$Z$}{o3}
  \fmffreeze
  \fmf{dashes}{v3,o2} 
  \fmflabel{$h$}{o2} \fmflabel{$h$}{o1}
\end{fmfgraph*}
\hspace{15mm}
\begin{fmfgraph*}(24,12)
  \fmfstraight
  \fmfleftn{i}{3} \fmfrightn{o}{3}
  \fmf{fermion}{i1,v1,i3}
  \fmf{boson,tens=3/2}{v1,v2}
  \fmf{dashes}{v2,o1} \fmflabel{$H$}{o1}
  \fmf{phantom}{v2,o3}
  \fmffreeze
  \fmf{boson}{v2,v3,o3} \fmflabel{$Z$}{o3}
  \fmffreeze
  \fmf{dashes}{v3,o2} 
  \fmflabel{$h$}{o2} \fmflabel{$h$}{o1}
\end{fmfgraph*}
\hspace{15mm}
\begin{fmfgraph*}(24,12)
  \fmfstraight
  \fmfleftn{i}{3} \fmfrightn{o}{3}
  \fmf{fermion}{i1,v1,i3}
  \fmf{boson,tens=3/2}{v1,v2}
  \fmf{dashes}{v2,o1} \fmflabel{$h$}{o1}
  \fmf{dashes}{v2,o2} \fmflabel{$h$}{o2}
  \fmf{boson}{v2,o3} \fmflabel{$Z$}{o3}
\end{fmfgraph*}
\end{fmfshrink}
}
\end{flushleft}
\caption{
Diagrams contributing to $gg\to hh$ (top) and $e^+e^-\to hhZ$ 
(bottom) in the MSSM. 
}
\label{fig:graphs}
\end{figure}

These reference reactions were chosen to be the gluon--fusion mechanism, 
$gg\to hh$, for the LHC (see top of Fig.~\ref{fig:graphs}) and the 
Higgs--strahlung process, $e^+e^-\to h h Z$, for the LC
(see bottom of Fig.~\ref{fig:graphs}), where $h$ is the lightest of the
MSSM scalar Higgs bosons.  The reason for this preference is simple.
Firstly, a stable upper limit exists on the value of $m_h$, of the
order of 130 GeV, now at two-loop level \cite{twoloop}, so that
its detection is potentially well within the reach of both the LHC and
a LC.  In contrast, the mass of all other Higgs bosons of the MSSM may
vary from the electroweak (EW) scale, ${\cal O}(m_Z)$, up to the TeV
region.  In addition, as noted in Ref.~\cite{pp}, the multi-$b$ final
state in $gg\to hh\to b\bar b b\bar b$, with two resonances and large
transverse momenta, may be exploited in the search for the $h$ scalar
in the large $\tan\beta$ and moderate $m_A$ region. This is a corner
of the MSSM parameter space that has so far eluded the scope of the
standard Higgs production and decay modes \cite{standard}.  (The
symbol $A$ here denotes the pseudoscalar Higgs boson of the MSSM, and
we reserve the notation $H$ for the heaviest scalar Higgs state of the
model.)
However, this paper will not investigate the LHC discovery potential
in this mode, given the very sophisticated treatment of the background
(well beyond the scope of this note) required by the assumption
that no $h$ scalar state has been previously discovered (see below).
This will be done in Ref.~\cite{more}.  Furthermore, the $gg\to hh$
and $e^+e^-\to hhZ$ modes largely dominate double Higgs production
\cite{pp,ee}, at least for centre-of-mass (CM) energies of 14 TeV
at the LHC and 500 GeV in the case of a LC, the default values of
our analysis. (Notice that we assume no
polarization of the incoming beams in $e^+e^-$ scattering.)
 Finally, when $m_H\OOrd2m_h$, the two reactions are
resonant, as they can both proceed via intermediate states involving
$H$ scalars, through $gg\to H$ and $e^+e^-\to HZ$, which in turn decay
via $H\to hh$ \cite{BRs}. Thus, the production cross sections are
largely enhanced \cite{pp,ee} (up to two orders of magnitude above
typical SM rates at the LHC \cite{pp}) and become clearly visible. 
This allows the possibility of probing the
trilinear $Hhh$ vertex at one or both these colliders.

The dominant decay rate of the MSSM $h$ scalar is into $b\bar b$
pairs, regardless of the value of $\tan\beta$ \cite{BRs}.  Therefore,
the final signatures of our reference reactions always involve four
$b$-quarks in the final state. (In the case of a LC environment, a
further trigger on the accompanying $Z$ boson can be exploited.)  

If one assumes very efficient tagging and high-purity sampling of
$b$-quarks, the background to $hh$ events at the LHC is dominated by
the irreducible QCD modes \cite{ATLTDR}. Among these, we focus here 
on the cases $q\bar q,gg\to b\bar b b\bar b$, as representative of
ideal $b$-tagging performances. These modes
consist of a purely QCD contribution
of ${\cal O}(\alpha_s^4)$, an entirely EW process of ${\cal
  O}(\alpha_{em}^4)$ (with no double
Higgs intermediate states) and an ${\cal O}(\alpha_s^2\alpha_{em}^2)$
component consisting of EW and QCD interactions.  (Note that in the EW
case only $q\bar q$ initiated subprocesses are allowed at tree-level.)
For a LC, the final state of the signal  is $ b\bar b b\bar b Z$, with
the $Z$ reconstructed from its decay products in some channel. Here, the EW
background is of $\ord(\alpha_{em}^5)$ away from resonances
(and, again, contains no more than one intermediate Higgs boson), whereas
the EW/QCD background is proportional to $(\alpha_s^2\alpha_{em}^3)$.

In general, EW backgrounds can be problematic due to the presence of
$Z$ vectors and single Higgs scalars yielding $b \bar b$ pairs, with
the partons being typically at large transverse momenta and well separated.  In
contrast, the QCD backgrounds involve no heavy objects decaying to $b
\bar b$ pairs and are dominated by the typical infrared (i.e., soft
and collinear) QCD behavior of the partons in the final state.
However, they can yield large production rates because of the strong
couplings.
\begin{fmffile}{fd}
\end{fmffile}

In this study, we investigate the interplay between the signal and
background at both colliders, adopting detector as well as dedicated
selection cuts. We carry out our analysis at both parton and hadron
level.  The plan of this note is as follows. The next Section details
the procedure adopted in the numerical computation.
Sect.~\ref{sec_results} displays our results and contains our
discussion. Finally, in the last section, we summarize our findings
and consider possible future studies.

\section{Calculation}
\label{sec_calculation}

For the parton level simulation, the double Higgs production process
at the LHC, via $gg$ fusion, has been simulated using the program
of Ref.~\cite{spira} to generate the interaction $gg \to hh$,
with the matrix elements (MEs) taken at leading-order (LO) for consistency
with our treatment of the background. We then
perform the two $h\to b\bar b$ decays to obtain the actual $4b$-final
state.  For double Higgs production at a LC, we use a source
code for the signal derived from that already used in Ref.~\cite{noi}. 
At both colliders,
amplitudes for background events were generated by means of MadGraph
\cite{tim} and the {\tt HELAS} package \cite{HELAS}. Note that
interferences between signal and backgrounds, and between the various
background contributions themselves, have been neglected.  This is a
good approximation for the interferences involving the signal because
of the very narrow width of the MSSM lightest Higgs boson.  Similarly,
the various background subprocesses have very different topologies,
and one would expect their interferences to be small in general.
 
The Higgs boson masses and couplings of the MSSM can be expressed at
tree-level in terms of the mass of the pseudoscalar Higgs state,
$m_A$, and the ratio of the vacuum expectation values of the two neutral
fields in the two iso-doublets, $\tan\beta$.
At higher order however, top and stop
loop-effects can become significant.  Radiative corrections in the
one-loop leading $m_t^4$ ap\-pro\-xi\-ma\-tion are parameterized by
\beq
\epsilon \approx \frac{3 G_F m_t^4}{\sqrt{2} \pi^2 \sin^2 \beta} 
\log \frac{m_S^2}{m_t^2} 
\eeq
where the SUSY breaking scale is given by the common squark mass,
$m_S$, set equal to $1$~TeV in the numerical analysis. If stop mixing
effects are modest at the SUSY scale, they can be accounted for by
shifting $m_S^2$ in $\epsilon$ by the amount $\Delta m_S^2 = \hat{A}^2
[1-\hat{A}^2/(12 m_S^2)]$
($\hat{A}$ is the trilinear common coupling). 
The charged and neutral CP-even Higgs boson
masses, and the Higgs mixing angle $\alpha$ are given in this
approximation by the relations:
\begin{eqnarray}
m_{H^\pm}^2 \!\!&=&\!\!  m_A^2 + 
m_Z^2 \cos^2 \theta_W, \non\\
m_{h,H}^2 \!\!&=&\!\! \textstyle{\frac{1}{2}}
[ m_A^2+m_Z^2+\epsilon \non\\
&\mp&
\sqrt{(m_A^2+m_Z^2+\epsilon)^2- 4m_A^2 m_Z^2 \cos^2 2\beta
   - 4\epsilon( m_A^2 \sin^2 \beta + m_Z^2 \cos^2 \beta)} ],
\non \\
\tan 2\alpha \!\!&=&\!\! \tan 2\beta
 \frac{m_A^2 + m_Z^2}{m_A^2 - m_Z^2 +\epsilon/\cos 2\beta} \qquad
\mbox{with} \qquad  - \frac{\pi}{2} \leq \alpha \leq 0,
\label{mass}
\end{eqnarray}
as a function of $m_A$ and $\tan\beta$.  The triple Higgs self-couplings of
the MSSM can be parameterized \cite{okada,djouadi} in units
of $M_Z^2/v$, $v=246$ GeV, as,
\beq
\lambda_{hhh} &=& 3 \cos2\alpha \sin (\beta+\alpha) 
+ 3 \frac{\epsilon}{m_Z^2} \frac{\cos \alpha}{\sin\beta} \cos^2\alpha,  
\non \\
\lambda_{Hhh} &=& 2\sin2 \alpha \sin (\beta+\alpha) -\cos 2\alpha \cos(\beta
+ \alpha) + 3 \frac{\epsilon}{m_Z^2} \frac{\sin \alpha}{\sin\beta}
\cos^2\alpha, \non \\
\lambda_{HHh} &=& -2 \sin 2\alpha \cos (\beta+\alpha) - \cos 2\alpha \sin(\beta
+ \alpha) + 3 \frac{\epsilon}{m_Z^2} \frac{\cos \alpha}{\sin\beta}
\sin^2\alpha, \non \\
\lambda_{HHH} &=& 3 \cos 2\alpha \cos (\beta+\alpha) 
+ 3 \frac{\epsilon}{m_Z^2} \frac{\sin \alpha}{\sin\beta} \sin^2 \alpha,
\non \\
\lambda_{hAA} &=& \cos 2\beta \sin(\beta+ \alpha)+ 
\frac{\epsilon}{m_Z^2} 
\frac{\cos \alpha}{\sin\beta} \cos^2\beta, \non \\
\lambda_{HAA} &=& - \cos 2\beta \cos(\beta+ \alpha) + 
\frac{\epsilon}{m_Z^2} 
\frac{\sin \alpha}{\sin\beta} \cos^2\beta.
\label{coup}
\eeq

Next-to-leading order (NLO) effects are certainly dominant, though the
next-to-next-to-leading order (NNLO) ones cannot entirely be neglected
(especially in the Higgs mass relations).  Thus, in the numerical
analysis, the complete one-loop and the leading two-loop corrections
to the MSSM Higgs masses and the triple Higgs self-couplings are
included.  The Higgs masses,
widths and self-couplings have been computed using the {\sc HDECAY}
program described in Ref.~\cite{hdecay1}, which uses a running $b$-mass
in evaluating the $h\ar b\bar b$ decay fraction.  Thus, for
consistency, we have evolved the value of $m_b$ entering the $hbb$
Yukawa couplings of the $h\ar b\bar b$ decay currents of our processes
in the same way.

For our analysis, we have considered $\tan \beta=3$ and $50$.  For the
LHC, high values of $\tan \beta$ produce a signal cross section much
larger than the $\tan\beta=3$ scenario, over almost the entire range
of $m_A$. However, this enhancement is due to the increase of the
down-type quark-Higgs coupling, which is proportional to $\tan\beta$
itself, and serves only to magnify the dominance of the quark box
diagrams of Fig.~\ref{fig:graphs}. Unfortunately, these graphs have no
dependence on either of the two triple Higgs self-couplings entering
the gluon-gluon process considered here, i.e., $\lambda_{hhh}$ and
$\lambda_{Hhh}$. Thus, although the cross section is comfortably
observable, all sensitivity to such vertices is lost. Therefore, the
measurement of the triple Higgs self-coupling, $\lambda_{Hhh}$, is only
feasible at the LHC for low $\tan \beta$ due to the resonant
production of the heavy Higgs boson (see Fig. 5a of Ref.~\cite{pp}). 

In contrast, the cross section for double Higgs production at the LC
is small for large $\tan\beta$ because there is no heavy Higgs
resonance (see Fig.~8 of Ref.~\cite{ee}). 
As soon as it becomes kinematically possible to decay the
heavy Higgs into a light Higgs pair, the $ZZH$ coupling is already too
small to generate a sizable cross section. Furthermore, the continuum
MSSM cross section is suppressed with respect to the SM cross section
since the MSSM couplings $ZZH$ and $ZZh$ vary with $\cos(\beta-\alpha)$ and
$\sin(\beta-\alpha)$, respectively, with respect to the corresponding SM
coupling. Notice that in this regime, at a LC, the $\lambda_{hhh}$
vertex could in principle be accessible instead, since
$\lambda_{hhh}\gg \lambda_{Hhh}$ (see Fig.~2 of Ref.~\cite{ee})
and because of the kinematic enhancement induced by $m_h\ll m_H$.
Unfortunately, we will see that the size of the $e^+e^-\to hhZ$ cross
section itself is prohibitively small.

We assume that $b$-jets are distinguishable from light-quark and gluon
jets and no efficiency to tag the four $b$-quarks is included in our
parton level results. We further neglect considering the possibility
of the  $b$-jet charge
determination. Also, to simplify the calculations, the $Z$ boson
appearing in the final state of the LC process is treated as
on-shell and no branching ratio
(BR) is applied to quantify its possible decays. In
practice, one may assume that it decays leptonically (i.e., $Z\to
\ell^+\ell^-$, with $\ell=e,\mu,\tau$) or hadronically into
light-quark jets (i.e., $Z\to q\bar q$, with $q\ne b$), in order to
avoid problems with $6b$-quark combinatorics.  Furthermore, in the LC
analysis, we have not simulated the effects of Initial State Radiation
(ISR), beamstrahlung or Linac energy spread.  Indeed, we expect them
to affect signal and backgrounds rather similarly, so we can neglect
them for the time being. Indeed, since a detailed phenomenological
study, including both hadronization and detector effects, already
exists for the case of double Higgs-strahlung in $e^+e^-$ \cite{Lutz},
whose conclusions basically support those attained in the
theoretical study of Ref.~\cite{noi}, we limit ourselves here to
update the latter to the case of the MSSM.

So far only resonant production $gg$ $\to$ $H$ $\to$ $hh$ $\to$ 
$b\bar bb\bar b$ has been investigated
\cite{ATLTDR}, with  full hadronic and detector simulation and
considering also the (large) QCD backgrounds, and a similar study
does not exist for continuum double Higgs production at the LHC. 
(See Ref.~\cite{ERW} for a detailed account of the $gg\to H$ $\to$
$hh$ $\to$ $\gamma\gamma b\bar b$ decay channel.)
The event simulation has been
performed by using a special version of {\sc PYTHIA} \cite{pythia1}, in
which the relevant LO MEs for double Higgs production of
Ref.~\cite{spira} have been implemented by M. El Kacimi and R. Lafaye.
These MEs take into account both continuum and resonant
double Higgs boson production and their interferences.
(The insertion of those for $e^+e^-$ processes is in progress.) The
{\sc PYTHIA} interface to {\sc HDECAY} has been exploited in order to
generate the MSSM Higgs mass spectrum and the relevant Higgs BRs, thus
maintaining consistency with the parton level approach.  As for the
LHC detector simulation, the fast simulation package
was used, with high luminosity (i.e., $\int{\cal L}dt=100$ fb$^{-1}$)
parameters. 

The motivation for our study is twofold. On the one
hand, to complement the studies of Ref.~\cite{ATLTDR} by also
considering the continuum production $gg\to hh\to b\bar b b\bar b$
at large $\tan\beta$. On the other hand, to explore the 
possibility of further kinematic suppression
of the various irreducible backgrounds to the resonant channel
at small $\tan\beta$.

\section{Results}
\label{sec_results}

\subsection{The LHC analysis}
\label{subsec_LHC}

In our LHC analysis, following the discussion in Sect.~\ref{sec_calculation},
we focus most of our attention on the case $\tan\beta=3$, with 
$m_A=210$ GeV, although other combinations of these two MSSM parameters
will also be considered.
We further set $A=-\mu=1$ TeV and take all sparticle masses (and other SUSY
scales) to be as large as $1$ TeV.

\subsubsection{$gg\to hh\to b\bar{b}b\bar{b}$ at parton level}
\label{subsubsec_parton_LHC}

In our parton level analysis, we identify jets with the partons from
which they originate (without smearing the momenta) and apply all cuts
directly to the partons.
We mimic the finite coverage of the LHC detectors by imposing a
transverse momentum threshold on each of the four $b$-jets, 
\begin{equation}\label{pTbcut_LHC}
p_T({b}) > 30~{\mathrm{GeV}}
\end{equation}
and requiring their pseudorapidity to be 
\begin{equation}\label{etabcut_LHC}
|\eta({b})| < 2.5.
\end{equation}
Also, to allow for their detection as separate objects, we impose
an isolation criterium among $b$-jets,
\begin{equation}\label{Rbbcut_LHC}
\Delta R({bb})>0.4,
\end{equation}
by means of  the usual cone variable
$\Delta R({ij})=\sqrt{\Delta\eta({ij})^2+\Delta\phi({ij})^2}$,
defined in terms of relative differences in pseudo-rapidity
$\eta_{ij}$ and azimuth $\phi_{ij}$ of the $i$-th and $j$-th $b$-jets.

As preliminary and very basic selection cuts (also to
help the stability of the numerical integration), we have required that
the invariant mass of the entire $4b$-system is at least twice the
mass of the lightest MSSM Higgs boson (apart from mass resolution and
gluon emission effects), e.g., 
\begin{equation}\label{Mbbbbcut_LHC}
m({bbbb})\ge 2m_h-40~{\rm{GeV}},
\end{equation}
and that exactly two $h$-resonances are reconstructed, such that
\begin{equation}\label{Mbbcut_LHC}
|m({bb})-m_h|<20~{\mathrm{GeV}}.
\end{equation}
In doing so, we implicitly assume that the $h$ scalar boson has
already been discovered and its mass measured through some other
channel, as we have already intimated in the Introduction. 

\begin{figure}[!ht]
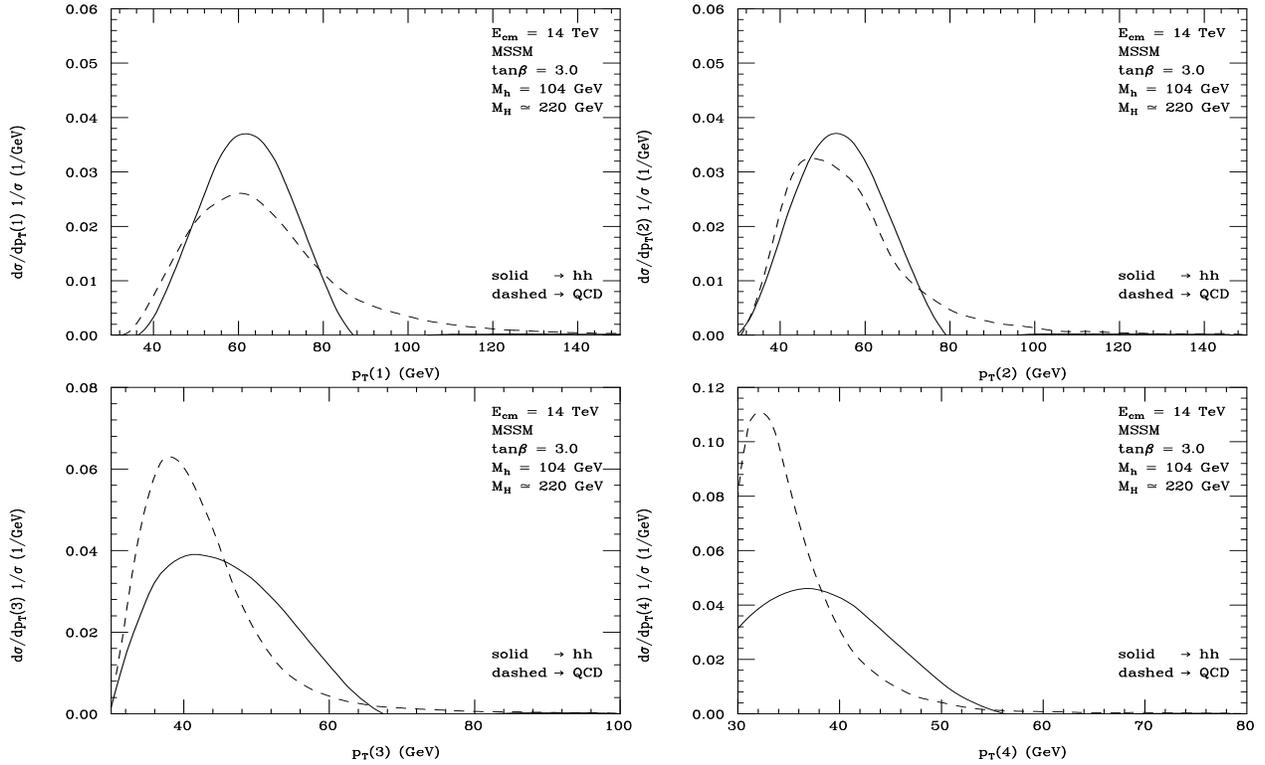

\begin{minipage}[b]{.495\linewidth}
\centering\epsfig{file=pT1_LHC.ps,angle=90,height=5cm,width=\linewidth}
\end{minipage}\hfill\hfill
\begin{minipage}[b]{.495\linewidth}
\centering\epsfig{file=pT2_LHC.ps,angle=90,height=5cm,width=\linewidth}
\end{minipage}\hfill\hfill
\begin{minipage}[b]{.495\linewidth}
\centering\epsfig{file=pT3_LHC.ps,angle=90,height=5cm,width=\linewidth}
\end{minipage}\hfill\hfill
\begin{minipage}[b]{.495\linewidth}
\centering\epsfig{file=pT4_LHC.ps,angle=90,height=5cm,width=\linewidth}
\end{minipage}

\caption{Distributions in transverse momentum of the
four $p_T$-ordered $b$-jets in $gg\to hh\to b\bar b b\bar b$ and
in the QCD background, after the cuts
(\ref{pTbcut_LHC})--(\ref{Mbbcut_LHC}) at the LHC, 
for $\tan\beta=3$, $m_h=104$ GeV and $m_H\simeq220$ GeV.
Normalization is to unity.}
\vspace*{-3mm}
\label{fig:pTaftercuts_LHC}
\end{figure}

After the above cuts have been implemented, we have found that the two
$4b$-backgrounds proceeding through EW interactions are negligible
compared to the pure QCD process.  In fact, the constraints described
in eqs.~(\ref{Mbbbbcut_LHC})--(\ref{Mbbcut_LHC}) produce the strongest
suppression, almost completely washing out the relatively enhancing
effects that the cuts in (\ref{pTbcut_LHC})--(\ref{Rbbcut_LHC}) have
on the EW components of the backgrounds with respect to the pure QCD
one, owning to the intermediate production of massive $Z$ bosons in
the former. In the end, the production rates of the three subprocesses
scale approximately as their coupling strengths: i.e., ${\cal
  O}(\alpha_s^4)$ : ${\cal O}(\alpha_s^2\alpha_{em}^2)$ : ${\cal
  O}(\alpha_{em}^4)$. Therefore, in the reminder of our analysis, we
will neglect EW effects, as they represent not more than a 10\%
correction to the QCD rates, which are in turn affected by much larger
QCD $K$-factors.  As for the pure QCD background itself, it hugely
overwhelms the double Higgs signal at this stage.  The cross section
of the former is about 7.85 pb, whereas that of the latter is
approximately 0.16 pb.

To appreciate the dominance of the $m_h$ cuts, one may refer to
Fig.~\ref{fig:pTaftercuts_LHC}, where the distributions in transverse
momentum of the four $p_T$-ordered $b$-quarks (such that $p_{T}(b_1)>
... >p_{T}(b_4)$) of both signal and QCD background are shown.  Having
asked the four $b$-jets of the background to closely emulate the
$gg\to hh\to b\bar b b\bar b$ kinematics, it is not surprising to see
a `degeneracy' in the shape of all spectra. Clearly, no
further background suppression can be gained by increasing the
$p_T(b)$ cuts. The same can be said for $\eta({b})$ and $\Delta
R({bb})$. Others quantities ought to be exploited.

\begin{figure}[!ht]
  ~\hskip2.5cm\epsfig{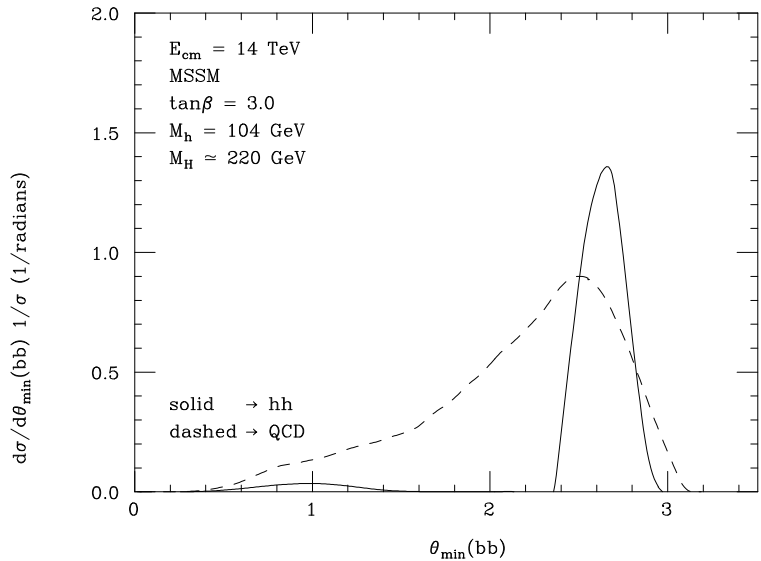}
\caption{Distributions in minimum  relative angle (in radians)
  in the $4b$-system rest frame between two $b$-jets
  reconstructing $m_h$ in $gg\to hh\to b\bar b b\bar b$ and in the QCD
  background, after the cuts (\ref{pTbcut_LHC})--(\ref{Mbbcut_LHC}) at
  the LHC, for $\tan\beta=3$, $m_h=104$ GeV and $m_H\simeq220$ GeV.
  Normalization is to unity.}
\label{fig:thetabb_LHC}
\vspace*{-3mm}
\end{figure}

In Fig.~\ref{fig:thetabb_LHC}, we present the signal and QCD
background distributions in the minimum angle formed between the
two $b$-quarks coming from the `same Higgs' (i.e., those fulfilling
the cuts in (\ref{Mbbcut_LHC})) in the $4b$-system rest frame (the
plot is rater similar for the maximum angle, thus also on average).
There, one can see a strong tendency of the two $2b$-pairs produced in
the Higgs decays to lie back-to-back, reflecting the $2\to2$
intermediate dynamics of Higgs pair production via $gg\to hh$. Missing
such kinematically constrained virtual state, the QCD background shows
a much larger angular spread towards small $\theta_{\rm{min}}(bb)$
values, eventually tamed by the isolation cut (\ref{Rbbcut_LHC}).

The somewhat peculiar shape of the signal distribution is due to
destructive interference.  Recall that the signal contains not only
diagrams proceeding via a heavy Higgs resonance (the upper-left hand
graph of Fig.~\ref{fig:graphs}), which results in the large peak in
Fig.~\ref{fig:thetabb_LHC}, but also contains a continuum contribution
mediated by box graphs (the upper-right hand graph of
Fig.~\ref{fig:graphs}). These two contributions destructively
interfere leading to the depletion of events between the large
back-to-back peak and the small remaining 'bump' of the continuum
contribution as seen in Fig.~\ref{fig:thetabb_LHC}.

In the end, a good criterium to enhance the signal-to-background ratio
($S/B$) is to require, e.g., $\theta({bb})>2.4$ radians, i.e., a
separation between the $2b$-jets reconstructing the lightest Higgs
boson mass of about 140 degrees in angle. (Incidentally, we also have 
investigated the angle that each of these $2b$-pairs form with the
beam axis, but found no significant difference between signal and QCD
background).

\begin{figure}[!ht]
~\hskip2.5cm\epsfig{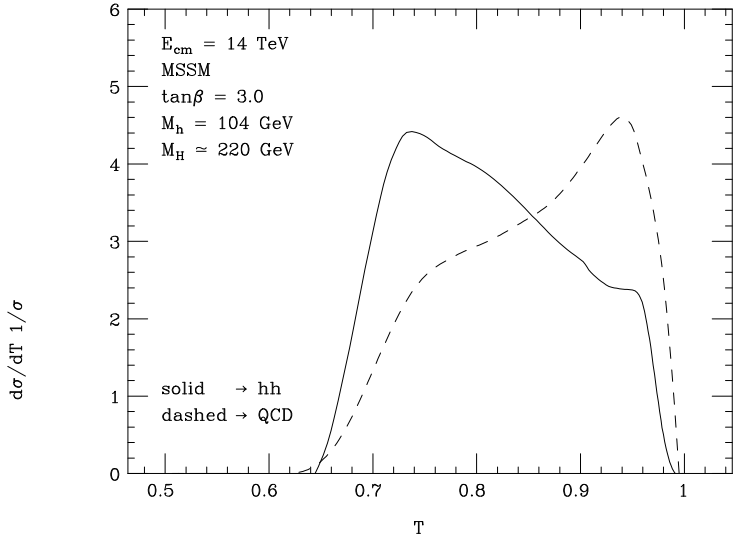}
\caption{
  Distributions in thrust in the rest frame of the $4b$-system
  in $gg\to hh\to b\bar b b\bar b$ and in the QCD background, after
  the cuts (\ref{pTbcut_LHC})--(\ref{Mbbcut_LHC}) at the LHC, for
  $\tan\beta=3$, $m_h=104$ GeV and $m_H\simeq220$ GeV.  Normalization
  is to unity.}
\label{fig:thrust_LHC}
\end{figure}

An additional consequence that one should expect from the presence
of two intermediate massive objects in $gg\to hh\to b\bar b b\bar b$
events is the spherical appearance of the jets in the final
state, in contrast to the usual planar behavior of the infrared
QCD interactions. These phenomena can be appreciated in
Fig.~\ref{fig:thrust_LHC}. Notice there the strong tendency
of the background to yield high thrust configurations, again controlled
by the separation cuts when $T$ approaches unity.
On the contrary, the average value of the thrust in the signal
is much lower, being the effect of accidental pairings of `wrong'
$2b$-pairs (the shoulder at high thrust values) marginal. An effective
selection cut seems to be, e.g., $T<0.85$.

\begin{figure}[!ht]
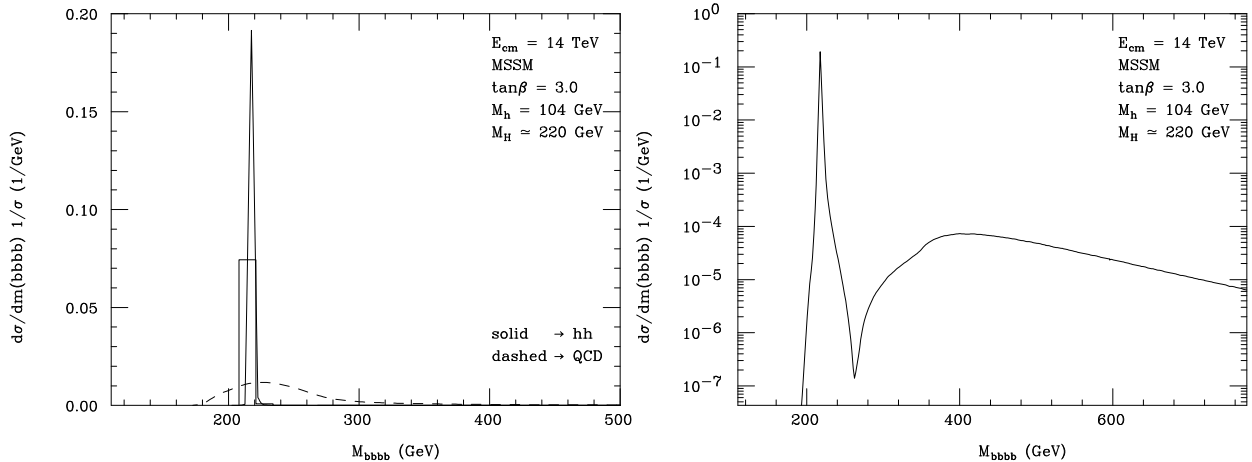

\begin{minipage}[b]{.495\linewidth}
\centering\epsfig{file=Mbbbb_LHC.ps,angle=90,height=6cm,width=\linewidth}
\end{minipage}\hfill\hfill
\begin{minipage}[b]{.495\linewidth}
\centering\epsfig{file=Mbbbblog_LHC.ps,angle=90,height=6cm,width=\linewidth}
\end{minipage}
\caption{ Distributions in invariant mass of the $4b$-system
  in $gg\to hh\to b\bar b b\bar b$ and in the QCD background, after
  the cuts (\ref{pTbcut_LHC})--(\ref{Mbbcut_LHC}) at the LHC, for
  $\tan\beta=3$, $m_h=104$ GeV and $m_H\simeq220$ GeV.  Normalization
  is to unity. The left hand plot shows both the signal (solid curve)
  and the QCD background (dashed curve), distributed in 5 GeV bins.
  The same signal is also shown as a histogram for a more
  experimentally realistic binning of 13 GeV. The right hand plot also
  shows the signal (collected in 5 GeV wide bins) on a logarithmic
  scale. Here the structure of the continuum contribution (and its
  destructive interference with the heavy Higgs decay contribution)
  can be seen.}
\label{fig:Mbbbb_LHC}
\vspace*{-3mm}
\end{figure}

Furthermore, if the heavy Higgs mass is sufficiently well measured at
the LHC then one can exploit the large fraction \cite{pp} of
$4b$-events which peak at $m_H$ in the signal, as dictated by the
$H\to hh$ decay, improving the signal-to-background ratio.  This peak
at $m_H$ can be clearly seen in the left hand plot of
Fig.~\ref{fig:Mbbbb_LHC}, where it dominates the QCD background, even
for bins 13 GeV wide.  In fact, not only could the QCD background be
considerably suppressed but also those contributions to $gg\to hh$ not
proceeding through an intermediate $H$ state should be removed, this
greatly enhancing the sensitivity of the signal process to the
$\lambda_{Hhh}$ coupling. This can be seen in the right hand plot of
Fig.~\ref{fig:Mbbbb_LHC} where the signal is shown on a logarithmic
scale. The continuum contribution due to the box graphs (and its
destructive interference with the heavy Higgs decay contribution) is
now evident although one should note that it is considerably suppressed
compared to the peak at $m_H$.

Now, if a less than 10\% mass resolution can be achieved on the light
and heavy Higgs masses, then one can tighten cut~(\ref{Mbbcut_LHC}) to
$|m({bb})-m_h|<10$ GeV and introduce the additional cut
$|m({bbbb})-m_H|<20$ GeV. These cuts taken together with those in
$\theta({bb})$ and $T$ already suggested, reduce the QCD background to
the same level as the signal. In fact, we have found that the cross
section of the background drops to approximately 174 fb whereas that
of the signal remains as large as 126 fb, this yielding a very high
statistical significance at high luminosity. Even for less optimistic
mass resolutions the signal-to-background ratio is still significantly
large. For example, selecting events with $|m(bb)-m_h|<20$ GeV and
$|m(bbbb)-m_H|<40$ GeV, the corresponding numbers are approximately
102 fb for the signal and 453 fb for the background. Notice that the
signal actually decreases as these Higgs mass windows are made larger.
This is due to our insistence that exactly two $b \bar b$ pairs should
reconstruct the light Higgs mass. As the light Higgs mass window is
enlarged from $m_h \pm 10$~GeV to $m_h \pm 20$~GeV, it becomes more
likely that accidental pairings reconstruct the light Higgs boson.
Since one is then unable to unambiguously assign the $b$ quarks to the
light Higgs bosons, the event is rejected and the signal drops.

Although we have discussed here an ideal situation which is difficult
to match with more sophisticated hadronic and detector simulations, it
still demonstrates that the measurement of the $\lambda_{Hhh}$
coupling could be well within the potential of the LHC, at least for
our particular choice of MSSM parameters.  Comforted by such a
conclusion, we now move on to more realistic studies.


\subsubsection{$gg\to hh\to b\bar{b}b\bar{b}$ at the LHC experiments}
\label{subsubsec_hadron_LHC}

Although the LHC experiments will be the first where one can attempt
to measure the Higgs self-couplings, the analysis is very challenging
because of the smallness of the production cross sections.  Even in
the most favorable cases, the production rate is never larger than a
few picobarns, already including one-loop QCD corrections, as computed
in Ref.~\cite{spira}. The cross sections at this accuracy are given in
Tab.~\ref{tab:cross}, for the resonant process (case 1 with $m_H = 220$ GeV) 
as well as
three non resonant scenarios: one at the same $\tan\beta$ but with
the $H\to hh$ decay channel closed (case 2), a second at very large
$\tan\beta$ and no visible resonance (case 3) and, finally, the SM
option (case 4, where $m_h$ identifies with the mass of the standard
Higgs state).

\begin{table}[!ht] 
\begin{center}
\begin{tabular}{|l||c|c|c|c|c|c|c|c|c|} \hline
case & model & $\tan\beta$ & $m_h$ (GeV) 
& $A$ (TeV) & 
$\mu$ (TeV) & $\sigma$ (fb) & dominant mode \\ \hline
1  & MSSM  & 3       & 104   
& $+1$   & $-1$   & 2000  & $gg \to H \to hh$ \\
2  & MSSM  & 3       & 100   
& $+1$   & $-1$   & 20  & $gg \to hh$ \\
3  & MSSM  & 50      & 105   
& $+1$   & $+1$   & 5000 & $gg \to hh$ \\
4  & SM    & -       & 105   
& -   & -   & 40 & $gg \to hh$ \\ \hline
\end{tabular}
\caption{Cross sections for double Higgs production $hh$ 
at the LHC via 
gluon-gluon fusion at NLO accuracy, for three possible configurations
of the MSSM and in the SM as well.
}
\label{tab:cross}
\vspace*{-7mm}
\end{center}
\end{table} 

\subsubsection{LHC trigger acceptance}
\label{subsubsect_trigger}

For $4b$-final states, possible LHC triggers are high $p_T$
electron/muons and jets. As an example, the foreseen ATLAS level 1 trigger
thresholds on $p_T$ and acceptance for a
$4b$-selection (with the four $b$-jets reconstructed in the detector)
are given in Tab.~\ref{tab:btag}, assuming the LHC
to be running at high luminosity.

\begin{table}[!ht]
\begin{center}
\begin{tabular}{|l||c|c|c|c|c|c|c|} \hline
trigger type: & 1 $e$ &  1 $\mu$  & 2 $\mu$ & 1 jet & 3 jets & 4 jets & total\\
$p_T$ in GeV & 30 & 20 & 10 & 290 & 130 & 90 & \\  \hline
case 1, $\epsilon(bbbb)$ in \% & 0.01 & 0.01 & 0.4 & 0.08 & 0.08 & 0.05 & 
0.53 \\ \hline
case 2 & $<0.01$ & $<0.01$ & 2.1 & 2.9 & 3.8 & 4.2 & 8.8\\ \hline
case 3 & $<0.01$ & $<0.01$ & 2.2 & 2.7 & 3.8 & 4.1 & 8.7\\ \hline
case 4 & $<0.01$ & $<0.01$ & 2.0 & 2.5 & 3.3 & 3.6 & 7.8\\ \hline
\end{tabular}
\caption{Kinematical
acceptance of the ATLAS detector to trigger four
$b$-jets 
 (including  detector acceptance) at high luminosity.
}
\label{tab:btag}
\end{center}
\end{table}

\begin{figure}
\vspace*{-9mm}
\begin{center}
  \includegraphics[width=6cm]{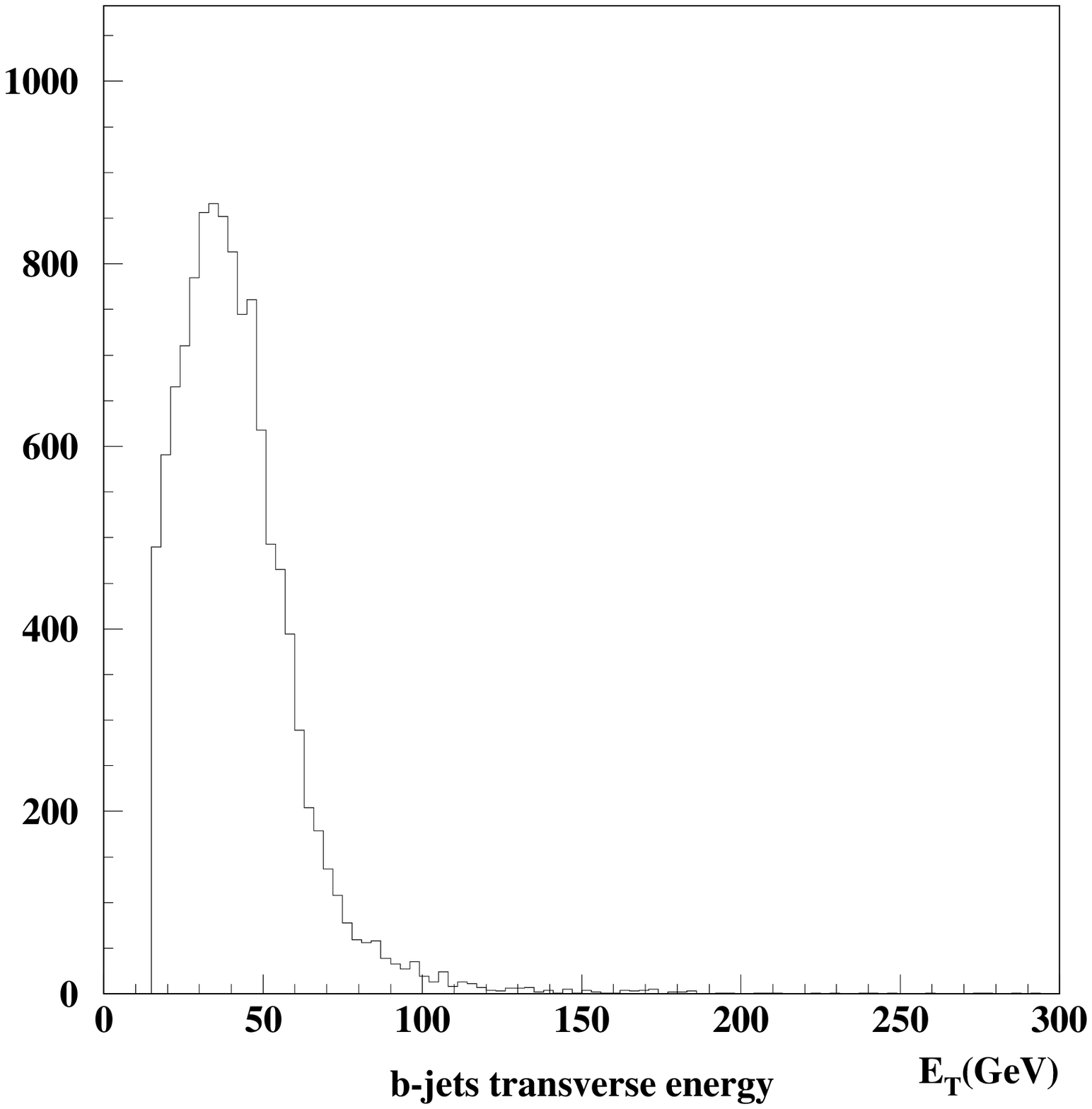}
  \includegraphics[width=6cm]{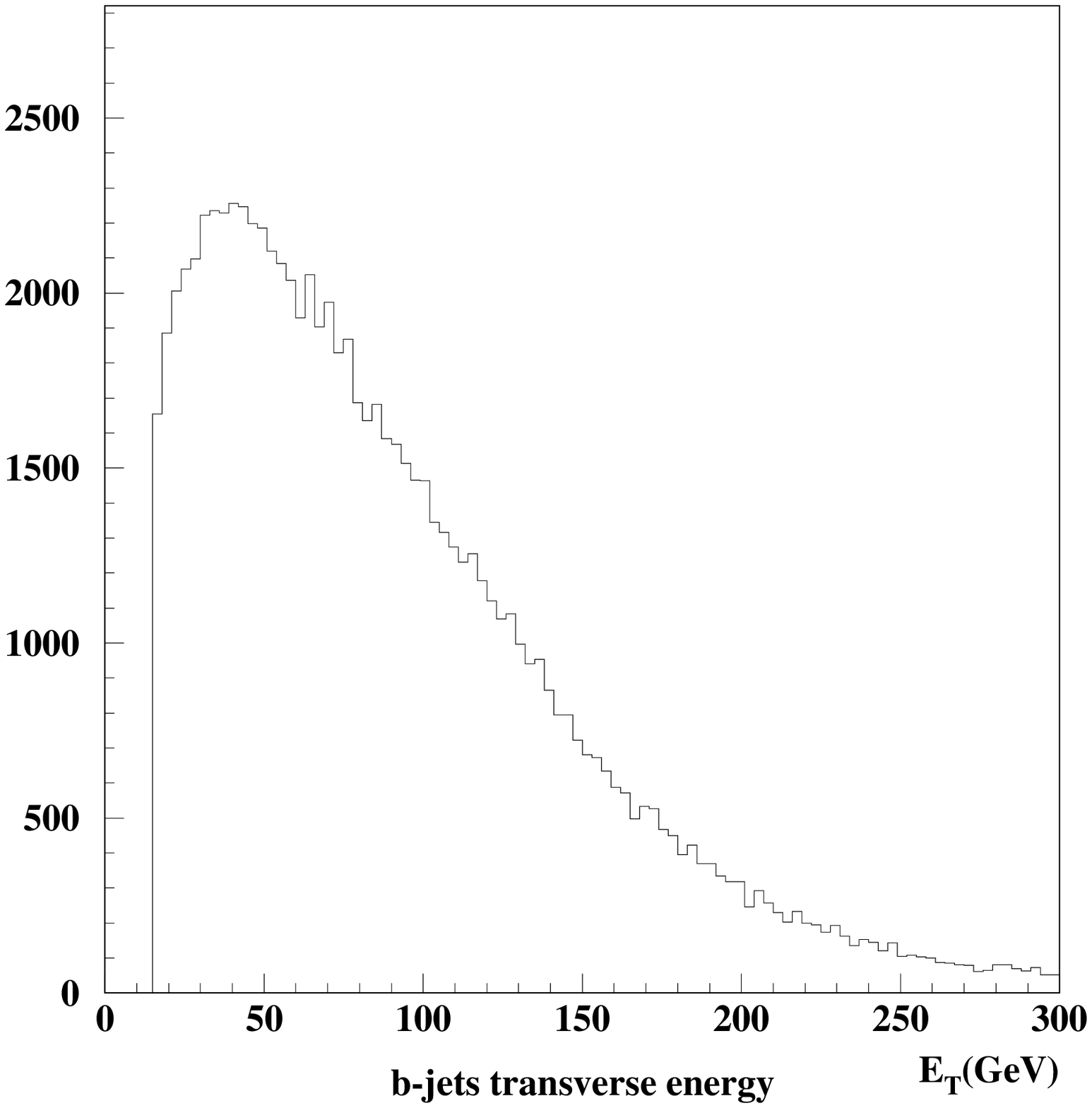}
 \caption{Reconstructed transverse energy/momentum for    
   $b$-jets in $gg\to hh\to b\bar{b}b\bar{b}$ events of case 1
   (left plot) and $b$-jets in $gg\to hh\to b\bar{b}b\bar{b}$ events
   of case 2 (right plot) with ATLAS fast simulation \cite{ATLFAST} at high
   luminosity. Normalization is arbitrary.}
  \label{ATLF-ptjet-2}
\end{center}
\vspace*{-5mm}
\end{figure}

\noindent
The overall trigger acceptance is at best 8--9\%,
 for cases 2,3,4. The very low efficiency for case 1 is
clearly a consequence of the small value of the difference $m_H-2m_h$,
translating into a softer $p_T(b)$ spectrum with respect to the other
cases (compare the left-hand with the right-hand
side of Fig.~\ref{ATLF-ptjet-2}).
One can further see in the left-hand plot of Fig.~\ref{ATLF-ptjet-2} that 
the bulk of the signal lies below the lowest $p_T(b)$ threshold
of Tab.~\ref{tab:btag} (i.e., $90$ GeV), so that adopting smaller 
trigger thresholds could result in a
dramatic enhancement of our efficiency. Of course, this 
would also substantially increase the low transverse momentum QCD
background, as we can see in the parton level analysis of
Fig.~\ref{fig:pTaftercuts_LHC}. 

For example, by lowering the thresholds to 180, 80
and 50~GeV for 1, 3 and 4 jets, respectively (compare
to Tab.~\ref{tab:btag}), the overall
trigger acceptance on the signal goes up to 1.8\%, i.e., by almost
a factor of 4. Meanwhile though,
the ATLAS level-1 jet trigger rates increase by a factor of 10
\cite{trig}.
Anyhow, even for our poor default value of $\epsilon(bbbb)$ in
Tab.~\ref{tab:btag}, we will see that case 1
still yields a reasonable number of events in the end.
Optimizations of the $b$-jet transverse momentum thresholds
are in progress \cite{more}.

\subsubsection{LHC events selection for $gg\to hh\to b\bar{b}b\bar{b}$}

Jets are reconstructed merging
tracks inside $\Delta R(bb)=0.4$. Only jets with transverse
energy/momentum greater than 30~GeV and with $|\eta(b)|<2.5$ are kept.
(Thus, the same cuts as in the parton level analysis, now applied
instead to jets.)  The effect from pile up is included in the
resolution.  A jet energy correction is then applied.

\begin{figure}[!ht]
\vspace*{-9mm}
\begin{center}
  \includegraphics[width=9cm]{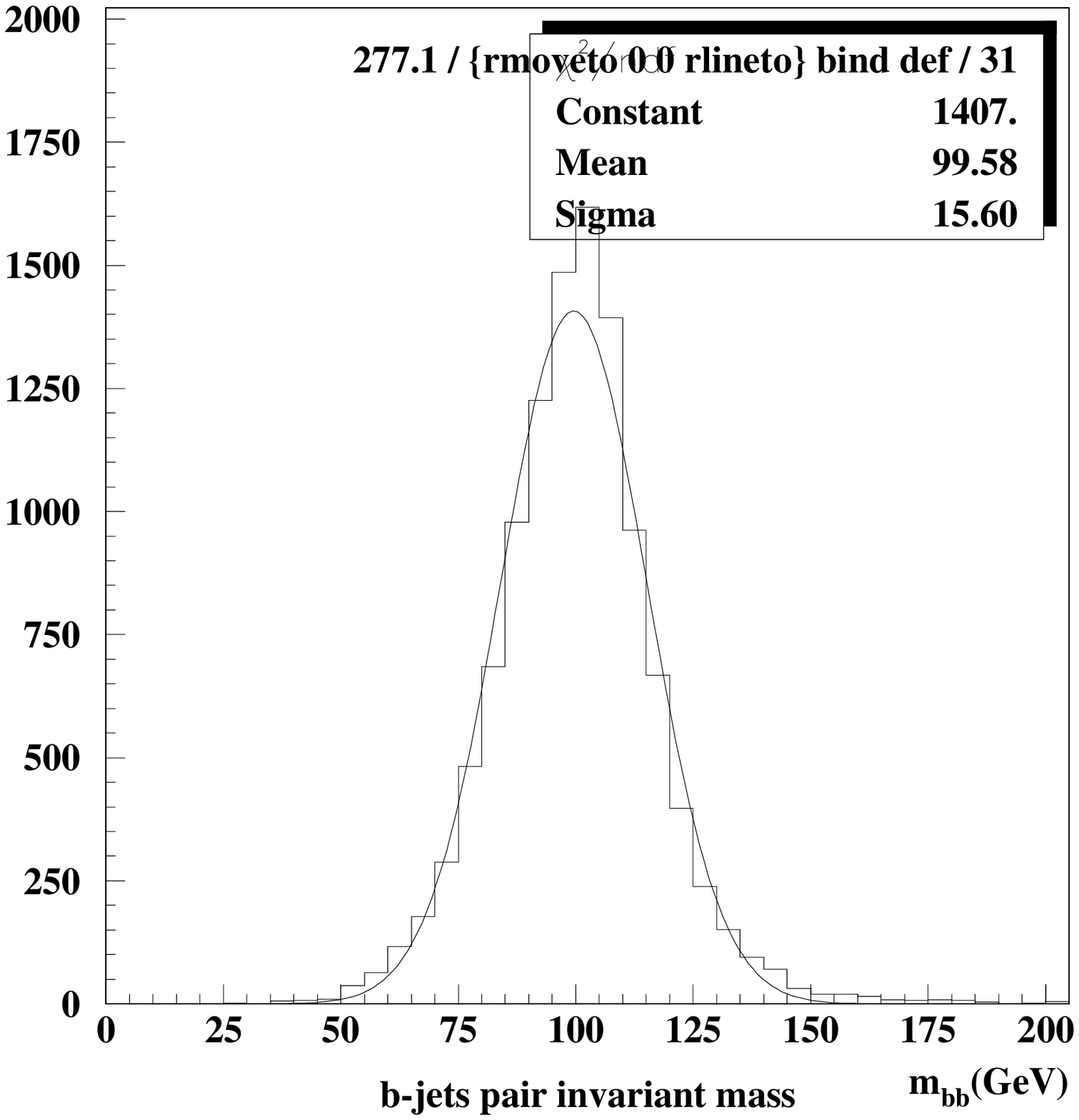} 
  \caption{ Reconstructed invariant mass distribution of
$2b$-jet pairs in continuum $gg\to hh\to b\bar{b}b\bar{b}$ events (case 2) with 
the fast simulation at high luminosity.
Normalization is arbitrary. (Results of a Gaussian fit are
also given.) }
  \label{ATLF-bbmass-2}
\end{center}
\vspace*{-9mm}
\end{figure}

\begin{figure}[!ht]
\vspace*{-9mm}
\begin{center}
  \includegraphics[width=9cm]{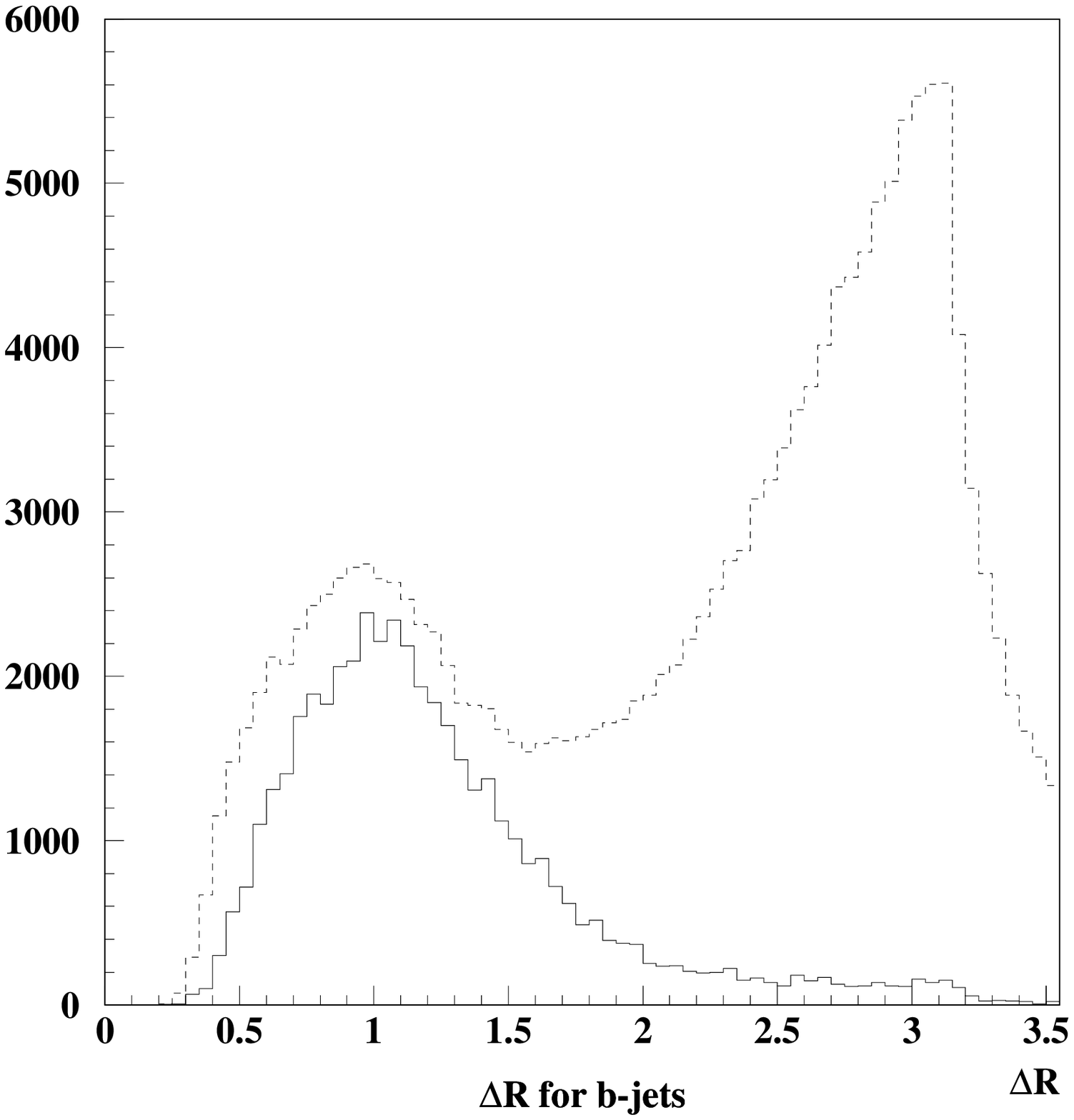} 
  \caption{Reconstructed $\Delta R(bb,bb)$ between
$2b$-jet systems from $h\to bb$ decays 
in continuum $gg\to hh\to b\bar b b\bar b$ events (case 2) with the fast
simulation at high luminosity. The dashed histogram shows the same 
distribution for all pairs of jets.  
Normalization is arbitrary.}
  \label{ATLF-bbDR-2}
\end{center}
\vspace*{-5mm}
\end{figure}

The invariant masses of each jet pair can then be computed.  Assuming
that the lightest Higgs boson mass is known, events with  $m(bb)$ sufficiently 
close to $m_h$ can efficiently be selected, see
Fig.~\ref{ATLF-bbmass-2}.  Another cut on the $\Delta R(bb,bb)$
between pairs of $b$-jets can also be applied to reduce the intrinsic
combinatorial background, since the latter concentrates at large
$\Delta R(bb,bb)$ values, see Fig.~\ref{ATLF-bbDR-2}.

\begin{figure}[!ht]
\vspace*{-9mm}
\begin{center}
  \includegraphics[width=10cm]{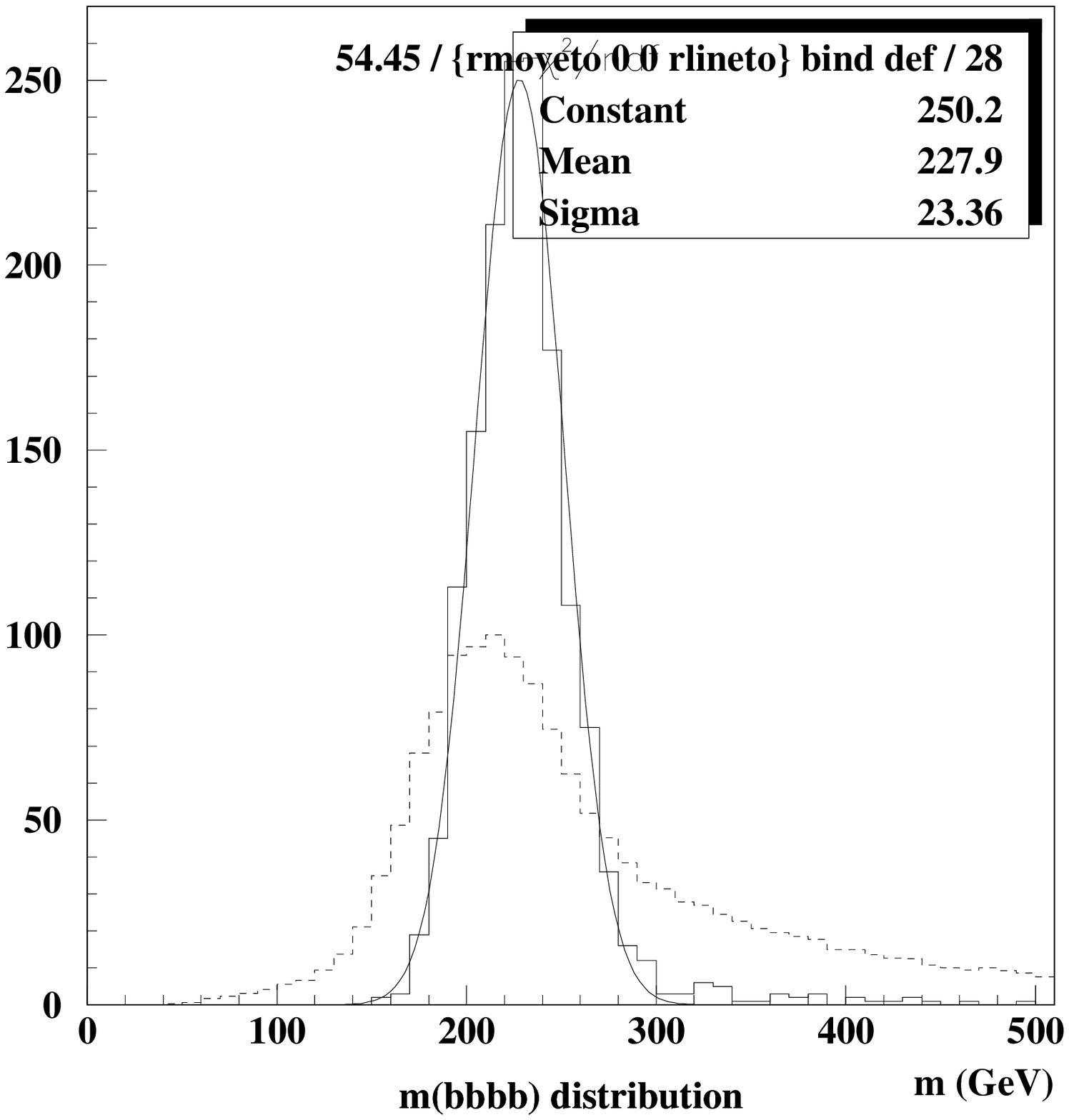} 
  \caption{Reconstructed $4b$-jet invariant mass for
    $b$-jets coming from the $hh$ pair in $gg\to hh\to b\bar bb\bar b$
    events (case 1) with the fast simulation at high luminosity.
    The dashed histogram shows the same distribution for all groups of
    four jets.  Normalization is arbitrary. (Results of a
    Gaussian fit to the first spectrum are also given.)}
  \label{ATLF-bbbbmass-r-4}
\end{center}
\vspace*{-5mm}
\end{figure}

For case 1, as already discussed, we can further impose that the
invariant mass of the four $b$-jets should be the heavy Higgs mass,
$m_H$, in order to select the $H\to hh$ resonance, as confirmed by
Fig.~\ref{ATLF-bbbbmass-r-4}.  In the other three cases, where the
$H\to hh$ splitting is no longer dominant (MSSM) or non-existent (SM),
one can still insist that the $4b$-jet invariant mass should be higher
than two times the lightest Higgs mass, see 
Fig.~\ref{ATLF-bbbbmass-4} and recall
eq.~(\ref{Mbbbbcut_LHC}).  Finally, following
Fig.~\ref{ATLF-bbbbmass-c-4}, by constraining the $b$-jets pairs
four-momenta around the known light Higgs mass value, $m_h$, one can
further reject the intrinsic background by means of the $m(bbbb)$ spectrum.

\begin{figure}[!ht]
\vspace*{-9mm}
\begin{center}
  \includegraphics[width=10cm]{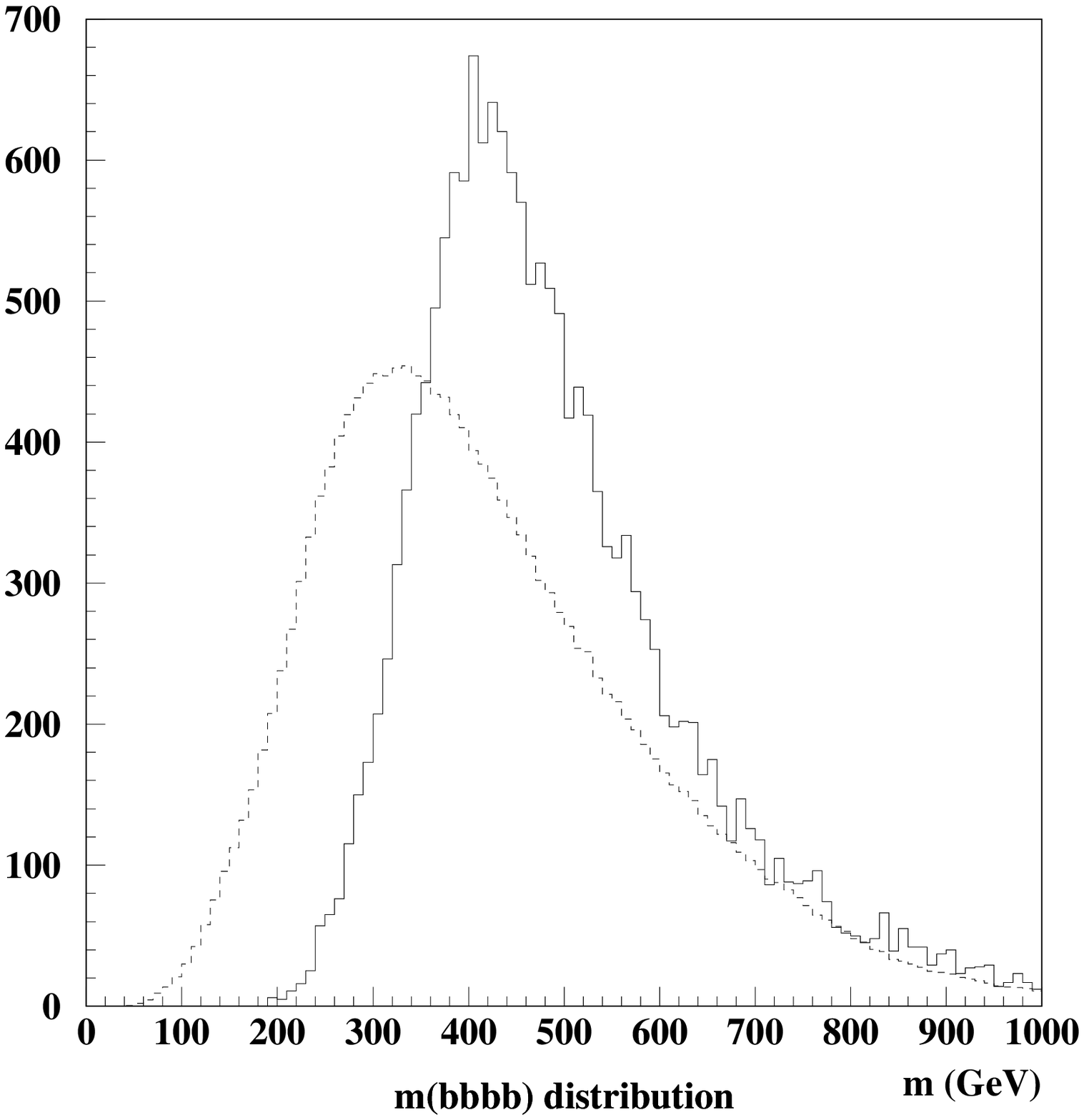} 
  \caption{Reconstructed $4b$-jet invariant mass for
$b$-jets coming from the $hh$ pair in $gg\to hh\to b\bar bb\bar b$ events
(case 4) with the fast simulation at high luminosity.
The dashed histogram shows the same distribution for all groups of four jets.
Normalization is arbitrary.}
  \label{ATLF-bbbbmass-4}
\end{center}
\vspace*{-5mm}
\end{figure}

\begin{figure}[!ht]
\vspace*{-9mm}
\begin{center}
  \includegraphics[width=10cm]{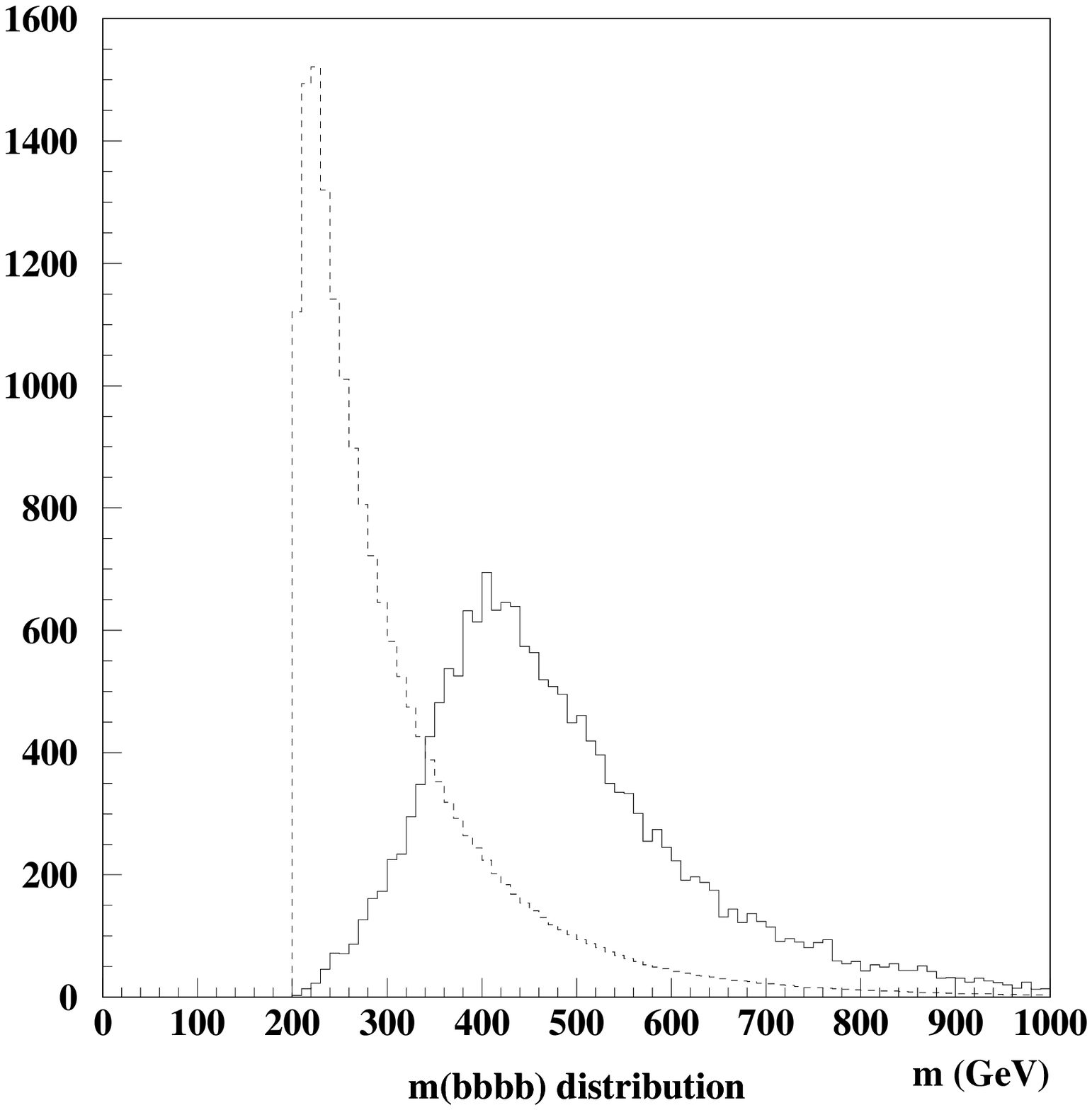} 
  \caption{Reconstructed $4b$-jet invariant mass for
$b$-jets coming from the $hh$ pair in $gg\to hh\to b\bar bb\bar b$ events 
(case 4) with the fast simulation at high luminosity. Here, the 
energy of the jet pairs is recalculated using the $m_h$ constraint.
The dashed histogram shows the same distribution for all groups of four jets.
Normalization is arbitrary.}
  \label{ATLF-bbbbmass-c-4}
\vspace*{-5mm}
\end{center}
\end{figure}

\subsubsection{LHC $b$-tagging in $gg\to hh\to b\bar{b}b\bar{b}$}

The $b$-tagging efficiency at high luminosity is set to 50\%,
with $p_T$ dependent correction factors for jets rejection. An average
rejection of 10 for $c$-jets and 100 for light-jets can be expected. We then
studied the effect on the selection efficiency of requiring from one
to four $b$-tags, although it is clear that, according to the parton
level studies, the huge background rate demands four $b$-tags, leading
to a 6\% tagging efficiency overall.

\subsubsection{Event rates at the LHC}

Taking into account all the efficiencies described above, and using
the NLO normalization of Tab.~\ref{tab:cross}, one can extract the
number of expected events per year at the LHC at high luminosity given
in Tab.~\ref{tab:rates}.  The selection cuts enforced here are the
following. For a start, we have kept configurations
where $|m(bb)-m_h|<30$ GeV (cases 1,3,4) or
$|m(bb)-m_h|<20$ GeV (case 2) and $\Delta R(bb,bb)<2.5$ (all four cases).
(If more than two $m_h$'s are reconstructed, the best two $2b$-pairs 
are selected according to the minimum value
of $\delta M^2=[m_h-m(bb)]^2+[m_h-m'(bb)^2]$.) Then, a cut on $m(bbbb)$
is applied: in presence of the $H\to hh$ resonance (case 1) we have
kept events within an $m_H$ mass window of $\pm 2\sigma$ (about 82\%
of the total number survive); otherwise
(cases 2,3,4) we have adjusted the $m(bbbb)\OOrd 2m_h$
cut so to keep 90\% of the sample. In the end, one finds
the numbers in Tab.~\ref{tab:rates}, that are encouraging indeed.

\begin{table}[!ht]
\vspace*{-3mm}
\begin{center}
\begin{tabular}{|l||c|c|c|c|} \hline
                             & case 1 & 2     & 3    & 4 \\ \hline
$\sigma$ in fb                    & 2000   & 20    & 5000 & 40 \\
trigger threshold acceptance       &  0.53\%   & 8.8\% & 8.7\%  & 7.8\% \\
mass windows & 60\%   & 50\%  & 40\% & 40\% \\
$4b$-tagging                  & {6\%}  & {6\%} & {6\%}& {6\%} \\ \hline
events/year (no tagging)     &  636  &  88   & 17400  & 125 \\
events/year (four $b$-tags)  &  38   & 5.3   & 1044   & 7.5 \\ \hline
\end{tabular}
\caption{Total rates for $gg\to hh\to b\bar bb\bar b$, 
after all efficiencies have been included and selection
cuts (\ref{pTbcut_LHC})--(\ref{Rbbcut_LHC}) enforced at hadron level,
with 100 fb$^{-1}$ per year of luminosity.}
\label{tab:rates}
\end{center}
\vspace*{-5mm}
\end{table}

In conclusion then, looking at the results in Tab.~\ref{tab:rates}
and bearing in mind the potential seen in reducing 
 the pure QCD background via $gg\to {\cal O}(\alpha_s^4)\to
b\bar b b\bar b$ (see
Figs.~\ref{fig:thetabb_LHC}--\ref{fig:Mbbbb_LHC}), one should be
confident in the LHC having the potential to measure the
$\lambda_{Hhh}$ coupling in resonant $H\to hh$ events (case 1). To
give more substance to such a claim, we have now initiated background
studies at hadron and detector level, following the guidelines
obtained by the parton level analysis \cite{more}.  As for other
configurations of the MSSM (such as case 2) or in the SM (case
4), the expectations are
more pessimistic. Case 3 deserves
further attention.  In fact, notice the large number of events
surviving and recall what mentioned in the Introduction concerning
 the potential of the non-resonant
$gg\to hh\to b\bar b b\bar b$ process as a discovery channel of the
light Higgs boson of the MSSM in the large $\tan\beta$ region at
moderate $m_A$ values, a corner of the parameter space where the $h$
coverage is given only by SM-like production/decay modes, thus not
allowing one to access information on the MSSM parameters. 
Results on this topic too will be presented in Ref.~\cite{more}.

\subsection{The LC analysis}
\label{subsec_LC}

Here, we closely follow the selection procedure advocated in
Ref.~\cite{noi}.  In order to resolve the four $b$-jets as four
separate systems inside the LC detector region, we impose the
following cuts. First, that the energy of each $b$-jet is above a
minimum threshold,
\begin{equation}\label{Ebcut_LC}
E(b)>10~{\rm{GeV}}.
\end{equation}
Second, that any $b$-jet is isolated from all others, by
requiring a minimum angular separation,
\begin{equation}\label{cosbbcut_LC}
\cos\theta({b,b})<0.95.
\end{equation}
Similarly to the hadronic analysis, one can optimize 
$S/B$ by imposing the constraints
\cite{noi},
\begin{equation}\label{Mbbbbcut_LC}
m({bbbb})\ge 2m_h-10~{\rm{GeV}},
\end{equation}
\begin{equation}\label{Mbbcut_LC}
|m({bb})-m_h|<5~{\rm{GeV}},
\end{equation}
on exactly two combinations of  $2b$-jets.  Here, note that the mass
resolution adopted for the quark systems is significantly better
than in the LHC case, due to the cleanliness of the $e^+e^-$
environment and the expected performance of the LC detectors in jet
momentum and angle reconstruction \cite{resolution}.  Thus, given such
high mass resolution power from the LC detection apparatus, one may
further discriminate between $h$ and $Z$ mass peaks by requiring that
none of the $2b$-jet pairs falls around $m_Z$,
\begin{equation}\label{MZcut_LC}
|m({bb})-m_Z|>5~{\mathrm{GeV}}.
\end{equation}
Moreover, in the double Higgs-strahlung process $e^+e^-\to hhZ$,
the four $b$-quarks are produced centrally,
whereas this is generally not the case for the background (see
the discussion in Ref.~\cite{noi}). This can be exploited by enforcing
\begin{equation}\label{cosbbbbcut_LC}
|\cos\theta({bb,bbb,bbbb})|<0.75,
\end{equation}
where $\theta({bb,bbb,bbbb})$ are the polar angles of all 
two-, three- and four-jet systems.
\begin{figure}[!ht]
~\hskip2.5cm\epsfig{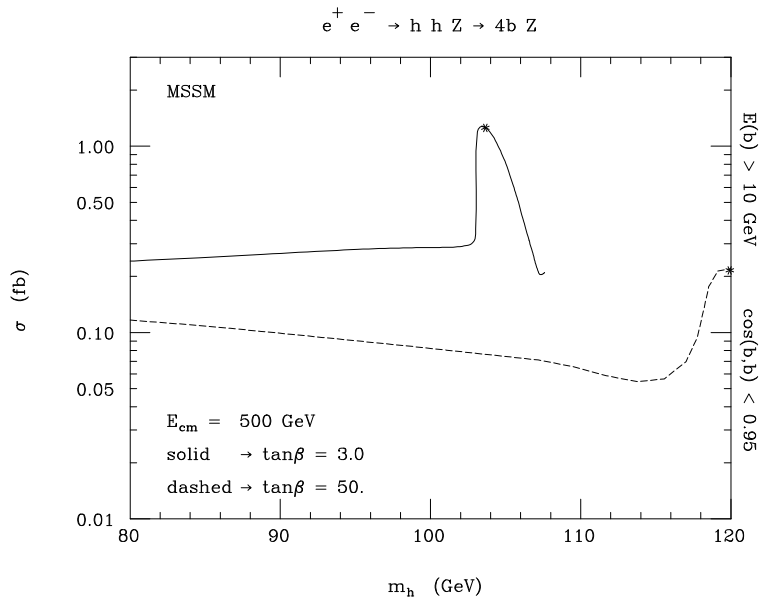}
\caption{Cross sections in femtobarns for the $e^+e^-\to hhZ$ signal
in the $h\to b\bar b b\bar b$ decay channel,
at a LC with 500 GeV as CM energy, as a function of $m_h$
for $\tan\beta=3$ and 50.
Our acceptance cuts in energy and separation of the four $b$-quarks
(\ref{Ebcut_LC})--(\ref{cosbbcut_LC}) have been implemented.
No beam polarization is included.}
\label{fig:signal_LC}
\end{figure}

Fig.~\ref{fig:signal_LC} shows the production and decay rates of the
signal process, $e^+e^-\to hhZ\to b\bar b b\bar b Z$, as obtained at
the partonic level, after the cuts
(\ref{Ebcut_LC})--(\ref{cosbbcut_LC}) have been implemented. The MSSM
setup here includes some mixing, having adopted $A=2.4$ TeV and
$\mu=1$ TeV, at both $\tan\beta=3$ and 50.  Notice the onset of the
$H\to hh\to b\bar b b\bar b$ decay sequence in the Higgs-strahlung
process $e^+e^-\to HZ$ at low $\tan\beta$.  The same does not occur
for large values, as previously explained. The
impact of the above jet selection cuts on the signal is marginal, as
the $b$-quarks are here naturally isolated and energetic, being the
decay products of heavy objects. In fact, the rates in
Fig.~\ref{fig:signal_LC} would only be 10--20\% higher if all the
$4b$-quark phase space was allowed (the suppression being larger for
smaller Higgs masses). At the height of the resonant peak around
$m_h\approx104$ GeV at $\tan\beta=3$, 
the signal rate is not large but observable,
yielding more than one event every 1 fb$^{-1}$ of data.  For a high
luminosity 500 GeV TESLA design \cite{TESLA}, this would correspond to
more than 300 events per year. Given the very high efficiency expected
in tagging $b$-quark jets, estimated at 90\% for each pairs of heavy
quarks \cite{btag}, one should expect a strong sensitivity to the
triple Higgs self-coupling.  The situation at large $\tan\beta$ is
much more difficult instead, being the production rates smaller by
about a factor of 10.

\begin{figure}[!ht]
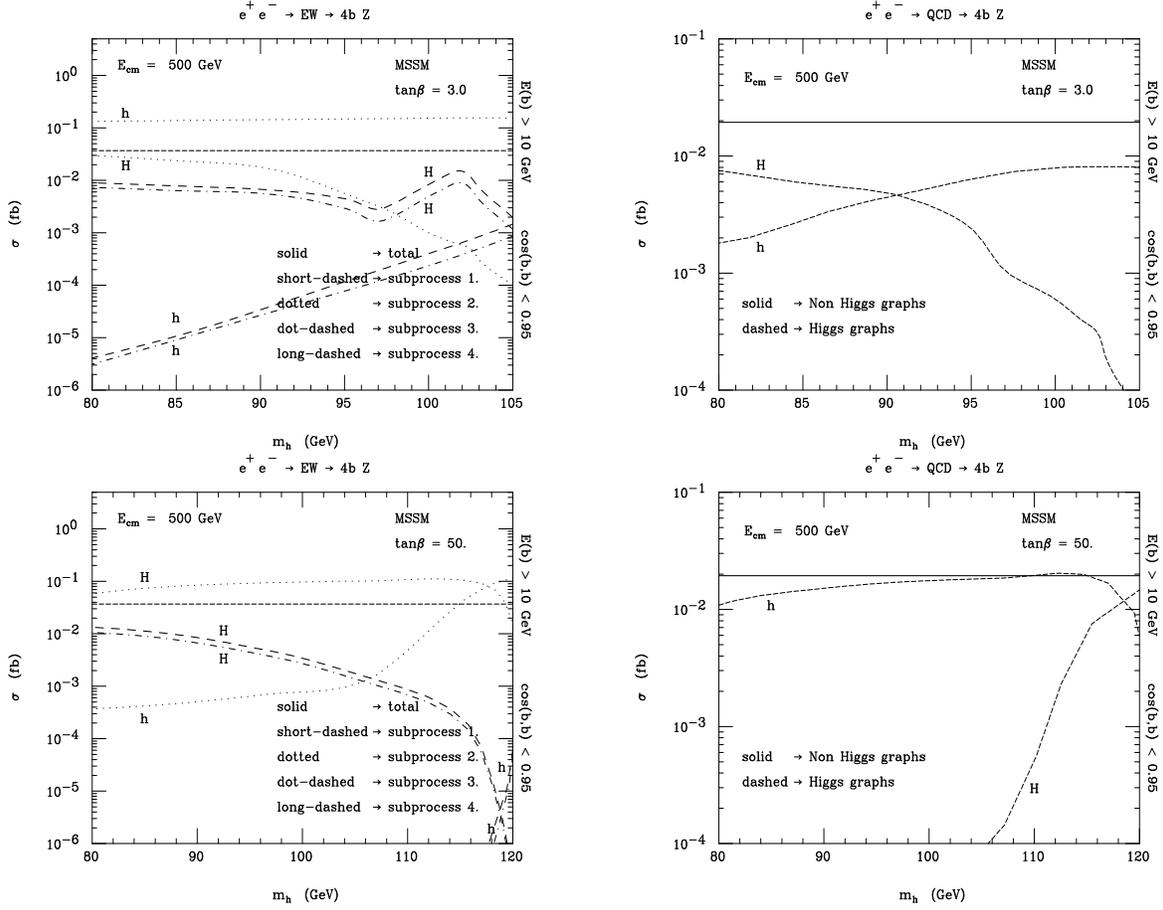

\begin{minipage}[b]{.495\linewidth}
\centering\epsfig{file=split_EW_MSSM_3.ps,angle=90,height=6cm,width=7cm}
\end{minipage}\hfill\hfill
\begin{minipage}[b]{.495\linewidth}
\centering\epsfig{file=split_QCD_MSSM_3.ps,angle=90,height=6cm,width=7cm}
\end{minipage}\hfill\hfill
\begin{minipage}[b]{.495\linewidth}
\centering\epsfig{file=split_EW_MSSM_50.ps,angle=90,height=6cm,width=7cm}
\end{minipage}\hfill\hfill
\begin{minipage}[b]{.495\linewidth}
\centering\epsfig{file=split_QCD_MSSM_50.ps,angle=90,height=6cm,width=7cm}
\end{minipage}
\caption{Cross sections in femtobarns for the dominant components
  of the EW (left) and EW/QCD (right) background to the $e^+e^-\to
  hhZ$ signal in the $h\to b\bar b b\bar b$ decay channel, at a LC
  with 500 GeV as CM energy, as a function of $m_h$ for $\tan\beta=3$
  (top) and 50 (bottom).  Our acceptance cuts in energy and separation
  of the four $b$-quarks (\ref{Ebcut_LC})--(\ref{cosbbcut_LC}) have
  been implemented.  No beam polarization is included.}
\label{fig:bkgd_LC}
\vspace*{-4mm}
\end{figure}

In the left-hand side of Fig.~\ref{fig:bkgd_LC} we present the EW
background, after the constraints in
(\ref{Ebcut_LC})--(\ref{cosbbcut_LC}) have been enforced, in the form
of the four dominant EW sub-processes.  These four channels are the
following.

\begin{enumerate}
\item $e^+e^-\to ZZZ\to b\bar b b\bar b Z$, first from the left in the
  second row of topologies in Fig.~3 of Ref.~\cite{noi}.  That is,
  triple $Z$ production with no Higgs boson involved.
\item $e^+e^-\to h/HZZ\to b\bar b b\bar b Z$, first(first) from the
  left(right) in the fifth(fourth) row of topologies in Fig.~2 of
  Ref.~\cite{noi} (also including the diagrams in which the on-shell
  $Z$ is connected to the electron-positron line).  That is, single
  Higgs-strahlung production in association with an additional $Z$,
  with the Higgs decaying to $b\bar b$. The cross sections of these
  two channels are obviously identical.
\item $e^+e^-\to h/HZ\to Z^*Z^*Z\to b\bar b b\bar b Z$, first from the
  right in the third row of topologies in Fig.~2 of Ref.~\cite{noi}.
  That is, single Higgs-strahlung production with the Higgs decaying
  to $b\bar b b\bar b$ via two off-shell $Z^*$ bosons.
\item $e^+e^-\to Zh/H\to b\bar b Z^*Z\to b\bar b b\bar b Z$,
  first(first) from the right(left) in the first(second) row of
  topologies in Fig.~2 of Ref.~\cite{noi}.  That is, two single
  Higgs-strahlung production channels with the Higgs decaying to
  $b\bar b Z$ via one off-shell $Z^*$ boson. Also the cross sections
  of these two channels are identical to each other, as in 2.
\end{enumerate}

The ${\cal O}(\alpha_s^2\alpha_{em}^3)$ EW/QCD background is dominated
by $e^+e^-\ar ZZ$ production with one of the two $Z$ bosons decaying
hadronically into four $b$-jets. This subprocess corresponds to the
topology in the middle of the first row of diagrams in Fig.~{4} of
Ref.~\cite{noi}. Notice that Higgs graphs are involved in this process
as well (bottom-right topology in the mentioned figure of \cite{noi}).
These correspond to single Higgs-strahlung production with the Higgs
scalar subsequently decaying into $b\bar bb\bar b$ via an off-shell
gluon. Their contribution is not entirely negligible, owing to the
large $ZH$ production rates, as can be seen in the right-hand side of
Fig.~\ref{fig:bkgd_LC}. The interferences among non-Higgs and Higgs
terms are always negligible.

In performing the signal-to-background analysis, we have chosen two
representative points only, identified by the two following
combinations: (i) $\tan\beta=3$ and $m_A=210$ GeV (yielding
$m_h\approx104$ GeV and $m_H\approx220$ GeV); (ii) $\tan\beta=50$ and
$m_A=130$ GeV (yielding $m_h\approx120$ GeV and $m_H\approx130$ GeV).
These correspond to the two asterisks in Fig.~\ref{fig:signal_LC},
that is, the maxima of the signal cross sections at both $\tan\beta$
values.  The first corresponds to resonant $H\to hh$ production,
whereas the latter to the continuum case.  If we enforce the
constraints of eq.~(\ref{Mbbbbcut_LC})--(\ref{cosbbbbcut_LC}), the
suppression of both EW and EW/QCD is enormous, so that the
corresponding cross sections are of ${\cal O}(10^{-3})$ fb, while the
signal rates only decrease by a factor of four at most.  This is the
same situation that was seen for the SM case in Ref.~\cite{noi}.
Indeed, in the end it is just a matter of how many signal events
survive, the sum of the backgrounds representing no more than a 10\%
correction (see Fig.~11 of Ref.~\cite{noi}). For example, after 500
fb$^{-1}$ of data collected,
one is left with 156 and 15 events for case (i) and (ii),
respectively. However, these numbers do not yet include $b$-tagging
efficiency and $Z$ decay rates.


\section{Summary}
\label{sec_conclusions}

To summarize, the `double Higgs production' subgroup has contributed
to the activity of the Higgs WG by assessing the feasibility of
measurements of triple Higgs self-couplings at future TeV colliders.
The machines considered were the LHC at CERN (14 TeV) and a future LC
running at 500 GeV.  In both cases, a high luminosity setup was
assumed, given the smallness of the double Higgs production cross
sections.  In particular, the $H\to hh$ resonant enhancement was the
main focus of our studies, involving the lightest, $h$, and the
heaviest, $H$, of the neutral Higgs bosons of the MSSM, in the
kinematic regime $m_H\OOrd2m_h$. This dynamics can for example occur
in the following reactions: $gg\to hh$ in the hadronic case and
$e^+e^-\to hhZ$ in the leptonic one, but only at low $\tan\beta$.
These two processes proceed via intermediate stages of the form $gg\to
H$ and $e^+e^-\to HZ$, respectively, followed by the decay $H\to hh$.
Thus, they in principle allow one to determine the strength of the
$Hhh$ vertex involved, $\lambda_{Hhh}$, in turn constraining the form
of the MSSM Higgs potential itself. The signature considered was
$hh\to b\bar b b\bar b$, as the $h\to b\bar b$ decay rate is always
dominant.

We have found that several kinematic cuts can be exploited in order to
enhance the signal-to-background rate to level of high significance,
particularly at the $e^+e^-$ machine. At the $pp$ accelerator, in
fact, the selection of the signal is made much harder by the presence
of an enormous background in $4b$ final states due to pure QCD. In
parton level studies, based on the exact calculation of LO scattering
amplitudes of both signals and backgrounds (without any showering and
hadronization effects but with detector acceptances), we have found
very encouraging results. At a LC, the double Higgs signal can be
studied in an essentially background free environment. 
 At the LHC, the signal and the QCD background are in the end
at the same level with detectable but not very large cross sections.

Earlier full simulations performed for the $e^+e^-$ case had already
indicated that a more sophisticated treatment of both signal and
backgrounds, including fragmentation/hadronization and full detector
effects, should not spoil the results seen at the parton level.  For the
LHC, our preliminary studies of $gg\to H\to hh\to b\bar b b\bar b$
in presence of
the $gg\to hh\to b\bar b b\bar b$ continuum (and relative interferences)
also point to the feasibility of the signal selection, after realistic
detector simulation and event reconstruction. 
As for double $h$ production in the continuum, 
although not very useful for Higgs self-coupling measurements,
this seems a promising channel, if not to discover the
lightest MSSM Higgs boson certainly to study its properties 
and those of the Higgs sector in general
 (because of the large production
and decay rates at high $\tan\beta$ and its sensitivity to such a parameter),
as shown from novel simulations also presented in this study. 
(The discovery potential of this mode will eventually be addressed
in Ref.~\cite{more}.)
Despite lacking a full
background analysis in the LHC case, we have no reason to believe that
a comparable degree of suppression of background events seen at parton level
cannot be achieved also at hadron level.  Progress in this respect is
currently being made \cite{more}.

\subsubsection*{Acknowledgements}

SM acknowledges financial support from the UK-PPARC.  The authors
thank P. Aurenche and the organizers of the Workshop for the
stimulating environment that they have been able to create. DJM and MM
thank M.~Spira for useful discussions. Finally, we all thank Elzbieta
Richter-Was for many useful comments and suggestions.

\setcounter{figure}{0}
\setcounter{table}{0}
\setcounter{section}{0}
\setcounter{equation}{0}
\newpage

\newcommand{\fhf}{{\tt FeynHiggsFast}}
\newcommand{\sx}{\\ \vspace*{-4mm}} 

\begin{center}

{\large\sc {\bf Programs and Tools for Higgs Bosons}}

\vspace{0.5cm}

{\sc E. Boos, A. Djouadi, N. Ghodbane, S. Heinemeyer,}

\vspace*{2mm}

{\sc  V. Ilyin, J. Kalinowski, J.L. Kneur and M. Spira} 
\end{center}

\begin{abstract}

The search strategies for Higgs bosons at LEP, Tevatron, LHC and future
$e^+e^-$ linear colliders (LC) and muon colliders exploit various Higgs boson
production and decay channels. The strategies depend not only on the
experimental setup [e.g. hadron versus lepton colliders] but also on the
theoretical scenarii, for instance the Standard Model (SM) or some of its
extensions such as the Minimal Supersymmetric Standard Model (MSSM). It is of
vital importance to have the most reliable predictions for the Higgs
properties, branching ratios and production cross sections.

There exist several programs and packages which determine the properties 
of Higgs particles, their decays modes and production mechanisms at various 
colliders. These programs are in general independent, have different inputs 
and treat different aspects of the Higgs profile. During this workshop, many 
discussions have been made and some work has been done on how to update these 
various programs to include the latest theoretical developments, and how to 
link some of them. 

This report summarizes the work which has been performed in this context. 
\end{abstract}

\section{HDECAY} 

The program HDECAY \cite{hdecay6} can be used to
calculate Higgs boson partial decay widths and branching ratios within
the SM and the MSSM and includes:  

\begin{enumerate}

\item[$\bullet$] All decay channels that are kinematically allowed and which
have branching ratios larger than $10^{-4}$, {\it y compris} the loop mediated,
the three body decay modes and in the MSSM the cascade and the supersymmetric
decay channels \cite{review6}.  

\item[$\bullet$] In the MSSM, the complete radiative corrections in the
effective potential approach with full mixing in the stop/sbottom sectors; it
uses the renormalization group improved values of the Higgs masses and
couplings and the relevant next--to--leading--order corrections are implemented
\cite{mhiggsRG}.  

\item[$\bullet$] All relevant higher-order QCD corrections to the decays into
quark pairs and to the loop mediated decays into gluons and photons are
incorporated in a complete form \cite{QCD6}; the small leading electroweak
corrections are also included.  

\item[$\bullet$] Double off--shell decays of the CP--even Higgs bosons [SM 
Higgs and the $h,H$ bosons of the MSSM] into massive gauge bosons which then 
decay into four massless fermions, and all important below--threshold 
three--body decays [decays into one real and virtual gauge bosons, cascade
decays into a Higgs and a virtual gauge boson, decays into a real and
virtual heavy top quark, etc,..] \cite{below}. 

\item[$\bullet$] In the MSSM, all the decays into SUSY particles [neutralinos,
charginos, sleptons and squarks including mixing in the stop, sbottom
and stau sectors] when they are kinematically allowed \cite{SUSY}. 

\item[$\bullet$] In the MSSM, the SUSY particles are also included in the loop
mediated $\gamma \gamma$ and $gg$ decay channels, with the leading parts of
the QCD corrections incorporated \cite{loop}. 

\end{enumerate} 

The source code of the program, {\tt hdecay.f} written in FORTRAN, has been
tested on computers running under different operating systems. It is
self--contained and all the necessary subroutines [e.g. for integration] are
included.  The program provides a very flexible and convenient usage, fitting
to all options of phenomenological relevance.  The program is lengthy [more
than 6000 lines] but rather fast, especially if some options [as decays into
double off-shell gauge bosons] are switched off.  

The basic input parameters, fermion and gauge boson masses and their total
widths, coupling constants and, in the MSSM, soft SUSY-breaking parameters can
be chosen from an input file {\tt hdecay.in}. In this file several flags allow
switching on/off or changing some options [{\it e.g.} choosing a particular
Higgs boson, including/excluding the multi--body or SUSY decays, or
including/excluding specific higher-order QCD corrections].  

The results for the many decay branching ratios and the total decay widths are
written into output files {\tt br.Xi} [with $X=H^0,h,H,A$ and $i=1,...$] with
headers indicating the various processes and giving some of the parameters.  

Since the release of the original version of the program several bugs have 
been fixed and a number of improvements and new theoretical calculations have 
been implemented. During this workshop, the following points have been included:

\begin{enumerate} 

\item[$\bullet$] Link to the {\tt FeynHiggsFast} routine which gives the masses 
and couplings of the MSSM up to two--loop order in the diagrammatic approach 
\cite{mhiggsFD}. 

\item[$\bullet$] Link to the {\tt SUSPECT} routine for the Renormalisation
Group evolution and for the proper electroweak symmetry breaking in the 
minimal Supergravity model \cite{suspect}.  

\item[$\bullet$] Implementation of Higgs boson decays to a gravitino and 
neutralino or chargino in gauge--mediated SUSY breaking models \cite{gg6}.

\item[$\bullet$] Inclusion of  gluino loops in Higgs boson decays to $q\bar{q}$
pairs \cite{hbb6}. 

\item[$\bullet$] Determination and inclusion of the RG improved two--loop 
contributions to the MSSM Higgs boson self-interactions.

\end{enumerate}

In addition, the  inclusion of the [possibly large] QCD corrections for the 
MSSM Higgs boson decays into squark pairs \cite{hsqsq} has started. \s

The log-book of all modifications and the most recent version of the
program can be found on the web page {\tt http://www.desy.de/$\sim$spira/prog}.

\section{Programs for Higgs production} 

Several programs for Higgs boson production at hadron colliders in the context 
of the SM and the MSSM, including the next--to--leading order (NLO) QCD 
corrections, are available at the web page: 
                  {\tt http://www.desy.de/$\sim$spira}. 
The purpose of these programs, and some improvements made during this Workshop,
are summarized below. For the physics context, see the contribution in Section 5
of these proceedings. \s

{\tt HIGLU} calculates the total cross sections for Higgs production in the
gluon--fusion mechanism, $gg \to$ Higgs, including the NLO QCD corrections in
the SM, MSSM and in a general two--Higgs doublet model [by initializing the
Yukawa couplings to quarks]. It includes both top and bottom quark loops which
generate the Higgs couplings to gluons. Moreover the program calculates 
the decay width of Higgs bosons into gluons at NLO. \s

{\tt V2HV} calculates the LO and NLO cross sections for the production in the
Higgs--strahlung mechanism, $qq \to V+ \Phi$ where $V=W/Z$ and $\Phi$ is a 
CP--even Higgs boson. The QCD corrections are those of the Drell--Yan
process; see Section 5. \s

{\tt VV2H} calculates the LO and NLO cross sections for the production in the
weak vector boson fusion mechanism, $qq \to V^*V^* \to qq \Phi$ where $\Phi$ is
a CP--even Higgs boson. The QCD corrections are included in the structure
function approach; see Section 5. \s

{\tt HQQ} calculates the LO cross sections for the production of neutral Higgs 
bosons in association with heavy quarks, $gg/q\bar{q} \to Q\bar{Q}+$ Higgs. 
The NLO QCD corrections are not yet completely available and are not included.\s

{\tt HPAIR} calculates the LO and NLO cross sections for the production of
pairs of neutral Higgs bosons in the the gluon--gluon fusion mechanism, $gg \to
\Phi_1 \Phi_2$, or in the Drell--Yan like process, $q\bar{q} \to \Phi_1 \Phi_2$.
The NLO corrections are included only in the heavy top quark limit for
the $gg$ process. \s

The source programs are written in FORTRAN and have been tested on computers
running under different operating systems. In most cases, the various 
relevant input parameters can be chosen from an input file including a flag
specifying the model. 

Since the first release of these programs, the following improvements have been 
made [some of them during this Workshop]: 

\begin{enumerate} 

\item[$\bullet$] A link to different subroutines calculating the MSSM Higgs 
boson masses and couplings has been installed for all the programs.

\item[$\bullet$] The contribution of squark loops has been included in {\tt 
HIGLU}. 

\item[$\bullet$] The SUSY--QCD corrections have been included in {\tt V2HV}
and {\tt VV2H}. 

\item[$\bullet$] The contribution of initial $b$--quark densities has been
included in {\tt HQQ}. 

\item[$\bullet$] The new version of {\tt HDECAY} for the neutral Higgs boson
total decay widths has been included in {\tt HPAIR}. 
  \end{enumerate}

\section{FeynHiggsFast}

In this section\footnote{This section is written with W. Hollik and G. 
Weiglein.} we present the Fortran code \fhf. Starting from low energy MSSM
parameters [$m_t$ the top quark mass, $\tan \beta$ the ratio of the vev's of 
the two Higgs
doublets, the pseudoscalar Higgs mass $M_A$, the soft SUSY breaking scalar
masses $M_{\tilde{t}_L}, M_{\tilde{t}_R}$, the trilinear coupling $A_t$ and the
higgsino mass parameter $\mu$], \fhf \ calculates the masses of the neutral
CP--even Higgs bosons, $M_h$ and $M_H$, as well as the corresponding mixing
angle $\alpha$, at the two--loop level \cite{mhiggsFD}. In addition the mass of
the charged Higgs boson, $M_{H^\pm}$, is evaluated at the one--loop level. The
$\rho$--parameter, which allows for constraints in the scalar fermion sector of
the MSSM, is evaluated up to ${\cal O}(\alpha \alpha_s)$, taking into account
the gluon exchange contribution at the two--loop level~\cite{drhosuqcd}. \s

\fhf\ is based on a compact analytical approximation formula, containing at the
two--loop level the leading corrections of ${\cal O}(\alpha \alpha_s)$ obtained
in the Feynman--diagrammatic approach~\cite{mhiggsFD} and of ${\cal O}
(G_F^2m_t^6)$ obtained with renormalization group (RG) methods~\cite{mhiggsRG}. 
Contrary to the full result in the FD approach~\cite{mhiggsFD} which has been
incorporated into the FORTRAN code {\tt FeynHiggs} \cite{feynhiggs}, the
approximation formula is much shorter. Thus, the program \fhf\ is about $3 
\times 10^4$ times faster than {\tt FeynHiggs}, while the agreement between 
the two codes is better than 2 GeV for the CP--even Higgs bosons masses in 
most parts of the MSSM parameter space. \s

The complete program \fhf\ consists of about 1300 lines FORTRAN code. The 
executable file fills about 65 KB disk space. The calculation for one set of 
parameters, including the $\Delta\rho$ constraint, takes about $2 \times 
10^{-5}$ seconds on a Sigma station [Alpha processor, 600 MHz processing speed,
512 MB RAM]. The program can be obtained from the {\tt FeynHiggs} home page:
\     {\tt http://www-itp.physik.uni-karlsruhe.de/feynhiggs}~ \ 
where the code itself is available, together with a short instruction, 
information about bug fixes, etc... \s

\fhf\ consists of a front--end, {\tt program FeynHiggsFast}, and the main part 
where the calculation is performed, starting with {\tt subroutine 
feynhiggsfastsub}. The front--end can be manipulated by the user at will, 
whereas the main part should not be changed. In this way \fhf\ can be 
accommodated as a subroutine to existing programs, thus providing an extreme 
fast evaluation for the masses and mixing angles in the MSSM Higgs sector.
As discussed previously, this has already been successfully performed for the 
program HDECAY during this workshop. \s

\fhf\ asks for the low energy SUSY parameter, listed in Table 1.
Concerning the stop sector, the user has the option
to enter either the physical parameters, i.e. the masses and the mixing angle
($m_{\tilde{t}_1}, m_{\tilde{t}_2}$ and $\sin \theta_{\tilde{t}}$) or the
unphysical soft SUSY breaking scalar mass parameters
$M_{\tilde{t}_L},M_{\tilde{t}_R}$ and the mixing parameter $M_t^{LR}=m_t
(A_t -\mu \cot \beta)$. From these input parameters \fhf\ calculates the 
masses and the mixing angle of the MSSM neutral CP--even Higgs bosons, as well 
as the mass of the charged Higgs boson and the $\rho$ parameter.  

\begin{table}[ht!]
\renewcommand{\arraystretch}{1.3}
\begin{center}
\begin{tabular}{|c||c||c|} \hline
 input parameter & MSSM expression & expression in program 
\\ \hline \hline
{\tt tan(beta)}          & $\tan \beta$      & {\tt ttb} \\
{\tt Msusy\_top\_L  }    & $M_{\tilde{t}_L}$    & {\tt msusytl} \\
{\tt Msusy\_top\_R}      & $M_{\tilde{t}_R}$    & {\tt msusytr} \\
{\tt MtLR }              & $M_t^{LR}$    & {\tt mtlr} \\
{\tt MSt2}               & $m_{\tilde{t}_2} $    & {\tt mst2} \\
{\tt delmst}             & $\Delta m_{\tilde{t}}= m_{\tilde{t}_2}-
                           m_{\tilde{t}_1}$     & {\tt delmst} \\
{\tt sin(theta\_stop)}   & $\sin \theta_{\tilde t}$   & {\tt stt} \\
{\tt MT}                 & $m_t$      & {\tt mmt} \\
{\tt Mue}                & $\mu$      & {\tt mmue} \\
{\tt MA}                 & $M_A$      & {\tt mma} \\
\hline
\end{tabular}
\renewcommand{\arraystretch}{1}
\caption[]{The meaning of the different MSSM variables to be entered into \fhf.}
\end{center}
\end{table}

\section{SUSPECT}

The fortran code\footnote{The program can be down-loaded from the node:
http://lpm.univ-montp2.fr:7082/~djouadi/gdr.html} {\tt SUSPECT} \cite{suspect}
calculates essentially the masses and some of the couplings of the SUSY and
Higgs particles within the framework of the MSSM. It includes several specific
options whose purpose is, hopefully, to gain more flexibility with the
generally non-trivial Lagrangian-to-physical parameter relationship in the
MSSM. In particular, besides the now widespread procedure of evolving the soft
parameters from some universal ``minimal SUGRA" high energy initial values 
down to obtain a corresponding low-energy spectrum, SUSPECT can also treat
almost arbitrary non--universal departures from this SUGRA model.  The latest
version 1.2 is a subroutine, so that it can be easily interfaced with any other
FORTRAN codes, as will be described below.  It also includes some new useful
tools like, for instance, the possibility of evolving the parameters ``
bottom--up", the possibility of choosing as input some of the parameters that
are usually obtained as output, etc. 

The latest version of the program consists of three parts: the subroutine
{\tt suspect12.f}, {\tt suspect12-call.f} an example of calling routine and
{\tt suspect12.in} a typical example of input file. To interface {\tt 
SUSPECT1.2} properly with your own main code, the easiest way is first to run 
the example code {\tt suspect12-call.f}. Once familiar with the calling 
procedure, you may simply implement in your calling code a few appropriate 
command lines stripped from the example file, that you can adapt to your 
purpose. \smallskip

The core of the SUSPECT algorithm is conveniently separated into three 
different tasks, that are indeed conceptually --and technically --relatively 
separated: $(i)$ Renormalization group evolution (RGE), $(ii)$ physical 
spectrum calculations (PS), $(iii)$ effective potential calculation with  
implementation of electroweak symmetry breaking (EWSB). The overall 
algorithm then reads as follows: choice of a model assumption/option $\to$ 
choice of initial scale $Q_{\rm in}$ (driven from input file {\tt suspect12.in}
or from user's main code) $\to$ RG evolution  $\to$ consistency of EWSB
which involves the effective potential at one--loop (iterating until stability 
is reached) $\to$ physical spectrum calculation: gauginos, sfermions, Higgses
$\to$ final masses and results (warning + comments as well) collected in file 
{\tt suspect.out}. 

An important aspect of {\tt SUSPECT} is a special attention given to the
consistency of EWSB, which makes that not all of the scalar sector parameters
are independent. [For the moment only the simplest constraints $\partial
(V_{\rm eff})/\partial v_{u,d} =0$ are included; the constraints from the
absence of Charge and Color Breaking (CCB) minima will be implemented in a
later version].  In particular, this is used to define different set of
input/output scalar parameters. Although this resulting flexibility in the
choice of input parameters is welcome, its actual implementation is quite  non
trivial, which is payed by a slower CPU time.  Moreover, one should keep in
mind that it is often a main source of possible discrepancies with other
similar task codes which implement EWSB in a different way.

Another important ingredient of {\tt SUSPECT} is the implementation of
RG evolution, in different (loop) approximations. The RGE can be implemented
(or not) by using different {\tt ichoice(1)} input parameters. For instance,
for {\tt ichoice(1)=0} one has the unconstrained MSSM with no RGE, i.e.
the relevant input parameter are assumed to be at LOW scale. For
{\tt ichoice(1) = 1}, RGE in the unconstrained MSSM with non--universality 
and the inputs are assumed at high scale, except $\tan \beta$ to be given
at low energies. {\tt ichoice(1) = 2}: unconstrained MSSM with RGE bottom--up;
the relevant input is set similarly as with {\tt ichoice(1)= 0}, but the final 
output consists of all the soft parameters at the high scale. 
{\tt ichoice(1) = 10}:  minimal SUGRA  model.
 
For interfacing {\tt SUSPECT1.2} with your main code, all the user has to 
control is the way to dialog between her/his "main" routine/program and 
the {\tt SUSPECT1.2} subroutine, together with the precise meaning of the 
different ``dialog" parameters, which are of two kinds: 

-- The ``physical" parameters, are those parameters that are either necessary
input for a given model and/or running option, or the desired output. All such 
parameters are passed from the calling code to {\tt SUSPECT} and back via
specific {\tt COMMONS}. By ``physical" we mean either truly physical 
parameters such as masses etc [and that are generally the output of {\tt 
SUSPECT} calculation], or MSSM basic parameters such as the SUSY and soft--SUSY 
breaking terms of the MSSM Lagrangian, that are generally input for the 
{\tt SUSPECT} calculation.

-- The ``control" parameters, whose different purpose is to choose various 
running options. There are three main ``control" parameters appearing as 
arguments of the {\tt SUSPECT} calling command: $(i)$ {\tt iknowl} sets some 
degree of control on various parts of the algorithm  [{\tt=0} blind use, i.e. 
no control on any ``algorithmic parameter, {\tt =1} more educated use, $(ii)$ 
{\tt input} setting control [{\tt =0} relevant parameters are read form {\tt 
suspect12.in} and {\tt =1} define the relevant inputs from your calling 
program] and ($iii)$ {\tt ichoice} for the choice of model parameters with 
{\tt ichoice(1)} discussed above for the RGE and {\tt ichoice(6)} for the 
scalar sector input [{\tt =0} for $\mu, M_A$ inputs and {\tt =1} for 
$M_{H_u}^2, M_{H_d}^2$ as inputs].  

All details on the main core {\tt SUSPECT} routines, input and output 
parameters as well as physical and control parameters can be found on
the web site and in Ref.~\cite{suspect}. 
\newpage

\section{SUSYGEN}

{\tt SUSYGEN2} \cite{kats:susygen2} is a Monte Carlo event generator for the
production and decay of supersymmetric particles and has been initially
designed for $e^+e^-$ colliders. It has been extensively used by all four LEP
experiments to simulate the expected signals.  It includes pair production of
charginos and neutralinos, scalar leptons and quarks. It offers also a
possibility to study the production of a gravitino plus a neutralino within
GMSB models and the production of single gauginos if one assumes R-Parity to be
broken.

All important decay modes of SUSY particles relevant to LEP
energies have been implemented, including cascades, radiative decays and
R-Parity violating decays to standard model particles.  The decay is included
through the exact matrix elements.  The lightest supersymmetric
particle (LSP) can either  be the neutralino $\tilde{\chi}^0_1$, the
sneutrino $\tilde{\nu}$ or the gravitino $\tilde{G}$ in R--parity conserving
models, or any SUSY particle if R--parity is violated.

The initial state radiative corrections  take account of $p_T / p_L$ effects 
in the Structure Function formalism.  QED final state radiation is implemented
using the \texttt{PHOTOS} \cite{was:photos} library.  An optimized
hadronization interface to \texttt{JETSET} \cite{sjos:jetset} is  provided,
which also takes into account  lifetimes of sparticles. Finally, a widely
used feature of \texttt{SUSYGEN2} is the possibility to perform automatic 
scans on the parameter space through user friendly ntuples.

Recently \texttt{SUSYGEN2}  has been upgraded to \texttt{SUSYGEN3}
\cite{nous:susygen3} in order  to adapt to the needs of the next generation of
linear colliders, but also in order to extend its potential to supersymmetric
particles searches at $e^-p$ colliders (e.g HERA) and hadronic colliders (e.g
Tevatron or LHC).  The main new features relevant for linear colliders are the
inclusion of beamstrahlung through an interface to \texttt{CIRCEE}
\cite{ohl:beamstrah}, the full spin correlation in initial and final states,
the inclusion of CP violating phases and the possibility to have an elaborate
calculation of the MSUGRA spectrum through an interface to \texttt{SUSPECT}
\cite{suspect}. \s

\noindent {\bf a) Mass spectrum calculation}: \s

\texttt{SUSYGEN2} offers different frameworks for the mass spectrum
calculation. One can first assume the different mass parameters entering in
the MSSM: $M_1$, $M_2$ and $M_3$ the gaugino mass parameters, $\mu$, the
Higgsino mass mixing parameter, the scalar fermions masses $M_{\tilde{f}_L}$
and $M_{\tilde{f}_R}$, the trilinear mixing parameters $A_t$ $A_b$ and $A_\tau$
to be free. This gives the so called ``unconstrained MSSM''.  Another
approach to the mass spectrum calculation is based on the supergravity inspired
models. In this case the soft breaking mass parameters are assumed to be
universal at the GUT scale reducing the number of parameters to $m_{1/2}$, the
common gaugino mass parameter, $m_0$, the common sfermion mass parameter, the
sign of $\mu$, $\tan\beta$, the ratio of the two vacuum expectation values of
the two Higgs doublets, $A_0$, the common  trilinear couplings.  All these
parameters are defined at the GUT scale.  

In \texttt{SUSYGEN3}, one can keep the approach used in \texttt{SUSYGEN2}.  In
this case, only  $m_0$ is defined at the GUT scale and the sfermion masses are
evolved from the GUT scale to the electroweak (EW) scale according to the
formulae given in appendix of Ref.~\cite{ambros.gut.evo}. The other parameters
$M_1$, $\mu$, $A_t$, $A_b$ and $A_\tau$ are defined at the EW scale and mixing
of the third generation sfermion is taken into account through the parameters
$A_t$, $A_b$ and $A_\tau$.  \texttt{SUSYGEN3} offers now the possibility to do
a better treatment of the mass spectrum calculation within mSUGRA through an
interface to the \texttt{SUSPECT} program \cite{suspect}.  In practice, if the
flag \texttt{SUSPECT} is set to \texttt{TRUE} in the input data card which
fixes the model, the entire mass parameters at the EW scale will be derived
from these at the GUT scale ($m_{1/2}$, $m_0$, sign of $\mu$, $A_0$ and
$\tan\beta$). \s

\noindent {\bf b) Beam polarization and spin correlations} \s

Since one expects high luminosities for the next generation of linear colliders
(e.g. $\sim 500$ fb$^{-1}$ for the TESLA project), one can use beam polarization
to reduce the standard model backgrounds and use the polarization dependence of
the cross sections to study specific SUSY parameters.  Moreover, as it has been
stressed by several authors \cite{gudi:spincorr}, spin correlations play a
major role in the kinematic distributions of final particles.  To fulfill these
two requirements, the ``helicity amplitude method" \cite{mana:helicity}
was used for the calculation of the different Feynman amplitudes for production
and decay, in order to obtain full spin correlation. 
Since such  amplitudes involve products and contractions of
fermionic currents, two basic functions, namely the $B$ and 
$Z$ functions were defined through:
\begin{eqnarray}
B_{\lambda_1,\lambda_2}^{\lambda} (p_1,p_2) &=& \bar{u}_{\lambda_1}(p_1,m_1) 
                      P_\lambda u_{\lambda_2}(p_2,m_2) \\ 
Z^{\lambda\lambda '}_{\lambda_1,\lambda_2,\lambda_3,\lambda_4}(p_1,p_2,p_3,p_4)
&=&\left[\bar{u}_{\lambda_1}(p_1,m_1)\gamma^\mu P_\lambda u_{\lambda_2}(p_2,m_2)
\right]\left[\bar{u}_{\lambda_3}(p_3,m_3)\gamma_\mu P_{\lambda '} \nonumber 
u_{\lambda_1}(p_4,m_4)\right]
\end{eqnarray}
where $P_\lambda$ stands for one of the two chiral projectors $P_L$ or $P_R$
and $u_{\lambda}(p,m)$ denotes the positive energy spinor solution of the Dirac
equation for a particle  of helicity $\lambda$, four momentum $p$ and mass $m$.
The decomposition of the bispinors $u_{\lambda}(p,m)$ in terms of the massless
helicity eigenstates $\omega_{\lambda}(k)$ yields simple analytical  
expressions for the $B$ and $Z$ functions. The amplitude is then factorized in
terms of these basic building blocks;  this fact permits compact and
transparent coding and speed of calculation.  The masses are not neglected in
any stage of the calculation.  For gaugino productions and decay, we use the
``widthless approximation". For instance, the calculation of the cross section 
associated to $e^+e^-\to\tilde{\chi}^0_2\tilde{\chi}^0_1\to \tilde{\chi}^0_1
\tilde{\chi}^0_1 e^+ e^- $ is done as follows: the total amplitude associated 
to a given
helicity configuration of the different particles is  approximated by the 
product of the amplitude associated to the production of the two neutralinos 
${\cal M} (e^+ e^- \to \tilde{\chi}^0_2 \tilde{\chi}^0_1$) with the amplitude 
corresponding to the decay of the next to lightest neutralino ${\cal M}(
\tilde{\chi}^0_2 \to \tilde{\chi}^0_1 e^+ e^-)$. The remnant of the propagator 
squared of $\tilde{\chi}^0_2$ is approximated by a factor 
given by $8\pi^4/(m_{\tilde{\chi}^0_2}\Gamma_{\tilde{\chi}^0_2})$. 
The phase space integration is done through the multichannel method 
\cite{pittau}. \s

\noindent {\bf c) Including phases in SUSY searches:} \s
 
In the MSSM, there are new potential sources of CP non--conservation
\cite{poko:phases}. Complex CP violating phases can arise from
several parameters present in the MSSM Lagrangian: the higgs mixing mass
parameter $\mu$, the gaugino masses $M_i$, the trilinear couplings $A_i$. 
Experimental constraints on these CP violating phases come from the electric
dipole moment of the electron and the neutron.  Since in \texttt{SUSYGEN3} all
the couplings, the different mass parameters $\mu$, $M_1$,
and the trilinear couplings $A_\tau$, $A_t$ and $A_b$ have been assumed to be
complex by default\cite{rosi:coupling}, the introduction of phases in the 
gaugino and sfermion
sector for masses as well for cross sections has been straightforward.

\newpage

\section{CompHEP}

{\tt CompHEP}\footnote{This section is written together with A. Pukhov 
and A. Semenov.} \cite{comphep6} is a package for automatic calculations of 
decay and production processes in the tree--level approximation in the 
framework of arbitrary
gauge models of particle interactions. The main idea prescribed into {\tt 
CompHEP}, is to make available passing on from the basic Lagrangian to the 
final distributions efficiently with a high level of automation. {\tt
CompHEP} is a
menu--driven system. The codes and the manual are available on the web site: \\ 
{\tt http://theory.npi.msu.su/\verb|~|comphep} (mirror on {\tt 
http://www.ifh.de/\verb|~|pukhov}). \sx

The present version has four built--in physical models. Two of them are the
Standard Model in the unitary and 't Hooft--Feynman gauges. The user can change
particle interactions and model parameters and introduce new vertices, thus
creating new models. Furthermore, in the framework of the {\tt CompHEP} project,
a program {\tt LanHEP} \cite{lanhep} was created to generate {\tt CompHEP} 
model files as will be discussed below. \sx

The {\tt CompHEP} package consists of two parts, a symbolic and a numerical 
one. The symbolic part is written in the {\tt C}  programming language and 
produces {\tt FORTRAN} and {\tt C} codes for squared matrix elements which are
used in the numerical calculation later on. There are two versions of the
numerical part,  a {\tt FORTRAN} and  a {\tt C} one, with almost equal
facilities. The {\tt C}  version has a more comfortable interface
but it does not possess an option to generate events and does not perform
calculations with quadruple  precision. \sx

The symbolic part of {\tt CompHEP} allows the user to: 

-- Select a process by specifying incoming and outgoing particles for
the decays $1 \rightarrow n$ ($<6$) and the production mechanisms  $2 
\rightarrow n$  ($<5$).
   
-- Generate Feynman diagrams,  display them, and generate squared 
diagrams.

-- Calculate analytical expressions corresponding to squared diagrams,
save them in {\tt REDUCE} and {\tt MATHEMATICA} forms for  further symbolic 
manipulations. 

-- Generate optimized {\tt FORTRAN} and {\tt C} codes for the 
squared matrix elements for further numerical calculations. \sx

The numerical part of {\tt CompHEP} allows to:

-- Convolute the squared matrix element with structure functions (for
proton and antiproton, electrons and photons). 

-- Modify physical parameters (energy, charges, masses etc.) involved 
in the processes. 

--  Select the scale for evaluation of $\alpha_S$ and parton structure 
functions.

-- Introduce various kinematical cuts.

-- Define the phase space parameterization and introduce a phase space 
mapping in order to smooth sharp peaks for effective Monte Carlo integration.

-- Perform Monte--Carlo integrations by VEGAS \cite{vegas} via the 
multichannel approach \cite{multichannel}. 

-- Generate events and make distributions with graphical and 
{\tt LaTeX} outputs. \sx

In the QCD part of these Proceedings, one can find more details on {\tt 
CompHEP} options, in particular the handling of the QCD aspects and the 
discussion of the automatic computation of processes with multiparticle 
final states. The {\tt CompHEP} package has been used in several studies 
performed at this Workshop, in particular in the Higgs working group. Examples 
are: Higgs boson searches in the $\gamma\gamma$+jet channel at the LHC 
[Sec.~2] and generation of events for associated production of light stops 
with Higgs bosons [Sec.~4].\sx

During this Workshop, a  new algorithm was proposed for the treatment of the
first and second generation quarks through the single generation of generalized
``up" and ``down" quarks \cite{BIS}. This algorithm neglects the masses of
these quarks and their mixing with third generation quarks. It is based on a
rotation of the S--matrix in flavor space and move the CKM matrix elements
from diagrams to distribution functions.  The complete set of new rules was
derived for a correct counting of the convolution with different parton
distributions for quarks of the first/second generations. Each rule corresponds
to a gauge invariant subset of diagrams; see also \cite{BO}. This technique
allows to reduce significantly the number of subprocesses contributing to the
same physical final state, especially for hadron colliders. It was realized in
the {\tt CompHEP} version installed at CERN ($\tt
/afs/cern.ch/cms/physics/COMPHEP$). \sx

Developments were also made during this Workshop for the implementation 
of SUSY models in {\tt CompHEP}; some of them concern the Higgs sector.
To derive the MSSM description for {\tt CompHEP} one can use the {\tt LanHEP} 
\cite{lanhep} program which allows to generate the Feynman rules from the 
Lagrangian input in compact forms close to the ones given in textbooks [e.g.
Lagrangian terms can be written with summation over indices and using 
compact expressions such as covariant derivatives and strength tensors for 
gauge fields]. There are given in terms of two--component spinors and with 
the superpotential formalism. The output for the Feynman rules is in {\tt 
LaTeX} format and in the form of {\tt CompHEP} model files.
For the MSSM Lagrangian, the complete description given in 
Ref.~\cite{rosi:coupling} is used, together with two extensions: vertices 
with R--parity violation and the light gravitino scenario in GMSB models. \sx

It is known that Higgs boson masses in the MSSM are significantly affected by
radiative corrections. To compute these corrections, the two--Higgs doublet
model potential \cite{haber} technique is exploited. This potential is
parametrized by 7 variables, $\lambda_1$...$\lambda_7$, for which analytical
formulae given by in M.~Carena et al. in Ref.~\cite{mhiggsRG} are implemented. 
{\tt CompHEP} allows to calculate arbitrary processes within the given physical
model. Thus, one has to deal with the $\lambda_i$ variables rather than with
the set of Higgs boson masses only. However, one can set the Higgs boson masses
as input parameters, but the $\lambda_i$ are derived after and the model is
changed correspondingly preserving gauge invariance. An interface is made 
with the {\tt FeynHiggs} program \cite{feynhiggs} [used as an external
library], thus providing an option to evaluate the CP--even Higgs boson masses 
in the most up--to--date way. \sx

The number of independent parameters in the MSSM can be reduced  by the 
implementation of the mSUGRA or GMSB models. More specifically, the soft 
SUSY--breaking parameters, gaugino and sfermion masses as well as trilinear 
couplings, are computed from the input parameters. It is possible to use the
{\tt ISASUSY} package \cite{ISASUSY} for the calculation of these soft 
SUSY--breaking parameters [as well as the CP--odd Higgs boson mass; the 
CP--even Higgs masses can be calculated by {\tt FeynHiggs}]. 
The masses of the sparticles are then calculated by {\tt CompHEP} from the 
formulae used in the unconstrained MSSM. 
SUSY models for 
{\tt CompHEP} with the {\tt FeynHiggs} and {\tt ISASUSY} options included, 
can be obtained from the web site: \
{\tt http://theory.npi.msu.su/\~{}semenov/mssm.html}

\newpage

\subsubsection*{Acknowledgements}

J.K. is partially supported by the KBN Grant No. 2P03B 052 16. M. Spira is
supported by the Heisenberg Fellowship programme.

\end{document}